\documentstyle[12pt,elsart,elsart12,epsfig]{article}
\begin{document}

\begin {center}
{\Large Four sorts of meson}

{D.V.~Bugg}, \\
{Queen Mary, University of London, London E1\,4NS, UK}
\end {center}
{\bf abstract}
An extensive spectrum of light non-strange $q\bar q$
states up to a mass of 2400 MeV has emerged from Crystal Barrel and
PS172 data on $\bar pp \to Resonance \to A + B$ in 17 final states.
These data are reviewed with detailed comments on the
status of each resonance. For $I=0$, $C=+1$, the spectrum is
complete and very secure.

Six `extra' states are identified.
Four of them have $I = 0$, $C = +1$ and spin-parities predicted for
glueballs. Their mass ratios agree closely with predictions from
Lattice QCD calculations.
However, branching ratios for decays are not flavour
blind and require mixing with neighbouring
$q\bar q$ states. This mixing remains controversial.

Two further states, $\eta _2(1870)$ and $\pi _2(1880)$, are
candidates for a pair of hybrids. Their decay branching ratios
$BR [\eta _2(1870) \to f_2(1270)\eta ]/
BR [\eta _2(1870) \to a_2(1320)\pi ]$ and
$BR [\pi _2(1880) \to a_2(1320)\eta ]/
BR [\pi _2(1880) \to f_2(1270)\pi ]$ are much larger than predicted by
SU(3) for $q\bar q$ states.

A fourth category of anomalously light $0^+$
mesons is made of
$\sigma (540)$, $\kappa (750)$, $a_0(980)$ and $f_0(980)$,
forming a non-$q\bar q$ nonet.
Their properties are reviewed using data from
both Crystal Barrel and elsewhere.

\vspace{1cm}


\section {Contents}
1. Introduction and survey of results
\newline
2. The Crystal Barrel experiment
\begin{verse}
   2.1 Rate dependence
\newline
~~~   2.2 Monte Carlo and data selection
\newline
\end{verse}
3. How to establish resonances
\begin{verse}
   3.1 Star ratings of resonances
\end{verse}
4. The partial wave analysis
\begin{verse}
   4.1 Formulae for amplitudes and resonances
\newline
~~~   4.2 Illustration of spectroscopic notation
\newline
~~~   4.3 General comments on procedures
\end{verse}
5. Review of $I = 0$, $C = +1$ resonances
\begin{verse}
   5.1 2-body channels
\newline
~~~   5.2 3-body data
\newline
~~~   5.3 $2^-$ states
\newline
~~~   5.4 Star ratings of resonances
\newline
~~~   5.5 $f_0(1710)$ and $f_0(1770)$
\newline
~~~   5.6 SU(3) relations for 2-body channels
\end{verse}
6. $I = 1$, $C = +1$ resonance
\begin{verse}
  6.1 2-body data
\newline
~~~   6.2 $\bar pp \to \eta \eta \pi ^0$
\newline
~~~   6.3 $\bar pp \to 3\pi ^0$
\newline
~~~   6.4 Evidence for $\pi _2(1880)$
\end{verse}
7. $I = 1$, $C = -1$ resonances
\begin{verse}
   7.1 Vector polarisation of the $\omega$ and its implications
\newline
~~~   7.2 $\omega \eta \pi ^0$ data
\newline
~~~   7.3 Observed resonances
\end{verse}
8. $I = 0$, $C = -1$ resonances
\newline
9. Comments on systematics of $q\bar q$ resonances
\newline
10. Glueballs
\begin{verse}
  10.1 $f_0(1500)$ and the $q\bar q$ nonet of $f_0(1370$,
$a_0(1450)$, $K_0(1430)$ and $f_0(1710)$
\newline
~~~   10.2 Mixing of the glueball with $q\bar q$ states
\newline
~~~   10.3 Features of central production data
\newline
~~~   10.4 $f_0(1950)$
\newline
~~~   10.5 $\bar pp \to \phi \phi$
\newline
~~~   10.6 Evidence for the $0^-$ glueball
\newline
~~~   10.7 Sum rules for $J/\Psi$ radiative decays
\end{verse}
11. Future prospects
\newline
Acknowledgements
\newline
References

\newpage
\setcounter{section}{0}
\section {Introduction and Survey of Results}
Many new resonances have appeared in Crystal Barrel  and PS172 data on
$\bar pp$ annihilation in flight.
They follow a remarkably simple pattern.
The objective of this review is to outline techniques
used in the analysis and also to assign a `star rating' to each state.
Hopefully the star rating will assist comparison with model
predictons, where it is important to know which states are identified
most reliably. Techniques of analysis may be of interest
to further experiments which will produce overlapping information.
Results from 35 journal publications in collaboration with the St.
Petersburg group are summarised.

The main idea is to study $s$-channel resonances:
$$\bar pp \to Resonance \to A + B,$$
by full partial wave analysis of 17 final states
$A + B$ listed in Table 1. Coupled channel analyses have been carried
out, following the classic techniques applied to the $\pi N$ system in
1960-70, imposing the fundamental constraints of analyticity and
unitarity.

\begin {table}[htp]
\begin {center}
\begin {tabular}{llll} \hline $I=0$, $C =
+1$  & $I = 1$, $C = +1$ & $I = 0$, $C = -1$ & $I =1$,
 $C = -1$  \\\hline
$\pi ^-\pi ^+$, $\pi ^0\pi ^0$, $\eta \eta$, $\eta \eta '$
& $\eta \pi ^0$, $\eta '\pi ^0$ & $\omega \eta$ & $\pi ^-\pi ^+$,
$\omega \pi ^0$ \\
$\eta \pi ^0 \pi ^0$, $\eta '\pi ^0 \pi ^0$, $\eta
\eta \eta$
 & $3\pi ^0$, $\eta \eta \pi ^0$ & $\omega \pi ^0 \pi ^0$ &
 $\omega \eta \pi ^0$ \\
$\eta \eta \pi ^0 \pi ^0$ & $\eta \pi ^0 \pi ^0 \pi ^0$ \\
\hline
\end {tabular}
\caption {Channels studied; decays $\eta \to \gamma \gamma$ and
$\eta \to 3\pi ^0$ are both used, and $\eta ' \to \eta \pi ^0 \pi ^0$;
also $\omega \to \pi ^0 \gamma$ and $\omega \to \pi ^+\pi ^- \pi ^0$.}
\end {center}
\end {table}

The majority of final states are made up only of neutral
particles.
The Crystal Barrel has excellent $\gamma$ detection in a barrel of
1380 CsI crystals with good energy and angular resolution (see
Section 2).
Crystals cover 98\% of the solid angle. Detection of charged
particles is limited to 71\% of the solid angle.
Neutral final states fall cleanly into four categories of isospin $I$
and change conjugation $C$, as shown in Table 1; $\pi ^-\pi
^+$ contains both $I = 0$ and 1.

The largest number of decay channels is available for $I = 0$, $C = +1$.
For the $\pi ^- \pi ^+$ final state, $d\sigma /d\Omega$ and polarisation
data are included from earlier experiments [1,2].
The polarisation data play a vital role in identifying helicities and
hence separating $\bar pp$ $^3F_2$ states from $^3P_2$. There is also a
second point:
\begin {equation} P\frac {d\sigma }{d\Omega } \propto Im
(F_1^*F_2)
\end {equation}
where $F_1$ and $F_2$ are helicity
combinations discussed  later. Differential cross sections
depend on moduli squared of amplitudes and real parts of interferences.
The combination of differential cross sections and polarisations
determines both magnitudes and phases of amplitudes uniquely. In any
one data set, significance levels of resonances are similar to those of
$\pi N$ analyses above 1700 MeV; for $I = 0$, $C = +1$, the analysis of
8 channels with consistent parameters leads to extremely secure
amplitudes and a complete set of the expected $q\bar q$ states in the
mass range 1910--2410 MeV, each with a statistical significance $\ge
25\sigma$.

The spectrum shown in Fig. 1 [3] is remarkably simple.
Almost within errors, observed states lie on parallel trajectories
of $M^2$ (where $M$ is mass) v. radial excitation number $n$.
The mean slope is $1.14 \pm 0.013$ GeV$^2$ per unit of $n$.
These trajectories resemble the well-known Regge trajectories;
however on Regge trajectories, states alternate in $I$ and $C$ (for
example $\rho _1$, $f_2$, $\rho _3$, $f_4$, $\rho _5$).

\begin {figure} [htp]
\begin {center}
\epsfig{file=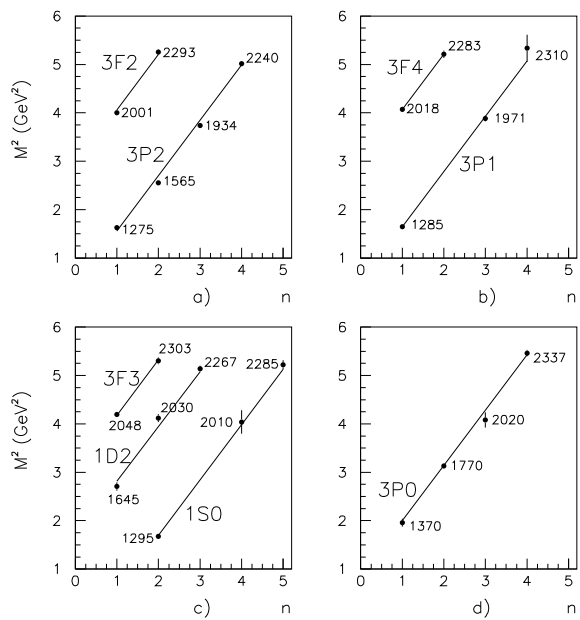,width=9.2cm}\
\caption{Trajectories of $I = 0$, $C = +1$ states.
Numbers indicate masses in MeV.}
\end {center}
\end {figure}

Similar spectra are observed for other $I$ and $C$, but are less
complete and less precise.
Fig. 2  shows the spectrum for $I = 1$, $C = +1$ [4,5].
Since quarks are nearly massless, one expects results
to be almost degenerate with $I = 0$, $C = +1$.
This is indeed the case.
However, the separation between $^3F_2$ and
$^3P_2$ is much poorer for $I = 1$, $C = +1$ because no
polarisation data are available.

\begin {figure} [htp]
\begin {center}
\epsfig{file=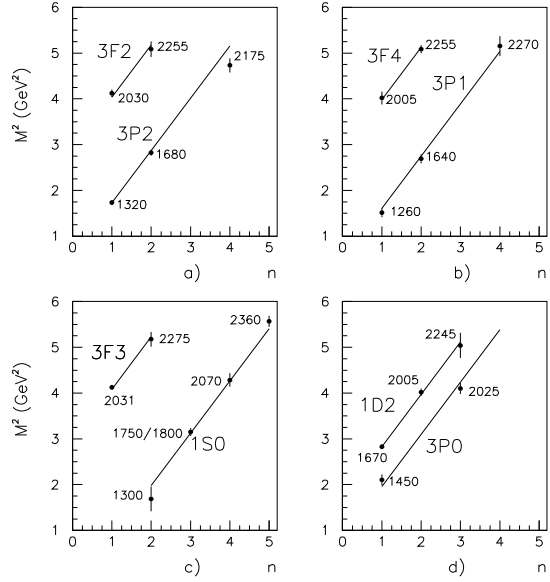,width=8.7cm}\
\caption{Trajectories of $I = 1$, $C = +1$ states.}
\end {center}
\end {figure}

For $I = 1$, $C = -1$, the observed spectrum is shown in Fig.
3(b) [6,7].
Here polarisation data for $\bar pp \to \pi ^- \pi ^+$ are
available and do help enormously in separating magnitudes and
phases of amplitudes. As a result, the observed spectrum is
complete except for one high-lying $J^{PC} = 1^{--}$ state
and the $^3G_3$ $\bar pp$ state.
The $^3D_3$ states are particularly well identified.

\begin {figure} [htp]
\begin {center}
\epsfig{file=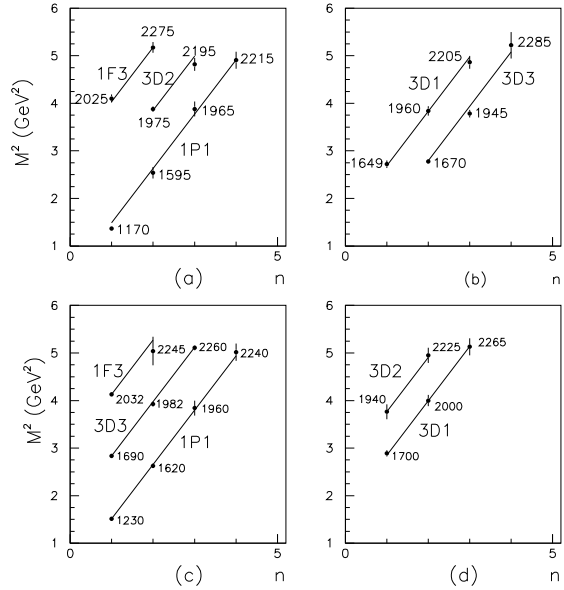,width=8.7cm}\
\caption{Trajectories of $C = -1$ states with (a) and (b) $I = 0$,
(c) and (d) $I = 1$. In (b), the $^3D_3$ trajectory
is moved one place right for clarity; in (d), $^3D_2$ is moved one
place left.}
\end {center}
\end {figure}

Figs. 3(a) and (b) show the spectrum of $I = 0$, $C = -1$ states [7].
Errors are the largest here.
Statistics for the $\omega \eta$ channel are typically a factor
6 weaker than for $\omega \pi ^0$.
The second available channel is $\omega \pi ^0 \pi ^0$ and is difficult
to analyse because of broad $0^+$ structure in $\pi ^0$;
those states which are observed are again almost degenerate with $I =
1$, $C = -1$.

All observed resonances cluster into fairly narrow mass ranges
(i) 1590-1700 MeV, (ii) 1930--2100 MeV, (iii) 2240--2340 MeV.
These `towers' of resonances are illustrated in Fig. 4.

\begin {figure} [htp]
\begin {center}
\epsfig{file=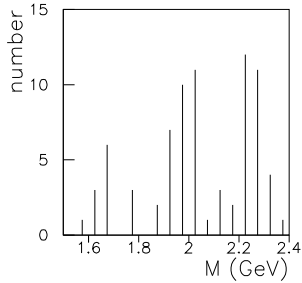,width=5cm}\
\caption {Clustering of resonances in mass.}
\end {center}
\end {figure}

Sections 1--9 review $q\bar q$ states in detail.
There are four `extra' states with $I = 0$, $C = +1$ having masses
close to Lattice QCD predictions for glueballs and the same $J^P$.
Section 10 reviews these glueball candidates and the mixing
of $f_0(1500)$ with the nearby nonet made of $f_0(1370)$, $a_0(1450)$,
$K_0(1430)$ and $f_0(1710)$.
Section 11 reviews the light scalar mesons: $\sigma (540)$, $\kappa
(750)$, $f_0(980)$ and $a_0(980)$.
Section 12 take a brief look at what is needed in future.

\section {The Crystal Barrel experiment}
This experiment ran at LEAR, the $\bar p$ storage ring
at CERN, using nine $\bar p$ beam momenta of 600, 900, 1050, 1200, 1350,
1525, 1642, 1800 and 1940 MeV/c.
The corresponding $\bar pp$ mass range is from 1962 to 2409 MeV.
Most of the data were taken during the last 4 months of LEAR
operation, August -- December 1996.
Data from the PS172 experiment wth a polarised target will also be
used.
This experiment ran during 1986 from 360 to 1550 MeV/c; the
extension of the mass range down to 1910 MeV is important.

The Crystal Barrel detector is illustrated in Fig. 5.
Details are to be found in a full technical report [8].
A 4.4 cm liquid hydrogen target was surrounded sequentially by
multiwire chambers for triggering (labelled 5 on the figure),
a jet drift chamber (4) for detection of charged particles,
a barrel (3) of 1380 CsI crystals covering 98\% of $4\pi$ solid
angle, and a magnet (1) producing a solenoidal field of 1.5T
(though this was run at 1.0T for all-neutral data).
For running after 1994, the multiwire chambers were
replaced by a silicon vertex detector. However, for detection of
neutral final states, the chambers and silicon vertex detector
acted simply as vetos over a solid angle 98\% of $4\pi$.

\begin{figure}
\begin{center}
\epsfig{file=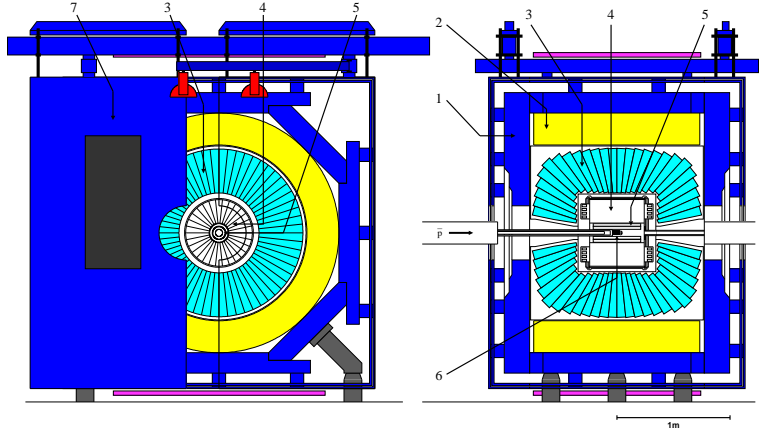,width=10cm}\
\caption{Overall layout of the Crystal Barrel detector, showing (1) magnet
yoke, (2) magnet coils, (3) CsI barrel, (4) jet drift chamber,
(5) multi-wire proportional chambers, (6) liquid hydrogen target.}
\end{center}
\end{figure}

The CsI crystals measured photon showers with an angular resolution of
20 mrad in both polar and azimuthal angles.
The photon detection efficiency was high down to energies below
20 MeV.
The energy resolution was given by $\Delta E/E = 0.025/E^{1/4}$, with
E in GeV.
In order to reject events which failed to conserve energy,
the total energy deposited in the CsI crystals was derived on-line;
the trigger selected events having a total energy within 200 MeV of
that in $\bar pp$ interactions [9].

Incident $\bar p$ were detected by a fast coincidence between a small
proportional counter $P$ and a Si counter $Si_C$ of radius 5 mm
at the end of the beam-pipe and only a few cm from the entrance
of the liquid hydrogen target.
Four further silicon detectors located at the same point and
segmented into quadrants were used to steer and focus the beam.
Two veto counters were positioned 20 cm
downstream of the target and inside the CsI barrel; they provided a
trigger for interactions in the target. The data-taking rate was $\sim
60$ events/s at a beam intensity of $2 \times 10^5~\bar p$/s.

\begin{figure} [htp]
\begin{center}
\epsfig{file=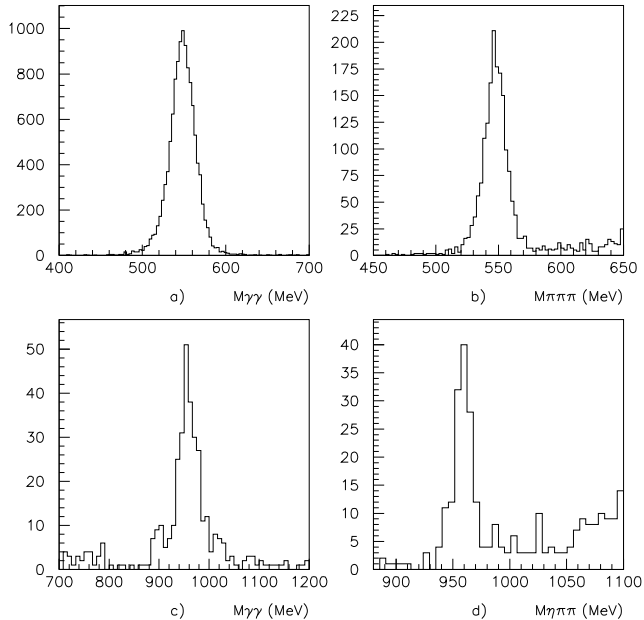,width=9.5cm}
~\
\caption{Distributions of (a) $M(\gamma \gamma )$ near the
$\eta$ from events fitting $\eta \gamma \gamma$, after cuts;
(b) $M(3\pi ^0)$ near the $\eta$ from $\eta 3\pi ^0$ events;
(c) $M(\gamma \gamma )$ near the $\eta '$ from $\eta \gamma \gamma$
data;
(d) $M(\eta \pi ^0 \pi ^0)$ near the $\eta '$ from
$\eta \eta \pi ^0 \pi ^0$ events. Data are at 1800 MeV/c.}
\end{center}
\end{figure}
One gets a general impression of the cleanliness with which
neutral particles were detected from Fig. 6.
Panels (a) and (b) show $\eta $ peaks in the $\eta \eta$
final state, from $2\gamma$ events in (a) and from $3\pi ^0$ events
in (b) [10].
As an illustration of selection procedures, consider $\bar pp \to
\eta \eta$.
Cuts were applied on confidence levels $(CL)$ as follows:
(i) $CL(\pi ^0 \gamma \gamma )$ to remove $\pi ^0 \pi ^0$,
(ii) $CL(\pi ^0 \gamma ) < 0.01\%$,
(iii) $CL(\eta \eta ) > CL(\eta \eta ')$,
(iv) $CL(\eta \eta ) > 0.1$,
In the few cases where there was more than one $\eta \eta $
solution, selected events were required to have
a $CL$ at least a factor 10 better than the second.
The entire process was followed through a Monte Carlo simulation
discussed below.
Observed backgrounds agreed closely with this simulation after
allowing for branching ratios of the background channels.
The $\eta$ peaks in both (a) and (b) were centred (purely from
kinematics of the production reaction) within
1-2 MeV of the Particle Data Group value (PDG) [11] at all
beam momenta; this checks systematics of the energy
calibration. Fig. 7 shows angular distributions derived from
$4\gamma$ and $8\gamma$ events, independently normalised to Monte
Carlo simulations.
The  close agreement in both shapes and absolute magnitudes is a major
cross-check.

\begin{figure} [htp]
\begin{center}
\epsfig{file=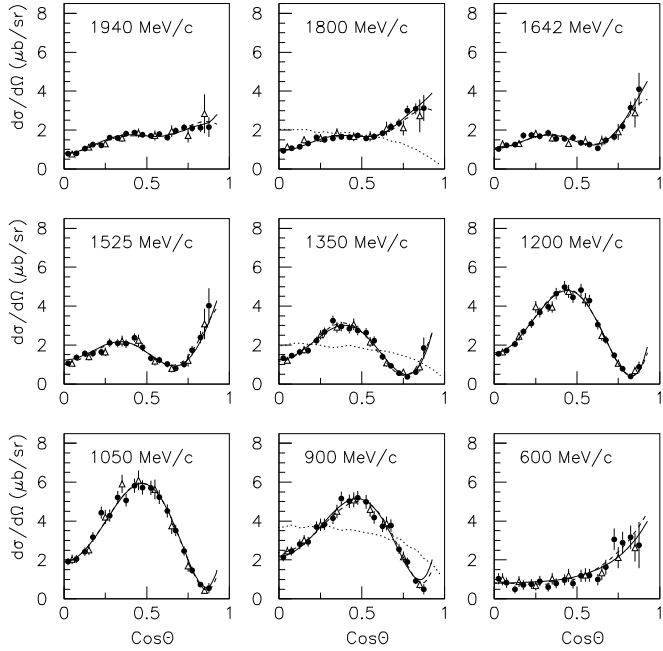,width=11cm}
~\
\caption{Comparison of angular distributions for $\eta \eta$ from
$4\gamma$ events (black circles) and $8\gamma $ (open triangles).
Full curves show the result of the amplitude analysis.
Dotted curves illustrate the acceptance for three momenta;
small undulations are due crystal sizes and overlap of showers between
crystals, well simulated by the Monte Carlo.}
\end{center}
\end{figure}

Figs. 6(c) and (d) show $\eta '$ peaks from $2\gamma$ and
$\eta \pi ^0 \pi ^0$ decays.
Statistics are lower than for $\eta \eta$, but angular
distributions and normalisations agree
between $4\gamma$ and $8 \gamma$ for $\bar pp \to \eta '
\eta$ within statistics [10], and also for $\bar pp \to \eta ' \pi
^0$ [12].

Showers were selected from adjoining groups of nine CsI crystals.
The relation between shower energy and pulse-height in crystals was
obtained [8] by fitting the $\pi ^0$ mass for $\gamma \gamma$ events,
using all crystal combinations.
Three variants of this calibration were tried.
The primary effects of different calibrations are
removed by the overall kinematic fit to energy-momentum conservation;
this uses the beam momentum, known to better than 0.1\%.
The result is that different calibrations altered
$\le 1\%$ of selected events.
However, the performance was better than that.
Between one calibration and another, the overall change in accepted
events was a factor 10 smaller; i.e. changing from one
selection to another, the few events which were lost were compensated
by an almost equal number entering the sample. A more important
consideration was the identification of crystals which malfunctioned.
For most running conditions, there were at most 1 or 2 crystals which
failed in some way (noisy or dead). These were identified and cut
completely from both the data and the Monte Carlo simulation. Otherwise
a faulty crystal could lead to a dip in the angular distribution over
the angles covered by 9 adjoining crystals.

\subsection {Rate dependence}
We now consider the absolute normalisation of cross sections.
This is important in an amplitude analysis covering many momenta.
A potentially serious rate dependence was found, but ultimately
affected normalisations by only $\sim 3-6\%$, varying slowly
with beam momentum.
It arose from pile-up in the CsI crystals.
Beam intensities were $2 \times 10^5$/s and interaction rates
in the target were typically $6 \times 10^3$/s.

\begin{figure} [htp]
\begin{center}
\epsfig{file=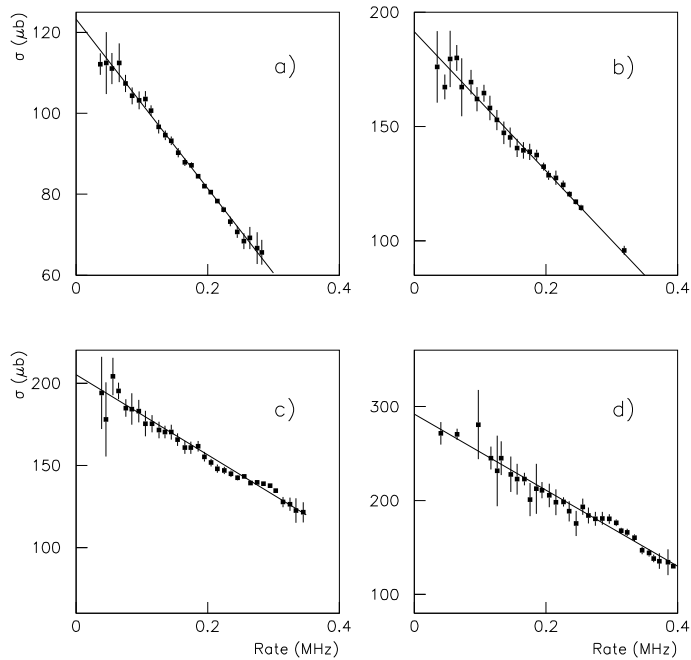,width=10cm}
~\
\caption{Rate dependence observed at (a) 1800, (b) 1525 (c) 1350 and
(d) 1050 MeV/c for the sum of $(4+5+6+7+8)\gamma$ events.
Lines show the fitted rate dependence.}
\end{center}
\end{figure}

The scintillation from CsI has two components, one with a
risetime of $6\mu s$, but with a fall-time of $100 \mu s$.
The discriminators used pole-zero cancellation which was supposed
to eliminate the long component. But in practice large energy
deposits in one crystal could produce some ringing in its discriminator
output. This led to significant pile-up between one event and the
next. A detailed discussion is to be found in Ref. [10].

The normalisation was derived from beam counts $P.Si_C$, the length
and density of the target, the observed number of events and the
Monte Carlo simulation of  detection efficiency.
The general off-line procedure was to reconstruct 45 exclusive final
states, each one kinematically fitted to channels with 4 to 10$\gamma$.
Rather few have 9 or 10 $\gamma$.
Fig. 8  shows as a function of beam rate the total of events fitting
with confidence levels $>10\%$ to any channel with 4 to 8 $\gamma$.
Typically $\sim 20\%$ of triggers fit such channels.
There is obviously a strong rate dependence and it was necessary to
extrapolate cross sections to zero beam rate.
The rate dependence varied with beam steering. At several momenta,
separate running conditions  gave different slopes, but
extrapolated values were consistent within $3\%$.

\begin{figure}
\begin{center}
\epsfig{file=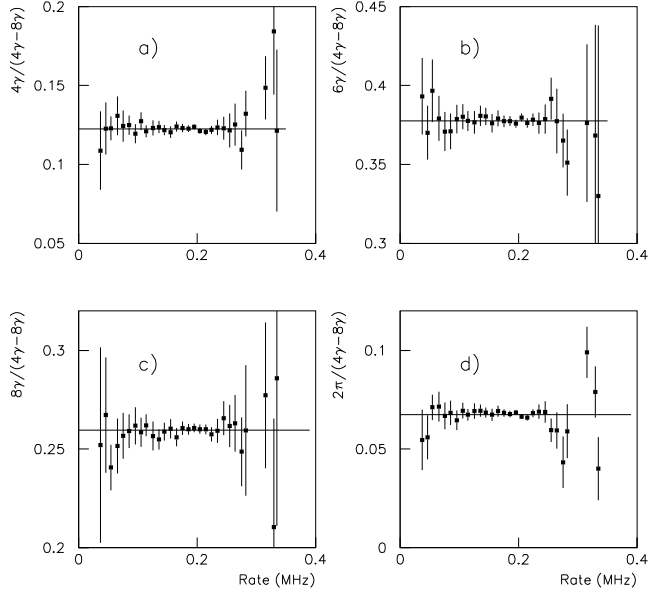,width=9cm}\
\caption {Fractions of (a) $4\gamma$, (b) $6\gamma$, (c) $8\gamma$
events and (d) $\pi ^0 \pi ^0$ events, out of the total of final
$(4 + 5 + 6 + 7 + 8)\gamma $ events at 1800 MeV/c v.
incident $\bar p$ rate.}
\end {center}
\end {figure}

Fortunately, the rate-dependence is found to
kill genuine events of all types in the same way.
Let $N(n\gamma)$ be the number of events fitting above 10\% confidence
level with $n$ photons, and let $N(4-8)\gamma$ be the sum of these
from $n = 4 $ to 8.
Fig. 9 displays ratios $N(n\gamma )/N(4-8)\gamma$ for $4\gamma$,
$6\gamma$, $8\gamma$ and for $\pi ^0 \pi ^0$ as a function of beam
rate. Any variation with rate is $<3\%$. Neither is there any observed
change with beam rate in $\pi ^0 \pi ^0$ angular distributions  or
Dalitz plots of $6\gamma$ events [13].
The conclusion is that angular distributions can be extracted
safely at any beam rate; but the normalisation must be obtained by
extrapolating to zero beam-rate. A cross-check is also satisfied
between $\pi ^0 \pi ^0$ data from Crystal Barrel and symmetrised
$\pi ^- \pi ^+$ differential cross sections from PS 172, see below.

\subsection {Monte Carlo and Data Selection}
Data are fitted kinematically to a large number of physics channels:
45 for $(4-10)\gamma $.
In order to assess branching ratios to every channel and cross-talk
between them, at least 20,000 Monte Carlo events are generated for every
one of the 45 fitted channels, using GEANT. In the first approximation,
events fitting the correct channel determine the reconstruction
efficiency $\epsilon _i$ in that channel. Events fitting the wrong
channel measure the probability of cross-talk $x_{ij}$ between
channels $i$ and $j$. More exactly, a set of 45 x 45
simultaneous equations is solved containing on the left-hand side the
observed number of fitted data events $D_i$, and on the right-hand side
reconstruction efficiencies $\epsilon _i$ and true numbers of created
events $N_i$ in every channel plus terms allowing for cross-talk
$x_{ij}$ between channels: \begin {equation} D_i = \epsilon _iN_i +
\sum _{j\ne i}~x_{ij}N_j. \end {equation} The solution is constrained
so that the numbers of real events $N_i$ in every channel are
positive or zero.

This procedure is carried out for several confidence levels
(1, 5, 10, 20\%) and using a wide variety of selection procedures.
A choice is then made, optimising the ratio of signal to background.
It turns out that this ratio is not very sensitive to confidence level
over the range 5--20\%. Hence
selection criteria are  not usually critical.

\section {How to establish resonances}
One objective of this review is to assign a star rating to
all resonances. This is to some extent subjective, but some
assessment is better than none, as a guide to further work.
Before coming to details, it is necessary to review the
general approach to identifying resonances.

It is sometimes argued that magnitudes and phases of all amplitudes
should be fitted freely, so as to demonstrae that phases follow
the variation expected for resonances.
Though that is a fine ideal, it is unrealistic and leads to problems.
Phases can only be determined with respect to some known amplitude.
As examples, phase variation of the $Z_0$ and the $J/\Psi$ may be found
from interferences with electrweak amplitudes.
In $\pi ^+p$ scattering at low energies, the $\Delta (1232)$ is
established as a resonance by the fact that its cross section
reaches the unitarity limit; interferences with S-waves then
establish its resonant phase variation, since there is not
enough cross section for S-waves to resonate too.
Further examples are rather rare.

The usual assumption is that isolated, well defined peaks far from
thresholds are assumed to be resonances.
The $\pi _2(1670)$ and $f_4(2050)$ are examples.
The phases of other amplitudes may be determined using these and
other established resonances as interferometers.
This is the procedure followed in Crystal Barrel analyses.

The $f_4(2050)$ appears strongly in data for $I = 0$, $C = +1$.
It acts as an interferometer and determines phases of other partial
waves. They have a similar phase variation with $s$ and must
therefore also resonate. In $\bar pp \to \pi ^-\pi ^+$, differential
cross sections establish real parts of interferences and polarisation
data measure the imaginary parts of the same interferences. Together,
these data establish quite precisely the phase variation of all partial
waves with respect to $f_4(2050)$. Further strong peaks appear as
$f_4(2300)$ and $\eta _2(2267)$. They act as further interferometers,
loosely coupled themselves in phase to the tail of $f_4(2050)$.

A strong $\rho _3(1982)$ appears in $\bar pp \to \pi ^- \pi ^+$.
In turn, this acts as an interferometer for $I = 1$, $C = -1$.
The amplitude analysis grows like a rolling snowball.
The phases of other partial waves at low mass are measured
with respect to $\rho _3(1982)$ and establish resonances in
other partial waves at low mass. Variations of magnitudes are
consistent with the observed phases.
Peaks appear also in the mass range 2200-2300 MeV, all correlated
in phase. Collectively, their phases are loosely coupled to the tails
of lower mass states.

For $I = 1$, $C = +1$ and $I = 0$, $C = -1$ there are no polarisation
data. Phase information from differential cross sections arise only
via real parts of interferences.
When relative phases are close to 0 or $180^\circ$, errors of
phases are large. Without some smoothing of the phases to
follow resonant behaviour (or some other analytic form),
noise in the analysis
can run out of control and can lead to a forest of narrow peaks.
Such a result is unphysical, since amplitudes should be analytic
functions of $s$.
Smoothing to analytic forms improves the determination of
partial wave amplitudes dramatically.

The technique used for $\pi N$ elastic scattering is to constrain
the analysis using fixed $t$ dispersion relations.
This works successfully up to 800 MeV.
Above that energy, amplitudes are parametrised
directly with analytic forms.

The approach used in Crystal Barrel analysis is to use analyticity
as a foundation, parametrising amplitudes from the outset in terms
of simple Breit-Wigner resonances of constant width, or constants,
or linear functions of $s$.
Even then, caution is necessary.
If a linear function develops a phase variation $>90^\circ$,
it is likely to represent a resonance.
It is well known to be dangerous to fit data with amplitudes
having similar characteristics; orthogonal functions are needed.
In reality, one can usually replace linear functions having rising
phases as the tails of resonances well below the $\bar pp$ threshold,
e.g. $\pi (1800)$ or $\rho _3(1690)$; their phase variations above
the $\bar pp$ threshold are small and under reasonable
control. Linear variations with falling phases are a
possibility but in practice are not observed.
A valuable constraint in the analysis is
that amplitudes for all partial waves except $^1S_0$ and $^3S_1$ go to
zero at the $\bar pp$ threshold; these two go to scattering lengths and
are the most difficulty to deal with.

\begin{figure}
\begin{center}
\epsfig{file=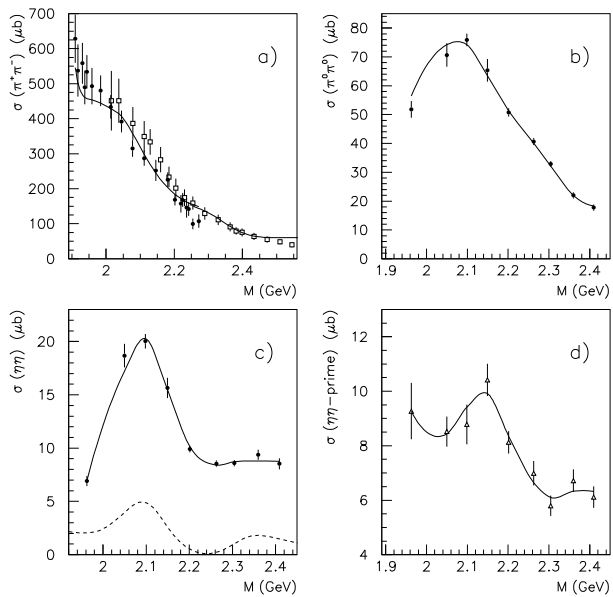,width=9cm}\
\caption {Integrated cross sections for 2-body scattering, compared
with the partial wave fit.
In (a), black circles are data of Hasan et al. and open squares are
those of Eisenhandler et al. In (b), (c) and (d) the integration is
from $\cos \theta = 0$ to 0.875 and is over only one hemisphere;
the $\eta \eta '$ data are averaged over $4\gamma$ and $8\gamma $ data.
For $\pi ^- \pi ^+$, the integration is over the full angular range.
In (c), the dashed curve shows the $0^+$ intensity. }
\end {center}
\end {figure}
There is a practical problem with the $\bar pp$
threshold.
Fig. 10 shows integrated cross sections for two-body channels.
For neutral final states $\pi ^0 \pi ^0$ and $\eta \eta$,
only $J^P = 0^+$, $2^+$, $4^+$ and $6^+$ are allowed
by Bose statistics.
There are no contributions from $\bar pp$ S-waves, which have
quantum numbers $J^{PC} = 0^{-+}$ ($\bar pp$ $^1S_0$) and $1^{--}$
($\bar pp$ $^3S_1$).
One sees in Fig. 10 that neutral cross sections go to low values near
the $\bar pp$ threshold.
However, the $\pi ^-\pi ^+$ cross section goes to large values,
following the familar $1/v$ law.
The $\bar pp$ total cross sections follows a similar behaviour.
The S-wave amplitudes near threshold need to contain this
feature, and it becomes difficult to separate the $1/v$ behaviour
from resonances at the $\bar pp$ threshold.

A further problem is that singlet $\bar pp$ states do not
interfere with $f_4(2050)$ (the triplet state $^3F_4$ in $\bar pp$).

An experimental problem is that Crystal Barrel data at the lowest
momenta 600 and 900 MeV/c straddle the lowest tower of resonances.
These momenta correspond to masses of 1962 and 2049 MeV.
In several cases, what can be said is that there are definite
peaks, inconsistent with the tails of resonances below threshold.
However, because of the limited number of available momenta,
masses and widths become strongly correlated; precise masses
are subject to some systematic uncertainty, which is correlated
over resonances of different $J^P$ via interference effects.
Each of these cases needs to be examined critically.
For $I = 0$, $C = +1$ and $I = 1$, $C = -1$, the PS172 data for
$\bar pp \to \pi ^- \pi ^+$ down to 360 MeV/c are very important in
extending the mass range to 1912 MeV in small steps of $\bar p$
momentum.

\subsection {Star Ratings of Resonances}
Resonances will be given ratings up to $4^*$.
The highest class requires observation of 3 or more strong,
unmistakable peaks and a good mass determination (with error $\delta
M\le 40$ MeV). Such states are equivalent to those in the summary table
of the Particle Data Group - better than some cases like $a_0(1450)$,
which until 2003 was observed only in a single channel.

States are demoted to lower star ratings if the mass determination
is poor, particularly if $\Gamma >300$ MeV. The $3^*$
resonances are reasonably well established, usually in two
strong channels with errors for masses $\le \pm40$ MeV; some
are established in 3 channels but with $\delta M$ in the range
40--70 MeV.

The $2^*$ resonances need confirmation elsewhere. They are
observed either (i) in one strong channel with $\delta M < 40$ MeV
or (ii) in 2 strong channels with sizable error in the mass.

Finally, $1^*$ states are tentative: weak channels and poor mass
determination.

\section {The Partial Wave Analysis}
The foundation for the analysis was laid in 1992-4 while fitting
PS172 data on $\bar pp \to \pi ^- \pi ^+$ [1].
Early in this analysis it became clear that there is a large
$J=4$ amplitude which can be identified with the well known
$f_4(2050)$.
At momenta $> 1$ GeV/c, a prominent feature of the polarisation
data, illustrated below in Fig. 13, is a polarisation close to
$+1$ over much of the angular range and over most beam momenta,
broken by some rapid fluctuations.
If there were a single $f_4$ resonance and non-resonant amplitudes
in other partial waves, the polarisation would necessarily show
a dramatic energy variation through the resonance.
This is in complete contradiction with the data. The only way to
fit the data successfully is to have resonances in many
partial waves.

If one invokes analyticity to relate real and imaginary parts of
amplitudes, the two available sets of data, (polarisations and
differential cross sections), can in principle lead to a unique
set of partial waves.
The first steps in analysing Crystal Barrel data were to add high
statistics data for $\bar pp \to \pi ^0 \pi ^0$, $\eta \eta$,
$\eta \eta'$ and $\eta \pi ^0 \pi ^0$,
all having $I = 0$, $C = +1$.
It rapidly became apparent that all these data could likewise
be fitted in terms of resonances and simple backgrounds.
A coupled channel analysis led to consistency between all sets of
data and rather precise determinations of resonance parameters
[13--15]. The final analysis for these quantum numbers is in
Ref. [3].
\subsection {Formulae for Amplitudes and Resonances}
It is easiest to condider one reaction as an example:
$\bar pp \to a_0(980)\pi ^0 \to \eta \pi ^0 \pi ^0$.
The amplitude for one partial wave is written
\begin {equation}
f = \frac {\rho _{\bar pp}}{p}\frac {g \e ^{i\phi } B_\ell(p)B_L(q)}
{M^2 - s - iM\Gamma }[F(s_{12})Z_{12} + F(s_{13})Z_{13}].
\end {equation}
Here $g$ is a real coupling constant, and $\phi$ the strong-interaction
phase for this partial wave. Blatt-Weisskopf centrifugal
barrier factors $B$ are included explicitly for production with orbital
angular momentum $\ell$ from $\bar pp$ and for coupling with angular
momentum $L$ to the decay channel (momentum $q$).
The first centrifugal barrier includes the correct threshold dependence
on momentum $p$ in the $\bar pp$ channel.
Formulae for $B$ are given by Bugg, Sarantsev and Zou [16].
The radius of
the centrifugal barrier is optimised at  $0.83 $ fm in Ref.
[3]. It is important only for the high
partial waves where the barrier is strong and well determined.
For $L = 4$ waves, the barrier radius is increased to 1.1 fm.
The factor $1/p$ in eqn. (3) allows for the flux in the $\bar pp$
entrance channel, and $\rho _{\bar pp}$ is the phase space for this
channel, $2p/\sqrt {s}$. Near the $\bar pp$ threshold, the S-wave cross
section is then proportional to $1/p\sqrt{s}$, the well-known $1/v$
law.

Strictly speaking, the Breit-Wigner amplitude should take the
general form
\begin {equation}
f = 1/[M^2 - s - m(s) -iM\Gamma (s)],
\end {equation}
where $\Gamma (s)$ is the sum of widths to decay channels.
For each channel, $\Gamma (s)$ is proportional to the available
phase space. This is the well known Flatt\' e form [17], which will be
used below for special cases such as $f_0(980)$ and $a0(980)$;
these resonances lie close to the $KK$ threshold and the
$s$-dependence of the $KK$ width is important. There is a dispersive
term $m(s)$ given by Tornqvist [18]. For present
purposes, these refinements are not relevant. The analysis concerns
high masses where many decay channels are open. Many of the decay
channels are unknown. The approximation is therefore that $\Gamma$ can
be taken as a constant; $m(s)$ is then zero. Backgrounds are
usually parametrised as constants or as
resonances below the $\bar pp$ threshold.

For the $\eta \pi ^0 \pi ^0$ final state, there are two possible
amplitudes for production of $a_0(980)$ in $\eta \pi$;
$F(s_{12})$ is the Flatt\' e formula for the $a_0$ resonance between
particles 1 and 2, and $F(s_{13})$ is the corresponding amplitude
between particles 1 and 3.
The functions $Z_{12}$ and $Z_{13}$ describe the dependence of the
production amplitude on the production angle $\theta$ in the
centre of mass and on decay angles $\alpha ,\beta$ for decay
in the resonance rest frame.
These functions are
written in terms of Lorentz invariant tensors, following the methods
described by Chung [19].

Partial waves with $J^P = 2^+$ and $4^+$ couple to $\bar pp$ with
$\ell  = J \pm 1$. Multiple scattering through the resonance is expected
to lead to approximately the same phase $\phi$ for both $\ell$ values.
The ratio of coupling constants $g_{J+1}/g_{J-1}$ is therefore fitted
to a real ratio $r_J$; this assumption can be checked in several cases
and data are consistent with this assumption.

A general comment concerns possible $t$-channel contributions.
Singularities due to those exchanges are distant
and therefore contribute mostly to low partial waves.
The high partial waves with $J > 2$ are consistent with no background.
Secondly, $t$-channel exchange amplitudes acquire the phases of
resonances to which they contribute and may be considered as
driving forces for those resonances; an illustration is the well-known
Chew-Low theory [20], where nucleon exchange
affects the line-shape of $\Delta (1232)$.
In the present analysis, line-shapes of resonances cannot be
distinguished in most cases from simple Briet-Wigner resonances
with constant widths.

\subsection {Illustration of spectroscopic notation}
 As an example of spectroscopic notation, consider $\bar pp \to \omega
 \pi$.
The initial $\bar pp$ state may have spin $s = 0$ or 1.
For singlet ($s = 0$) states, the total angular momentum is
$J = \ell$.
For triplet states $(s = 1)$, $J = \ell$ or $\ell \pm 1$.
The parity is $P = (-1)^{\ell }$ and $C = (-1)^{\ell + s}$.
In the $\omega \pi $ exit channel, the spin $s$ is 1 for the
$\omega$.
As examples, $J^{PC} = 1^{--}$ may couple to initial $\bar pp$
$^3S_1$ and $^3D_1$ partial waves having $\ell = 0 $ or 2;
the exit $\omega \pi $ channel has $L = 1$.
In the analysis, $^3S_1$ and $^3D_1$ partial waves are fitted with
a ratio of coupling constants
$r_J= g_{\ell = J+1}/g_{\ell = J - 1}$.
A second example is $J^{PC} = 3^{+-}$.
This couples to $\bar pp$ singlet states and decays to $\omega \pi ^0$
with $L = 2$ or 4.
These two partial waves are again fitted with a ratio of coupling
constants $r_J = g_{L = J+1}/g_{L = J-1}$.
Multiple scattering through the resonance is expected to lead
to approximately the same phase $\phi$ for both values of $\ell$ or
$L$.
It is indeed found that all phase differences between $\ell = J \pm 1$
lie within the range 0 to $\pm 15 ^\circ$.
Where possible, ratios $r_J$ are taken to be real, in order to stiffen
the analysis.

The $\omega \eta \pi ^0$ data are fitted to
sequential 2-body processes: $\omega a_2(1320)$, $\omega a_0(980)$,
$b_1(1235)\eta$ and $\omega (1650)\pi ^0$.
As regards notation,  consider $\omega a_2$.
Spins 1 of the $\omega$ and 2 for the $a_2$ combine to
total spin $s' =  1$, 2 or 3.
The initial $\bar pp$ $^1F_3$ states with $J^{PC} = 3^{+-}$ may
decay to $\omega a_2$ with $L = 1$, 3 or 5.
In practice, the lowest $L$ value is always dominant;  in
every case $L $ values above the second may be omitted because of
the strong centrifugal barrier.
It is however necessary to consider all possible values of total spin
$s'$.

\subsection {General comments on procedures}
The selection of amplitudes is kept as simple as possible, consistent
with the data; introducing amplitudes whose significance is
marginal introduces noise.
A rule-of-thumb is that one can normally
remove amplitudes which improve log likelihood by $<20$. For $I =  0$,
$C =+1$ where 8 data sets are available and the solution is strongly
over-constrained, amplitudes contributing $< 40$ are omitted; they
contribute $<0.3\%$ of the total intensity in all cases.

Programmes can optimise $M$ and $\Gamma$ as well as coupling constants
and the barrier radius $R$.
In practice it is important to scan all $M$ and $\Gamma$
through at least 7 steps (freeing other $M$ and $\Gamma$) so as to
check that there is a well defined parabolic optimum in log
likelihood. Movements of other $M$ and $\Gamma$ reveal correlations.
For a few poorly determined resonances near the $\bar pp$
threshold, log likelihood flattens out for masses below threshold.
Likewise, poorly determined widths may tend to flatten off or run
away at large values; this can arise via cross-talk with other
resonances having the same quantum numbers at higher or lower masses.
Resonances with poor behaviour of this
type are assigned a low star rating or are absorbed into background
terms where possible.

It is bad practice to introduce broad isotropic backgrounds. These
interfere over the whole Dalitz plot and can perturb resonances.
A symptom of instability is the development of large destructive
interferences between components. This is checked by printing out
the integrated contributions from every amplitude and every
pair of interferences.

Well known resonances such as $f_2(1270)$,
$a_2(1320)$, $a_0(980)$, $b_1(1235)$, etc are fixed at masses
and widths quoted by the PDG; the D/S ratio of amplitudes for
decay of $b_1(1235)$ (and other resonances) are fixed at
PDG values.

A general point is that programmes must be as fast and as simple as
possible, since one needs to run hundreds of variants
of the fit with varying sets of amplitudes.
A simple though laborious procedure to look for alternative
solutions is to flip the signs of the smaller amplitudes one
by one and in pairs and check whether the solution returns to
its starting point or a different solution.
Signs of $r$ parameters are also flipped.
In practice Barrelet ambiguities
do not survive in the coupled-channel fits. Only one
ambiguous solution is found, for $\eta \pi ^0$ as described below.
The bigger problem arises with poor solutions where several
amplitudes are strongly correlated. The addition of each new set
of data improves the stability of the solution greatly; one new
channel is worth a factor 4 statistics in existing data.
Generally speaking, 50,000 events per channel are sufficient;
further statistics do not improve the solution and one has to
beware that errors become systematic rather than statistical.

Cross-talk between high partial waves arises because
of limitations on acceptance for small production angles.
For Crystal Barrel data, cross-talk begins to develop between
$\ell  = 4$ and $\ell = 5$ partial waves. A general guide is that it
is necessary to have an acceptance which extends to halfway
between $\cos \theta = \pm 1$ and the first zero of the highest
Legendre polynomial. To identify $\ell = 5$ partial waves
requires acceptance to $\cos \theta = \pm 0.96$ or $\theta < 6^\circ$,
where $\theta$ is the centre of mass production angle.
In practice this is available in the backward hemisphere because
of the Lorentz boost, but is not available in the forward hemisphere.
An obvious point is that bins where
acceptance varies rapidly with scattering angle should be removed
or flagged.
For all-neutral final states, angular distributions are symmetric
forward-backward because of conservation of $C$-parity; physically,
neutral products do not distinguish between initial $p$ and $\bar p$.
A feature of the PS172 data is that the angular acceptance is
complete to $\cos \theta = \pm 1$ since the beam went through the
multi-wire detectors.

Adjoining resonances which overlap by more than their half-width
are poorly separated. This makes separation of $^3S_1$ and $^3D_1$
resonances difficult, since they alternate at steps of
$\sim 150$ MeV and have widths typically 250 MeV.

One sometimes finds that resonance parameters are determined
sensitively in simple channels, e.g. $\omega \eta$,
despite low statistics. Higher statistics may exist in more
complex channels, e.g. $\omega \pi ^0 \pi ^0$.
However, analysis of these more complex channels may give poorer
sensitivity, for example due to the broad $\sigma$ in the
$\omega \sigma$ channel.
[Here $\sigma$ denotes the whole $\pi \pi$ S-wave amplitude].
In this case it may be desirable to fix resonance parameters for
these more complex channels. There is an alternative technique
for dealing with this problem. It may also be used when
searching for weak decay modes of resonances.
A penalty function is introduced. This penalty function
takes the form of contributions to $\chi ^2$:
$$
\chi ^2_i = \left[ \frac {\Lambda - \Lambda _0}{\delta \Lambda }
\right]^2_i. $$
Here $\Lambda$ is a parameter and $\Lambda _0$ a target value.
The denominator $\delta \Lambda$ is adjusted so that each questionable
ingredient contributes $\chi ^2$ = 9 (i.e. $3\sigma$).
Those amplitudes which are really needed feel little influence from
this small penalty function compared to the influence of data points;
amplitudes which are unstable or not needed are well constrained.
This is particularly useful for searching for small branching
ratios; here $\Lambda _0$ is set to zero.
This procedure is applied for $I = 1$, $C = +1$,
where some instabilities are discussed below; the penalty
function contributes $\sim 650$ to $\chi ^2$ compared with
$\sim 650,000$ for log likelihood. Incidentally, the program
optimises both $\chi ^2 $ for some data sets (e.g. binned
2-body angular distributions) and log likelihood (for 3-body
channels).

There is a useful trick in checking $J^P$ of resonances.
It is difficult to exhibit the angular dependence of some
amplitudes, for example high $J^P \to a_2\pi$; the angular
dependence $Z$ of eqn. (3) involves complicated correlations
between production angle $\theta$ and decay angles $\alpha,~\beta $
of the $a_2$.
Putting any new resonance into the analysis always gives some
improvement. Some of this improvement comes from the correct
line-shape and some from the correct angular dependence.
It is useful to substitute different (wrong) angles and examine
how sensitive the analysis is to the right choice of angles
and correlations.

\section {Review of $I = 0$, $C = +1$ resonances}
Most resonance from the Crystal Barrel in-flight analysis
are listed by the Particle Data Group in a section called
`Other Light  Unflavoured Mesons', pp 010001-539 of Ref. [11].
The reason is that the Particle Data Group requires confirmation
from a separate experiment, rather than from different channels
of data in one experiment. There is very little other coverage of
this mass range from other experiments.
Some results are combined by the PDG with those of other groups in the
usual way, some are not.
Many of the resonances described here are observed in several
decay channels and are better established that other entries from
single groups, so this creates a confusing situation.
Resonance parameters from the final $I =  0$, $C = +1 $
coupled-channel analysis [3] are given in Table 2.
\begin{table} [htp]
\begin{center}
\begin{tabular}{ccccccc}
$J^{PC}$ & * & Mass $M$ & Width $\Gamma$ & $r$ &
$\Delta S $ & Observed\\
      &   & (MeV)    & (MeV) &  & $(\eta \pi \pi )$ & channels  \\\hline

$6^{++}$ & - & $2485 \pm 40 $ & $410 \pm 90$ & 0 & 245 & \\
$4^{++}$ & $4^*$ & $2283 \pm 17 $ & $310 \pm 25$ & $2.7 \pm 0.5$ & 2185
& $\pi \pi$,$\eta \eta$,$f_2\eta$,VES\\
$4^{++}$ & $4^*$ & $2018 \pm 6 $ & $182 \pm 7$ & $0.0 \pm 0.04$ & 1607
& $\pi\pi$,$\eta \eta$, $\eta \eta '$,$a_2\pi$,($f_2\eta$),PDG \\
$4^{-+}$ & $2^*$ & $2328 \pm 38 $ & $240 \pm 90$ & & 558
& $a_0\pi$,($a_2\pi $) \\
$3^{++}$ & $4^*$ & $2303 \pm 15$ & $214 \pm 29$ & & 1173
& $[f_2\eta ]_{L=1,3}$,$f_2\eta '$ \\
$3^{++}$ & $4^*$ & $2048 \pm 8 $ & $213 \pm 34$ & & 1345
& $[f_2\eta ]_{L=1,3}$,$a_2\pi$,$a_0\pi$ \\
$2^{++}$ & $4^*$ & $2293 \pm 13 $ & $216 \pm 37$ & $-2.2 \pm 0.6$ & 1557
& $\pi\pi$,$\eta \eta$,$\eta \eta '$,$f_2\eta$,$a_2\pi$ \\
$2^{++}$ & $4^*$ & $2240 \pm 15 $ & $241 \pm 30$ & $0.46 \pm 0.09$ & 468
& $\pi \pi$,$\eta\eta$,$\eta \eta '$,$f_2\eta '$ \\
$2^{++}$ & $4^*$ & $2001 \pm 10 $ & $312 \pm 32$ & $5.0 \pm 0.5$ & 168
& $\pi\pi$,$\eta\eta$,$\eta \eta '$,$f_2\eta$ \\
$2^{++}$ & $4^*$ & $1934 \pm 20 $ & $271 \pm 25$ & $0.0 \pm 0.08$ & 1462
& $\pi\pi$,$\eta\eta$,$f_2\eta$,$a_2\pi$,GAMS,VES \\
$2^{++}$ & $3^*$ & $2010 \pm 25 $ & $495 \pm 35$ & $1.51 \pm 0.09$ & 694
& $\eta\eta \times 2$,WA102 \\
$2^{-+}$ & $4^*$ & $2267 \pm 14 $ & $290 \pm 50$ & & 1349
&  $f_2\eta '$,$[f_2\eta ]_{L=2}$,$a_0\pi$,
$f_0(1500)\eta$\\
$2^{-+}$ & $3^*$ & $2030 \pm 16 $ & $205 \pm 18$ &  &
& $[a_2\pi ]_{L=0,2}$,$a_0\pi$ \\
$2^{-+}$ & $4^*$ & $1870 \pm 16 $ & $250 \pm 35$ & &
& $f_2\eta$,$a_2\pi$,$a_0\pi$,WA102 \\
$1^{++}$ & $2^*$ & $2310 \pm 60 $ & $255 \pm 70$ & & 882
& $f_2\eta$, $f_2\eta '$, $\sigma \eta$ \\
$1^{++}$ & $3^*$ & $1971 \pm 15 $ & $240 \pm 45$ & & 1451
& $f_2\eta$, $a_2\pi$ $a_0\pi$, ($f_0(980)\eta$,$\sigma
\eta$,$\sigma\eta'$) \\
$0^{++}$ & $3^*$ & $2337 \pm 14 $ & $217 \pm 33$ & &
& $\pi \pi$,$\eta\eta$\\
$0^{++}$ & $4^*$ & $2102 \pm 13 $ & $211 \pm 29$ & &
& $\eta \eta \times 2$,$\eta \eta '$,Mark III,BES I \\
$0^{++}$ & $3^*$ & $2020 \pm 38 $ & $405 \pm 40$ & & & $\pi \pi$,
WA102($\pi \pi ,4\pi$)\\
$0^{-+}$ & $3^*$ & $2285 \pm 20 $ & $325 \pm 30$ & & 1342
& $f_0(1500)\eta \times 2$,($\sigma \eta$),($\sigma \eta '$) \\
$0^{-+}$ & $2^*$ & $2010 ^{+35}_{-60}$ & $270 \pm 60$ & & 1189
& $a_2\pi$,$f_2\eta$,$\sigma\eta$, BES II \\\hline
\end{tabular}
\caption {$I=0$, $C = +1$ resonances . Errors cover
systematic variations in a variety of fits with different
ingredients. The sixth
column shows changes in $S = $ log likelihood when each component is
omitted in turn from the fit to $\eta \pi ^0 \pi ^0$ data and remaining
components are re-optimised.
In the last column, channels in parentheses are weak compared with
the others, or less reliable.}
\end{center}
\end{table}

\begin{figure} [htb]
\begin{center}
\epsfig{file=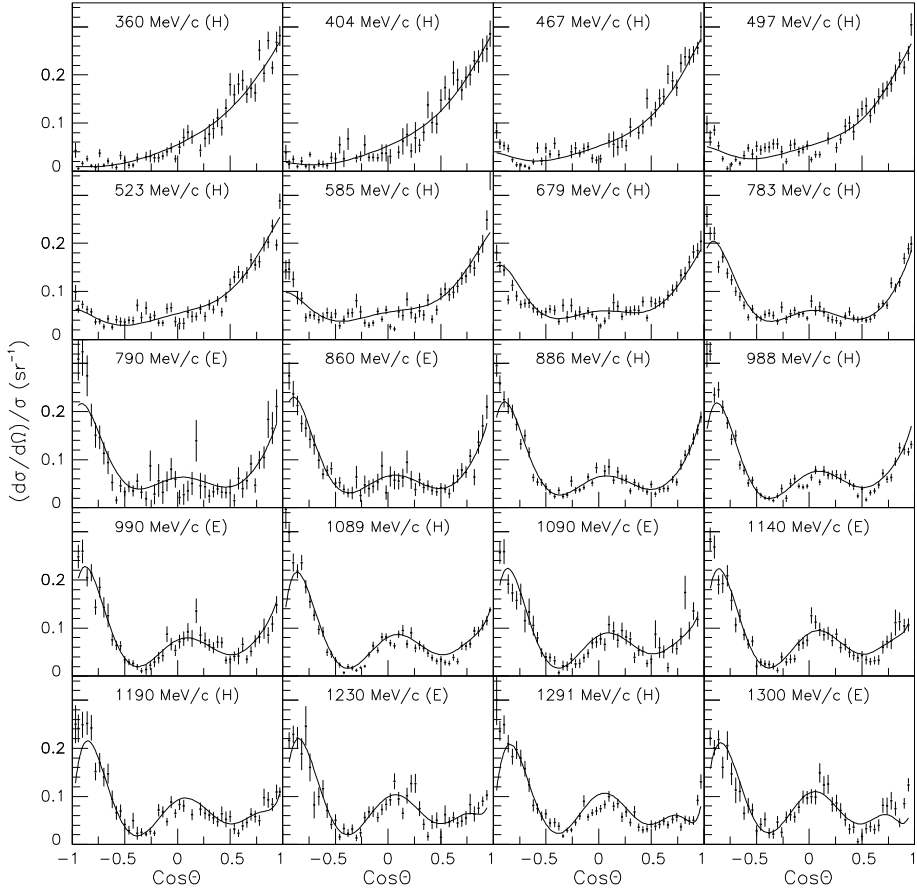,width=14cm}  
~\
\caption{Differential cross sections for $\bar pp \to \pi ^-\pi ^+$
from 360 to 1300 MeV/c, compared with the fit; panels are labelled
H for data of Hasan et al. and E for those of Eisenhandler et al.
Each distribution is normalised so as to integrate to $2\pi$.}
\end{center}
\end{figure}
\subsection {2-body channels}
Resonances with $J^P = 0^+$, $2^+$ and $4^+$ are observed in 2-body
channels, where polarisation data are available. The first
step was to determine these amplitudes.
The extensive differential cross section and polarisation data come
from Eisenhandler et al. [2] and PS172, Hasan et al. [1].
They are compared with the fit in Figs. 11--13.
\begin{figure} [htb]
\begin{center}
\epsfig{file=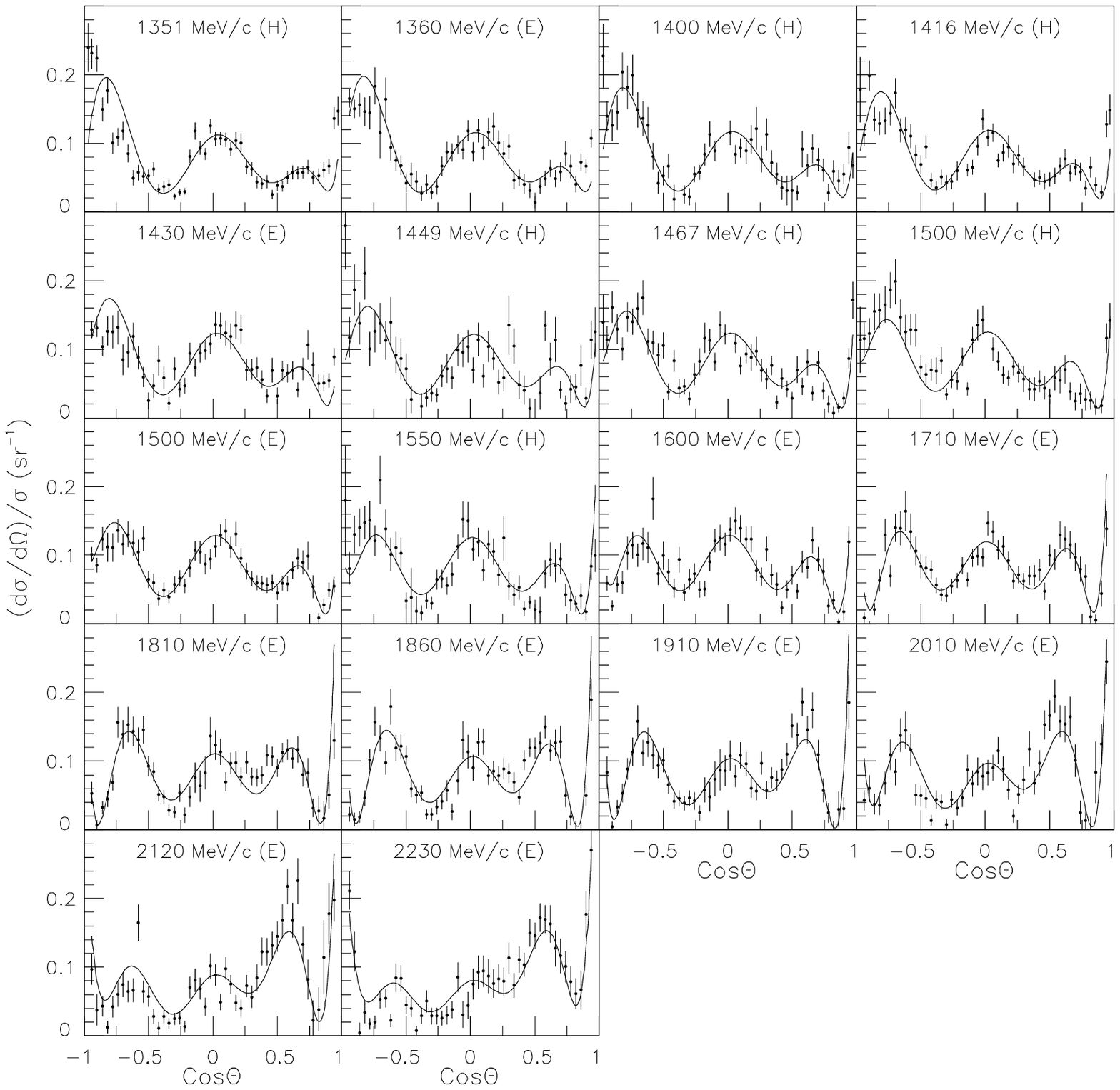,width=14cm}  
~\
\caption{As Fig. 11, 1351 to 2230 MeV/c.}
\end{center}
\end{figure}
\begin{figure} [htb]
\begin{center}
\epsfig{file=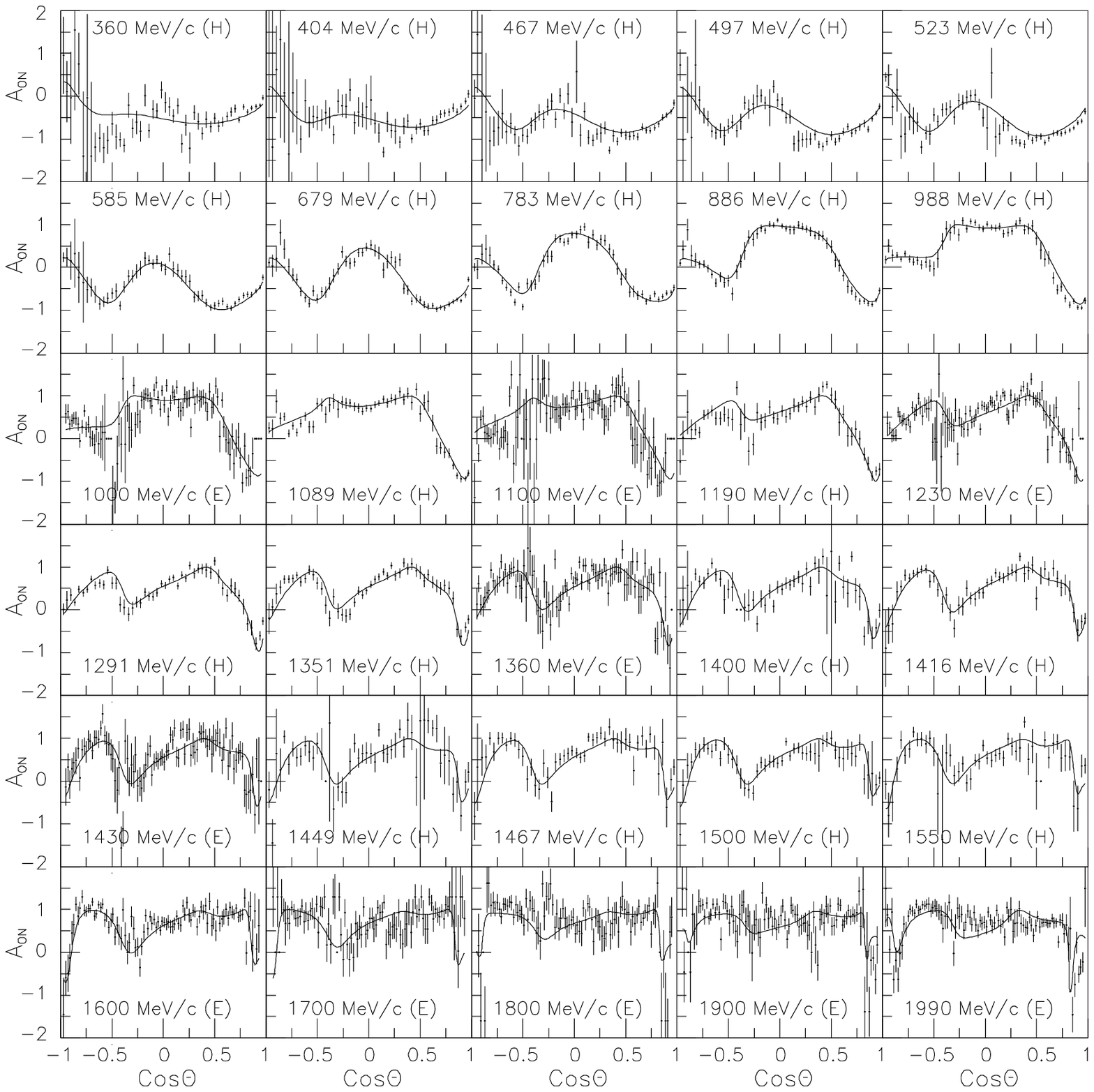,width=15cm} 
~\
\caption{Polarisation data for $\bar pp \to \pi ^- \pi ^+$,
compared with the fit.}
\end{center}
\end{figure}

Why are polarisation data needed?
For $\bar pp \to \pi ^- \pi ^+$, there are only triplet amplitudes
with $J = L \pm 1$: $^3S_1$, $^3P_0$, $^3D_1$, $^3D_3$,
$\ldots$.
It is convenient to use two amplitudes which measure transfer
of helicity +1/2 of the proton in the initial state to
$\pm 1/2$ in the final state.
These are denoted $F_{++}$ (helicity non-flip) and $F_{+-}$ (helicity
flip).
[$F_{--} = F_{++}$ and $F_{-+} = F_{+-}$].
In terms of partial wave amplitudes and Legendre polynomials,
\begin {eqnarray}
F_{++} &=& \frac {1}{4}\sum _J (2J+1)f^J_{++}P_J(\cos \theta ) \\
F_{+-} &=& \frac {1}{4}\sum _J \frac {2J+1}{\sqrt {J(J+1)}}f^J_{++}
P^1_J(\cos \theta ), \\
\sqrt {2J+1}f_{++} &=& \sqrt {J}T_{J-1,J} - \sqrt {J+1}T_{J+1,J} \\
\sqrt {2J+1}f_{+-} &=& \sqrt {J+1}T_{J-1,J} + \sqrt {J}T_{J+1,J}.
\end {eqnarray}
Note the difference in sign in the last two of these expressions for
T-matrix elements $T_{J+1,J}$.
Differential cross sections and polarisations are given by
\begin {eqnarray}
A_{0N}d\sigma /d\Omega &=& Tr(F^*\sigma _Y F) = 2Im(F^*_{++}F_{+-}) \\
d\sigma /d\Omega &=&|F_{++}|^2 + |F_{+-}|^2 .
\end {eqnarray}
The polarisation is very sensitive to the difference between
amplitudes for $L = J \pm 1$, for example $^3P_2$ and $^3F_2$
amplitudes. It is also phase sensitive.

The analysis of 2-body data alone [13] leads directly to a unique
and accurate solution for all $0^+$, $2^+$ and $4^+$ states.
Intensities of fitted partial waves are shown in Fig. 14.
A small detail is that the low-mass tail of a higher $6^+$
resonance is included, though the mass of that state is above
the available mass range and is therefore not well determined.
The $f_4(2050)$ is observed strongly in Fig. 14 in all of $\pi ^0 \pi
^0$, $\eta \eta$ and $\eta \eta '$, as well as in $\eta \pi ^0 \pi ^0$
data discussed below.
The 2-body data required the presence of two $^3P_2$ states,
two $^3F_2$ and two $0^+$ states.
The strong peaking of the $^3P_2$ amplitude at the lowest
mass is associated with $f_2(1934)$, or $f_2(1910)$ of the PDG.
Fig. 9. of Ref. [13] shows the effect of dropping each of
these states one by one from the analysis. Removing any of
them causes clear visual disagreements with the data as well
as large changes in log likelihood.
\begin{figure} [htb]
\begin{center}
\epsfig{file=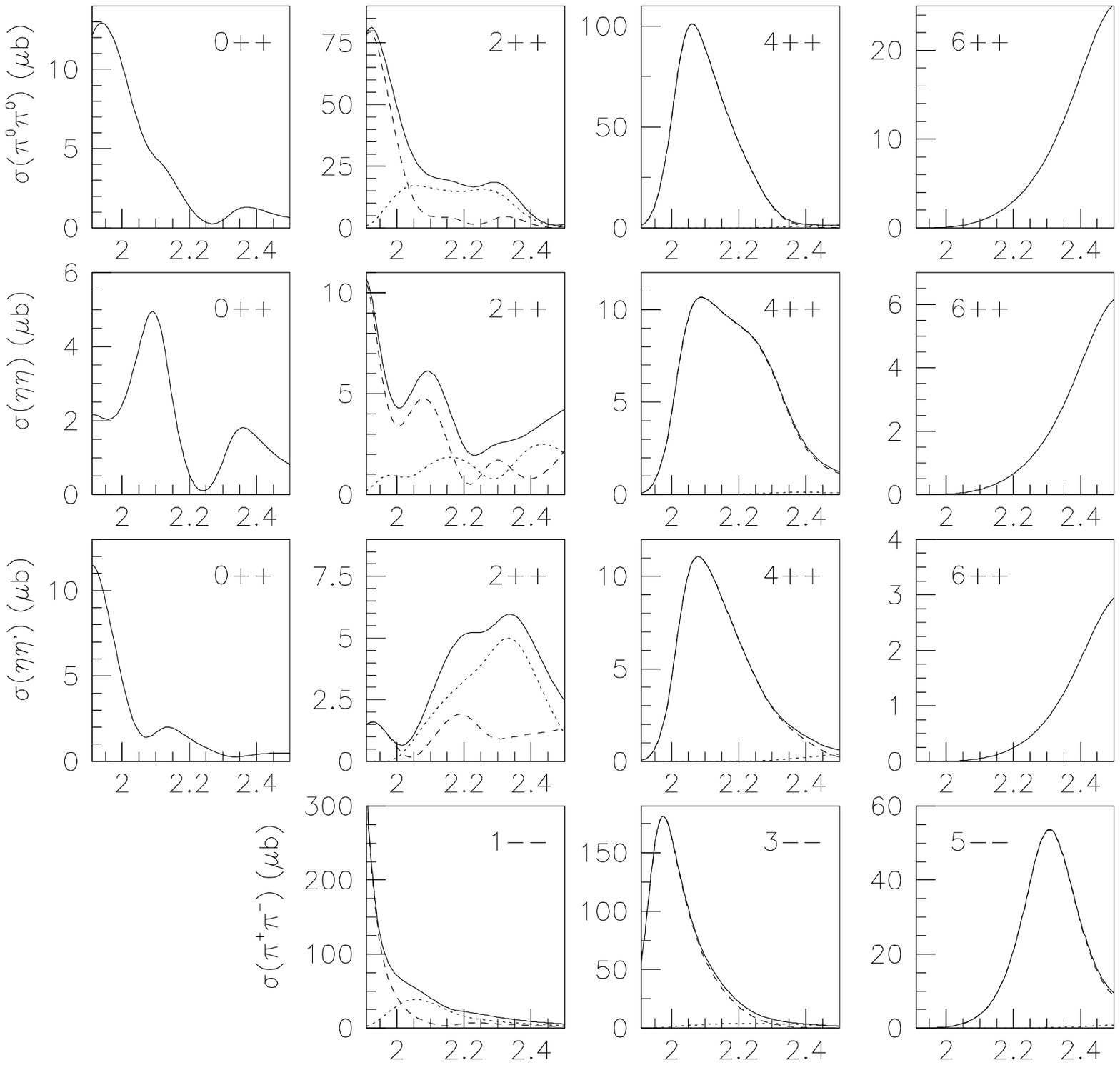,width=11cm}
~\
\caption{Intensities for partial waves for $\pi ^0 \pi ^0$,
$\eta \eta$ and $\eta \eta '$.
Dashed curves show contributions from $L = J-1$ and dotted curves
from $L = J+1$.}
\end{center}
\end{figure}

A cross-check on the analysis is possible between $\pi ^0 \pi ^0$
data and $\pi ^- \pi ^+$. States with $I = 1$, $C = -1$ contribute
to $\pi ^- \pi ^+$ but are determined separately from $\omega \pi $
data discussed below.
If one allows for their small intensity in  $\pi ^- \pi ^+$
and folds the angular distribution forward-backwards about $90^\circ$,
in order to remove interferences between $I = 0$ and $I = 1$,
the resulting angular distribution agrees accurately in shape
with $\pi ^0 \pi ^0$ and their absolute magnitudes agree within
$1.1 \pm 2.3\%$. This is a valuable check on the entire process of
selecting and reconstructing Crystal Barrel data, and the
correction for rate dependence.

\subsection {3-body data}
Resonances with $J^P = 0^-$, $1^+$, $2^-$, $3^+$ and $4^-$ appear only
in 3-body channels.
However, statistics for $\eta \pi ^0 \pi ^0$ are very high, typically
$\ge 70~K$ events per momentum.
There are two dominant channels, $f_2(1270)\eta$ and $a_2(1320)\pi$
which are both well determined.
Mass projections at some momenta are shown in Fig. 15.
Small $f_0(980)$, $f_0(1500)$ and $a_0(980)$ signals are also visible.
\begin{figure}
\begin{center}
\epsfig{file=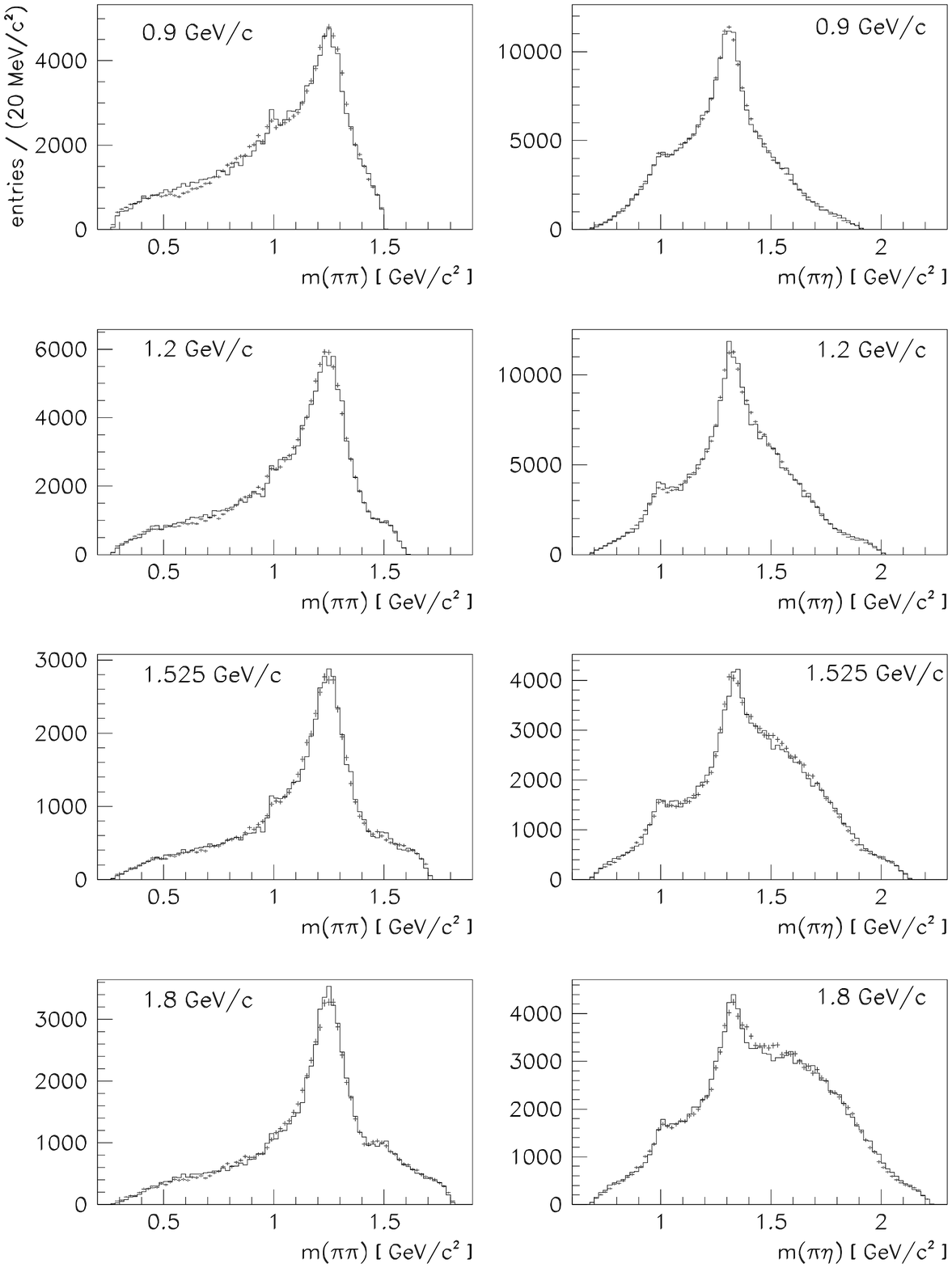,width=11cm}~\
\caption{Mass projections on to $M(\pi \pi )$ and $M(\pi \eta )$ for
$\eta \pi ^0 \pi ^0$
data at (a) and (b) 900 MeV/c, (c) and (d) 1200 MeV/c, (e) and (f)
1525, (g) and (h) 1800 MeV/c; histograms show the result of the fit.
Data are uncorrected for (small) variations of acceptance; these are
included in the maximum likelihood fit. }
\end{center}
\end{figure}

\begin{figure}[htbp]
\begin{center}\hspace*{-0.cm}
\epsfysize=15cm
\epsffile{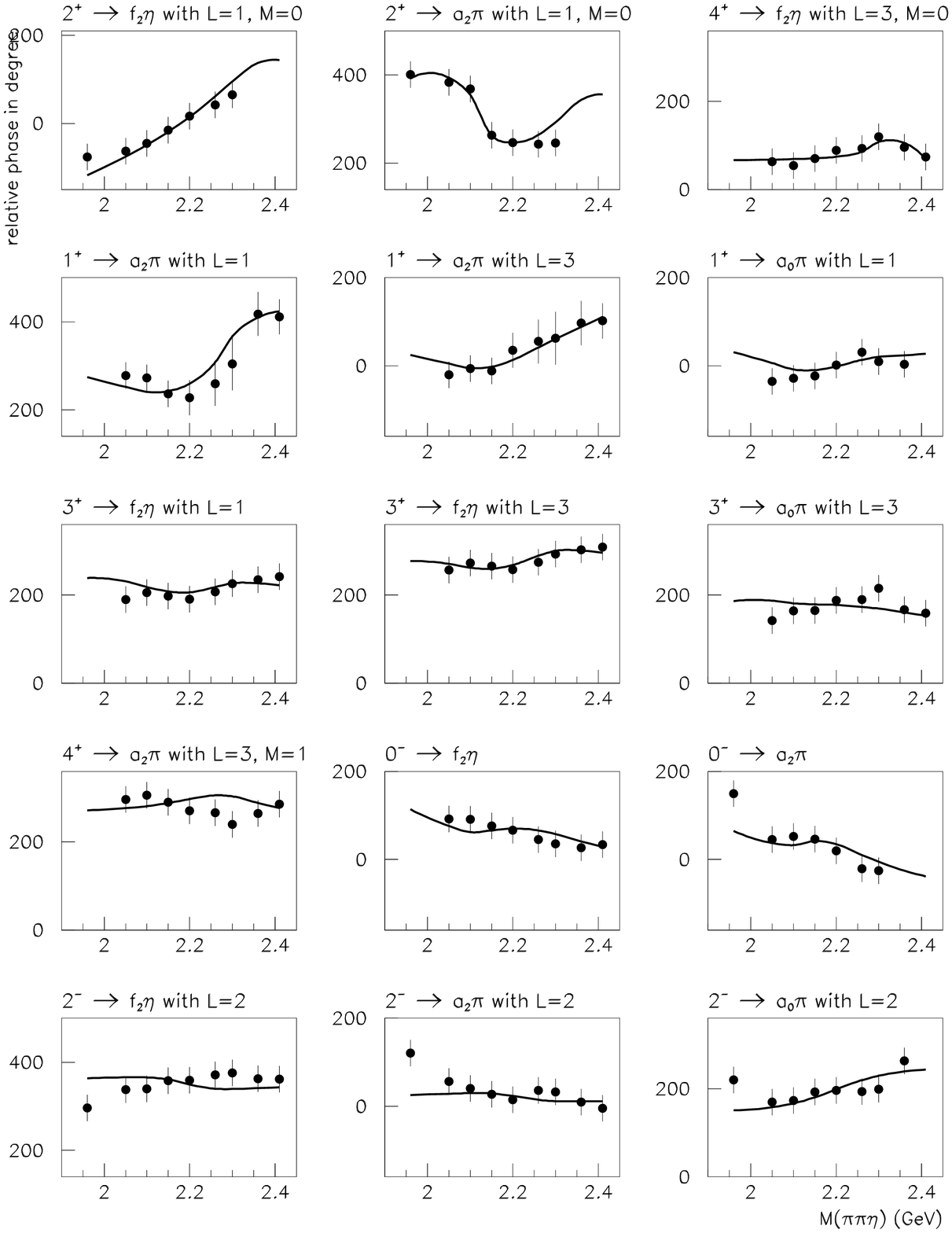}
\end{center}
\caption{Relative phases of partial waves for $\eta \pi ^0 \pi ^0$.
The phases for $2^+$ and $4^+$ with M=0 are relative to $4^+\to a_2\pi$
with L=3, M=0; the phases for $1^+$, $3^+$ and $4^+$ with M=1 are
relative to $4^+\to f_2\eta$ with L=3, M=1; the phases for $0^-$ and
$2^-$ are relative to $2^-\to f_2\eta$ with L=0. }
\end{figure}
The $f_4(2050)$ acts as an
interferometer at the lower momenta, and likewise $f_4(2300)$ at higher
momenta.
Phases of many other partial waves show rather small mass
dependence with respect to these $f_4$.
This is illustrated in Fig. 16; there, $M$ refer to the
helicity in the initial $\bar pp$ state.
It is unavoidable that resonances need to be introduced into
$3^+$, $1^+$ and $0^-$ partial waves.
The behaviour of $2^+$ partial waves is somewhat more complex.
The 2-body data define very precisely the existence of
states which are dominantly $^3P_2$ at 1934 and 2240 MeV
and $^3F_2$ at 2001 and 2293 MeV.
The first of these may be identified with the $f_2(1910)$
observed by GAMS [21] and VES [22] collaborations.
The small spacings of $2^+$ states leads to rather rapid
variations of phases with mass, but the 3-body data are
entirely consistent with 2-body results.

There are then two lucky breaks.
The first is that data for the final state $\eta '\pi ^0 \pi ^0$
are dominated by a single resonance in $f_2(1270)\eta '$ [23].
Dalitz plots are shown in Fig. 17. There is an obvious band
due to $f_2(1270)$ along the lower left-hand edge of these
Dalitz plots.
A very clear peak is observed in the integrated cross section
for $f_2\eta '$ at 2267 MeV, see Fig. 18.
It is consistent with a resonance required near this
mass in $\eta \pi ^0 \pi ^0$, but gives a better definition of
resonance parameters. In turn, this $2^-$ state then acts as a powerful
interferometer to determine other partial waves at high masses.

\begin{figure} [h]
\begin{center}
\epsfig{file=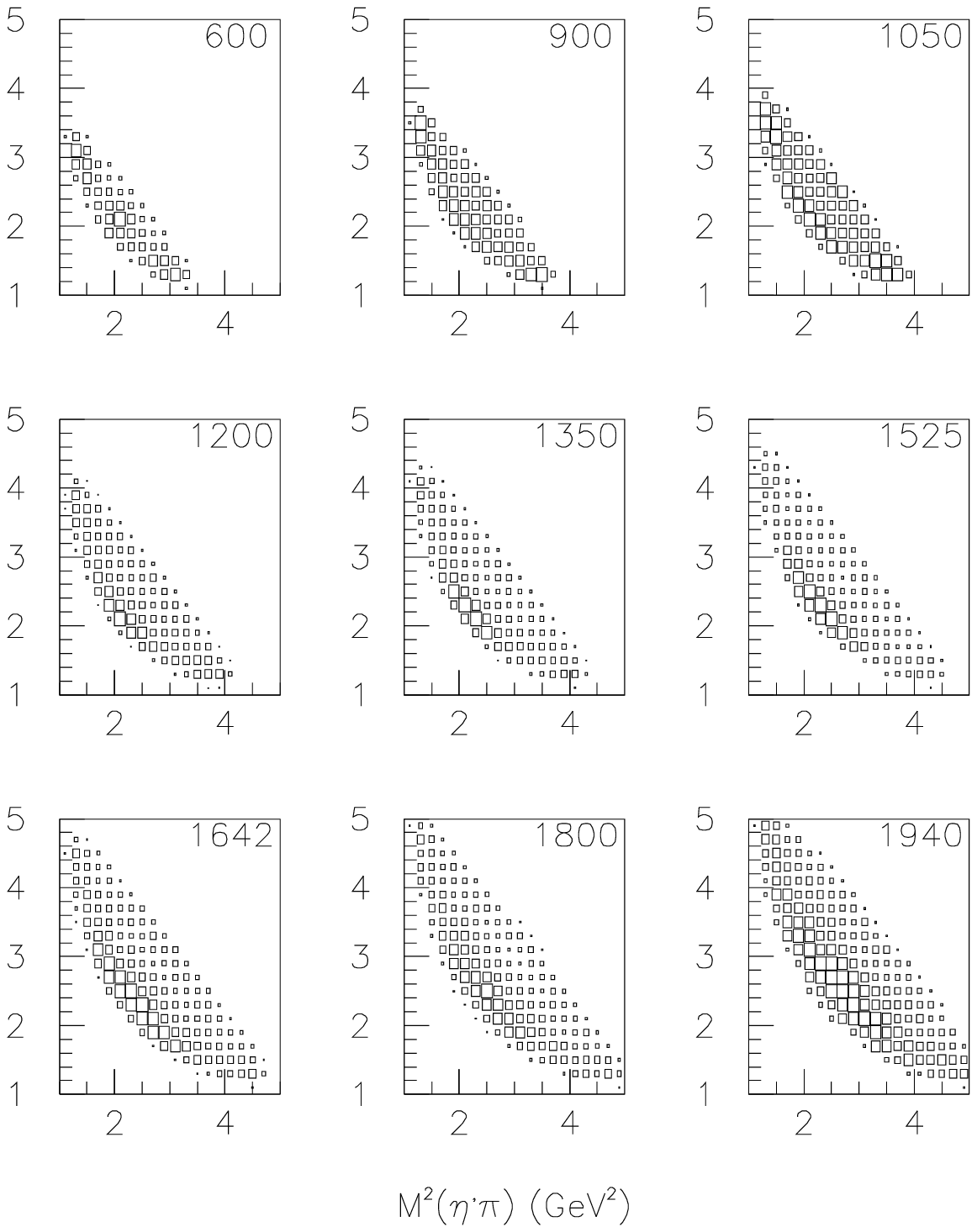,width=10cm}\
\caption{Dalitz plots at all beam momenta for $\eta '\pi ^0
\pi ^0$ data. Numerical values indicate beam momenta in MeV/c. }
\end{center}
\end{figure}
\begin{figure} [h]
\begin{center}
\epsfig{file=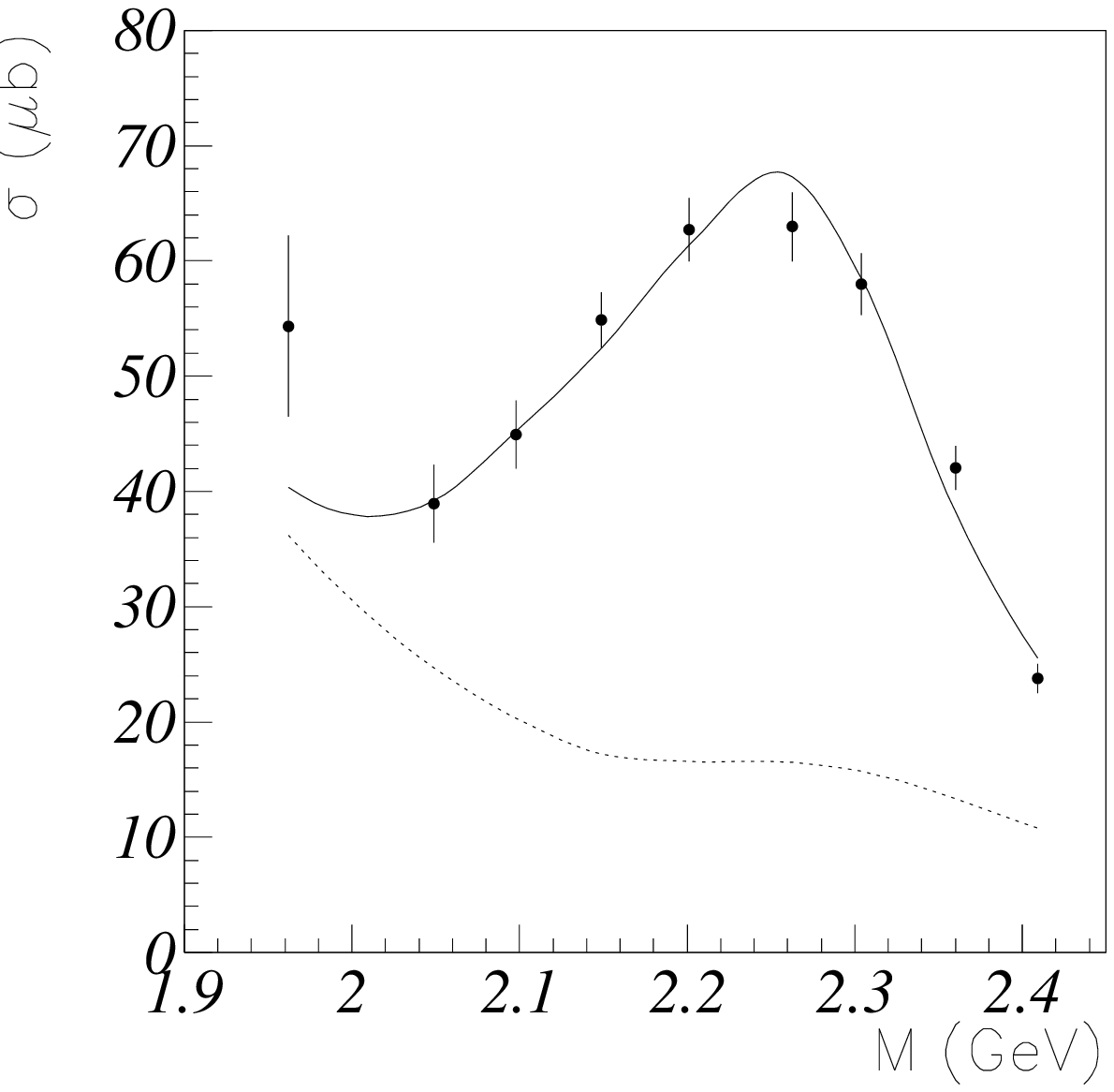,width=5cm}\
\caption{The integrated cross sections for
the final state $\eta '\pi ^0\pi ^0$, after correction for backgrounds
and for all decay modes of $\eta '$, $\eta$ and $\pi ^0$;
the full curve shows the fit from the amplitude analysis;
the dotted curve shows the $\sigma \eta '$ contribution.}
\end{center}
\end{figure}

\newpage
The second lucky break comes from $\eta \eta \eta$ data [24].
Although statistics are low, there is a strong peak
at high mass due to the well identified
$f_0(1500)\eta$ S-wave final state with quantum
numbers $0^-$. This is shown by the curve labelled $0^-$ in Fig. 19;
in plotting this curve, interference between the three $f_0(1500)$
bands has been removed.
There is a small but distinctive contribution from the $2^-$ state
at 2267 MeV in the $f_0(1500)\eta$ D-wave.
The analysis is very simple and leads to a cleanly identified $0^-$
resonance at 2285 MeV.
\begin{figure} [htbp]
\begin{center}
\epsfig{file=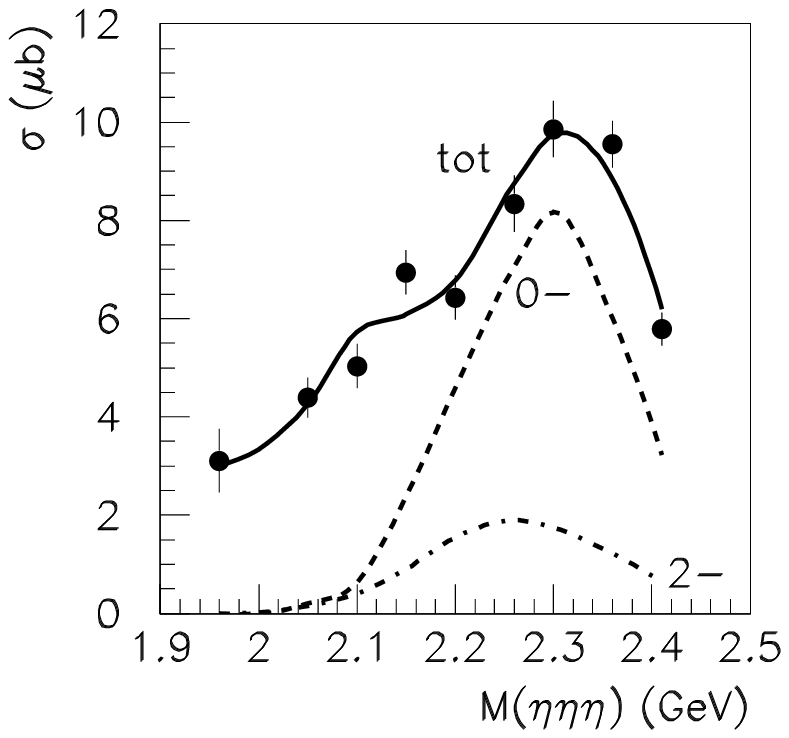,width=6cm}\
\caption {The fit to the mean integrated cross section for
$\bar pp \to 3\eta$ from $6\gamma$ and $10\gamma$ data,
(b) $\eta (2285) \to [f_0(1500)\eta]_{L=0}$ (dashed),
and $\eta _2(2267) \to [f_0(1500)\eta ]_{L = 2}$ (chain curve).}
\end{center}
\end{figure}

Returning to $\eta \pi \pi$ data, there remain a $4^-$ state, two $1^+$
states and a lower $0^-$ state which are needed in the full combined fit.

\subsection {$2^-$ states}
It is necessary to discuss separately $2^-$ states at 1645, 1860
and 2030 MeV. They appear in $\eta \pi ^0 \pi ^0 \pi ^0$ data in
the production reaction $\bar pp \to \pi ^0 (\eta \pi \pi )$.
This was first studied [25] in data at just two momenta: 1200 and 1940
MeV/c, prior to the running of the other momenta .
Two $2^-$ $I = 0$ states were observed.
The lower one was initially observed decaying purely to
$a_2(1320)\pi$ at 1645 MeV. This fits nicely as the expected
partner of the well known $I = 1$ $\pi _2(1670)$.
However, the same data contained a second strong signal in
the $f_2(1270)\eta$ final state at 1870 MeV.
The second signal could not be explained as the high mass
tail of $\eta _2(1645)$ decaying to $f_2\eta$.
Furthermore, the relative strengths of the two signals
were quite different in data at 1200 and 1940 MeV/c.
Two separate states were required.

Both resonances were subsequently confirmed by the WA102
collaboration. They observed the $\eta _2(1645)$ in decays
to $\eta \pi \pi$ and $\eta _2(1870)$ in decays to
both $\eta \pi \pi$ and $K\bar K \pi$ [26].
However, the $K\bar K\pi$ decay mode is not strong
enough to make an interpretation likely as the
$s\bar s$ partner of $\eta _2(1645)$.
A feature of $\bar pp$ annihilation is
that well known $s\bar s$ states such as $f_2'(1525)$ are
produced very weakly, if at all; the $\eta _2(1870)$ is
produced rather strongly.
A possible explanation is that it is the radial
excitation of $\eta _2(1645)$, but that would be
expected to lie around 2000-2050 MeV, near
$f_4(2050)$.

In later data, a much more extensive study was made
of $\eta \pi ^0 \pi ^0 \pi ^0$ data with statistics
a factor 7 larger [27].
A further $2^-$ state appears as expected at 2030 MeV,
decaying dominantly to the $a_2(1320)\pi $ D-wave.

\begin{figure} [b]
\begin{center}
\epsfig{file=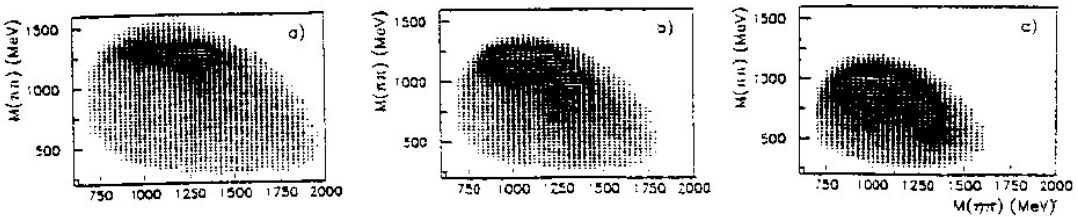,width=15cm}\
\caption {Scatter plots of $M(\pi \pi )$ v. $M(\eta \pi )$ for three
ranges of $M(\eta \pi \pi )$: (a) 1945--2115 MeV, centred on $\eta
_2(2030)$, (b) 1775--1945 MeV over $\eta _2(1870)$ and
(c) 1560--1750 MeV, centred on $\eta _2(1645)$.}
\end{center}
\end{figure}
The main features of the data are shown in Fig. 20 as
scatter plots of $M(\pi \pi )$ v. $M(\pi \eta )$ for three $\eta \pi
\pi$ mass ranges centred on $\eta _2(1645)$, $\eta _2(1870)$ and $\eta
_2(2030)$. In (c), there is a vertical band due to $a_2(1320)$ from the
decay of $\eta _2(1645)$; the horizontal band at 1000 MeV is due to the
low mass tail of the stronger $\eta _2(1870) \to f_2\eta$. In (b), the
horizontal band from $\eta _2(1870)$ is dominant and the vertical
$a_2(1320)$ band becomes weaker. In (a), the $f_2(1270)$ becomes weaker
and there is a strong peak at the intersection of $f_2(1270)$ and
$a_2(1320)$; the amplitude analysis shows that the $\eta _2(2030)$
decays dominantly to the $a_2(1320)\pi$ D-wave. The essential point of
the analysis is shown by the branching ratios in Table 3. [These refer
directly to data and are not corrected for other charge states or for
other decay modes of the $a_2$ and $\eta$.] Branching ratios change
dramatically with mass, requiring three separate states. The
$f_2(1870)$ decays dominantly to $f_2\eta$ and $a_0\pi$; the $\eta
_2(2030)$ has dominant $a_2(1320)\eta$ decays with $L = 0$ and 2, and
the $f_2\eta$ decay is weak.
\begin{figure} [t]
\begin{center}
\epsfig{file=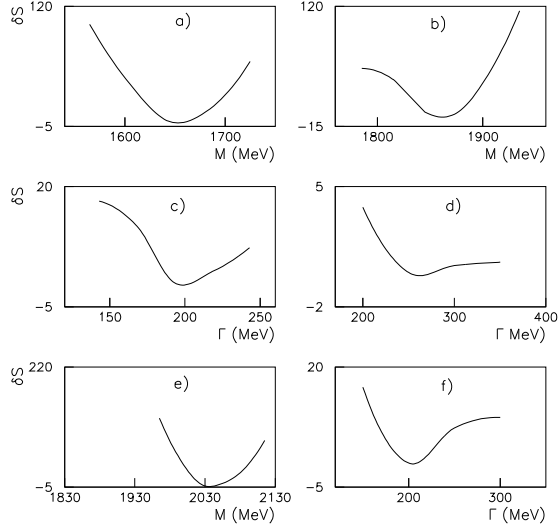,width=8cm}\
\caption {$S = $ --log likelihood v. masses and widths of
(a) and (b) $\eta _2(1645)$, (c) and (d) $\eta _2(1875)$,
(e) and (f) $\eta _2(2030)$.}
\end{center}
\end{figure}

\begin{table}  [b]
\begin{center}
\begin{tabular}{cccc}
\hline
Resonance & $\eta _2(1645)$ & $\eta _2(1870)$   & $\eta _2(2030)$
\\\hline
$BR[a_0\pi]/BR[a_2\pi ]$ & $0.21 \pm 0.13$ & $3.9 \pm 0.6$ &
$0.21 \pm 0.02$ \\
$BR[f_2\eta]/BR[a_2\pi ]$ & -  & $4.5 \pm 0.6$ & $0.25 \pm 0.09$
\\\hline
\end {tabular}
\caption {Decay branching ratios for $\eta
_2(1645)$, $\eta _2(1870)$ and $\eta _2(2030)$.}
\end{center}
\end{table}

Fig. 21 shows log likelihood for the three $2^-$ states
for mass and width.
There is some evidence for decay to $\eta \sigma$, though
that is a difficult final state to identify because the
$\sigma$ is broad.

The $\eta _2$ states couple to $\bar pp$ $^1D_2$.
From the straight-line trajectories of Fig. 1, it appears
that $\eta _2(1870)$ is an `extra' state. Isgur, Kokoski and
Paton [28] predicted the existence of hybrids around 1800--1900
MeV.

The hybrid consists of $q$ and $\bar q$ connected by a gluon
having one unit of orbital angular momentum about the axis
joining the quarks.
It rotates like a skipping rope between $q$ and $\bar q$.
It is predicted that this orbital angular momentum will be transferred to one
pair of quarks in the final state [29,30]. The expected decay modes are
therefore to one $^3P$ combination of quarks and one S-state
combination. Favoured decay modes are $f_2(1270)\eta$
and $a_2(1320)\pi$, in agreement with $\eta_2(1870)$ decays. However,
the $f_2\eta$ decay mode is strikingly strong. One can use SU(3) to
calculate the expected ratio $BR[a_2\pi]_{L=0}/BR[f_2\eta ]$,
integrating their available phase space over $a_2$ and $f_2$
line-shapes.
If it is a $q\bar q$ state, the predicted ratio for $\eta _2(2030)$ is
9.7 after correcting for all charged states and decay modes of $a_2$
and $\eta$; this compares well with the observed $10.0 \pm 1.8$.
However, for $\eta _2(1870)$, the prediction is 8.5 compared with
the observed $1.27 \pm 0.7$.
The conclusion is that the decay to $f_2\eta$ is distinctive and
far above the value expected for a $q\bar q$ state. It
will be shown below that there is an $I = 1$ partner at 1880 MeV
decaying to $a_2(1320)\eta$. If
these make a hybrid pair, it is the first pair to be identified.
Further hybrids containing hidden strangeness are then to be
expected at $\sim 2000$ MeV decaying to $K_2(1430)\eta$ and at
$\sim 2120 $ MeV decaying to $f_2'(1525)\eta$ and/or $K_2(1430)\bar K$.

\begin{figure} [htbp]
\begin{center}
\epsfig{file=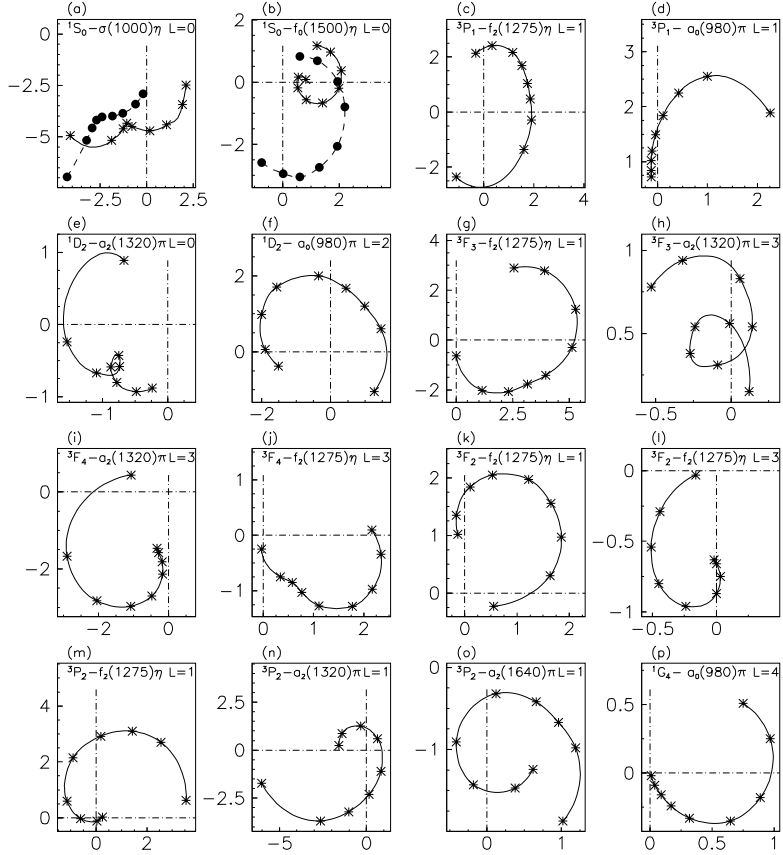,width=12cm}\  
\caption {Argand diagrams for prominent partial waves in $\eta \pi
^0\pi ^0$ data; crosses mark beam momenta; the amplitudes
move anti-clockwise with increasing beam momentum; in (a) and (b)
dashed curves show the effect of removing the lower $0^-$ resonance.}
\end{center}
\end{figure}

\subsection {Star ratings of states}
All states of Table 2 have a statistical significance
$> 25$ standard deviations.
The even spin states are even more significant in 2-body decays.
Argand diagrams for $\eta \pi \pi$ are shown in Fig. 22
and for $\pi \pi$ decays in Fig. 23.
\begin{figure} [htbp]
\begin{center}
\epsfig{file=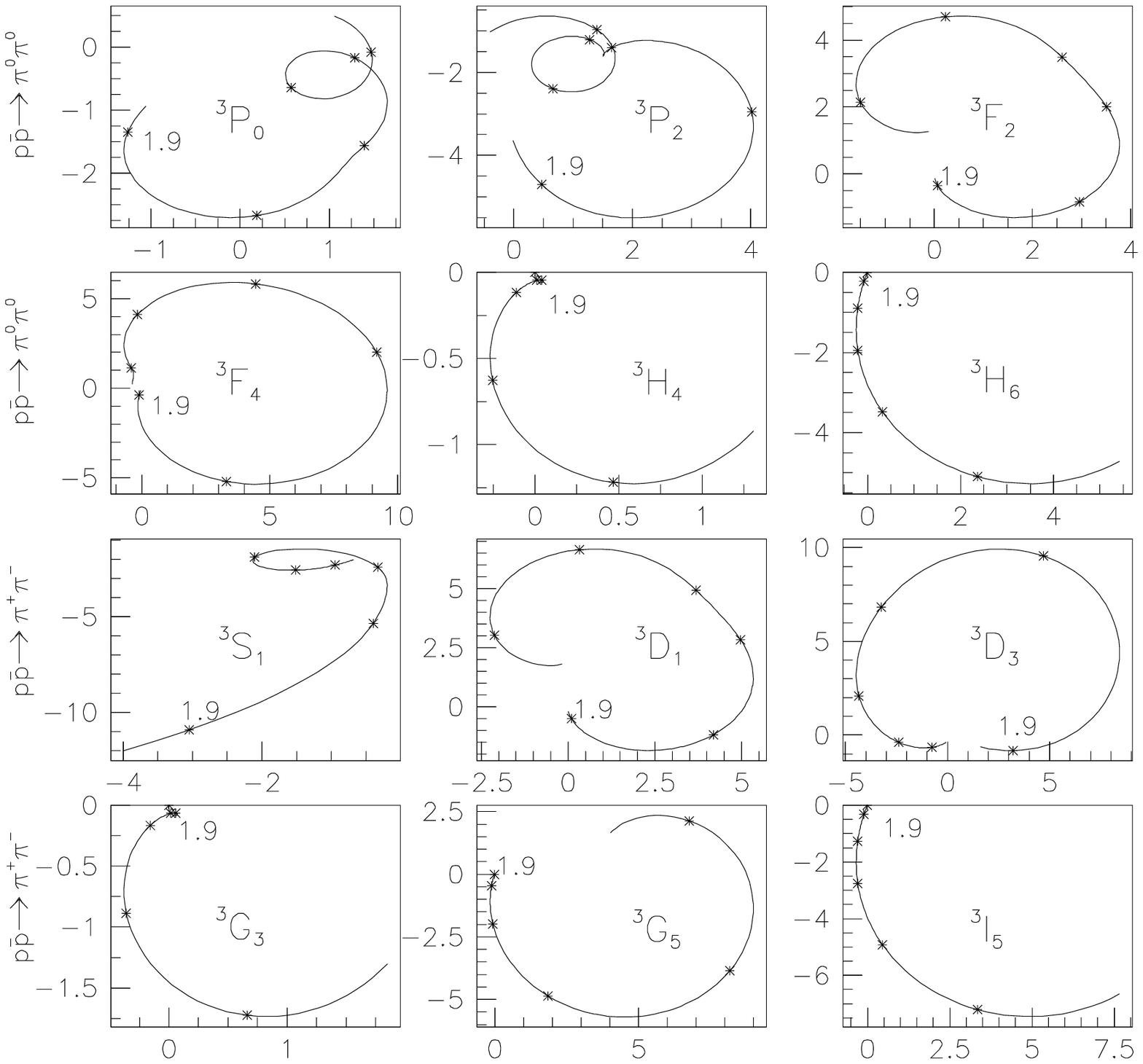,width=12cm}\
\caption {Argand diagrams for $\pi ^+ \pi ^- $amplitudes; crosses mark
masses at 100 MeV intervals of mass beginning at 1900 MeV.}
\end{center}
\end{figure}

The data require a larger and more linear phase variation than can
be produced by a single resonance.
In electronics, it is well known that a linear phase variation
with frequency may be obtained with a Bessel filter [31], which
has a sequence of poles almost linearly spaced in frequency.
Physically, such a system generates a time delay which is
independent of frequency, since group velocity is proportional
to the gradient of phase v. frequency.
It appears that a sequence of meson resonances likewise
conspires to produce approximately a linear phase variation with
mass squared $s$.

Intensities of the main individual partial waves in $\eta \pi ^0 \pi
^0$ are displayed in Fig. 24. In each panel, titles indicate in
descending order full, dashed, chain and dotted curves.
\begin{figure}[htbp]
\begin{center}
\hspace*{-0.cm}
\epsfysize=14cm
\epsffile{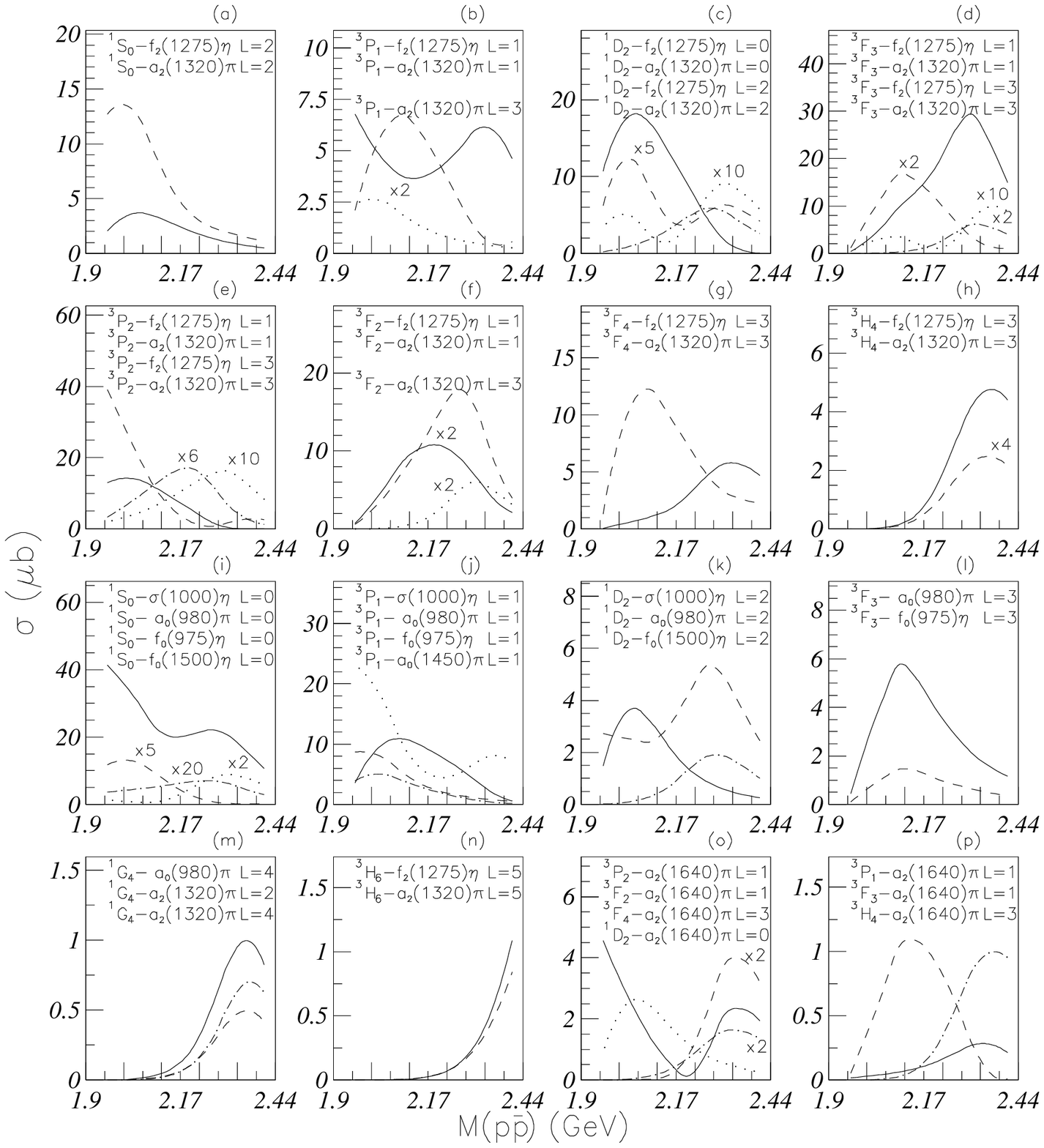}
\end{center}
\caption{Intensities fitted to individual partial waves, and corrected
for all $\eta$ and $\pi ^0$ decays.
The titles in each panel show in descending order full, dashed, chain
and dotted curves; $\sigma (1000)$ denotes the broad $f_0(400-1200)$ in
the $\pi \pi$ amplitude. Numerical values indicate factors by which
small intensities have been scaled so as to make them visible.}
\end{figure}

The $f_4(2050)$ is seen as a clear peak in $\pi \pi$, $\eta \eta$,
and $a_2(1320)\pi$, Fig. 24(g)  and is clearly
$4^*$.
It peaks at 2070 MeV because of the effect of the centrifugal barrier
for production from $\bar pp$, but the mass of the Breit-Wigner
resonance is substantially lower, at 2018 MeV.
The $f_4$ state at 2283 MeV is the $f_4(2300)$ of the PDG.
It is not conspicous as a peak in 2-body data, but has been
found by several earlier partial wave analysis of $\pi ^- \pi ^+$ data
by Hasan [1], A.D. Martin and Pennington, [32], B.R. Martin and
Morgan [33] and Carter [34]. It need not appear as a peak in $\bar pp
\to \pi\pi$ because of interference with $f_4(2050)$. It does
however appears clearly as a peak in $f_2\eta$ in  Fig. 24(g)
at $2320 \pm 30$ MeV. Recent VES data also display conspicuous $4^+$
peaks in $\omega \omega$ at $2325 \pm 15 $ MeV with $\Gamma = 235 \pm
40$ MeV [35] and in $\eta \pi \pi$ data with $M = 2330 \pm 10(stat) \pm
20(syst)$ MeV with $\Gamma = 235 \pm 20 \pm 40$ MeV [36]. Peak
positions agree well with Crystal Barrel. This is also a $4^*$
resonance.

The $4^-$ state ($\bar pp$ $^1G_4$) appears only in $\eta \pi \pi$
data.
Amplitudes are small, but rather stable in all fits.
The dominant decay is to $a_0(980)\pi$, Fig. 24(m).
This resonance is classified as $2^*$. Log likelihood
changes by 558 when it is omitted. General experience is that any
contribution improving log likelihood by more than 500 has well
determined parameters; any affecting log likelihood by $>300$ is
definitely needed.

The upper $3^+$ state decays to $f_2\pi$ with both $L = 1$ and
3; it produces the strongest intensity at high mass in $\eta \pi \pi$
data, Fig. 24(d). It also appears  as a clear peak in $\eta '\pi ^0 \pi
^0$ data. The Argand diagram of Fig. 22(h) shows two loops requiring
two $3^+$ resonances. The lower one appears  clearly as peaks in
Fig. 24(d) and (l) in $f_2(1270)\eta$ and $a_0(980)\pi$ and more weakly
in $a_2(1320)\pi$. Phases are  well determined by interference with
$f_4(2050)$, so it really should be given $4^*$ status. Its mass is
well determined with an error of only $\pm 8$ MeV.

The upper $2^-$ state at 2267 MeV appears as a clear peak in the raw
$\eta '\pi \pi$ data of Fig. 18. It is also observed clearly
in $\eta \pi \pi$ in decays to $[f_2\eta ]_{L=2}$ in Fig.
24(c)(chain curve) and to both $a_0(980)\pi$ and $f_0(1500)\eta$, Fig.
24(k). It deserves to be classified as $4^*$. The Argand diagram of
Fig. 22(e) requires a lower $2^-$ state also. It appears as a peak in
Fig. 24(c) in both $[f_2\eta ]_{L=0}$ and $a_2\pi$; it is also seen in
several decay modes in $\eta \pi ^0\pi ^0\pi ^0$ data. So its existence
is not in doubt. Because of its overlap with $\eta _2(1870)$, its mass
and width may have some systematic uncertainty, so it falls into the
$3^*$ category.

The first four $2^+$ states  of Table 2 are all extremely well
determined in 2-body data and $\eta \pi \pi$ and are all $4^*$.
The $f_2(1980)$ will be discussed later as $f_2(1950)$ of the PDG.

Coming to low spins, states are less well defined.
The lower $1^+$ state ($^3P_1$) is large in several channels
in Figs. 24(b) and (j), dominantly $f_2\eta$, $a_2\pi$ and
$a_0(980)$. Despite the fact that it lies at the bottom of the
mass range, parameters are quite stable: $M = 1971 \pm 15$ MeV,
$\Gamma = 241 \pm 45$ MeV. It just deserves $3^*$ status.
However, the upper $1^+$ state is the least well determined state in
$\eta \pi \pi$ data.
The reason is that it makes only small contributions
to the distinctive channels $a_2(1320)\pi$ and $f_2(1270)\eta$,
and there are large interferences with  other amplitudes. As the
resonance mass is changed, these interferences are able to absorb the
variation with small changes in log likelihood. Nonetheless it does
improve log likelihood by 882, a very large amount. Since it also
appears in $\eta '\pi \pi$, it qualifies as $2^*$.

The upper $0^-$ state ($\bar pp$ $^1S_0)$ at 2285 MeV appears as a clear
peak in the raw $3\eta$ data of Fig. 19. It appears again in the
$f_0(1500)\eta$ channel in $\eta \pi \pi$ with a magnitude consistent
with the known branching ratio of $f_0(1500)$ between $\pi\pi$ and
$\eta\eta$.
It also contributes to $\sigma \eta$.
It is classified as $3^*$. The lower
$0^-$ state is close to the $\bar pp$ threshold and its
amplitude goes to a scattering length at threshold.
This makes it difficult to separate from a threshold effect.
On the basis of Crystal Barrel data it is only $1^*$.
However, new BES II data on $J/\Psi \to \gamma 4\pi$ identify
this state clearly at 2040 MeV in $4\pi $ [37], raising its
status to $2^*$.

Finally we come to $0^+$ states. The upper $0^+$ state at
2337 MeV is remarkably well defined in all of $\pi ^- \pi ^+$,
$\pi ^0 \pi ^0$ and $\eta \eta$.
This qualifies it as $3^*$.
The $f_0(2100)$ appears as an $\eta \eta$ peak in $\eta \eta \pi$
data at 1940 MeV/c [38], see Fig. 14 of Section 5.1. It is also
conspicuous in $\eta \eta$ as shown in Fig.
14 and contributes to $\eta \eta '$.
If it is dropped from the
2-body analysis, $\chi ^2$ increases by 4030, an enormous amount.
Allowing for its low multiplicity $(2J + 1)$, it is one of the most
strongly produced states in $\bar pp$ annihilation. It was discovered
in $J/\Psi \to \gamma 4\pi$ [39] and also appears as a very clear peak
in E760 data on $\bar pp \to \eta \eta \pi ^0$ [40], but has no spin
determination there; these data are illustrated below in Fig. 25.
Its mass and width are very well determined by
the E760 data.
Crystal Barrel data agree within a few MeV with that  determination.
It is classified as $4^*$.

An $f_0(2020)$ with a width of 400 MeV has been reported by the
WA102 collaboration [41] in decays to both $2\pi$ and $4\pi$.
Adding this to the analysis of Crystal Barrel $\pi ^0\pi ^0$ data
gives a significant improvement in $\chi ^2$ of 370 and a visible
improvement in fits to some $d\sigma /d\Omega$ data.
However, overlap and interference with the strong $f_0(2100)$ make
a determination of its mass and width difficult.
It also obscures $f_0(2100)$ in $\pi \pi$ in Fig. 14.
From Crystal Barrel data alone it is a $1^*$ resonance, but its
observation in two channels of WA102 data raises its status to $3^*$.

In summary, all states of Table 2 are securely determined in mass
except for the upper $1^+$ state at 2310 MeV and the lower $0^-$
state at 2010 MeV; both are certainly required, but have rather
large uncertainties in mass, $\pm 60$ MeV.

\subsection {$f_0(1710)$ and $f_0(1770)$}
Data from MARK III [42] and DM2 [43] on $J/\Psi \to \gamma K\bar K$
establish the existence of $f_0(1710)$ decaying to $K\bar K$
It is also observed clearly in data on  $J/\Psi \to \omega K\bar K$
[44-46].
From these and the absence of evidence for decays to $\pi \pi$, one
can deduce a branching ratio $[f_0(1710)
\to \pi \pi ]/[f_0(1710) \to K\bar K] < 0.11$ with 95\% confidence.
The $f_0(1710)$ is an obvious candidate for an $s\bar s$ state
completing a nonet with $f_0(1370)$, $a_0(1450)$ and $K_0(1430)$.

\begin{figure}[htbp]
\begin{center}
\hspace*{-0.cm}
\epsfysize=5cm
\epsffile{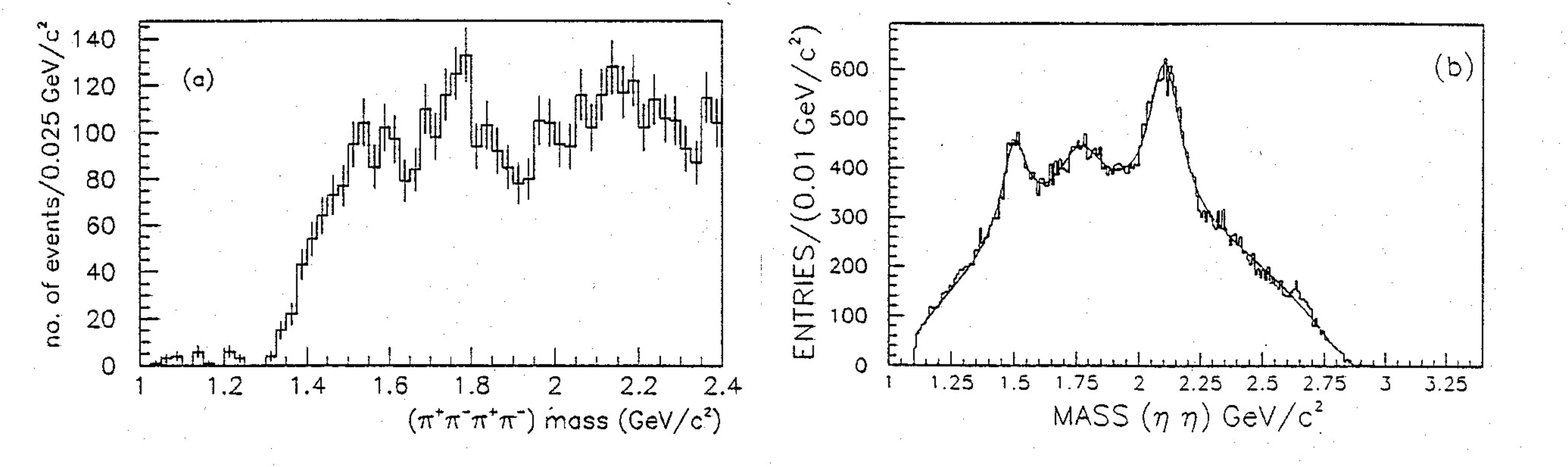}
\caption {(a) Mark III data for $J/\Psi \to \gamma 4\pi $,
(b) the $\eta \eta$ mass spectum from E760 data on $\bar pp \to
\eta \eta \pi ^0$ at 3.0 GeV.}
\end{center}
\end{figure}
A radial excitation of $f_0(1370)$ is expected in the same general
mass range.
There is ample evidence for this state, and the PDG used to
call it $X(1740)$.
Today, all entries are lumped under $f_0(1710)$.
However, these need to be examined critically.
In an analysis of Mark III data for $J/\Psi \to \gamma \pi ^+ \pi ^-
\pi ^+ \pi ^-$, Bugg et al. identified a clear peak in $4\pi$ at
1750 MeV as a $0^+$ resonance [39].
The Mark III data are shown in Fig. 25.
If it is identified as $f_0(1710)$ it would require a branching
ratio for $J/\Psi \to \gamma 4\pi$ of $(10.9 \pm 1.3)\times
10^{-4}$, larger than that to $\gamma K\bar K$, $(8.5 ^{+1.2}_{-0.9}
) \times 10^{-4}$.
That does not fit nicely with an $s\bar s$ state, and suggests the
presence of a second state.

\begin{figure}[htbp]
\begin{center}
\hspace*{-0.cm}
\epsfysize=12cm
\epsffile{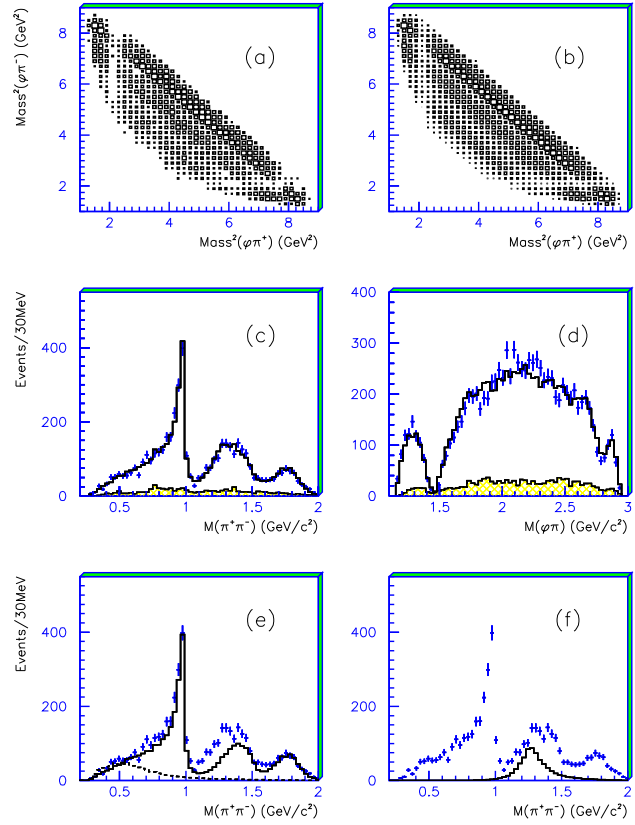}
\caption {BES II data from $J/\Psi \to \phi \pi ^+\pi ^-$.
(a) and (b) measured and fitted Dalitz plots; (c) and (d) $\pi
\pi$ and $\phi \pi$ mass projections; the upper histogram shows the
fit to data and the lower shaded histogram show background derived
from a sidebin to the $\phi$; in (d) the dip at 1.5 GeV arises from a
cut to remove $K^*\bar K^*$;
(e) and (f) fitted $0^+$ and $2^+$ contributions; the dashed
histogram in (e) shows the $\sigma $ contribution.}
\end{center}
\end{figure}

In $J/\Psi \to \phi \pi \pi$ there is a peak at 1765 MeV
but no corresponding signal in $J/\Psi \to \phi K\bar K$ [46].
New data from BES confirm and strengthen this observation [47].
These data are shown in Figs. 26 and 27. There is a clear peak in Fig.
26(c) and (e) at 1780 MeV; $J^P = 0^+$ is strongly favoured over $2^+$.
In Fig. 27 there is no corresponding peak in $\phi K\bar K$, but
a small shoulder in Fig. 27(c) at 1650 MeV; Mark III results are
similar.
This small shoulder is fitted by  $f_0(1710)$
interfering with $f_0(1500)$ and $f_2(1525)$.
The branching
ratio $\pi \pi/K\bar K$ for the 1765 MeV peak in $\phi \pi \pi$
is at least a factor 25 larger than for
$f_0(1710)$, so two separate states are definitely required.
The natural interpretation of the peak in $\phi \pi \pi$ is in
terms of a second $f_0(1770)$ decaying dominantly into $\pi \pi$
and $4\pi$ but not significantly into $K\bar K$. This fits in with
the second $0^+$ state expected in this mass range. There is
however a long-standing puzzle why the $f_0(1710)$, which seems
to be dominantly $s\bar s$, should be produced in $\omega KK$ while
the $f_0(1770)$, apparently mostly non-strange, should be
produced in the final state $\phi KK$.
\begin{figure}[htbp]
\begin{center}
\hspace*{-0.cm}
\epsfysize=12cm
\epsffile{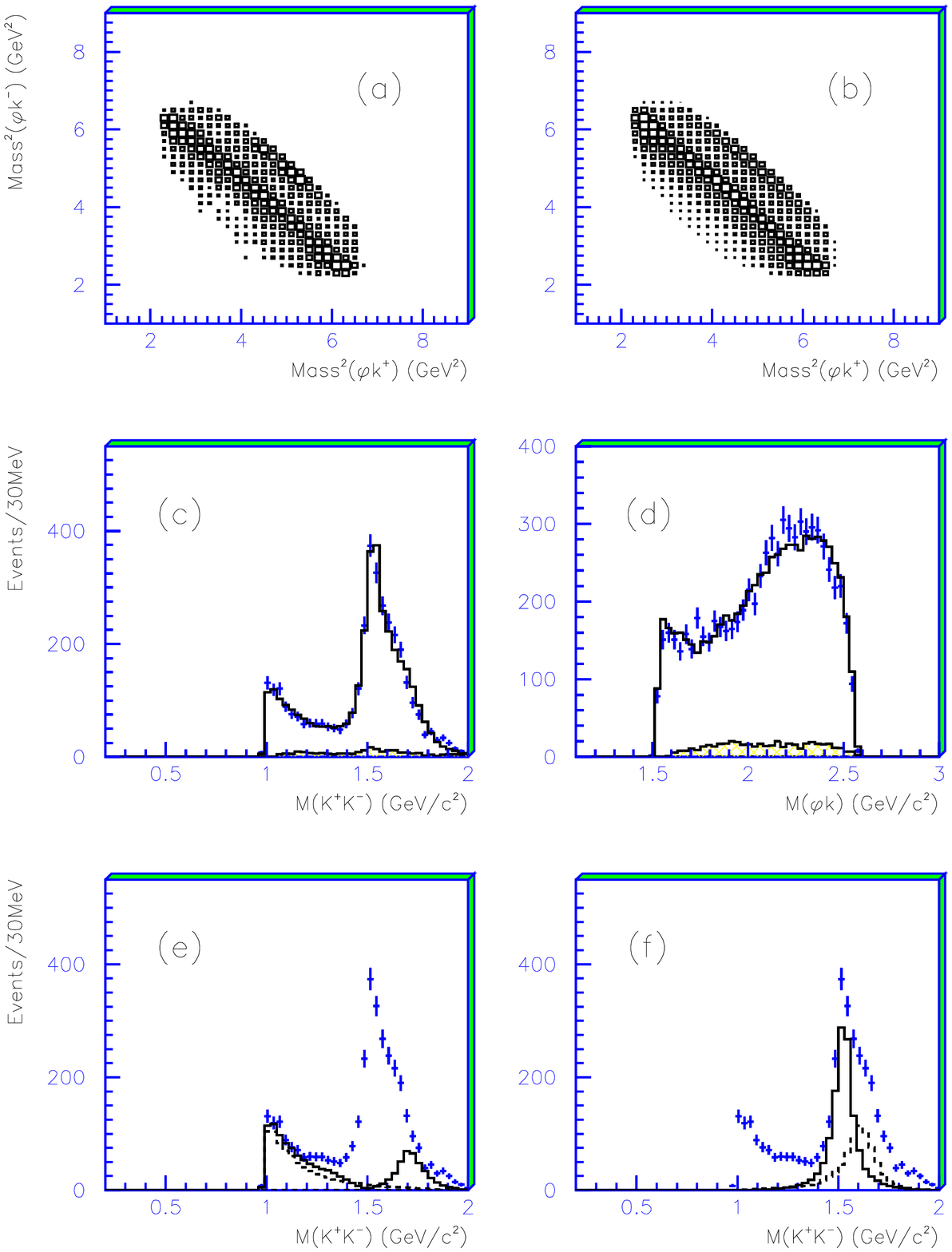}
\caption {As Fig. 28 for $J/\Psi \to \phi K^+K^-$.}
\end{center}
\end{figure}

In Crystal Barrel data on $\bar pp \to \eta \eta \pi ^0$,
there is evidence for the same $f_0(1770)$ decaying to
$\eta \eta $ [48].
Fig. 28 shows the $\eta \eta$ mass projection for data
at 900, 1050 and 1200 MeV/c.
The histogram shows the best fit without inclusion of
$f_0(1770)$.
At all three momenta there is a small surplus of events,
which may be fitted well by including a
further contribution from $f_0(1770)$ with $M = 1770 \pm 12$ MeV,
$\Gamma = 220 \pm 40$ MeV; these values are in good agreement with data
discussed above. The first three panels of Fig. 29 show log likelihood
v mass at the three momenta. Full curves are for the spin 0 assignment
and dashed curves for spin 2; spin 0 is obviously preferred. The last
panel shows log likelihood v. width.

The $f_0(1710)$ could decay to $\eta \eta$ via their $s\bar s$
components.
However, $s\bar s$ final states are conspicuous by their absence in
$\bar pp$ annihilation.
In the same $\eta \eta \pi ^0$ data, a scan of the $2^+$ amplitude v.
mass shows no peaking at $f_2'(1525)$, so $f_0(1710)$ appears an
unlikely source of the $\eta \eta$ signal.
Furthermore, mass and width fitted to $f_0(1770)$ appears significantly
different to those of $f_0(1710)$.

\begin{figure}[htbp]
\begin{center}
\hspace*{-0.cm}
\epsfysize=6cm
\epsffile{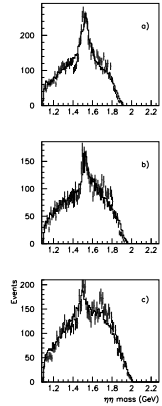}
\caption {Data for $\bar pp \to \eta \eta \pi ^0$.
Histograms show the fit without $f_0(1770)$ at (a) 900 MeV/c, (b) 1050
MeV/c and (c) 1200 MeV/c.}
\end{center}
\end{figure}
A somewhat different fit to these data has been
reported by Amsler et al. [49]. This fit omits $f_0(1770)$ and
generates the 1770 MeV shoulder as the tail of an interference between
$f_0(1370)$ and $f_0(1500)$; since they lie close to one another, they
behave like a dipole with a long tail at high masses. The
difficulty with this fit is that the $f_0(1370)$ branching ratio
to $\eta \eta$ is known to be small compared to that for $\pi \pi$;
the large $f_0(1370)$ signal required is considerably larger than is
fitted to $f_0(1370)$ in $3\pi$ data at all momenta, see Section 6
below.

\begin{figure}[htbp]
\begin{center}
\hspace*{-0.cm}
\epsfysize=8cm
\epsffile{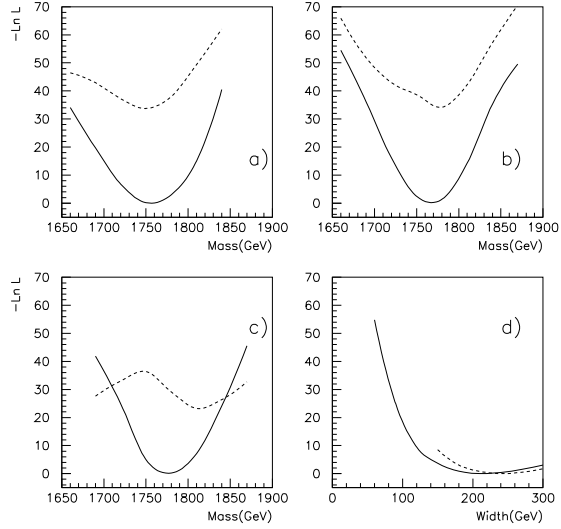}
\caption {Log likelihood v. mass of $f_0(1770)$ full lines $J=0$,
dashed lines $J=2$, (a) at 900 MeV/c, (b) 1050 MeV/c and
(c) 1200 MeV/c; (d) log likelihhod v. width at 1050 MeV/c (full
curve) and (d) 1200 MeV/c (dashed).}
\end{center}
\end{figure}
The first evidence for a signal in $\eta \eta$ at $1744 \pm 15$ MeV
was reported by the GAMS collaboration [50,51].
It was observed to have an isotropic decay angular distribution,
consistent with $J^P = 0^+$.

\subsection {SU(3) relations for 2-body channels}
Amplitudes for $\bar pp \to \pi ^0 \pi ^0$, $\eta \eta$ and
$\eta \eta '$ throw valuable light on the quark content of
resonances.
It is well known that $\eta$ and $\eta '$ may be expressed
as linear combinations of $|u\bar u + d\bar d>/\sqrt {2}$ and $s\bar s$.
Denoting the former by $n\bar n$, their quark content may be
written in terms of the pseudoscalar mixing angle $\Theta$ as
\begin {eqnarray}
|\bar nn> &=& \cos \Theta |\eta > + \sin \Theta |\eta '> \\
|\bar ss> &=& -\sin \Theta |\eta > + \cos \Theta |\eta '>,
\end {eqnarray}
where $\cos \Theta \simeq0.8$ and $\sin \Theta \simeq 0.6$.
Amplitudes for decay to $\pi ^0 \pi ^0$, $\eta \eta$ and $\eta \eta
'$
may then be written compactly as [13]
\begin {eqnarray}
f(n\bar n) &=&  \frac {|\pi ^0 \pi ^0 >}{\sqrt {2}} +
\frac {\cos ^2 \Theta }{\sqrt {2}}|\eta \eta >
+ \cos \Theta \sin \Theta |\eta \eta '>, \\
f(s\bar s) &=& \sqrt {\lambda }[\sin ^2 \Theta |\eta \eta >
+ \cos ^2 \Theta |\eta \eta > - \sqrt {2}\cos \Theta \sin \Theta
|\eta \eta '>].
\end {eqnarray}
The second of these equations includes a factor $\lambda$.
This factor has been fitted empirically by Anisovich to a large
variety of data on strange and non--strange final states at
high energies [52].
Its value is $\sqrt {\lambda } = 0.8$--0.9 and the central
value 0.85 is adopted in the Crystal Barrel analysis.

Resonances $R$ observed in data may be linear combinations of
$n\bar n$ and $s\bar s$:
\begin {equation}
R = \cos \Phi |q\bar q> + \sin \Phi |s\bar s>.
\end {equation}
Amplitudes for the decay of $R$ to these three channels are given by
\begin {eqnarray}
f(\pi ^0 \pi ^0) &=& \frac {\cos \Phi}{\sqrt {2}} \\
f(\eta \eta) &=& \frac {\cos \Phi }{\sqrt {2}}(\cos ^2 \Theta +
\sqrt {2\lambda } \sin ^2 \Theta \tan \Phi ) \\
f(\eta \eta ') &=& \frac {\cos \Phi }{\sqrt {2}}\cos  \Theta \sin
\Theta (1 - \sqrt {2\lambda } \tan \Phi ).
\end {eqnarray}

\begin{figure}[htbp]
\begin{center}
\hspace*{-0.cm}
\epsfysize=15cm
\epsffile{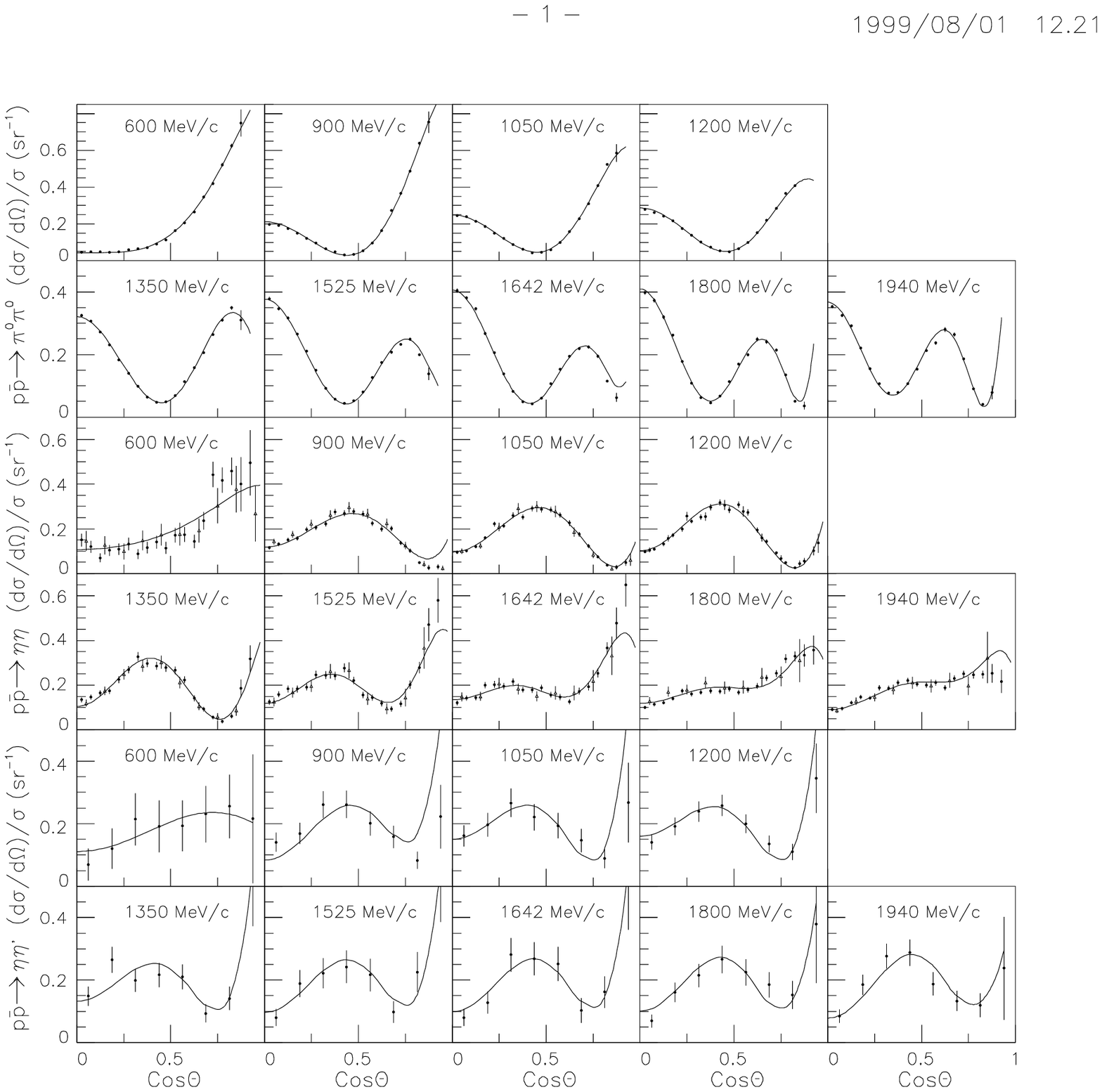}
\caption {Differential cross sections for $\bar pp \to \pi ^0 \pi ^0$
(first two rows),
$\eta \eta$ (rows 3 and 4) and $\eta \eta '$ (bottom 2 rows), compared
with the combined fit (full curves). For $\eta \eta '$, the average of
$4\gamma$ and $8\gamma$ data is used.}
\end{center}
\end{figure}
Most
flavour mixing angles $\Phi$ optimise close to zero, i.e. close to
pure $n\bar n$ states. Some mixing is essential, since $\eta \eta$
differential cross sections are visibly different to $\pi ^0 \pi ^0$.
The small mixing angles allow modest changes in branching ratios
between $\pi ^0 \pi ^0$, $\eta \eta$ and $\eta \eta '$. However, there
is a striking exception for $f_0(2100)$. This requires $\Phi = (65 \pm
6)^\circ $. The reason for this result is that the ratio of observed
amplitudes of $f_0(2100)$ is
\begin {equation} \pi ^0 \pi ^0 : \eta \eta : \eta \eta ' =
0.71 \pm 0.17 : 1 : -0.85 \pm 0.45.
\end {equation}
For an unmixed $n\bar n$ state the branching ratio $BR[\pi
^0 \pi ^0 ]/BR [\eta \eta ]$ would be $0.8^{-4} = 2.44$. The $f_0(2100)$
is one of the strongest $0^+$ resonances. It makes up $(38 \pm 5)\%$ of
the $\eta \eta$ signal. Intensities of $\eta \eta $ and $\eta \eta '$
cross sections v. mass are shown in Fig. 10(c) and (d). The
$f_0(2100)$ has the anomalous property that it is produced very
strongly in $\bar pp$ yet it decays dominantly to $s\bar s$. Below it
will be considered as a candidate for a $0^+$ glueball predicted close
to this mass. A glueball would have a mixing angle of $+35.6^\circ $.
The departure of the flavour mixing angle $\Phi$ from this value can be
accomodated by some mixing with a nearby $s\bar s$ state.
An overall picture of the fit to $\pi ^0 \pi ^0$, $\eta \eta$ and
$\eta \eta '$ data in shown in Fig. 30.

\section {$I = 1$, $C = +1$ resonances}
The channels available are $\pi ^0 \eta$, $\pi ^0 \eta '$, $3\pi ^0$
and $\eta \eta \pi ^0$.
Statistics are very high for $3\pi ^0$, typically $>250~K$ events
per momentum; backgrounds are very low [53].
Since $u$ and $d$ quarks are nearly massless, one expects the
spectrum of $I = 1$, $C = +1$ mesons to be close to that for
$I = 0$, $C = +1$.
Although results are indeed similar, they are very much less
accurate.
Some resonances are well defined, but several are not.
The well defined resonances agree reasonably well with $I = 0$,
$C = +1$. The remainder fall into line with $I = 0$ if one
chooses one of two alternative solutions to the data.

The fundamental difficulty is that there are two discrete
solutions for $\pi ^0 \eta$ and $\pi ^0 \eta '$ [12].
To get from one solution to the other, two $2^+$ amplitudes must
be flipped in sign;  a substantial change also  appears in
$^3H_4$.
The reason for the ambiguity is the lack of polarisation data,
which are needed to separate $^3P_2$ and $^3F_2$ cleanly; the
phase sensitivity of the polarisation data is also
missing.

The $3\pi ^0$ channel is dominated by $f_2(1270)\pi$.
Two $4^+$ resonances appear reasonably clearly as does a particularly
strong $3^+$ resonance at 2031 MeV. Further analysis requires
resonances in all partial waves up to $4^-$, in close analogy to $I =
0$, $C = +1$. However, determinations of resonance parameters are
rather fuzzy.
An initial analysis was made of the $f_2(1270)\pi$
channel alone [53].

This analysis was made at single energies,
and resonance parameters were deduced by fitting curves to
magnitudes and phases.
This technique is hazardous, since magnitudes and phases
contain non-analytic noise at individual momenta.
Where relative phases between partial waves are close to 0 or
$180^\circ$, phases can depart seriously from physical
values and can distort conclusions.
Removing this noise is important.
The later analysis of Ref. [4] was made
to all momenta simultaneously, fitting resonances directly to
data. There were substantial changes to the solution,
particularly at 600 and 900 MeV/c.

There are several reasons for this instability:
\begin {itemize}
\item {(i) As well as $f_2\pi$ in $3\pi ^0$ data, there is an almost
    flat, featureless
    physics background. It may be fitted with the $\sigma$ pole at
    550 MeV and a broad, slowly varying $\pi \pi$ amplitude over the
    mass range 1--2 GeV, well known in $\pi \pi$ elastic scattering
    [16,54]. The entire $\pi \pi$ S-wave amplitude will be denoted by
    $\sigma$ as a shorthand. The $\sigma \pi$ amplitudes appear
    mostly in $\bar pp$ singlet states. They interfere with $f_2\pi$
    singlet amplitudes and the interference leads to a poor definition
    of resonance parameters; substantial shifts of masses and widths
    may be accomodated by changes to phase angles.} \\
\item {(ii) Lack of polarisation data leads to uncertainty in $r_2$
    and $r_4$ parameters, i.e. separation between $^3P_2$ and $^3F_2$
    and between $^3F_4$ and $^3H_4$.} \\
\item {(iii) There is a rapid change in angular distributions between
    the two lowest beam momenta: 600 MeV/c ($M = 1962$ MeV)
    and 900 MeV/c ($M =  2049$ MeV); this is shown on Fig. 2 of Ref.
    [4]. The experiment was supposed to run also at 450 and 750 MeV/c,
    but these momenta were lost because of
    the closure of LEAR. For $I = 0$, the data from PS172 extend the
    range down to 360 MeV/c ($M = 1912$ MeV) in steps of 100 MeV/c.} \\
\end {itemize}
The ultimate conclusion is that $3\pi ^0$ data do not define the
partial waves cleanly, despite the high statistics.

Amongst the 3-body data are typically 5000--9000 $\eta \eta \pi ^0$
events per momentum.
It turns out that these data are actually more valuable than
$3\pi ^0$.
In $\eta \eta \pi ^0$, there are strong contributions from
spinless channels $a_0(980)\eta$, $f_0(1500)\pi$,
$f_0(1770)\pi$ and $f_0(2100)\pi$.
For these channels, amplitudes depend only on Legendre polynomials for
the production angle. The resulting analysis is very simple.
Singlet states dominate.
As a result, $0^-$, $1^+$, $2^-$, $3^+$ and $4^-$ resonances emerge
well defined. There is also a further distinctive $^3F_2$ state at 2255
MeV.
This information feeds back to stabilise the fit to $3\pi ^0$.

The final element in the analysis is the observation of a $\pi
_2(1880)$ state, $J^P = 2^-$, in the production process
$\bar pp \to \pi _2(1880)\pi$, $\pi _2 \to \eta \eta \pi$
in $\eta \eta \pi ^0 \pi ^0$ data.
This state overlaps partially with a further $2^-$ state at
2005 MeV. They can actually be separated by the fact that
$\pi_2 (2005)$ decays to $a_0(980)\eta$ whereas the $\pi _2(1880)$
does not. However, for other decay channels there is potential
ambiguity between these two close resonances.

\newpage
\begin{figure}
\begin {center}
\epsfig{file=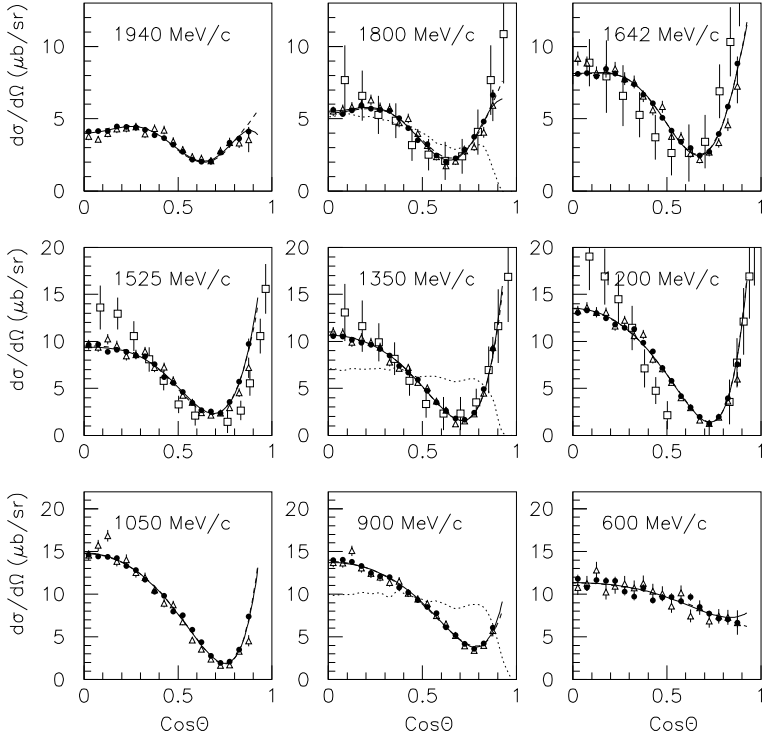,width=12.5cm}~\
\caption {Angular distributions for $\pi ^0 \eta$ corrected for
acceptance. Black circles show results from $4\gamma$ and open
triangles results from $8\gamma$. Full and dashed curves show fits from
solutions 1 and 2 respectively. Dotted curves illustrate the
acceptance. Open squares show earlier results of Dulude et al at
neighbouring momenta, after renormalisation to obtain the optimum
agreement with Crystal Barrel normalisation.}
\end {center}
\end {figure}
\subsection {2-body data}
Angular distributions for $\pi \eta$ are shown in Fig. 31 [12].
There is again excellent agreement between $4\gamma$ data (black
circles) and $8\gamma$ (open triangles) in both shapes and
absolute normalisations.
Open squares compare with earlier data of Dulude et al. [55]
after renormalising them to agree with the normalisation of
Crystal Barrel data.
Angular distributions for $\pi ^0 \eta '$ are shown in Fig.
32. There are typically 1000 events per momentum.
Here one sees differences between two alternative solutions, shown
by full and dashed curves.
Intensities of partial waves from these two solutions are shown in Fig.
33. The main difference lies in $2^+$.
\begin{figure}
\begin {center}
\epsfig{file=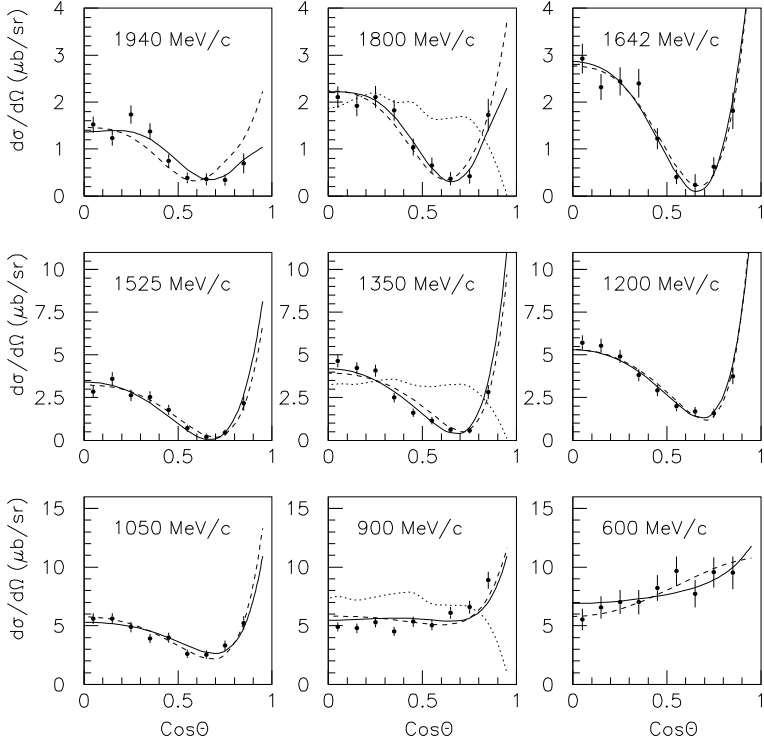,width=12.5cm}~\
\caption {Angular distributions for $\pi ^0 \eta '$ corrected for
acceptance. Full and dashed curves show solutions 1 and 2
respectively. Dotted curves illustrate the acceptance.}
\end {center}
\end{figure}
\begin{figure}
\begin {center}
\epsfig{file=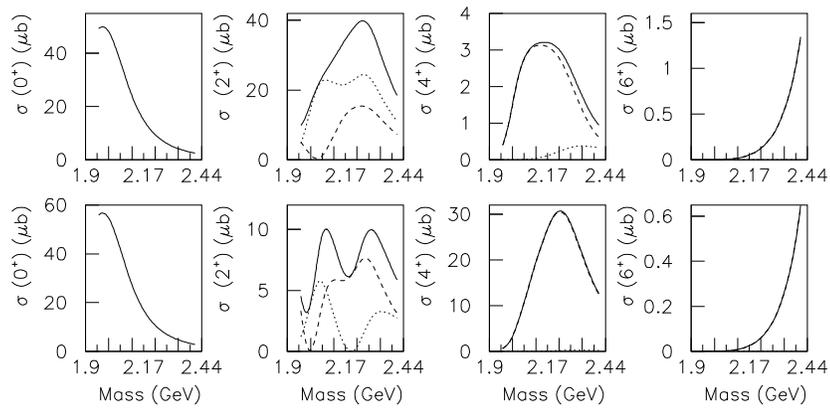,width=12cm}~\\
\caption {Intensities of $0^+$, $2^+$, $4^+$ and $6^+$ partial waves
for $\bar pp \to \pi ^0 \eta$ for the better solution (upper row)
and poorer solution (lower row).
Dashed curves show intensities for $^3P_2$ and $^3F_4$, dotted curves
for $^3F_2$ and $^3H_4$, full curves their sum.}
\end {center}
\end{figure}

Both solutions require the $^3F_2$ resonances at 2255 and
2030 MeV and some $^3P_2$ contribution.
In the earliest analysis of Ref. [12] there was no input from
$3\pi ^0$ data and the upper $^3P_2$ resonance at 2175 MeV
was not included.
Even in the final combined analysis, this is the least significant
contribution.

Both solutions require a $0^+$ state near 2025 MeV, though its
mass and width have sizable errors.
A lucky break is that independent evidence for this state
appears in $\eta \pi ^0 \pi ^0 \pi ^0$ data.
A small but definite $\eta (1440)$ signal is observed there in $\eta \pi
\pi$ [56]. It is illustrated in Fig. 34. This channel may be fitted to
production amplitudes using simple Legendre polynomial for $L = 0$, 2
and 4. The result is complete dominance by $L = 0$, with no significant
evidence for any $L = 2$ or 4 amplitudes.
This requires production from an initial $I = 1$, $J^{PC} = 0^{++}$
initial state. The cross section for production shows a distinct
peak at 2025 MeV, Fig. 35. It may be fitted with a resonance
at 2025 MeV, $\Gamma = 330$ MeV after inclusion of the
$\bar pp$ $L = 1$ centrifugal barrier for production.

However, even after this lucky break, the 2-body data fail to locate
the further state expected in the mass range 2300-2400 MeV as
partner to $f_0(2337)$. Note that the $a_0$ expected between
1450 and 2025 MeV is also missing. As a result, the $0^+$ sector
for $I = 1$ is poorly known.
\begin{figure}
\begin{center}\hspace*{-0.cm}
\epsfysize=11cm
\epsffile{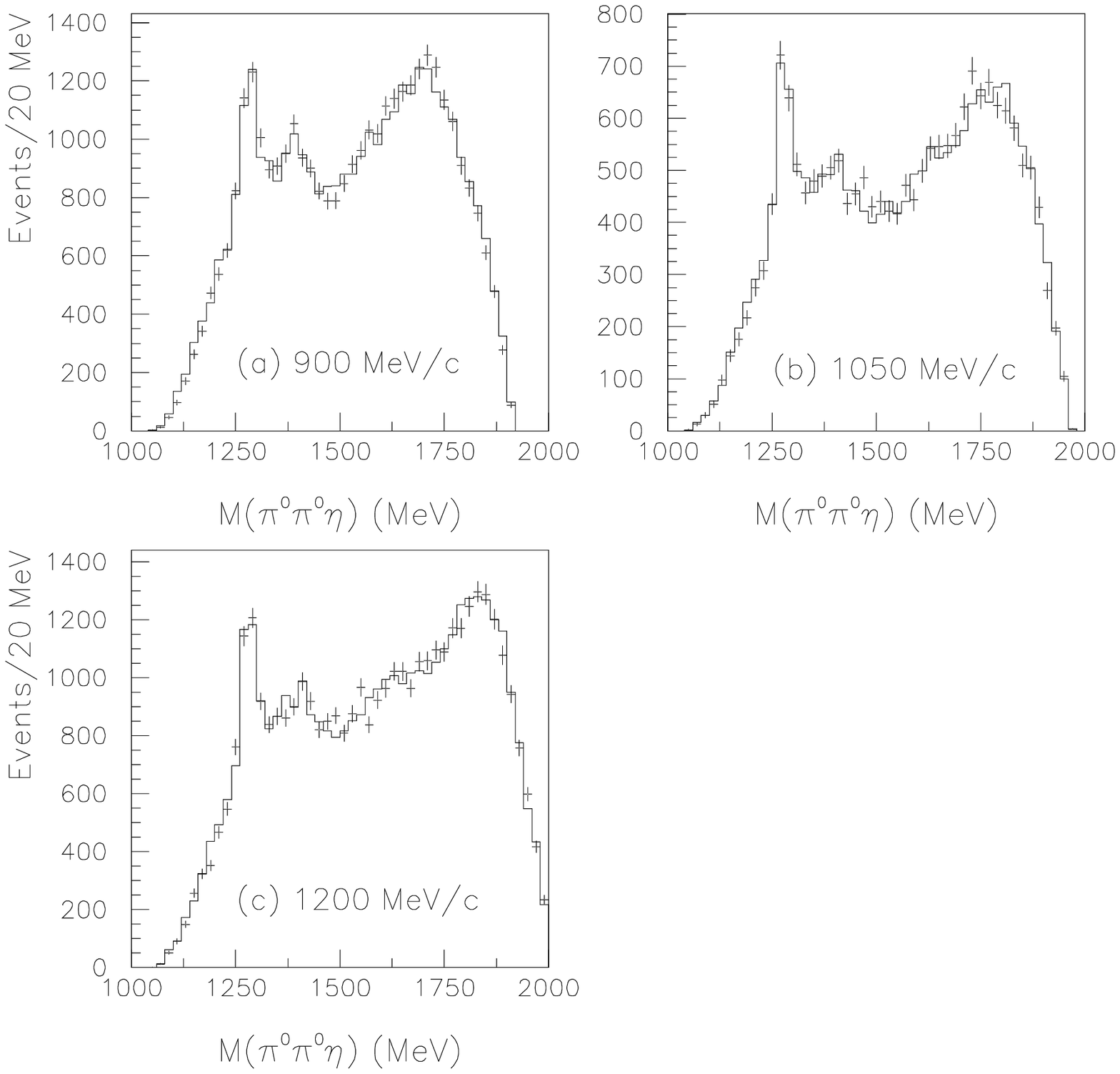}
\end{center}
\caption{Mass projections for all $\eta \pi ^0 \pi ^0$ combinations
in $\eta 3\pi ^0$ data after selecting events where $M(\eta \pi )$
lies within $\pm 50$ MeV of $a_0(980)$ or alternatively $M^2(\pi \pi
)>0.6$ GeV$^2$. Histograms show fits from the amplitude analysis. }
\end{figure}
\begin{figure}
\begin{center}\hspace*{-0.cm}
\epsfysize=6cm
\epsffile{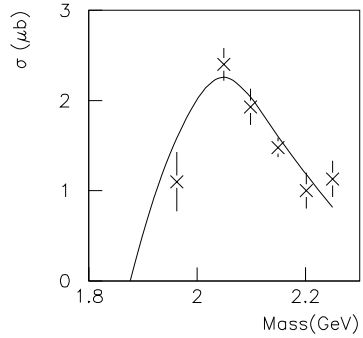}
\end{center}
\caption{Cross sections for $\bar pp \to \eta (1440)\pi$ with $L = 0$.
The curve shows the cross section for a resonance with $M = 2025$ MeV,
$\Gamma = 330$ MeV, including the $\bar pp$ $\ell = 1$ centrifugal
barrier at threshold.} \end{figure}

\newpage
\begin{figure}
\begin{center}\hspace*{-0.cm}
\epsfysize=10cm
\epsffile{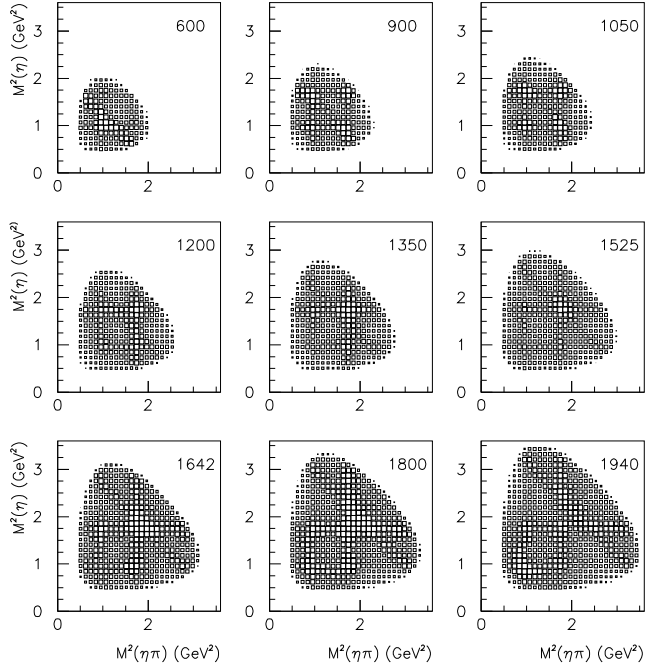}
\end{center}
\caption{Dalitz plots for $\bar pp \to \eta \eta \pi$; numbers indicate
beam momenta in MeV/c.}
\end{figure}
\subsection {$\bar pp \to \eta \eta \pi ^0$}
These data are discussed in detail in Ref. [5].
Dalitz plots are shown in Fig. 36 at all momenta.
There are obvious vertical and horizontal bands at $M^2(\eta \pi ) = 1$
and 1.7 GeV$^2$ due to $a_0(980)$ and $a_2(1320)$.
The is also a diagonal band at the highest four momenta due to
$f_0(1500)$.

The $\pi \pi$ mass projection is shown in Fig. 37.
The strong $f_0(1500)$ is clearly visible. There is
also a shoulder due to $f_0(1770)$
at beam momenta 1200-1642 MeV/c.
There is one  other important feature of Fig. 36.
At the intersection of vertical and
horizontal bands at $M^2(\eta \pi ) = 1$ GeV$^2$, there is a distinct
enhancement at the highest three momenta in the bottom row of the
figure. It turns out that this enhancement cannot be fitted adequately
by constructive interference alone between two $a_0(980)$
bands; such an enhancement is not there at lower momenta.
The enhancement in fact arises from a rapidly growing
signal due to $f_0(2100)$, whose kinematic threshold is at
1525 MeV/c. This feature is important in two respects.
Firstly it shows the presence of $f_0(2100)$ in $\eta \eta$.
This was reported in Ref. [38].
Secondly the $f_0(2100)$ is so close to threshold that it
appears only in the $f_0(2100)\pi$ S-wave, which has $J^P = 0^-$
in $\bar pp$.
This observation helps identify the highest $0^-$ resonance.
\begin{figure}[htbp]
\begin{center}\hspace*{-0.cm}
\epsfysize=9cm
\epsffile{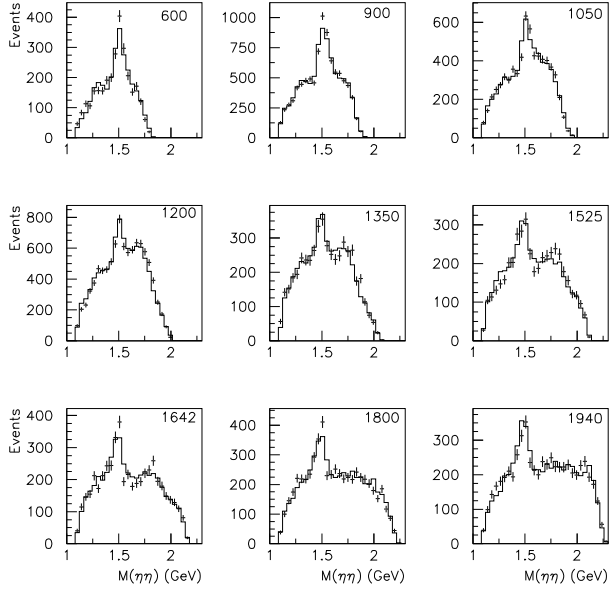}
\end{center}
\caption{Projections of $\eta \eta \pi ^0$ data on to $\eta \eta$
mass; histograms show the fit and beam momenta are indicated in MeV/c.}
\end{figure}

Table 4 shows the conspicuous resonant contributions in $\eta \eta \pi
^0$.
The final column shows changes in log likelihood when each
resonance is omitted from the fit.  All are
statistically at least 14 standard deviation signals, though
uncertainties in smaller amplitudes introduce systematic errors which
reduce the significance level to 8--12 standard deviations.
\begin{table}[htp]
\begin{center}
\begin{tabular}{cccccc}
\hline
Name & $J^P$ & $M$   & $\Gamma$ & Decay & $\Delta S $ \\
     &       &(MeV)  & (MeV)    & Channel  &  \\\hline
$\pi $ & $0^-$  & $2070 \pm 35$  &  $310 ^{+100}_{-50}$ &
$a_0(980)\eta$ & 272\\
    & & & & $f_0(1500)\pi$ & 324 \\
$\pi $ & $0^-$  & $2360 \pm 25$  &  $300 ^{+100}_{-50}$ &
$a_0(980)\eta$ & 197 \\
    & & & & $f_0(2100)\pi$ & 434 \\
$\pi _2$ & $2^-$  & $1990 \pm 30$  &  $200 \pm 40 $ &
$a_0(980)\eta$ & 531 \\
$\pi _4$ & $4^-$  & $2255 \pm 30$  &  $185 \pm 60 $ &
$a_0(980)\eta$ & 213 \\
$a _2$ & $2^+$  & $2255 \pm 20$  &  $230 \pm 15 $ &
$f_2(1980)\pi$ & 363 \\
$a _3$ & $3^+$  & (2031)  &  (150)  &
$a_2(1320)\eta$ & 185 \\
    & & & & $f_0(1500)\pi$ & 164 \\
$a _3$ & $3^+$  & $2260 \pm 50$  & $250 ^{+100}_{-50}$ &
$a_0(980)\eta$ & 141 \\\hline
\end {tabular}
\caption {Resonances parameters fitted to $\eta \eta \pi ^0$ data;
values in parentheses are fixed from $3\pi ^0$ data.
The final column shows changes in log likelihood when each channel
is dropped from the fit and all others are re-optimised.}
\end{center}
\end{table}

Singlet amplitudes are determined simply via
Legendre polynomials for production amplitudes.
Angular distributions shown in Ref. [5] verify directly the presence
of $0^-$, $2^-$ and $4^-$ states.
There is evidence also in $3\pi ^0$ data for both $0^-$ resonances,
though they are determined best by the $\eta \eta \pi$ data.
Both should be regarded as $2^*$.

The $2^-$ state at $\sim 1990$ MeV is conspicious also in $3\pi ^0$
data, Fig. 38(b) below. The combined fit to both channels gives
an optimum mass of $2005 \pm 15$ MeV and a width of $200 \pm 40$ MeV,
as shown below in Table 5.
The observation of this $2^-$ state in $a_0(980)\eta$ is  rather
important.
This resonance overlaps significantly with $\pi _2(1880)$ discussed
below. However, $\pi _2(1880)$ has a very weak decay to $a_0(980)\eta$
if any at all. The second order Legendre polynomial associated with the
$2^-$ state at 1990 in $\eta \eta \pi ^0$ data is very distinctive at
600, 900 and 1050 MeV/c and leaves no doubt of its presence. This state
should be regarded as $3^*$. The $4^-$ signal is entirely consistent
with a rather strong and distinctive $4^-$ signal in $f_2\pi$ in $3\pi
^0$ data and should be regarded as $3^*$.

The $3^+$ state at 2031 MeV is a confirmation of a very
strong $f_2\pi$ signal in $3\pi ^0$ data, making this state
$3^*$.
The upper $3^+$ state is not well identified in $3\pi ^0$ data,
so this should be regarded as $2^*$.
The $2^+$ state at 2255 MeV confirms a $^3F_2$ state in $3\pi ^0$
and $\eta \pi ^0$ and should be regarded as $3^*$.
Its distinctive decay into $f_2(1980)$ will be discussed in detail
in Section 10.4.

\subsection {$\bar pp \to 3\pi ^0$}
\begin{table}[htp]
\begin{center}
\begin{tabular}{ccccccccc}
\hline
 & $J^P$ & *& $M$   & $\Gamma$ & $r$ & $\Delta S $ & $M(I=0)$
& Observed \\
 & & & (MeV) & (MeV)    &  & (MeV) & & channels \\\hline
$\pi _4$ & $4^-$ & $3^*$ & $2250 \pm 15$  &  $215 \pm 25$ &  & 11108 &
$2328 \pm 38$ & $a_0\eta$,$f_2\eta$,($\sigma\eta$) \\
$a_4$    & $4^+$ & $3^*$ & $2255 \pm 40$  &  $330 ^{+110}_{-50}$ & $0.87
\pm 0.27$ & 2455 & $2283 \pm 17$ & $\pi\eta$,$\pi\eta '$,$f_2\pi$\\
$a_4$    & $4^+$ & $2^*$ & $2005 ^{+25}_{-45}$  &  $180 \pm 30$ &
$0.0 \pm 0.2$ & 2447 & $2018 \pm 6$ & $\pi\eta$,$f_2\pi$ \\
$a_3$    & $3^+$ & $2^*$ & $2275 \pm 35$  &  $350 ^{+100}_{-50}$ & & 3154
& $2303 \pm 15$ & $a_0\eta$,$f_2\eta$\\
$a_3$    & $3^+$ & $3^*$ & $2031 \pm 12$  &  $150 \pm 18$ & & 18410 &
$2048 \pm 8$ & $f_0(1500)\pi$,$a_2\eta$,$f_2\pi$ \\
$a_2$    & $2^+$ & $3^*$ & $2255 \pm 20$  &  $230 \pm 15$ & $-2.13 \pm
0.20$ & 2289 & $2293 \pm 13$ & $\pi\eta$,$f_2\eta $,
$f_2(1980)\pi$ \\
$a_2$    & $2^+$ & $2^*$ & $2175 \pm 40$  &  $310 ^{+90}_{-45}$ & $-0.05
\pm 0.31$ & 1059 & $2240 \pm 15$ & $f_2\pi$ \\
$a_2$    & $2^+$ & $2^*$ & $2030 \pm 20$  &  $205 \pm 30$ & $2.65 \pm
0.56$ & 1308 & $2001 \pm 10$ & $\pi\eta$,$\pi\eta '$,$f_2\pi$ \\
$\pi _2$ & $2^-$ & $2^*$ & $2245 \pm 60$  &  $320 ^{+100}_{-40}$ & &
2298 & $2267 \pm 14$ & $f_2\pi$,($\sigma\eta$) \\
$\pi _2$ & $2^-$ & $3^*$ & $2005 \pm 15$  &  $200 \pm 40$  & & 1633 &
$2030 \pm 15$ & $a_0\eta$,$f_2\pi$\\
$a_1$    & $1^+$ & $2^*$ & $2270 ^{+55}_{-40}$  &  $305 ^{+70}_{-35}$ & &
2571 & $2310 \pm 60$ & $f_2\pi$ \\
$\pi$    & $0^-$ & $2^*$ & $2360 \pm 25$  &  $300 ^{+100}_{-50}$ &
&1955 & $2285 \pm 20$ & $a_0\eta$,$f_0(2100)\pi$,$f_2\pi$ \\
$\pi$    & $0^-$ & $2^*$ & $2070 \pm 35$  &  $310 ^{+100}_{-50}$ &  & 1656
& $2010 ^{+35}_{-60}$ & $a_0\eta$,$f_0(1770)\pi$,$f_2\pi$ \\
$a_0$    & $0^+$ & $2^*$ & $1995 \pm 30$  &  $300 \pm 35$ & & 2134 &
$2040 \pm 38$ & $\pi\eta$,$\pi\eta '$,$\eta(1440)\pi$ \\\hline
$a_2$    & $2^+$ &  & $(1950) $  &  $(180)$ & $-0.05 \pm 0.30$ & 2638 &
$1934 \pm 20$ \\
$\pi _2$ & $2^-$ & $4^*$  & $(1880)$  &  (255) & & 7315 & $1860 \pm
15$\\
$a_1$    & $1^+$ & & $(1930)$  &  (155) & & 2609 & $1971 \pm 15$ \\
$a_1$    & $1^+$ &  & $(1640)$  &  (300) & & 232 \\
$\pi $   & $0^-$ & & $(1801)$  &  (210) & & 2402 \\\hline

\end {tabular}
\caption {Masses and widths of fitted $I = 1$, $C = +1$ resonances.
Values in parentheses are fixed from other data. Entries in the lower
half of the table are below the available mass range.
Column 7 shows changes $\Delta S$ in log likelihood for $3\pi ^0$ when
each resonance is removed from the fit and all remaining parameters are
re-optimised. Column 8 shows masses for $I = 0$, $C = +1$
resonances from Table 2 for comparison.}
\end{center}
\end{table}
Table 5 shows resonance parameters from the better of the two
solutions mentioned above. This solution has log
likelihood better by 1182. This is comparable to the effect of
removing one of the weaker resonances.
The essential change in going to
the poorer solution, shown in Table 6, is that
the masses of the upper $4^+$ state and both of the upper two $2^+$
states slide downwards by $\sim 40$ MeV. There  is also a small
increase in the mass of the lower $4^+$  state from 2005 to 2030
MeV. What is happening is that the two $4^+$ states, which have
similar values of $r_4$, are tending to merge.
Elsewhere it is observed that such merging of resonances of the
same $J^P$ almost always gives small improvements in log likelihood
through interference effects. Generally this should be regarded
with suspicion.
The masses of the highest $4^+$ and
$2^+$ states are 60--70 MeV lower than for $I=0$.
If one takes the $I = 0$ solution as a guide,
this solution looks physically less likely, as well as having
the poorer log likelihood.

Despite differences of detail between the two solutions, the
general pattern of masses and widths for $I = 0$ and $I = 1$
is similar. It is noticeable that most $I = 1$ masses tend to
lie 20 MeV lower than for $I = 0$.
Since the poorer solution tends to drag the masses of the upper
$2^+$ and $4^+$ states down, there is the possibility that
this ambiguity is everywhere having the effect of lowering masses,
through correlations between partial waves.
\begin{table}[htp]
\begin{center}
\begin{tabular}{cccccc}
\hline
Name & $J^P$ & $M$   & $\Gamma$ & $\Delta S$ & r \\
    &  & (MeV) & (MeV)    &  \\\hline
$a_4$ & $4^+$  & $2220 \pm 20$  &  $345 \pm 65$ & 2852 & $0.30 \pm 0.31$
\\
$a_4$ & $4^+$  & $2035 \pm 20$  &  $135 \pm 45$ & 2063 & $0.0 \pm
0.2$\\
$a_2$ &  $2^+$  & $2235 \pm 35$  &  $200 \pm 25$ & 3658 & $-1.74
\pm 0.36$\\
$a_2$ & $2^+$  & $2135 \pm 45$  &  $305 ^{+90}_{-45}$ &
1685 & $-0.56 \pm 0.53$  \\\hline
\end {tabular}
\caption {Masses and
width of resonances in the alternative solution. The last column shows
changes $\Delta S$ in log likelihood when each resonance is removed
from the fit and all remaining parameters are re-optimised.}
\end{center}
\end{table}

One of the problems in the analysis is that angular
distributions of 3-body data change rapidly
between the two lowest beam momenta, 600 and 900 MeV/c.
In order to accomodate this feature, both the lower
$3^+$ and $4^+$ states must be rather narrow.
Popov [57] reports an $I = 1$, $J^P = 3^+$ peak in E852 data for
$[b_1(1235)\pi ]$ at $\sim 1950 $ MeV, but fits to a much
lower mass of $M = 1873 \pm 1 \pm 18$ MeV with $\Gamma =
259 \pm 18 \pm 21$ MeV.
He also reports an $a_4$ at $1952 \pm 6 \pm 14$ MeV in
$[\omega \rho ]_{L=2}$ with $\Gamma = 231 \pm 11 \pm 21$ MeV.

Fig. 38 shows as full curves the intensities fitted to individual
$f_2\pi$ partial waves in the better solution.
Intensities for single energy solutions are shown by black squares.
This gives some idea of the stability of the fit.
\begin{figure}
\begin {center}
\epsfig{file=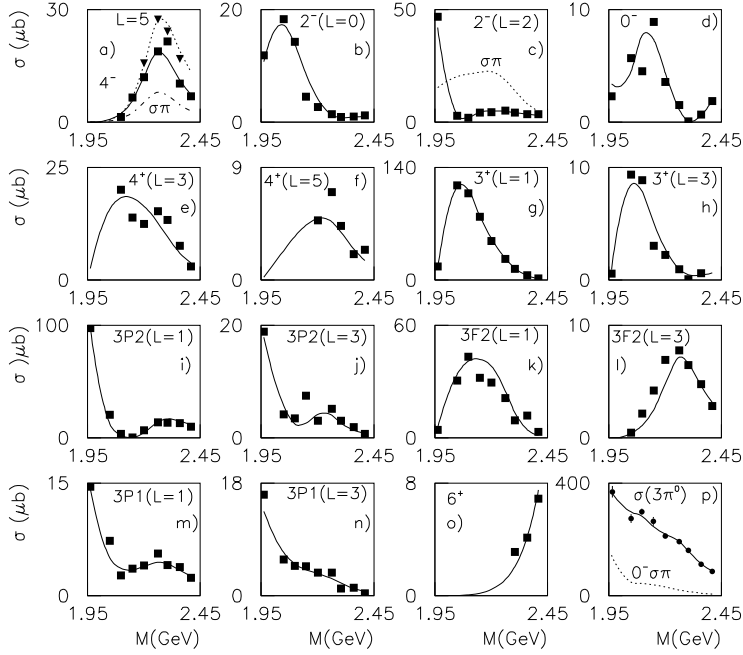,width=12cm}~\\
\caption {Intensities of individual $f_2\pi$ partial waves for the
preferred solution. Squares and triangles show free fits at single
energies. Full curves are for $f_2\pi$ final states.
Dotted and chain curves in (a) show respectively the $4^-$ $f_2\pi$
amplitude with $L = 5$ (triangles) and the $\sigma \pi$ contribution.
Dotted curves in (c) and (p) show $\sigma \pi$ contributions.
In (p), points and the full curve show integrated cross sections
for $3\pi ^0$. }
\end {center}
\end{figure}

Because of the ambiguity between two solutions, none of these
states can be regarded as $4^*$.
Both $4^+$ states are certainly required by
$\pi \eta$ and $f_2\pi$ data; the upper one is required also by
$\pi \eta '$ data.
The  lower $3^+$ state is exceptionally strong, Fig. 38(g)
and must be regarded as well established.
The $2^+$ states are poorly separated between $^3P_2$ and $^3F_2$,
but a solution is not possible without three of them in the mass
range 2000 MeV upwards.
Except for the $f_2(2255)$, which appears distinctively in $\eta \eta
\pi$, they should be regarded as $2^*$ because of the ambiguity in
the solution.

The upper $2^-$ state is required but poorly determined
and can only be classed as $2^*$. A lower $2^-$ state
is definitely required by the peak in $f_2\pi$ in
Fig. 38 (b).
Recently the E852 collaboration has presented evidence
for a similar resonance in $f_1(1285)\pi$
with $M = 2003 \pm 88 \pm 148$ MeV, $\Gamma = 306 \pm 132 \pm 121$
MeV [58].

For $J^P = 1^+$, both $f_2\pi$ data and
$\eta \eta \pi$ require a $1^+$ state around 2270 MeV,
but it can only be regarded as $2^*$.
There is a strong $1^+$ intensity at 600 MeV/c.
The $0^-$ states are better identified by $\eta \eta \pi$
data.

As remarked above, there is a large change at the lowest
momenta between the earliest analysis of $f_2\pi$ [53] and the
final analysis of Ref. [4]. The problem is an ambiguity
between $^3P_1$, $^3P_2$ and $^3F_4$ amplitudes and the rapid
change of angular distributions with mass.
The ambiguity between $^3P_1$ and $^3P_2$ is the principal problem,
so both can only be regarded as tentative identifications at
present: something to watch for in other experiments.
The very strong $^3F_3$ state is well established independently
of the others.
Good information from any other source resolving the
ambiguity would greatly stabilise the complete solution for
$I = 1$, $C = +1$.
One should be alert for a $2^+$ state decaying
dominantly to $\rho \omega$, as partner to $f_2(1910-1934) \to
\omega \omega$.
The strong rise in Fig. 38(i) at the lowest momentum suggests a
contribution from a low mass  $^3P_2$ state around 1950 MeV.
However, this is right at the bottom of the available mass range
and cannot be regarded as established at present.

\subsection {Evidence for $\pi _2(1880)$}
Evidence was presented for $\eta _2(1870)$ in Section 4.3.
An $I = 1$ partner is to be expected and is indeed observed
in $\bar pp \to \pi _2\pi$, $\pi _2 \to a_2(1320)\eta$,
i.e. in the $\eta \eta \pi ^0 \pi ^0$ final state [59].
The entire pattern of production cross sections is remarkably
similar to that for $\eta \pi ^0 \pi ^0 \pi ^0$.

The evidence for $\pi _2(1880)$ is found at beam momenta
900, 1050 and 1200 MeV/c (as for $\eta _2(1860))$.
Fig. 39 shows data at 1200 MeV/c.
In (a), dashed histograms show the fit without $\pi _2(1880)$.
There is a hint of a peak in $\eta \eta \pi$.
In (c), there is clear evidence that this is associated with the
$a_2(1320)$.
The essential evidence comes from (c), (d) and (e).
Without $\pi _2(1880)$, the dashed histogram in (c) fails to fit
the observed $a_2(1320)$ signal. Note that the normalisation
of the curve cannot be scaled up arbitrarily to fit the data,
since it is governed by fitted amplitudes.
In (d), the $a_2$ signal in the $\eta \eta \pi$ mass range
1850 MeV (approximate threshold) to 2050 MeV is clear.
In (e) the $\eta \eta$ enhancement below 1350 MeV arises as a
reflection of the decay channel $a_2(1320)\eta$.
\begin{figure}
\begin {center}
\epsfig{file=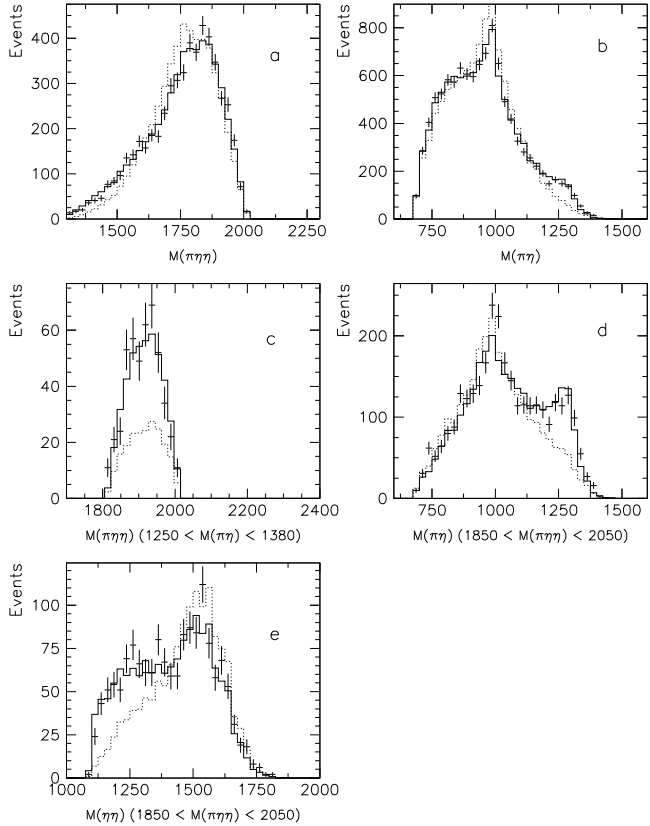,width=9cm}~\\
\caption {Full histograms show fits to $\eta \eta \pi ^0 \pi ^0$
data at 1200 MeV/c compared with data; dotted histograms show fits
without $\pi _2(1880)$; (a) the $\eta \eta \pi$ mass distribution,
(b) the $\pi \eta$ mass distribution, (c) the $\eta \eta \pi$
mass
distribution for events with a $\pi \eta$ combination within
$\pm 65$ MeV of $a_2(1320)$, (d) and (e) $\pi \eta$ and $\eta \eta$
mass distributions for events with $M(\eta \eta \pi )$ 1850--2050 MeV.
Units of mass are MeV.}
\end {center}
\end{figure}

Optimum parameters are $M = 1880 \pm 20$ MeV, $\Gamma = 255 \pm 45$
MeV.
The signal cannot be explained as the high mass tail of $\pi _2(1660)$:
it is too strong and the $\pi _2(1670)$ is too narrow to fit observed
cross section.
From Fig. 39(a), the signal clearly does not arise from $\eta \eta
\pi$ masses near 2005 MeV.
The angular distribution for decay to $a_2(1320)\eta$ is consistent
with S-wave decay, as expected so close to threshold; it is
inconsistent with $L = 1$ decays.

The VES collaboration also reports a
threshold enhancement in $a_2(1320)\eta$ with $J^P = 2^-$ in their
$\eta \eta \pi ^-$ data [60] and also in $\eta \pi ^+ \pi ^- \pi ^0$
data where $a_2(1320) \to \pi ^+\pi ^-\pi ^0$. They observe a strong
peak at the same mass in $[f_2(1270)\pi ]_{L=2}$ [61].
Daum et al. [62] also observed a peak at 1850 MeV in $[f_2(1270)\pi
]_{L=2}$ with a width of $\sim 240$ MeV.
Their result is reproduced in Fig. 40.
There is a clear peak and a phase motion with respect to $\pi _2(1670)$
entirely consistent with a second $2^-$ resonance associated
with the peak. However, they interpreted the data conservatively as
arising from interference between $\pi _2(1670)$ and a higher $\pi
_2(2100)$; that interpretation fails for Crystal Barrel data.
\begin{figure}
\begin {center}
\epsfig{file=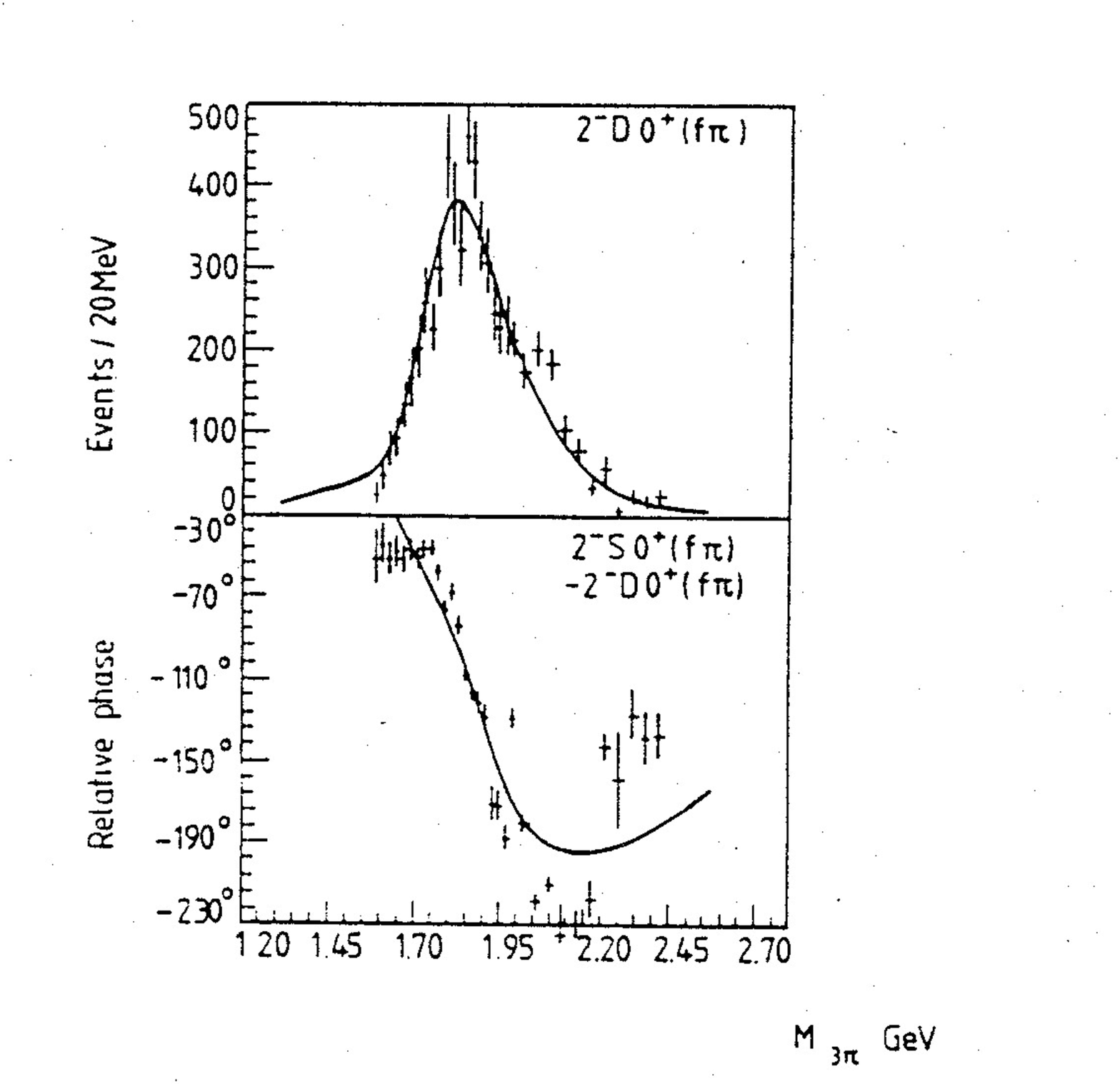,width=5cm}~\\
\caption {Data of Daum et al. [62].The upper figure shows the
$f_2\pi$ D-wave intensity; the lower figure shows the phase
difference between $\pi_2(1670)$ and $[f_2\pi ]_{L=2}$.}
\end {center}
\end{figure}

Popov [57] reports a $2^-$ peak in E852 data for $[\omega \rho ]_{L=1}$
for the combination of $\omega$ and $\rho$ spins $S = 1$.
He quotes $M = 1890 \pm 10 \pm 26$ MeV, $\Gamma= 350 \pm 22 \pm 55$
MeV. The mass agrees well with Crystal Barrel; the width is slightly
higher.
Eugenio [63] reports a similar $2^-$ peak in E852 data for
the $\eta \eta \pi ^-$ final state.
In view of all these observations, the $\pi _2(1880)$ should
be regarded as a $4^*$ state.

In section 5.3, it was reported that $\eta _2(1870)$ has a branching
ratio to $f_0(1270)\eta$ much stronger than to $a_2(1320)\pi$
and inconsistent with SU(3) for a $q\bar q$ state.
The $\pi _2(1880)$ is observed strongly in decays to
$a_2(1320)\pi$ by VES but no decay to the $f_2\pi$ S-wave is observed;
although no actual branching ratio is reported at present, it seems
likely that $\pi _2(1880)$ is also inconsistent with a $q\bar q$ state
and SU(3) relations for decays.

\section {$I = 1$, $C = -1$ resonances}
Data are available from $\pi ^- \pi ^+$, $\omega \pi ^0$  and
$\omega \eta \pi ^0$ final states.
The latter two channels have been measured for $\omega \to \pi ^0
\gamma$ at all momenta and for $\omega \to \pi ^+ \pi ^- \pi ^0$
at 6 momenta: 600, 900, 1200, 1525, 1642 and 1940 MeV/c.

The analysis of $\bar pp \to \pi ^-\pi ^+$ is reported in Ref. [13].
The $I = 1$ states with $J^{PC} = 1^{--}$, $3^{--}$ and $5^{--}$
cause forward-backward asymmetries in both $d\sigma /d\Omega$
and $Pd\sigma /d\Omega$ via interferences with $I = 0$ states; these
asymmetries are conspicuous and lead to good determinations of $C
= -1$ amplitudes.
A particular feature at low momenta is a very strong $\rho _3(1982)$
which interferes with $f_4(2050)$ over the whole momentum range around
750 MeV/c.
Its mass and width are determined very accurately by the $\pi ^- \pi
^+$
data.
This resonance then serves as a useful interferometer in $\omega \pi ^0$
and $\omega \eta \pi ^0$ data in the low mass range.
A second $J^{PC} = 3^{--}$ state is required also at 2260 MeV and
serves as a similar interferometer at high masses.
The $\rho _5(2295)$ appears as a large signal in $\pi ^- \pi
^+$, see Fig.  14 of Section 5.1.

Attention needs to be drawn to one point concerning the $1^{--}$ state
at 2000 MeV.
The best mass determination of this state comes from $\pi ^-\pi ^+$,
because of the PS172 data available in small momentum steps
down to a mass of 1912 MeV. As emphasised above, the combination
of differential cross sections and data from a polarised target
determines phases well for all partial waves.
The Argand diagram for $^3D_1$ on Fig. 23 of Section 5.4 requires
a phase variation approaching $180^\circ$ between 1900 and 2100 MeV,
hence a $1^{--}$ resonance.
The $^3S_1$ amplitude shown on the same figure has a $1/v$ behaviour
close to the $\bar pp$ threshold.
When one factors this $1/v$ dependence out, the
$^3S_1$ amplitude also requires the same $1^{--}$ resonance.
For $\omega \pi ^0$ and $\omega \eta \pi ^0$ data discussed below,
data from a polarised target are not available, so determinations
of  its mass and width are less reliable.
Once it is established, the $1^{--}$ amplitude again serves as a
useful interferometer in
$\omega \pi ^0$ and $\omega \eta \pi ^0$ data at low masses.

\begin{figure}
\centerline{\hspace{0.2cm}\epsfig{file=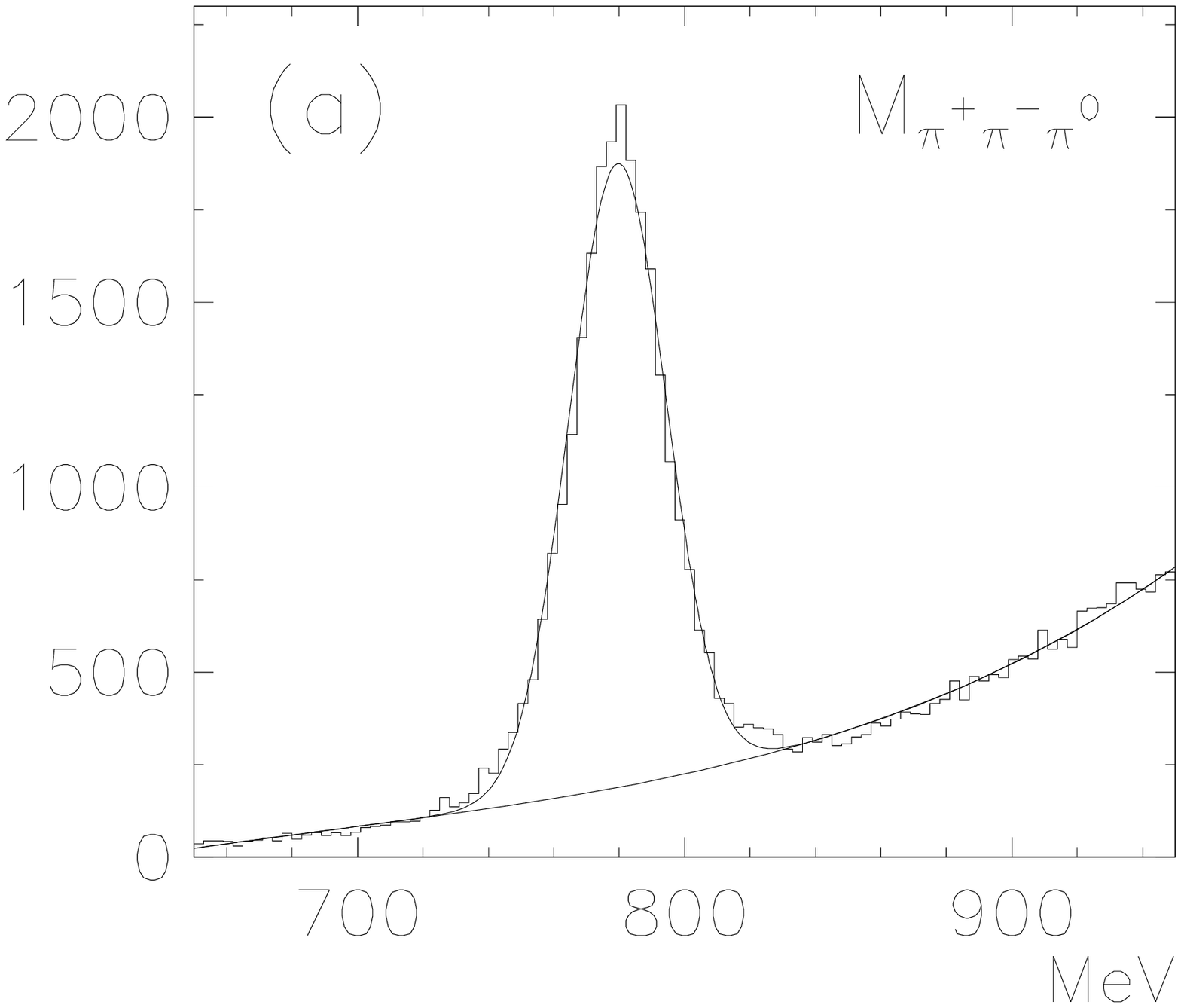,width=5.1cm}
            \epsfig{file=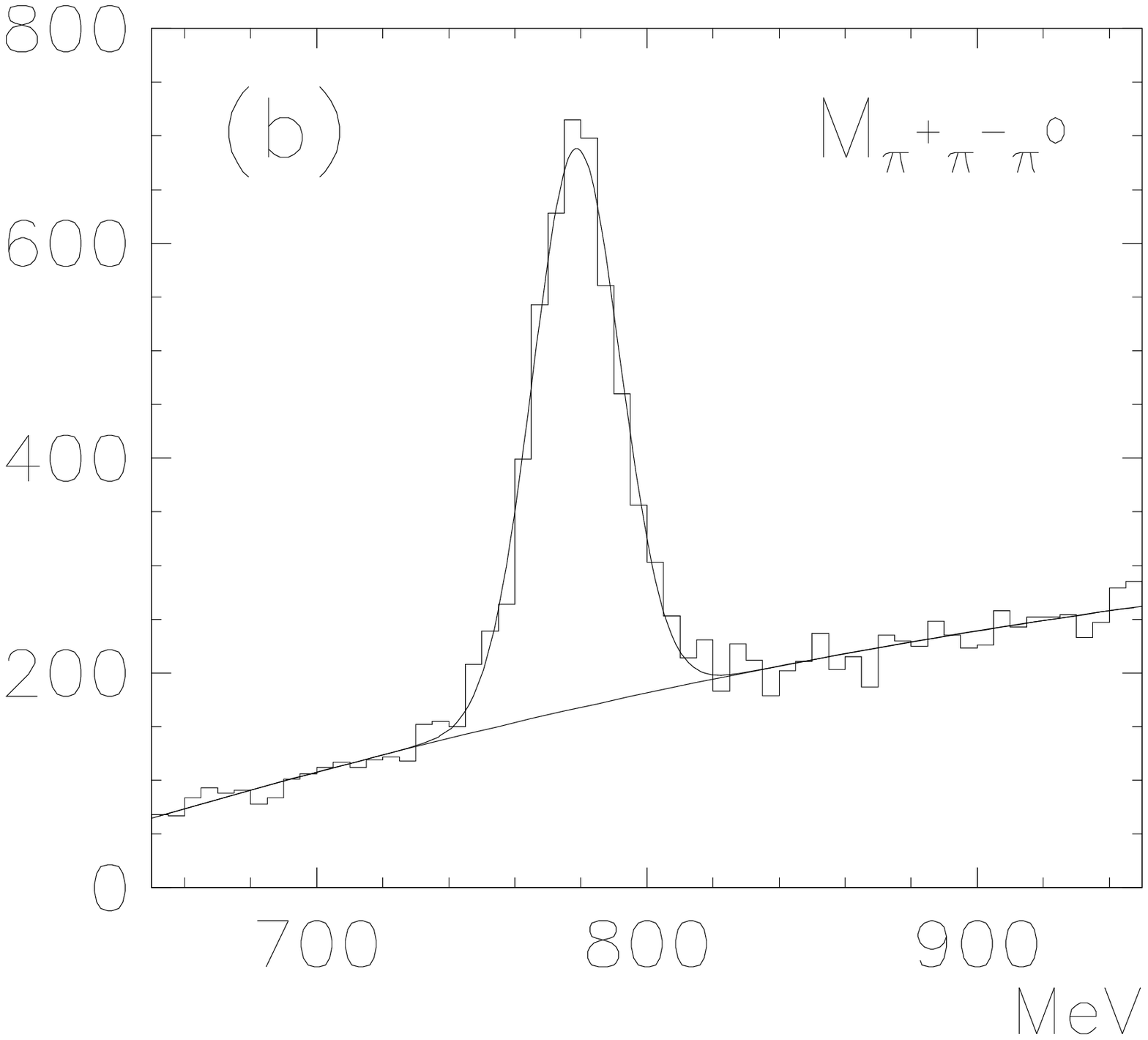,width=5.1cm}}
\vspace{-1.1cm}
\centerline{\hspace{0.3cm}\epsfig{file=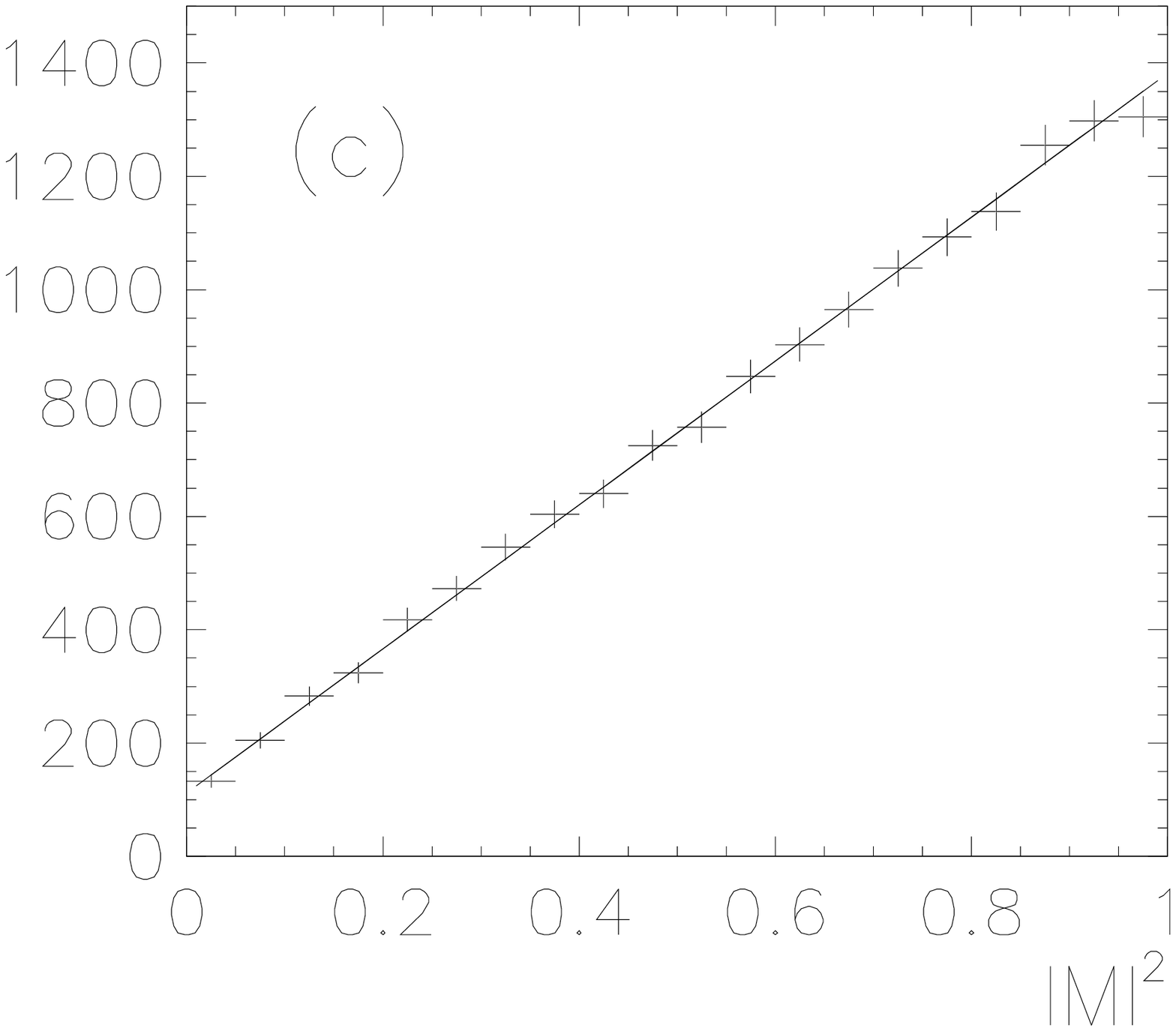,width=5.2cm}
            \epsfig{file=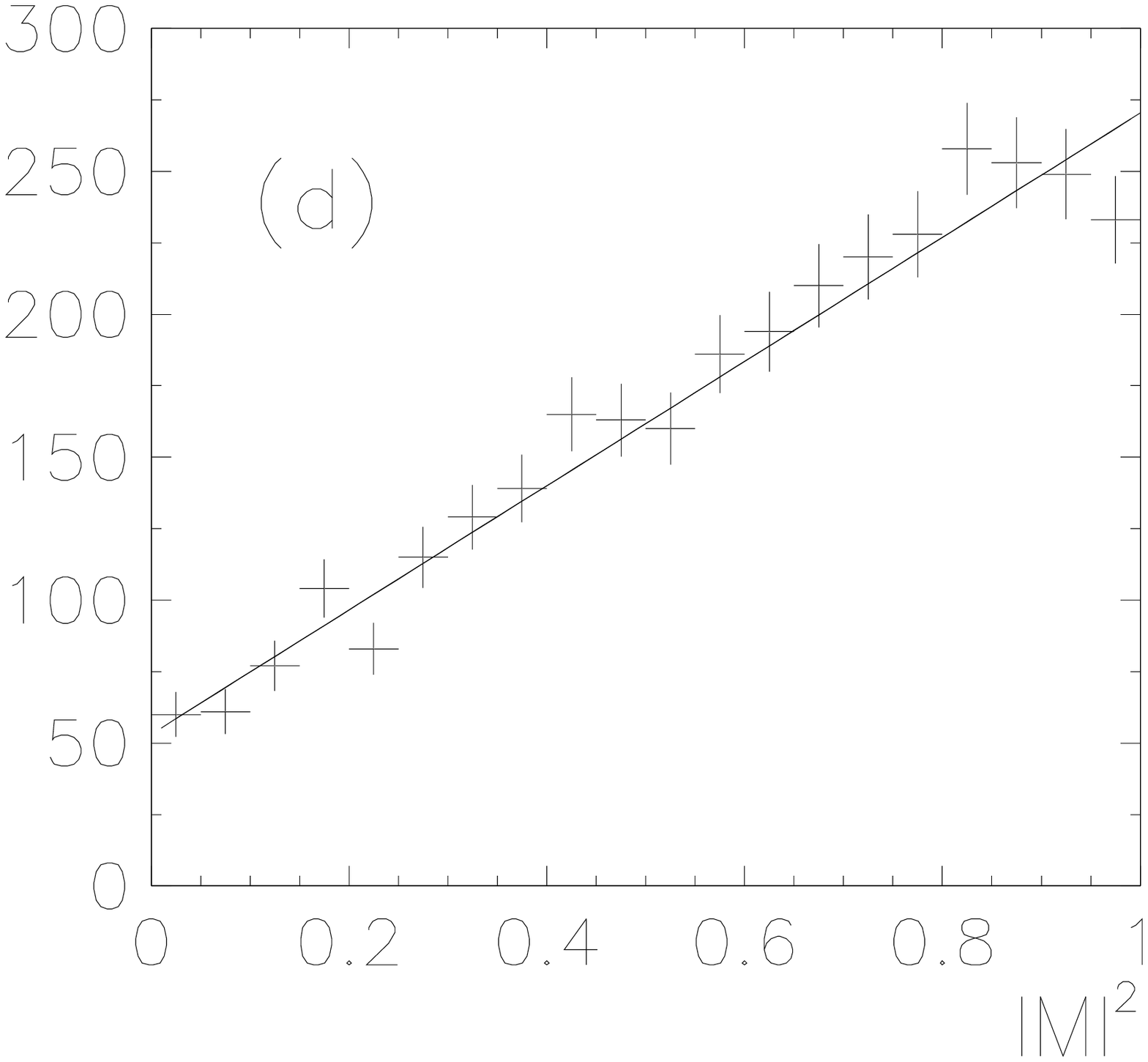,width=5.2cm}}
\vspace{-1.5cm}
\centerline{\epsfig{file=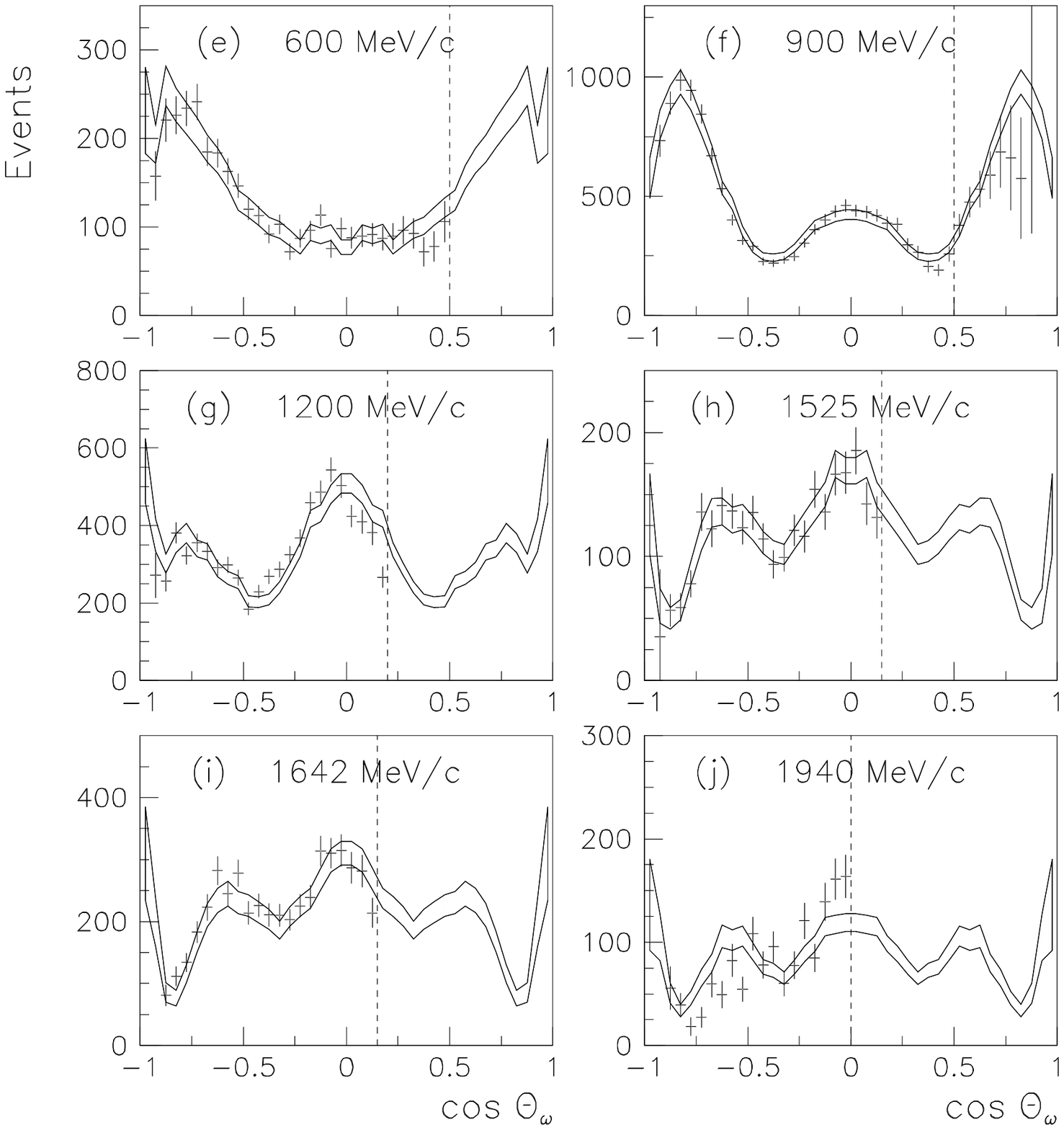,width=13cm}}
\vspace{-0.5cm}
\caption{$M(\pi ^+\pi ^- \pi ^0)$ from the data selection
for (a) $\omega \pi ^0$ data, (b) $\omega \eta \pi ^0$ at 900 MeV/c;
the number of events v. the square of the matrix element for
$\omega$ decay in (c) $\omega \pi ^0$, (d) $\omega \eta \pi ^0$;
(e)-(j): curves show the error corridor for $\omega \pi ^0$ differential
cross sections where $\omega$ decays to $\pi ^0 \gamma$;
they are corrected for angular acceptance; points with
errors show results for $\omega$ decays to $\pi ^+\pi ^-\pi ^0$;
the vertical dashed lines show the cut-off which has been used.}
\end{figure}

Analysis of $\bar pp \to \omega \pi ^0$, $\omega \to \pi ^0 \gamma$
is reported in Ref. [64];
$\omega \eta \pi ^0$ data with $\omega \to \pi ^0 \gamma$
are reported in Ref. [65].
Although this $\omega$ decay mode is identified cleanly, polarisation
information for the $\omega$ is
carried away by the photon, whose polarisation is not measured. This
was the reason for studying  $\omega \to \pi ^+\pi ^- \pi ^0$
[66]; the combined analysis of all $I = 1$, $C = -1$ channels is
reported there. The matrix element for
 $\omega \to \pi ^+ \pi ^- \pi ^0$ is relativistically $\epsilon
_{\alpha \beta \gamma \delta } p^\beta p^\gamma p^\delta$ where $p$
are 4-momenta of decay pions. Non-relativistically, this matrix element
becomes $p_i \wedge p_j$ in the rest frame of the $\omega$, where $p_i$
are 3-momenta of any two pions. The polarisation vector of the $\omega$
is therefore described by the normal $\vec {n}$ to its decay plane. A
measurement of this decay plane provides complete information on the
polarisation of the $\omega$.
That improves the amplitude analysis greatly.
Conclusions should therefore be taken from Ref. [66].

The measurement of both decay channels also provides a valuable
experimental cross-check, illustrated in Fig. 41.
Angular distributions are measured mostly in the backward hemisphere
for the $\omega$, because of the Lorentz boost from centre of mass to
the lab.
Figs. 41 (e)-(j) show as points with errors the angular distributions
for $\omega \pi ^0$ from $\omega \to \pi ^+ \pi ^- \pi ^0$ decays.
The corridors show angular distributions from the partial
wave analysis of $\omega \to \pi ^0 \gamma$.
Both are corrected for acceptance.
The observed agreement is a major cross-check on calibrations
of the detector, data selection and Monte Carlo simulations.
All of these are completely different between all-neutral data and
charged data.
One aspect of this is the treatment of the vertex.
For charged data, the vertex is determined by the intersection of the
two charged tracks.
For neutral data, the vertex is instead taken to be at the centre
of the target.
This assumption distorts the kinematic fit slightly, but the
correction is only $1\%$ to $d\sigma /d\Omega$ for all-neutral data.
This procedure is safer than fitting the vertex freely; that
alternative leads to strong and dangerous variations of acceptance
with $\cos \theta _\omega$.

Backgrounds are clearly visible in Figs. 41 (a) and (b) under the
$\omega$.
The backgrounds are checked in (c) and (d).
There, the number of events is plotted against the square of the
matrix  element for $\omega$ decay. The linear dependence checks that
the $\omega$ signals are clean. The intercepts at $|M|^2 = 0$ check the
magnitudes of the backgrounds.

\begin{figure} [b]
\centerline{\epsfig{file=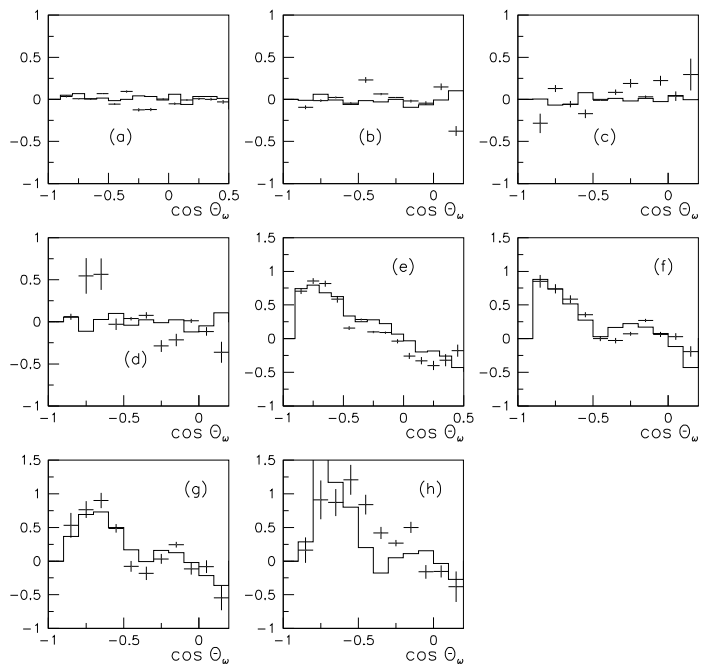,width=9.6cm}}
\caption{ Vector polarisation $P_y$ for $\omega \pi ^0$ data at (a) 900,
(b) 1200, (c) 1525 and (d) 1940 MeV/c compared with the partial wave
fit (histogram); (e)--(h) $Re~T_{21}$ at the same momenta. }
\end{figure}
\subsection {Vector polarisation of the $\omega$ and its implications}
A remarkable feature is observed for $\omega$
polarisation, with major physics implications. Both vector polarisation
$P_y$ and tensor polarisations $T_{20}$, $T_{21}$ and $T_{22}$ are
measured. For details and formulae, see Ref. [66]. Vector polarisations
are shown in the upper half of Fig. 42. They are consistent with zero
almost everywhere. Large tensor polarisations are observed at all
momenta in the lower half of Fig. 42 ($Re~T_{21})$ and in Fig. 43
($T_{20}$ and $T_{22})$. Vector polarisation depends on the imaginary
\begin{figure}
\centerline{\epsfig{file=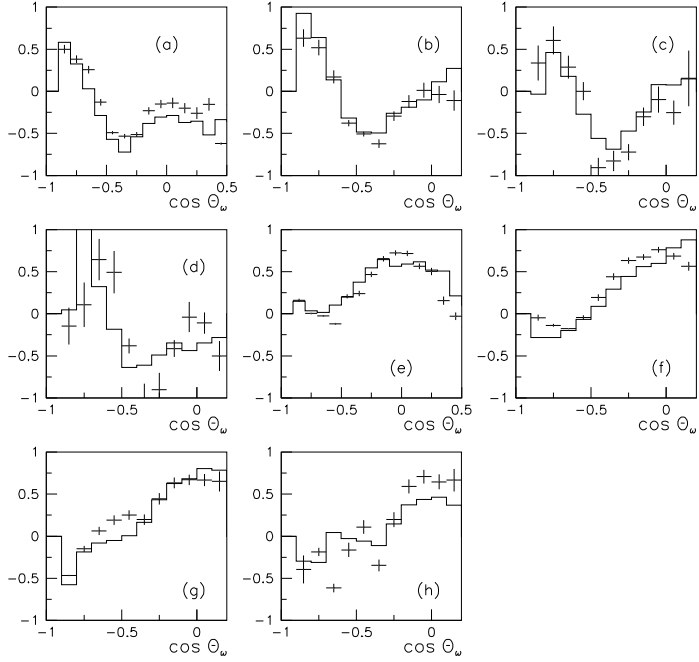,width=9.6cm}}
\caption{(a)--(d) $Re~T_{22}$ and (e)--(h) $T_{20}$ for $\omega \pi
^0$ data at 900, 1200, 1525 and 1940 MeV/c, compared with the partial
wave fit (histograms).}
\end{figure}
parts of inteferences between partial waves. Tensor polarisations
depend on moduli squared of amplitudes
and real parts of interferences. In particular, $Re~T_{21}$
contains real parts of precisely the same interference terms as appear
in $P_Y$. Formulae are given by Foroughi [67] for $pp \to d\pi ^+$, but
the same formulae apply to $\bar pp \to \omega \pi ^0$. The implication
of zero vector polarisation is that relative phases of amplitudes are
close to 0 or $180^\circ$; the physics implications will now be
discussed.

Consider first a given $J^P$ such as $1^{+}$.
This partial wave decays to $\omega \pi$ with orbital angular momenta
$L = 0$ or 2.
It is not surprising that partial waves to these final states have the same
phase, since a resonance implies multiple scattering through all coupled
channels.
Indeed, it is found that relative phases are consistent with zero
within errors for all resonances.
[The final analysis can then be stiffened by setting the relative
phase angle to zero.]

Vector polarisation $P_y$ can arise from interference between
singlet waves
$1^{+}$, $3^{+}$ and $5^{+}$.
Likewise, it can arise from interference between triplet waves
$1^{-}$, $2^{-}$, $3^{-}$, $4^{-}$ and $5^{-}$.
It is remarkable that these further interferences also lead to nearly
zero polarisation.
It implies coherence between different $J^P$.
There is  no full explanation for this result, but an interpretation
is suggested here.

\begin{figure} [b]
\centerline{\epsfig{file=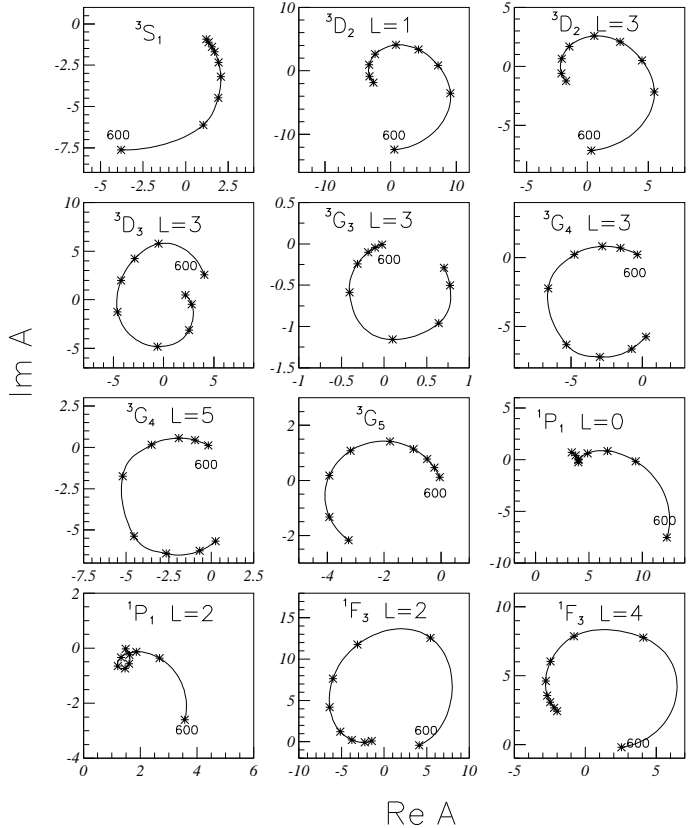,width=10cm}}
\caption{Argand diagrams for partial waves fitted to $\omega \pi ^0$
in the combined analysis with $\omega \eta \pi ^0$ and $\pi ^+\pi ^-$.
Crosses show beam momenta 600, 900, 1050, 1200, 1350, 1525, 1642, 1800
and 1940 MeV/c; all move anti-clockwise with increasing beam momentum.
$L$ is the orbital angular momentum in the $\omega \pi$ channel.}
\end{figure}
The incident plane wave
may be expanded in the usual way in terms of Legendre functions:
\begin {equation}
e ^{ikz} = \sum _{\ell} (2\ell + 1)i^{\ell} P_{\ell}(\cos \theta ).
\end {equation}
As a result, all partial waves of the initial state are related in phase
through the factor $i^{\ell}$.
Fig. 44 shows Argand diagrams for all partial waves.
Let us take as reference for triplet states the large $^3G_4 (\ell = 3)$
amplitude.
If one compares this by eye with diagrams for $^3D_2$ and $^3D_3$,
it is clear that there is a phase rotation of the diagam by
$\sim 180 ^{\circ}$ for $^3D_2$;
$^3D_3$ is similar in phase to $^3G_4$.
The Argand loop for $^3S_1$ is again rotated by $\sim 180 ^{\circ }$
with respect to $^3G_4$.
Singlet and triplet states do not interfere in $d\sigma /d\Omega$,
$P_y$, $T_{20}$, $T_{21}$ and $T_{22}$.
Therefore the phase of singlet states with respect to triplet
are arbitrary in Fig. 44.
The Argand loops for $^1P_1$ and $^1F_3$ are broadly similar,
though there is a large offset in the $^1P_1$ $L = 0$ amplitude.

A similar phase coherence is reported for $\bar pp \to \pi ^- \pi ^+$
[1].
There, both $I = 0$ and $I = 1$ amplitudes are present.
The amplitudes which differ by 1 in orbital angular momentum for $\bar pp$
are $\sim 90 ^{\circ}$ out of phase, as suggested by eqn. (21).

An analogy is given in Ref. [66].
Suppose a bottle is filled with irregular shaped objects, such as
screws.
If the bottle is shaken, the objects inside settle
to a more compact form.
Returning to resonances, they have well defined masses, widths and
coupling constants which make them individual.
But a suggestion is that, in the interaction, phases adjust to retain as
much coherence as possible with the incident plane wave.
This is a natural way of inducing order between the observed
resonances.
It helps explain why all Argand diagrams show a smooth variation
of phases with mass.

\begin{figure}
\centerline{\epsfig{file=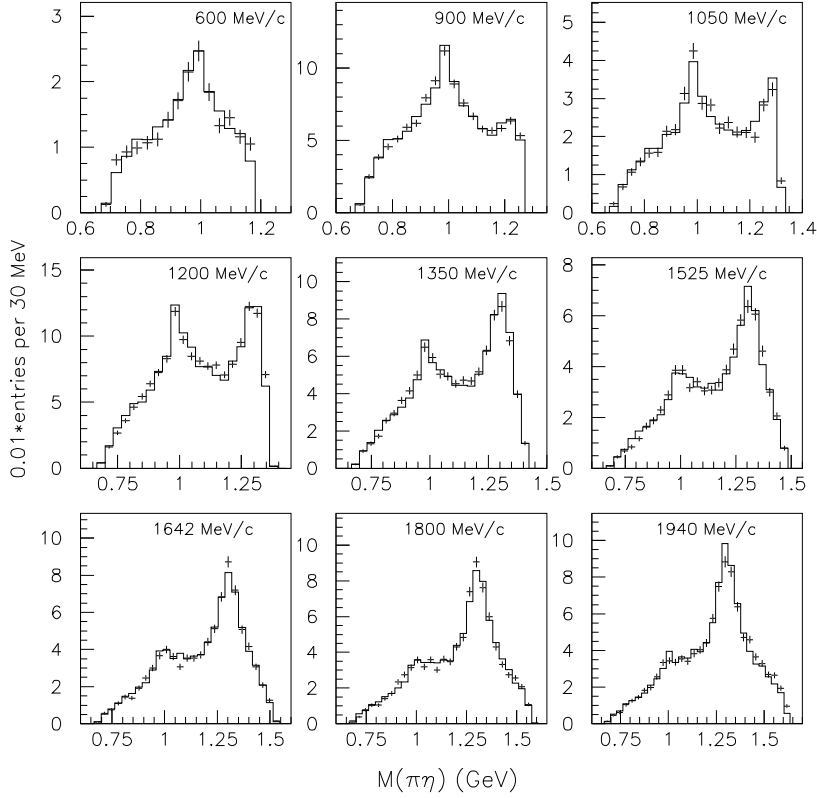,width=12cm}}
\caption{Projection of $\omega \eta \pi ^0$ data on to
$M(\pi ^0 \eta )$; full histograms show the fit.}
\end{figure}
\begin{figure}
\centerline{\epsfig{file=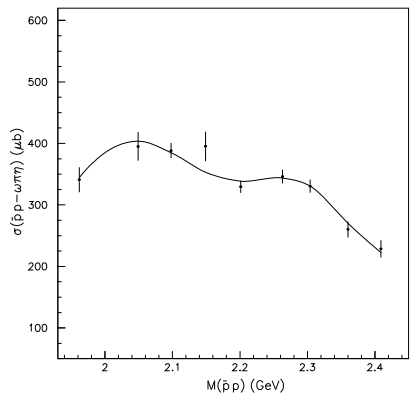,width=6cm}}
\caption{Integrated cross section for $\bar pp \to \omega \eta
\pi ^0$ compared with the fit.
It is corrected for acceptance and for branching ratios
of $\omega \to \pi ^0 \gamma$, $\eta \to \gamma \gamma$ and $\pi ^0 \to
\gamma \gamma$.}
\end{figure}
\subsection {$\omega \eta \pi ^0$ data}
Fig. 45 shows the $\pi ^0 \eta$ mass projection from
$\omega \eta \pi ^0$ data.
There are clear peaks due to $a_0(980)$ and $a_2(1320)$.
The strong increase in the $a_2$ signal at the higher masses
is due to the opening of the $a_2\omega$ threshold at 2100 MeV.
This threshold is also observed through its direct effect on the
integrated cross section, shown in Fig. 46.
In contrast, integrated cross sections for most other channels
peak at low mass, e.g. Fig. 10(a) for $\pi ^- \pi ^+$ in Section 3.

\begin{figure} [t]
\centerline{\epsfig{file=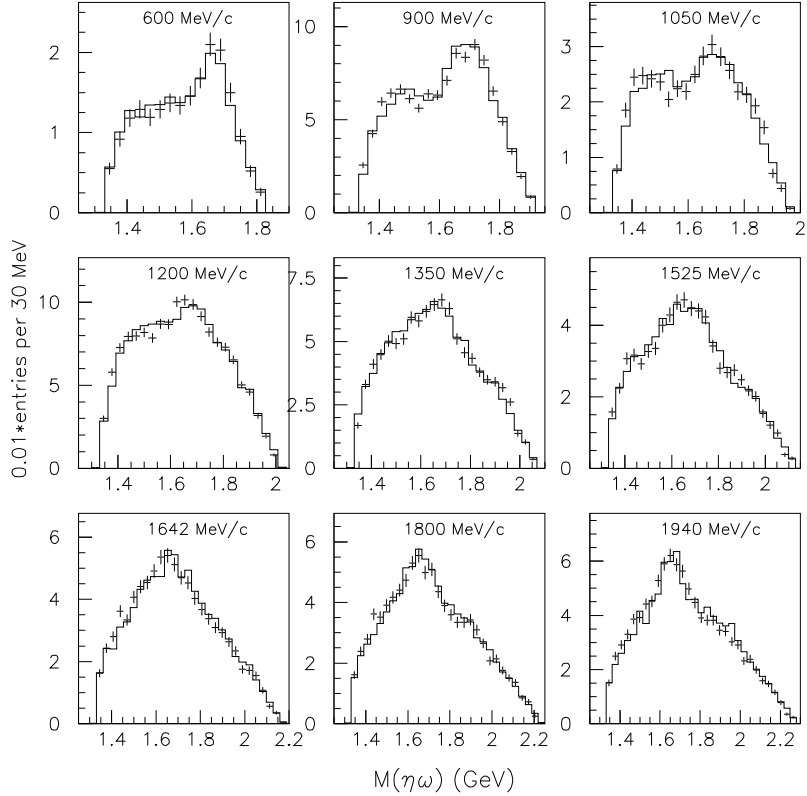,width=12cm}}
\caption{Projection of $\omega \eta \pi ^0$ data on to
$M(\eta \omega )$; full histograms show the fit including an
$\omega (1665)$ with $\Gamma = 260$ MeV.}
\end{figure}
A feature of the $\omega \eta \pi ^0$ data is shown in Fig. 47.
There is a clear peak at 1650 MeV in $\omega \eta$.
The amplitude analysis favours $J^{PC} = 1^{--}$ strongly over
$J^{PC}=1^{+-}$, $2^{--}$ or $3^{--}$.
The fitted mass and width $M = 1664 \pm 17$ MeV, $\Gamma = 260 \pm
40$ MeV agree well with PDG averages for $\omega _1(1650)$.
The final state $\omega _1(1650)\pi$ contributes strongly to the
$1^{+-}$ ($^1P_1 ~\bar pp$) resonance at 1960 MeV.
\subsection {Observed resonances}
Table 7 shows resonances from the final combined analysis
of all channels.
The complete spectrum of expected $q\bar q$ states is observed except
for
the $^3G_3$ state and the highest $^3S_1$ state expected near 2400 MeV.
The $^3G_3$ state may easily be hidden close to the strong upper
$^3D_3$ state; indeed, that resonance has a ratio $r_3$
of $^3G_3/^3D_3$ amplitudes significantly different
from 0; this hints at the presence of either a $^3G_3$ state
or quantum mechanical mixing with it.
There is even a hint of the last $^3S_1$ state.
A scan of the mass of the $1^-$ state at 2265 MeV reveals two
distinct optima in log likelihood, one at 2265 MeV and a second weaker
one at 2400 MeV;
however there is not sufficient evidence to isolate this state
so close to the top of the available mass range.

\begin{table} [htp]
\begin{center}
\begin{tabular}{lccccccc}
$J^{PC}$ & * & Mass $M$ & Width $\Gamma$ & $r$  & $\Delta S$ &
$\Delta S$ & Observed \\
   &  & (MeV)  & (MeV) &  &  $(\omega \pi )$ &
     $(\omega \eta \pi )$ & Channels \\\hline
$5^{--}$ & $4^*$ & $2300 \pm 45$ &$ 260 \pm 75$ & (0)  &
   473  & 133 & $\pi\pi $,$\omega \pi$,GAMS \\
$4^{--}$ & $4^*$ & $2230 \pm 25$ & $ 210 \pm 30$ & $0.37 \pm 0.05$ &
   1254 &  153 & $[\omega \pi ]_{L = 3,5}$,$a_0\omega$,$\omega '\pi$\\
$3^{--}$ & $4^*$ & $2260 \pm 20$ & $160 \pm 25$ & $1.6 \pm 1.0$ &
   341 &  578 &  $\pi\pi$,$\omega \pi$,$a_2\omega$,$b_1\eta$,
$\omega '\pi$ \\
$3^{--}$ & $4^*$ & $1982 \pm 14$  & $188 \pm 24$ & $0.006 \pm 0.008$ &
   314 &  138 & $\pi \pi$,$\omega \pi$,$a_0\omega$ \\
$2^{--}$ & $3^*$ & $2225 \pm 35$ & $335 ^{+100}_{-50}$ & $1.39 \pm 0.37$
  & 356  & 502 & $\omega \pi$,$a_2\omega$,$a_0\omega$ \\
$2^{--}$ & $2^*$ & $1940 \pm 40$  & $155 \pm 40$ & $1.30 \pm 0.38$ &
    433 &  93 & $\omega \pi$,$a_0\omega$ \\
$1^{--}$ & $2^*$ & $2265 \pm 40$ & $325 \pm 80$ & $-0.55 \pm 0.66$  &
   - &  313 & $\pi \pi$,$a_2\omega$,($\omega '\pi$) \\
$1^{--}$ & $3^*$ & $2110 \pm 35$ & $230 \pm 50$ & $-0.05 \pm 0.42$  &
  - &   834 &  $\pi \pi$,$a_2\omega $,($\omega '\pi$),GAMS \\
$1^{--}$ & $3^*$ & $2000 \pm 30$ & $260 \pm 45$ & $0.70 \pm 0.23$  &
   562 &  133 & $\pi \pi$,$\omega \pi$,$a_0\omega$ \\\hline
$5^{+-}$ & - & $(2500)$ & $\sim 370 $ & (0)  &
   195 &  - &  $\omega \pi$ \\
$3^{+-}$ & $3^*$ & $2245 \pm 50$ & $320 \pm 70$ & $0.78 \pm 0.47$ &
   264 &  187 & $\omega \pi$,$a_2\omega$,$\omega '\pi$,$b_1\eta$ \\
$3^{+-}$ & $4^*$ & $2032 \pm 12$ & $117 \pm 11$ & $2.06 \pm 0.20$ &
    3073 &  178 & ($\omega \pi$),$a_0\omega$,$\omega '\pi$,($b_1\eta$)
    \\
$1^{+-}$ & $2^*$ & $2240 \pm 35$ & $320 \pm 85$ & $1.2 \pm 0.5$ &
   62 &  542 & ($\omega\pi$),$a_2\omega$,$b_1\eta$ \\
$1^{+-}$ & $3^*$ & $1960 \pm 35$ & $230 \pm 50$  & $0.73 \pm 0.18$ &
   503 &    514 & $\omega \pi$,$a_0\omega$,$\omega '\pi$ \\
\hline
\end{tabular}
\caption {Resonance parameters from a combined fit to $\omega \pi ^0$,
$\omega \eta \pi ^0$ and $\pi ^- \pi ^+$, using both
$\omega \to \pi ^0 \gamma$ and $\omega \to \pi ^+\pi ^- \pi ^0$ decays.
Values of $r$ are ratios of coupling constants
$g_{J+1}/g_{J-1}$ for coupling to $\bar pp$ or $\omega \pi ^0$.
Columns 6 and 7 show changes in
$S = $ log likelihood when each resonance is omitted and others
are re-optimised. In the final column, $\omega '$ denotes $\omega
_1(1650)$; channels in parentheses are weak.}
\end{center}
\end{table}

Intensities of partial waves fitted to $\omega \pi ^0$ are shown in
Fig. 48.
A remarkable feature is a very strong $^3G_4$ state at 2230 MeV.
In all other data, $G$-states are weak, presumably because of the
strong $\ell = 4$ centrifugal barrier in the $\bar pp$ channel.
The analysis has been checked in many ways to ensure that there is
no cross-talk with $^3D_2$, which is quite weak at high momenta.
As an example, analyses have been carried out at single energies.
The fitted $^3G_4$ intensity follows the behaviour expected from
the centrigugal barrier.
It falls close to zero at low momenta where
the $^3D_2$ intensity is large; this is illustated in Fig. 4 of
Ref. [64].
So there is no doubt that the huge $^3G_4$ amplitude is a real
effect.
There is no obvious explanation.
It must indicate that for some reason this state decays dominantly
through $\omega \pi$.
The low mass of the $^3G_4$ state is also surprising; it is lower
than the nearby $F$-states.
This is in contrast to the well established behaviour of $I = 0$
states where masses of $G$-state resonances are higher than those
of $F$-states.
\begin{figure} \centerline{\epsfig{file=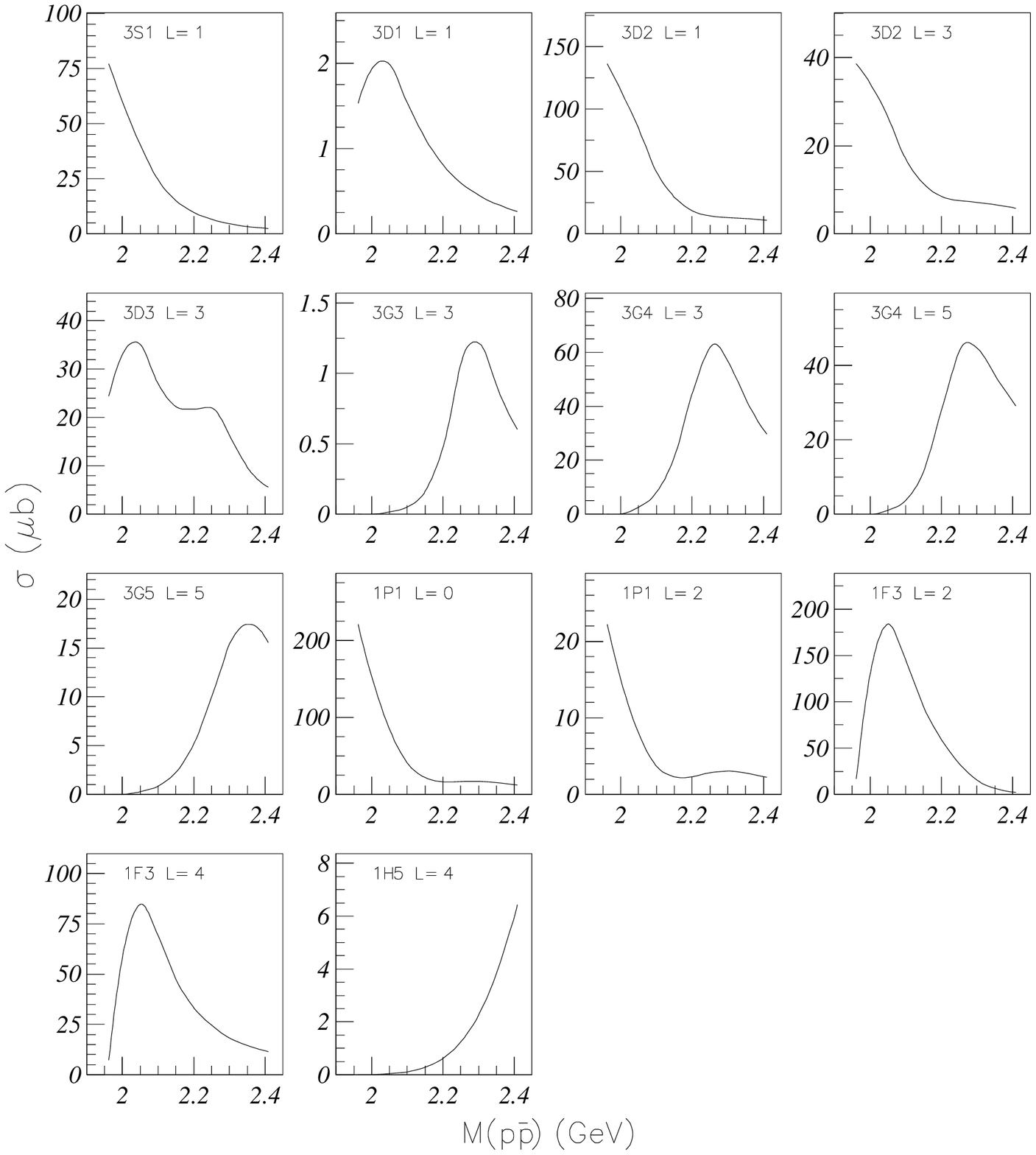,width=13.5cm}}
\caption{Intensities of partial waves fitted to $\omega \pi ^0$ data;
$L$ is the orbital angular momentum in the decay.}
\end{figure}

Table 7 shows in the right-hand entries the firmly established
decay channels where each resonance is observed.
The star rating depends on the number of channels in which each
resonance is observed and the stability of resonance parameters.
Figs. 49 and 50 show triplet and singlet partial waves for
$\bar pp \to \omega \eta \pi ^0$ from the final analysis of
Ref. [66]; these figures are an update of those from Ref. [65],
but changes are small.
Brief comments will now be given, starting with the highest spins.

The $\rho _5(2300)$ is already known from GAMS data [68].
The determination from $\bar pp \to \pi ^- \pi ^+$ is more
accurate, from the large $5^{--}$ amplitude observed there.
GAMS omitted any centrifugal barrier for the production process;
this barrier lead to a peak position in Crystal Barrel data of
2340 MeV in agreement with the mass of $2330 \pm 35$ MeV quoted by
GAMS.

The $4^-$ state at 2230 MeV is observed independently in $\omega
\pi$ decays with $L = 3$ and 5 and also in $a_0(980)\omega$
in Fig. 49.
Since it is the dominant signal at high masses in $\omega \pi$, it
is very clearly established and classified as $4^*$.

\begin{figure} \centerline{\epsfig{file=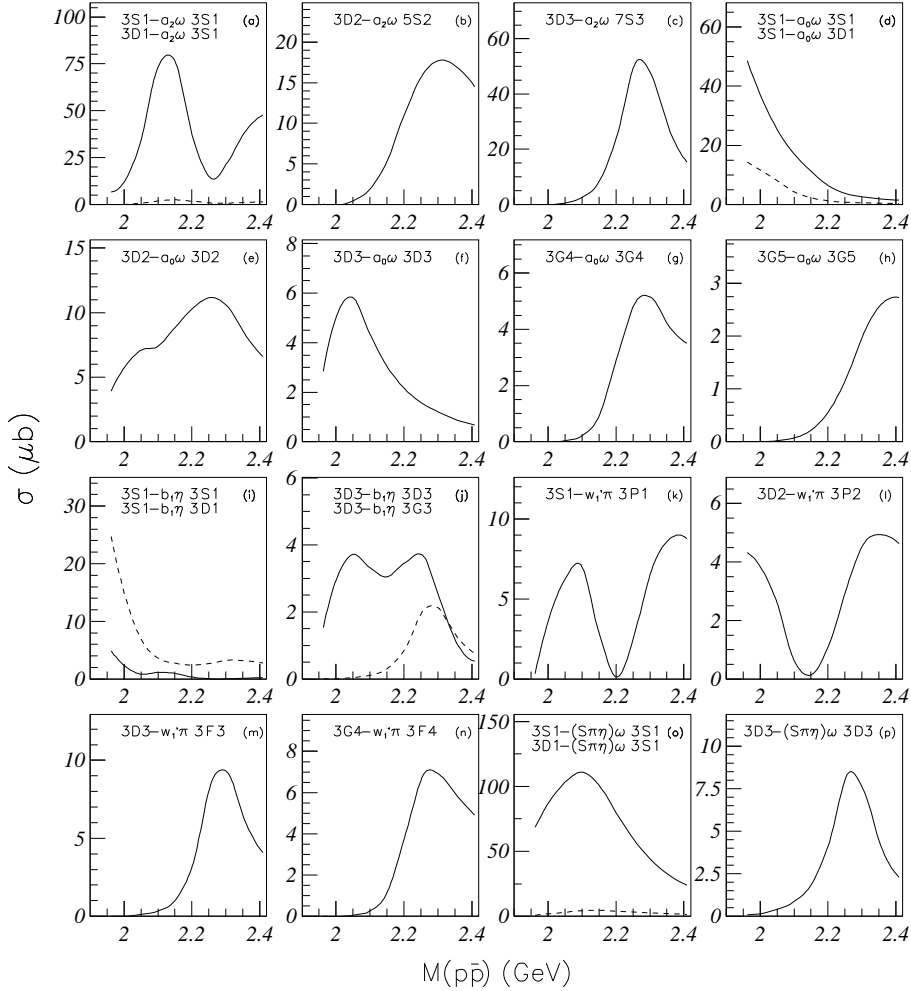,width=13.5cm}}
\caption{Intensities of dominant triplet partial waves in  $\omega
\eta \pi ^0$ data; dashed lines refer to the second title in panels.}
\end{figure}

The upper $3^-$ state is observed in 5 channels; the lower state
at 1982 MeV is  strong in $\pi ^-\pi ^+$ and accurately determined.
Both are therefore classified as $4^*$. Indeed, the trajectory of
$3^{--}$ states ($\bar pp ~ ^3D_3)$ in Fig. 3(c), including the very
well known $\rho _3(1690)$, provides one of the best determinations
of trajectory slopes.

The upper $2^-$ state at 2225 MeV is not given $4^*$ status because
of correlations with $3^-$ and $1^-$ partial waves.
The lower $2^-$ state at 1940 MeV is one of the two most weakly
established states. However, its Argand plot on Fig. 44 requires
a rapid phase variation at low mass because the
amplitude must go to zero at the $\bar pp$ threshold.
So a resonance is required, even if it is at the
bottom of the available mass range and therefore somewhat uncertain
in mass.

All three $1^{--}$ states are observed in three
channels. The $\rho (2150)$ was found by GAMS [69,70]
and now deserves $3^*$ status.
The mass scan for the upper one gives a distinct
optimum at 2265, but some indication of a second  at 2400 MeV. It
is possible that there are two states in the high mass range
and they cannot be fully resolved. The state at 2265 MeV is therefore
classified as $2^*$.
\begin{figure} \centerline{\epsfig{file=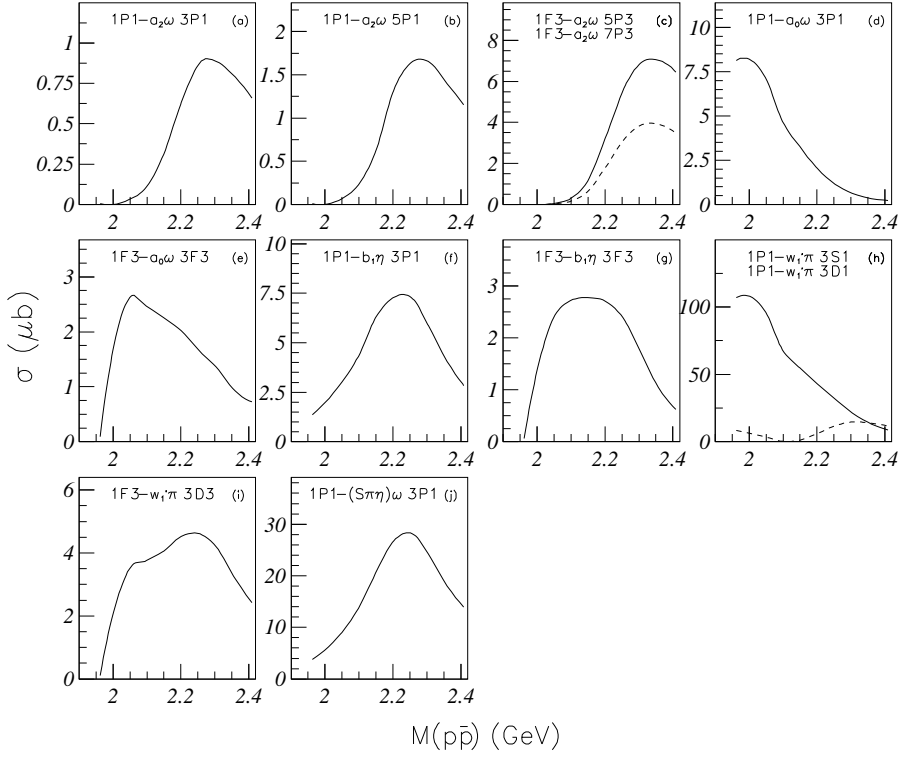,width=13.5cm}}
\caption{Intensities of singlet partial waves in  $\omega
\eta \pi ^0$ data; dashed lines refer to the second title.}
\end{figure}

Now consider singlet states.
The $3^{+-}$ state ($^1F_3$) at 2245 MeV is observed in 4 channels;
intensities of singlet partial waves for $\omega \eta \pi ^0$
are shown in Fig. 50.
Its presence is required in $\omega \pi$ by the $360^\circ$ loop in the
Argand diagram of Fig. 44.
However, there is a rather large error of $\pm 50$ MeV in its mass.
The lower $3^{+-}$ state is the strongest contribution to
$\omega \pi$ and deserves $4^*$ status.

The partial wave analysis requires a rather large
$1^{+-}$ intensity in the $\omega (1650)\pi$ S-wave, as shown in
the bottom left panel of Fig. 50.
It peaks above 2 GeV because of the $\ell = 1$ centrifugal barrier
for production and also the increasing phase space for production
of $\omega (1650)$.
It is identified directly by the peak in the $\omega \eta$
mass distribution.
The $\omega \pi ^0$ data also require a rather strong $^1P_1$
intensity at low masses.
It cannot be eliminated without a large change of 289 in log likelihood.
The Argand loop of Fig. 44 for $\omega \pi ^0$ requires a
strong phase variation when one remembs that the amplitude must go
to zero at the $\bar pp$ threshold; this demands a low mass $1^+$
resonances, even though there is some uncertainty in its precise
mass.

A final remark is  that Matveev reports
a new $b_1'$ at  $M = 1620 \pm 15$ MeV with  $\Gamma = 280 \pm
50$ MeV at the recent Aschaffenburg conference [71].
This is included on  Fig. 3(a).

\section {$I = 0$, $C = -1$ resonances}
There are three publications of Crystal Barrel data with these
quantum numbers.
The first was for $\omega \pi ^0\pi ^0$ data with $\omega$ decaying
to $\pi ^0 \gamma$ [72].
This was followed by an analysis of $\omega \eta$ where $\omega$
again decays to $\pi ^0 \gamma$ [73].
The final combined analysis added data for both channels where
$\omega \to \pi ^+ \pi ^- \pi ^0$ [74].
The measurement of $\omega$ polarisation in these data improves
the stability of the analysis considerably, so conclusions will
be taken from that work.

\begin{figure}
\centerline{\hspace{0.2cm}
\epsfig{file=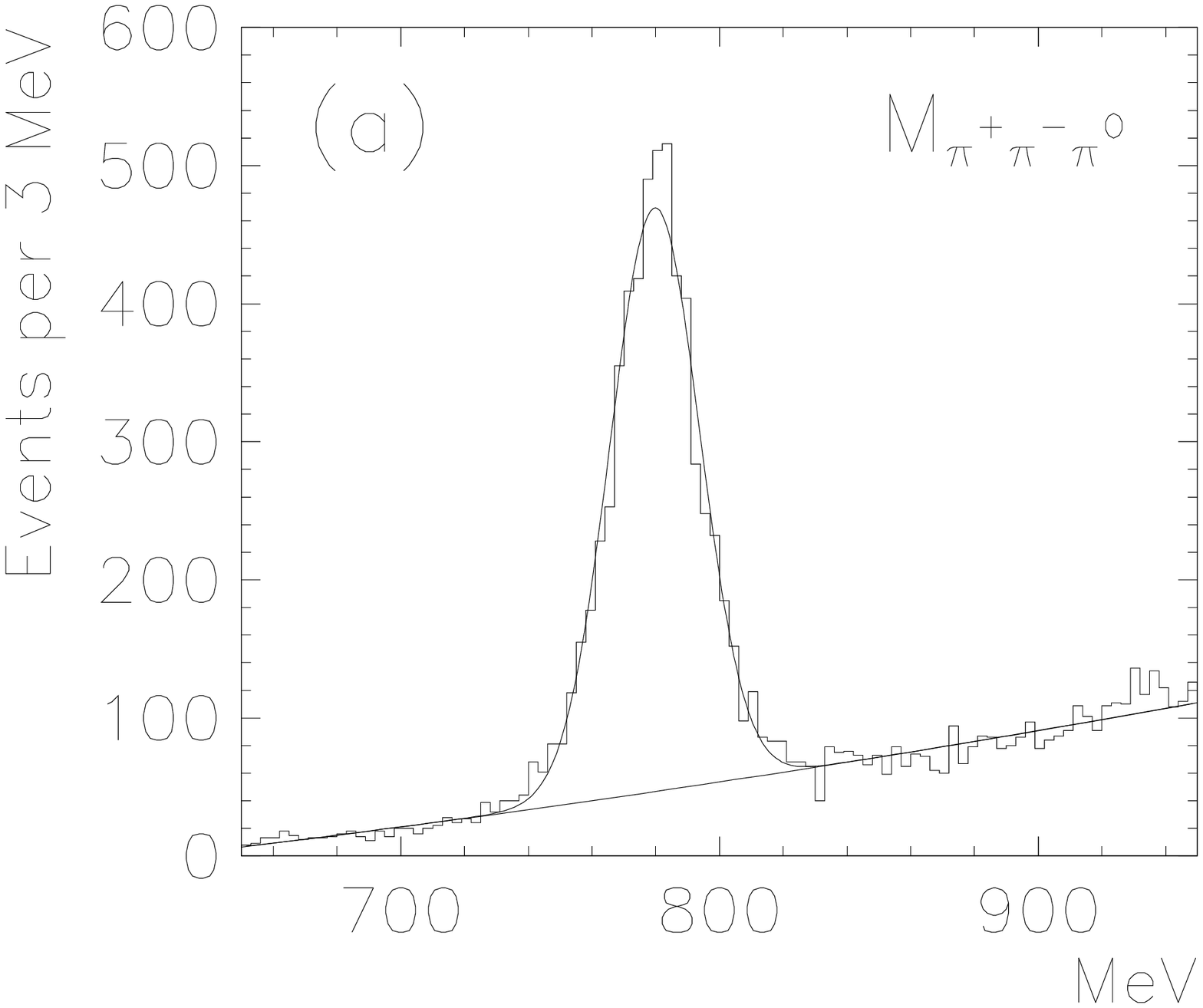,width=5.1cm}
\epsfig{file=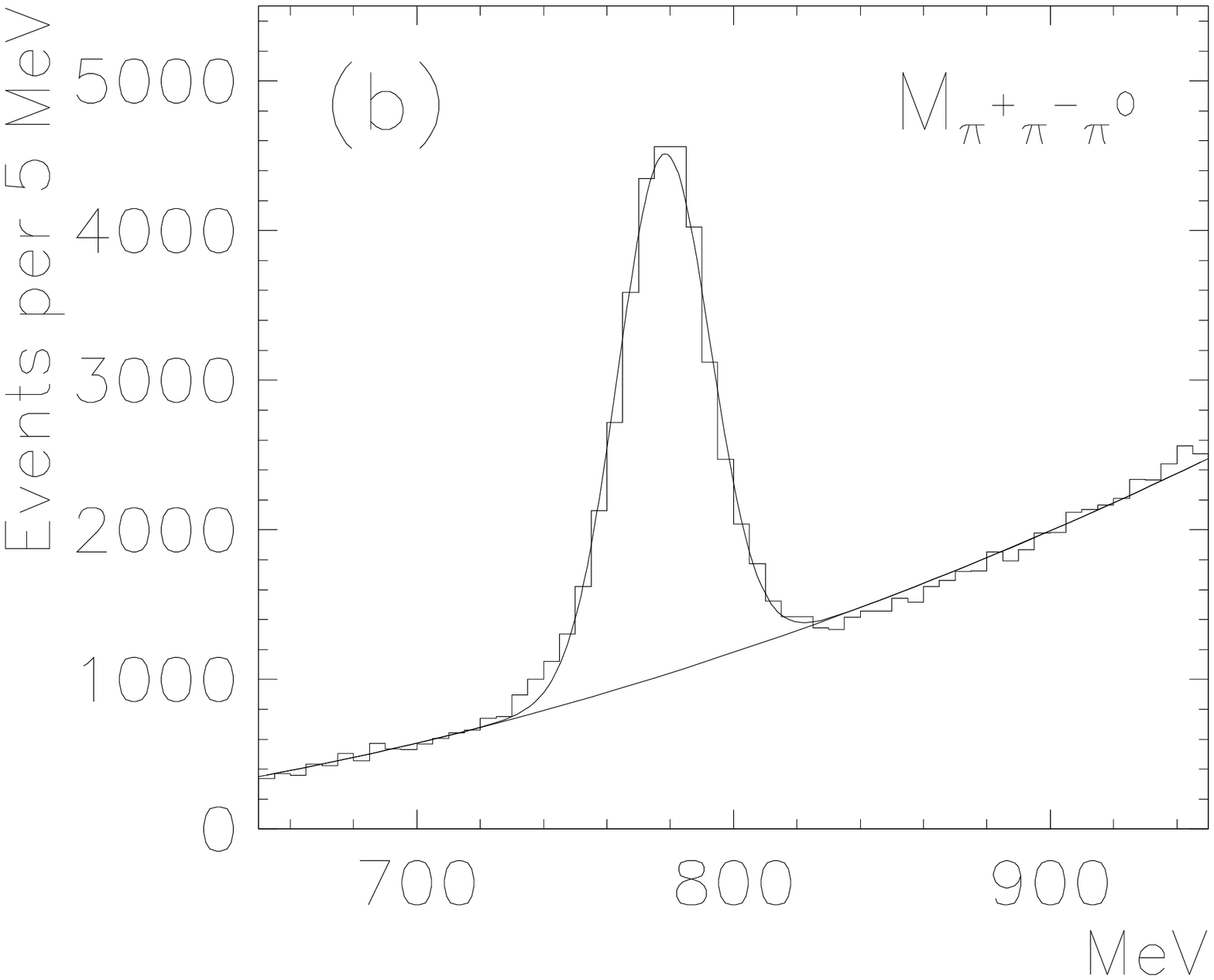,width=5.1cm}}
\vspace{-1.1cm}
\centerline{\hspace{0.3cm}
\epsfig{file=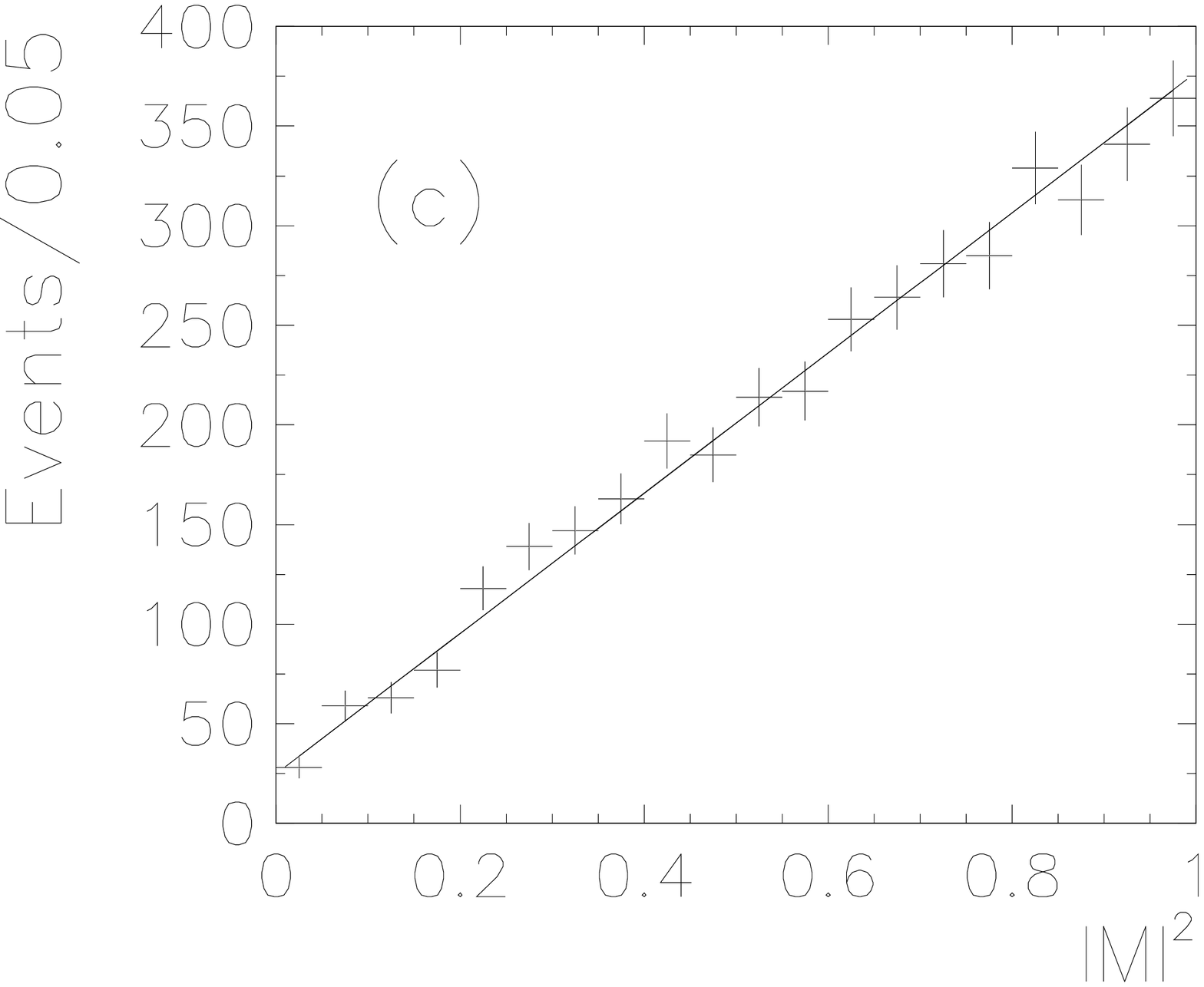,width=5.2cm}
\epsfig{file=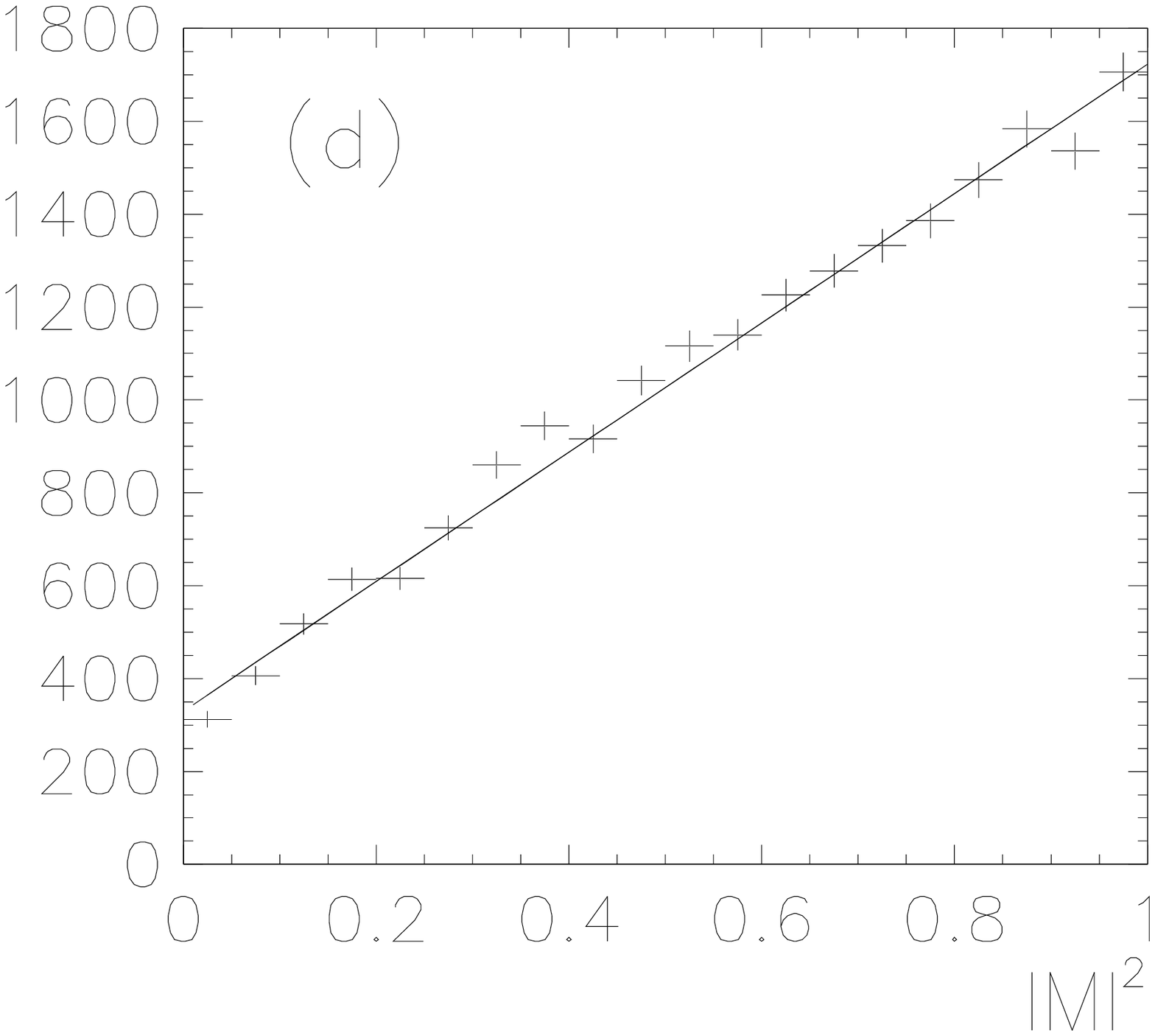,width=5.2cm}}
\vspace{-1.5cm}
\centerline{\epsfig{file=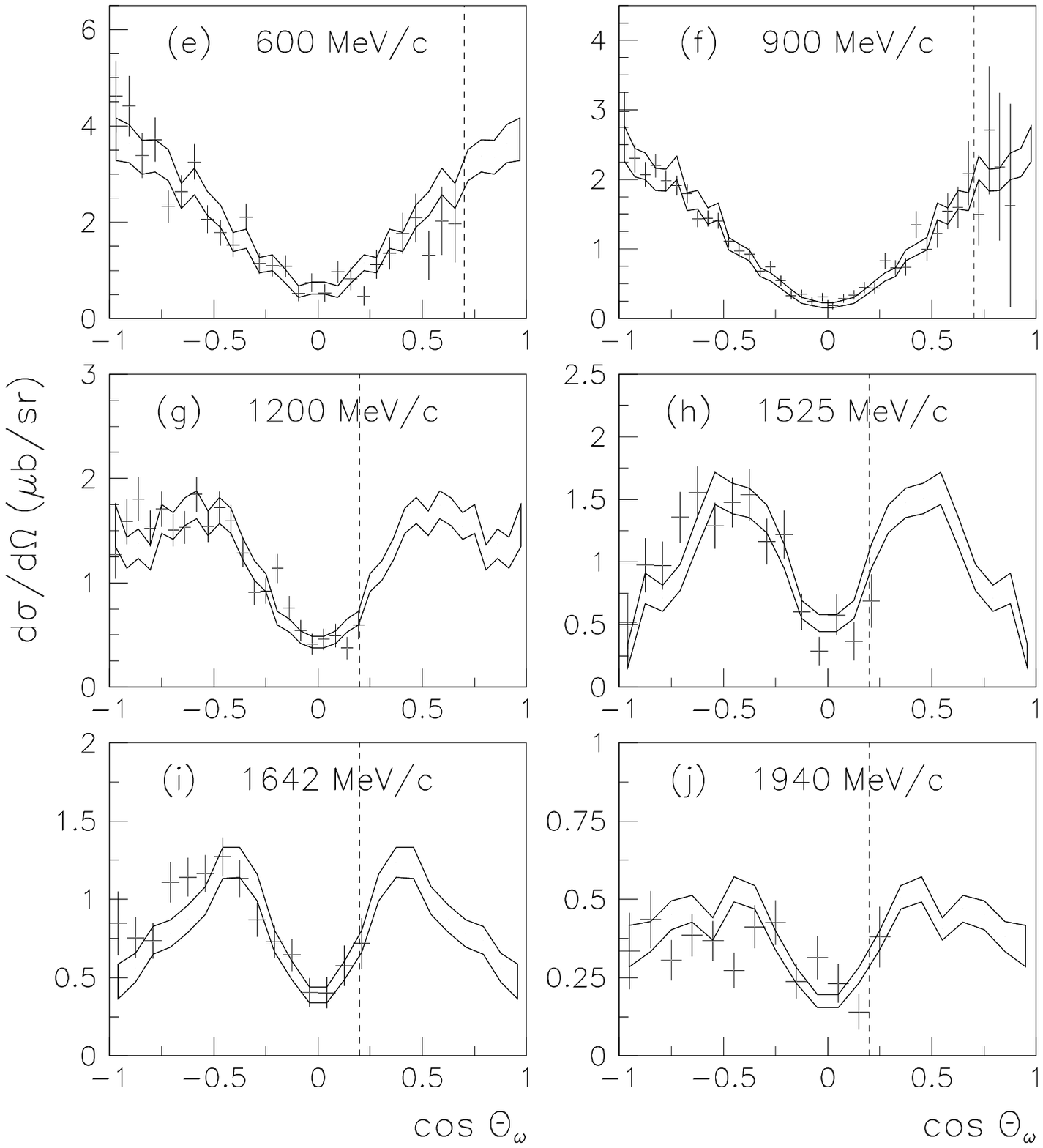,width=13cm}}
\vspace{-0.5cm}
\caption{$M(\pi ^+\pi ^- \pi ^0)$ from the data selection
for (a) $\omega \eta$ data, (b) $\omega \pi^0 \pi ^0$ at 900 MeV/c;
the number of events v. the square of the matrix element for
$\omega$ decay in (c) $\omega \eta$, (d) $\omega \pi ^0 \pi ^0$;
(e)-(j): curves show the error corridor for $\omega \eta$ differential
cross sections where $\omega$ decays to $\pi ^0 \gamma$;
they are corrected for angular acceptance; points with
errors show results for $\omega$ decays to $\pi ^+\pi ^-\pi ^0$;
the vertical dashed lines show the cut-off which has been used.}
\end{figure}
Features of the data are very similar to those for $I = 1$, $C = -1$
and need little comment.
Figs. 51(e)-(j) compare angular distributions for $\omega \eta$ in
decays via $\pi ^0 \gamma$ (shown by the corridors) and via
$\omega \to \pi ^+ \pi ^- \pi ^0$ (points with errors).
There is agreement within errors, as there was for $\omega \pi ^0$.
This checks the data selection and Monte Carlo simulations, but
with lower accuracy than for $\omega \pi ^0$ because of lower
statistics.
Fig. 52 shows vector and tensor polarisation $Re ~T_{21}$.
Other tensor polarisations are large, as for $\omega \pi ^0$, and
are shown in Ref. [74].
Again the vector polarisation is consistent with zero, indicating
the same phase coherence amongst the amplitudes as for $\omega \pi$.
\begin{figure}
\centerline{\epsfig{file=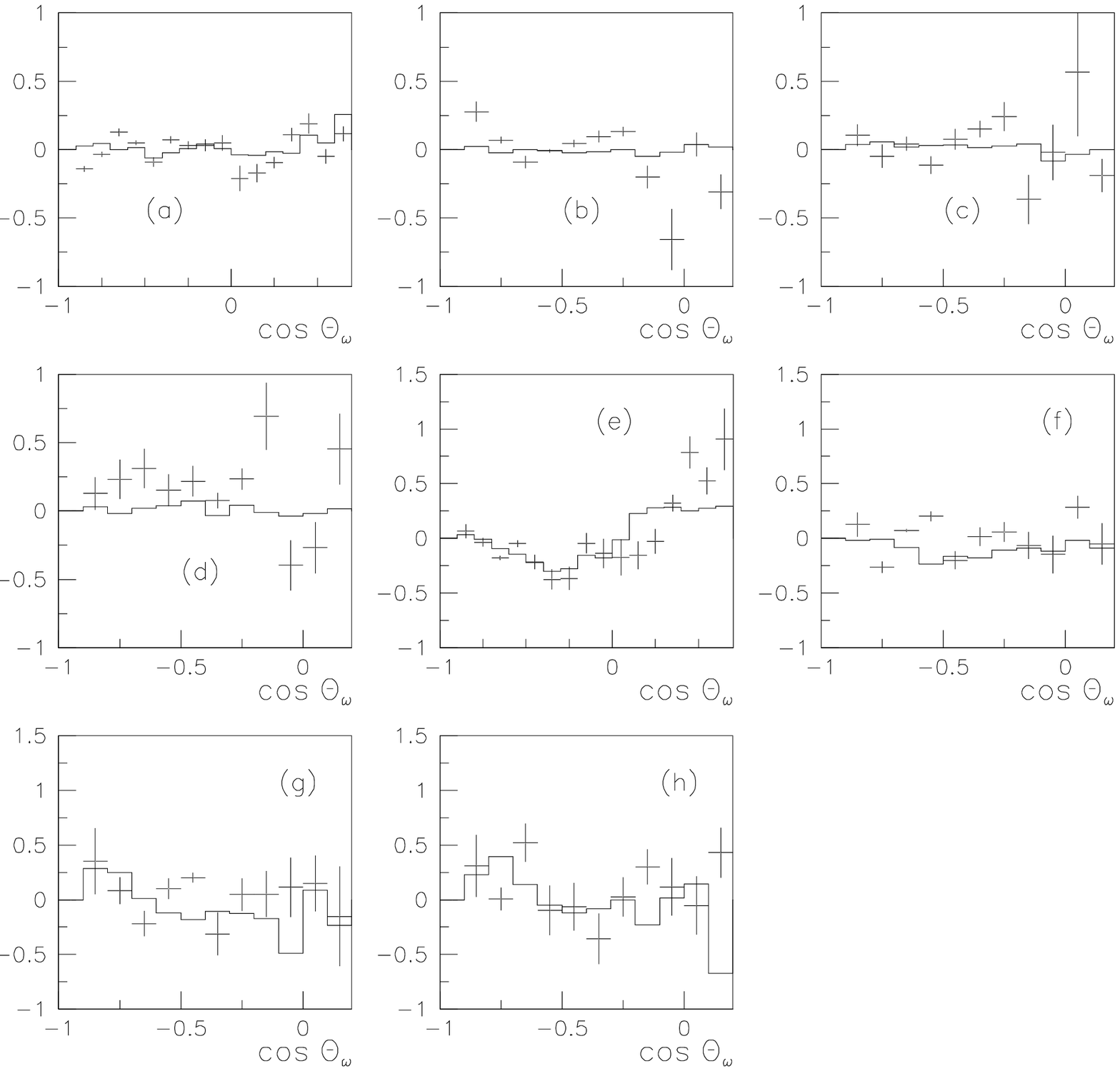,width=10cm}}
\caption{ Vector polarisation $P_y$ for $\omega \eta$ data at (a) 900,
(b) 1200, (c) 1525 and (d) 1940 MeV/c compared with the partial wave
fit (histogram); (e)--(h) $Re~T_{21}$ at the same momenta. }
\end{figure}

\newpage
\begin{figure}
\centerline{\epsfig{file=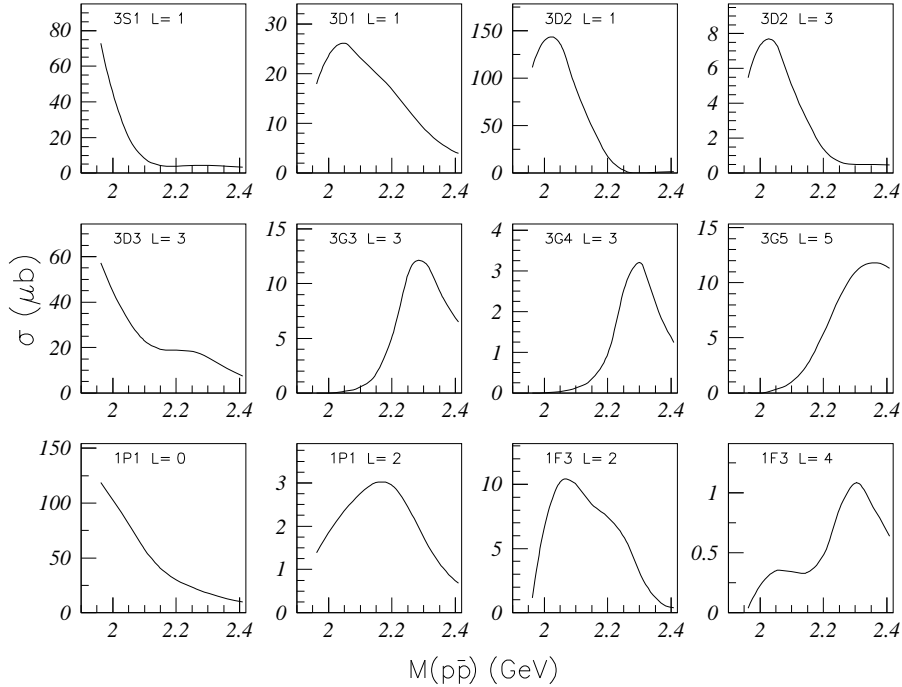,width=13.5cm}}
\caption{Intensities of partial waves fitted to $\omega \eta$ data;
$L$ is the orbital angular momentum in the decay.}
\end{figure}
Fig. 53 shows intensities of partial waves fitted to $\omega \eta$
and Fig. 53 shows Argand diagrams.
Fig. 55 shows intensities of the large partial waves fitted to
$\omega \pi ^0 \pi ^0$ data.

\newpage
\begin{figure}
\centerline{\epsfig{file=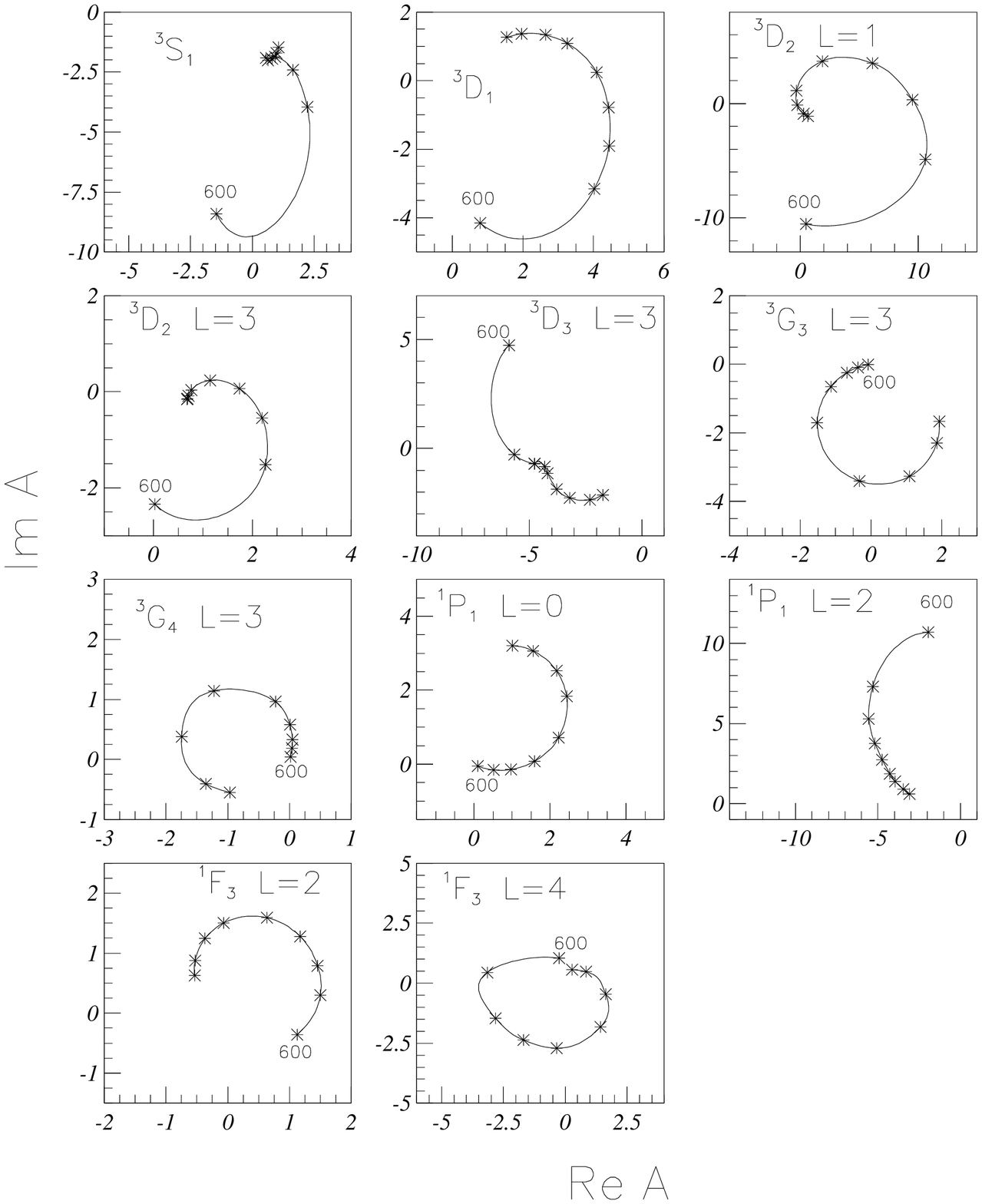,width=10cm}}
\caption{Argand diagrams for partial waves fitted to $\omega \eta$
in the combined analysis with $\omega \pi ^0  \pi ^0$.
Crosses show beam momenta 600, 900, 1050, 1200, 1350, 1525, 1642, 1800
and 1940 MeV/c;
$L$ is the orbital angular momentum in the $\omega \eta$ channel.}
\end{figure}

\newpage
\begin{figure}
\centerline{\epsfig{file=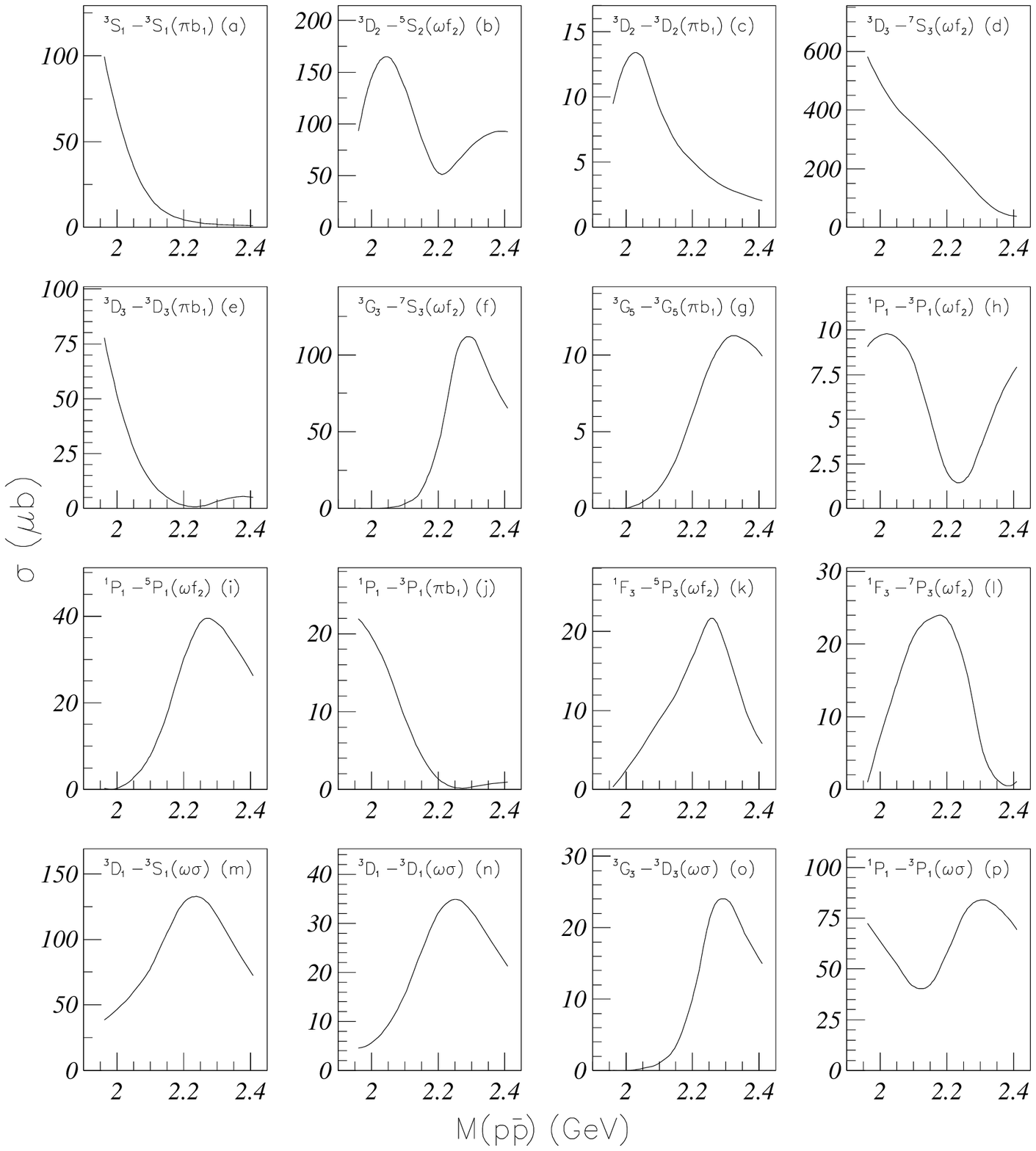,width=12cm}}
\caption{Intensities of the large partial waves fitted to $\omega \pi
^0\pi ^0$ data.}
\end{figure}

\newpage
The analysis of this family of $I = 0$, $C = -1$ states is the
weakest of all four families.
Statistics for the $\omega \eta$ channel are only $\sim 5000$
events per momentum.
In the $\omega \pi ^0 \pi ^0$ data, there is a strong $f_2(1270)$
signal but it is superimposed on an  $\omega \sigma$ physics
background. Uncertainties in parametrising the broad $\pi \pi$
S-wave amplitude above 1 GeV introduce possible systematic errors
into this physics background and, via interferences, into the
determination of resonance masses in the $\omega f_2$ channel.
However, without the $\omega \sigma$ channel, the data cannot be
fitted adequately, so there is no choice but to include it.

\begin {table}
\begin {center}
\begin{tabular}{cccccccc}
$J^{PC}$ & * &  Mass $M$ & Width $\Gamma$ & $r$ & $\Delta S$
& $\Delta S$ & Observed \\
      &  & (MeV)  & (MeV) &  & $(\omega \pi )$
      & $(\omega \eta \pi )$ & channels \\ \hline
$5^{--}$ & $2^*$ & $2250 \pm 70$ &$ 320 \pm 95$ & (0)  &
    677 &  244 &  $\omega \eta$, $\pi b_1$ \\
$4^{--}$ & $1^*$ & $2250 \pm 30$ & $ 150 \pm 50$ & (0)          &
     70  &  -  & $\omega \eta$ only \\
$3^{--}$ & $2/3^*$ & $2255 \pm 15$ & $175 \pm 30$ & $\sim 50$     &
    145  & 1436 & $\omega \eta$, $\omega f_2$, ($\omega \sigma$)  \\
$3^{--}$ & $2/3^*$ & $2285 \pm 60$ & $230 \pm 40$ & $1.4 \pm 1.0$ &
     98  & 1156 & $\omega \eta$, $\omega f_2$, ($\pi b_1$) \\
$3^{--}$ & $3^*$ & $1945 \pm 20$  & $115 \pm 22$ & $0.0 \pm 1.0$ &
    752 & 5793 &  $\omega \eta$, $\omega f_2$, $\pi b_1$\\
$2^{--}$ & $2^*$ & $2195 \pm 30$ & $225 \pm 40 $ & $0.28 \pm 0.59$ &
    160  & 356 & $\omega \eta$, $\omega f_2$ \\
$2^{--}$ & $3^*$ & $1975 \pm 20$  & $175 \pm 25$ & $0.60 \pm 0.10$ &
    948  & 1203 & $\omega \eta$, $\omega f_2$, $\pi b_1$ \\
$1^{--}$ & $1^*$ & $2205 \pm 30$ & $350 \pm 90$ & $-3.0 \pm 1.8$  &
    180  & 1410 & ($\omega \eta$), $\omega \sigma$ \\
$1^{--}$ & $3^*$ & $1960 \pm 25$ & $195 \pm 60$ & $2.4 \pm 0.45$  &
   1258  & 393 & $\omega \eta$, $\pi b_1$ \\\hline
$3^{+-}$ & $3^*$ & $2275 \pm 25$ & $190 \pm 45$ & -$0.74 \pm 0.60$ &
    253  & 140 & $\omega \eta $, $\omega f_2$ \\
$3^{+-}$ & $2^*$ & $2025 \pm 20$ & $145 \pm 30$ & $0.60 \pm 0.49$ &
    341 &  234 & $\omega \eta$, $\omega f_2$ \\
$1^{+-}$ & $2^*$ & $2215 \pm 40$ & $325 \pm 55$ & $4.45 \pm 1.16$ &
    272  & 1020 & $\omega \eta$, $\omega f_2$, ($\omega \sigma$) \\
$1^{+-}$ & $2^*$ & $1965 \pm 45 $ & $345 \pm 75$  & -$0.20 \pm 0.17$ &
   1808   & 700 & $\omega \eta$, $\omega f_2$, $\pi b_1$,
   ($\omega \sigma$) \\\hline
\end{tabular}
\caption {Resonance parameters from a combined fit to $\omega \eta$ and
$\omega \pi ^0 \pi ^0$ using both
$\omega \to \pi ^0 \gamma$ and $\omega \to \pi ^+\pi ^- \pi ^0$ decays.
Values in parentheses are fixed.
Values of $r$ are ratios of coupling constants
$g_{J+1}/g_{J-1}$ for coupling to $\bar pp$ or $\omega \eta$.
Columns 6 and 7 show changes in
$S = $ log likelihood when each resonance is omitted and others
are re-optimised.}
\end{center}
\end{table}
There is evidence in the final analysis for all expected $q\bar q$
states except for two $1^-$. However, several of them have
a $2^*$ rating because of (i) low statistics for $\omega \eta$,
(ii) the difficulties caused by the $\omega\sigma$ physics background
in $\omega \pi ^0 \pi ^0$, (iii) the lack of an accepted
interferometer to establish phase variations.
Table 8 shows fitted resonances and their star ratings.
None can be classified as $4^*$ because of the limited number of
decay channels.

It is best to start with the low mass range and triplet states.
The $^3D_3$ partial wave in $\omega \pi ^0 \pi ^0$ data of
Fig. 55(d) is very strong in the $\omega f_2$ final state
down to the lowest beam momentum.
This
cannot be confused with a threshold effect in $^3S_1$ because $^3S_1$
and $^3D_3$ give rise to quite different tensor polarisations of the
$f_2$, hence distinctive decay angular distributions. For $\omega \to
\pi ^+ \pi ^- \pi ^0$ decays, the correlation between the normal to the
$\omega$ decay plane and the decay angular distribution of the $f_2$ is
very sensitive to the different Clebsch-Gordan coefficients for
$^3D_3$, $^3D_2$, $^3D_1$ and  $^3S_1$ amplitudes. The conclusion is
that the selection of the $^3D_3$ initial state is very distinctive.
This can be demonstrated by running a crude initial
analysis at low momenta with just one partial wave at a time.
A small  but significant $b_1 \pi$ D-wave amplitude provides further
evidence  for the $^3D_3$ initial state.

Results for $\omega \eta$ are similar.
Again the normal to the $\omega $ decay plane in $\omega \to
\pi ^+\pi ^-\pi ^0$ decays provides good separation between
$^3D_3$, $^3D_2$ and $^3D_1$.
The momentum dependence separates $\bar pp$ S and D-waves.
The differential cross sections for $\omega \eta$
look like $(1 + 2\cos ^2 \theta )$ after correction for acceptanc;
here $\theta$ is the production angle.
The strong $\cos ^2\theta$ term
requires initial S or D-states (and final P-states).
The data cannot be reproduced by singlet partial waves.
The $^1P_1$ amplitude gives the isotropic term.
In the earliest  analysis of Ref. [73], the $^1F_3$ amplitude was
negligible.

Argand diagrams of Fig. 54 for $^3D_3$,
$^3D_2$ and $^3D_1$ $\omega \eta$ final states
require phase variations approaching $180^\circ$ between the
$\bar pp$ threshold and 2100 MeV (a beam momentum of 1050 MeV/c).
These partial waves dominate the $\omega \eta$ differential cross
section up to 1200 MeV/c.
The rapid decrease with momentum of the $\omega \eta$ cross section
forces all of the $^3D$ partial waves to agree with quite narrow
widths: 115 MeV for $^3D_3$, 175 MeV for $^3D_2$ and 195 MeV for
$^3D_1$. It is not possible to make them wider and fit the  observed
cross sections.
Observed phase relations between $^3D_3$,  $^3D_2$ and $^3D_1$
make resonances in all three partial waves unavoidable in the same
mass range.
The $^3S_1$ partial wave is difficult to separate from a $\bar pp$
threshold effect, but its phase falls naturally into line with that
observed for $^3D_1$.

In view of the well established $I = 1$ $^3D_3$ state at 1982
MeV, these results are not surprising.

An isotropic component of the $\omega \eta$ differential
cross section is fitted naturally by $^1P_1$; again the
normal to the $\omega$ decay plane in $\omega \to \pi ^+\pi ^-
\pi ^0$ isolates this partial wave.
A difficulty here is that there is no interferometer to
define the phase of the $^1P_1$ amplitude. However, its
mass dependence fits naturally to a Breit-Wigner amplitude
multiplied by the factor $\rho _{\bar pp}/p$  of
eqn. (3), section 4.1.
The charged decay mode of the $\omega$ also successfully locates
a low mass $^1F_3$ state at 2025 MeV; this state evaded
detection in the analysis of $\omega \to \pi ^0\gamma$.

However, it is clear that all these resonances are correlated
in masses and widths.
There is no established resonance to act as interferometer,
so resonance masses and widths are derived from the simple
assumption of eqn. (3) for Breit-Wigner resonances.
Results for masses come out in fair
agreement with $I = 1$, $C = -1$; there the $\rho _3(1982)$
is derived precisely from interferences with the known $f_4(2018)$.
In view of the assumptions in the $I=0$, $C = -1$ analysis, it is not
surprising that $I = 0$ S D-states have masses
which are systematically slightly lower than $I = 1$, $C = -1$.
This may not be a real effect.

The conclusion is that all of the strong $1^-$, $2^-$ and
$3^-$ states at low masses make a pattern consistent with $I=1$; they
deserve $3^*$ states for their existence, subject to the reservation
that there may well be a systematic error in  their mass scale. The
weak $^1P_1 $ and $^1F_3$ states at low mass  are classified  as
$2^*$.

Next consider high mass states.
The Argand loops of Fig. 54 for $^3D_1$, $^3D_2$ and $^3D_3$
all move through approximately $360^\circ$ between the
$\bar pp$ threshold and the highest mass, 2410 MeV.
This is consistent with a group of higher $1^-$, $2^-$ and
$3^-$ states. The structure in the Argand loop for
$^3D_3$ indicates two separate states. This structure is
apparent also in the $^3D_3$ intensity of Fig. 53.
Intensities of $^3D_1$ and $^3D_3$ partial waves drop to
low values at high masses in Fig. 53, but the loops
in Argand diagrams are distinctive.

The situation concerning the $3^-$ amplitudes at high mass
is somewhat surprising. The measured tensor polarisation in
the $\omega f_2$ final state in $\omega \pi ^0 \pi ^0$ data
appears to require a strong peak in Fig. 55(g).
The data fit naturally to a strong $^3G_3$ state at 2255
MeV and a weaker $^3D_3$ state at 2285 MeV.
However, this result should be treated with caution.
The polarisation of the $\omega$ in the final state
is sensitive to the orbital angular momentum $L$ in
the final state. However, it is much less sensitive to the
angular momentum $\ell$ in the production process;
$^3G_3$ and $^3D_3$ states differ only through Clebsch-Gordan
coefficients for coupling of $\bar pp$ helicity states.
To be sure of the separation between $^3D_3$ and $^3G_3$,
data from a polarised target are needed and are not
available at present.
A balanced view of the results is that there is evidence for
both $^3G_3$ and $^3D_3$ states, but they are not well
separated by present data.
In Table 8, they are shown as $2^*/3^*$; the $3^*$ refers
to their existence, but the $2^*$ refers to the imprecise
identification of masses.

The upper $2^{--}$ state is weak in $\omega \eta$ and determined
only through its interference with other neighbouring states
and with the lower $2^-$ state. It is therefore only $2^*$.

The upper $^3D_1$ state at 2205 MeV appears weakly in $\omega \eta$
and is determined mostly by $\omega \sigma$ data. Because of the
uncertainties in parametrising the broad $\sigma$, it can be
classified only as $1^*$.
For $I = 1$, there are two neighbouring states at 2110 and 2265 MeV.
This suggests that the $I = 0$ state may be a blurred combination
of two states.

The $5^-$ state at 2250 MeV appears quite strongly in both
$\omega \eta$ and $\pi b_1$, Fig. 55(g), but the mass has a
rather large error, so it only deserves $2^*$ status.
The $4^-$ state appears only in $\omega \eta$ and is $1^*$.

Turning to singlet states, the $^1F_3$ resonance at 2275 MeV appears
strongly in $\omega f_2$ and is also required by $\omega \eta$ with
$L = 4$.
The Argand diagram for $^1F_3$ with $L = 4$ completes a full
$360^\circ$ loop which can only be fitted with two $3^+$ states.
Since the mass is well determined, this state is given $3^*$ status.
The upper $1^+$ state appears weakly in $\omega \eta$ and is required by
$\omega f_2$. It deserves $2^*$ status.

In summary, the pattern of $I = 0$, $C = -1$ states is very similar
to that of $I = 1$, but is less well defined because of (i) the
lack of data from a polarised target, (ii) the uncertainties in fitting
the $\omega \sigma$ channel. Several of the states lying at the bottom
of the mass range may be subject to systematic errors in mass. That
situation is better for $I = 1$ because of the strong $3^-$ state at 1982
MeV. This state is accurately determined by $\pi ^- \pi ^+$ data and
then acts as an interferometer with other partial waves. No such
interferometer is available for $I = 0$, $C = -1$.

\section {Comments on systematics of $q\bar q$ resonances}
The first obvious comment is that there is no proof that these
are $q\bar q$ states rather then hybrids.
The regularity of the spectra suggests an origin common with
accepted $q\bar q$ states at low mass.
There is firm evidence for just one hybrid with
exotic quantum numbers $J^{PC} = 1^{-+}$ from the E852 collaboration
at 1593 MeV [75].
There is also evidence for structure with the same quantum numbers
near 1400 MeV from the E852 collaboration and
from Crystal Barrel data at rest [76,77].
Hybrids are expected closer to the 1800 MeV mass range.
So it seems possible that the 1400 MeV structure is connected with the
opening of the first S-wave inelastic thresholds with these quantum
numbers: $b_1(1235)\pi$ and $f_1(1285)\pi$.
Further hybrids are expected in the mass range above 1800 MeV,
but they are predicted to be broad and may be difficult to locate.

\subsection {Slopes of trajectories}
Figs. 1--3 of Section 1 summarise the observed spectra. Straight lines
are drawn through them.
An obvious comment is that this may well be an approximation.
It is to be expected, for example, that the opening of strong
thresholds may perturb the precise masses of resonances.
An obvious example is $f_2(1565)$, which sits precisely
at the $\omega \omega$ threshold and couples strongly to that
channel [78]. Therefore Figs. 1--3 should be taken as a
general indication of the systematics.
The straight lines do have predictive power.
For example,  the $b_1'(1620)$ was actually predicted very
close to this mass from the known mass of $b_1(1230)$
and was found just where expected [71]. A second similar
example is the recently discovered $h_1(1595)$.

There are still some missing states.
In the mass range 1600-1800 MeV, an $f_1$ is expected around
1670 MeV and an $\eta $  close to
1680 MeV. These low spin states are difficult to identify
in production processes in the presence of strong signals of
higher spins.
The $^3D_2$ states of both isospins are also missing around
1660 MeV.
The $0^+$ sector for $I = 1$ is poorly represented at present.
Further states are expected around 1760 and 2380 MeV; the
latter is right at the top of the Crystal Barrel range.
Also the whole situation concerning $^3S_1$ states is
obscure at present.

Table 9 summarises the slopes of $I = 0$, $C = +1$ resonances,
which are the best established; it also includes a very well
determined slope from the $I = 1$, $J^{PC} = 3^{--}$ family,
which have accurately known masses.
The mean slope is $1.143 \pm 0.013$ GeV$^2$.
At high masses, it is expected that spin-orbit, tensor and
spin-spin splitting will decrease, since these all arise from
short-range $1/r^2$ effects.
Therefore it is to be expected that slopes will become parallel
for high excitations, where there are many nodes in the
radial wave functions.
At the lowest masses, the $\pi$, $\eta$ and $\eta '$ have been
omitted from the trajectories because of the likelihood that
their masses are affected by the instanton interaction.

\begin {table}
\begin {center}
\begin{tabular}{cccccccc} $J^P$ &
Slope\\
      & (GeV$^2$) \\\hline
$4^{++}$ & $1.139 \pm 0.037$ \\
$3^{++}$ & $1.107 \pm 0.078$ \\
$^3F_2$  & $1.253 \pm 0.066$ \\
$^3P_2$  & $1.131 \pm 0.024$ \\
$2^{-+}$ & $1.217 \pm 0.055$ \\
$1^{++}$ & $1.120 \pm 0.027$ \\
$0^{++}$ & $1.164 \pm 0.041$ \\
$0^{-+}$ & $1.183 \pm 0.028$ \\
$3^{--}$ & $1.099 \pm 0.029$ \\\hline
\end{tabular}
\caption {Slopes of trajectories.}
\end{center}
\end{table}

For $2^+$ states, it is rather obvious from Fig. 1 that
$^3F_2$ states lie systematically higher than $^3P_2$.
A splitting of $\sim 80$ MeV is observed.
This is natural. For the F-states, there is a centrifugal
barrier at short range. The need to overcome this barrier
by attraction from the confining potential pushes the
masses of F-states up.
According to this argument, masses of $D$ states should
lie $\sim 40$ MeV below F-states.
The mass of the best established D-state, $\rho _3(1982)$ is
indeed $36 \pm  15$ MeV below that of $f_4(2018)$.

The centroids of $^3F_2$, $^3F_3$ and $^3F_4$ states with $I = 0$,
weighted by their multiplicities, are at
$2292 \pm 15$ and $2024 \pm 8$ MeV.
The $^1F_3$ states with $I = 0$ are at $2275 \pm 25$ MeV and
$2025 \pm 20$ MeV. Within the sizable errors, these are consistent
with the absence of spin-spin splitting.
The well established $\eta _2 (2267)$ lies $25 \pm 21$ MeV
below the centroid of the triplet F-states; and the
somewhat less well determined $\eta _2(2005)$ lies
$19 \pm 18$ MeV below the centroid of the F-states.
However, there is the possibility that the $\eta _2(2005)$
may be moved upwards by the well known level repulsion
by $\eta _2(1880)$.
In any case, these determinations are not as definitive as the
difference measured by $\rho _3(1982)$.

One would then expect G-states
to lie $\sim$ 40 MeV above F-states. The situation is not clear here
because most G-states are located with rather large errors.
In the absence of tensor and spin-orbit splitting,
$^3F_4$, $^3F_3$ and $^3F_2$ states are degenerate in mass.

Tensor and spin-orbit splitting may be assessed from the relations [79]
\begin {eqnarray}
\Delta M_{LS} &=& [-20M(^3F_2) - 7M(^3F_3) + 27M(^3F_4)]/54, \\
\Delta M_{T} &=& [-4M(^3F_2) + 7M(^3F_3) - 3M(^3F_4)]/14.
\end {eqnarray}
For both multiplets with $I = 0$, $C = +1$, $^3F_3$ states lie highest
in mass, requiring significant tensor splitting. This splitting is in
the sense predicted by one-gluon exchange. Using a linear confining
potential plus one-gluon exchange, Godfrey and Isgur [80] predict
$\delta M_T = 9$ MeV for the lower multiplet.
From Table 5, one finds $\delta M_T = 20 \pm 4.5$ MeV for the lower
multiplet and $7.1 \pm 7.3 $ MeV for the upper one; errors take into account
observed correlations in fitted masses.
Because the same centrifugal
barrier is used for $^3F_4$, $^3F_3$ and $^3F_2$ states, there is almost
no correlation of $\Delta M_T$ or $\Delta M_{LS}$ with the radius of
the barrier. The observed spin-orbit splittings are $\delta M_{LS} =
2.4 \pm 4.4$ MeV for the lower multiplet and $-6.3  \pm 5.1$ MeV for
the upper one; these average approximately to zero.

Glozman [81,82] has associated the small spin splitting at high
masses with Chiral Symmetry restoration and the string
picture of hadrons. He argues that this leads to parity doubling
high in the spectrum and cites a number of examples.

It is well known that Regge trajectories arise naturally
from the string picture of hadrons: quarks tied together
by a flux-tube of gluons.
The argument, originally due to Nambu, is in outline as
follows [83].
Suppose the mass originates from a uniform flux tube
joining two gluons.
Suppose also that for light quarks the system rotates
and the quarks move at the speed of light at radius $R$.
Then the mass associated with the flux tube is
\begin {equation}
M = 2 \int ^R_0 \frac {\sigma~dr}{\sqrt {1 - v^2(r)}}
  = 2 \int ^R_0 \frac {\sigma~dr}{\sqrt {1 - r^2/^2}}
  = \pi \sigma R .
\end {equation}
The angular momentum is given by
\begin {equation}
J = 2 \int ^R_0 \frac {\sigma rv(r)~dr}{\sqrt {1 - v^2(r)}}
  = \frac {2}{R} \int ^R_0 \frac {\sigma r^2~dr}{\sqrt {1 - r^2/R^2}}
  = \pi \sigma R^2/8.
\end {equation}
Then $ J = M^2/2\pi \sigma $,
where $\sigma$ is the density per unit length.

Fig. 56 shows Regge trajectories constucted from
mesons quoted in this review. On Regge trajectories, $M^2$ is plotted
against $J$. Here the centrifugal barrier comes into play, so the
slopes of these trajectories will be slightly larger than for
those plotted against radial excitation.
It is found to be 1.167 GeV$^2$.
\begin{figure}
\centerline{\epsfig{file=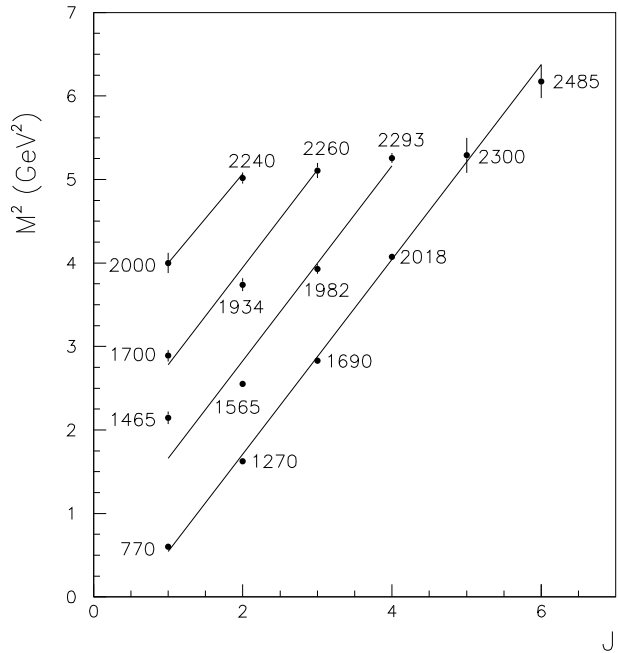,width=10cm}}
\caption{Regge trajectories. }
\end{figure}

It is not clear how to assign $1^-$ states to these trajectories.
The $\rho (1450)$ does not seem to fit naturally as the first
radial excitation of the $\rho (770)$.
It has been suggested by two groups that the mass of the $\rho '$
should be lower, $\sim 1.29$ GeV [84,85];
this has been a well known controversy for some years.
The lower mass would fit much better to a trajectory through the well
defined $\rho (1982)$.
Then $\rho (1700)$ and $\rho (2000)$ fit on to the next two
trajectories.
However, the $\rho (1700)$ is approximately  degenerate with
$^3D_3$ and is often taken to be the $^3D_1$ state.
The  $\rho (2000)$ is found experimentally to have a significant
$^3D_1$ component, but could be a mixed state.
The $1^{--}$ states present a dilemma.

Better data are needed for $1^{--}$ states.
A way forward would be to study diffractive dissociation of
photons to distinctive channels like $b_1(1235)\pi$.
This decay can go via $L = 0$ or 2.
It is likely that matrix elements for these two decays
will be distinctively different from initial states $^3D_1$
and $^3S_1$, because of the D-wave tied up in the $^3D_1$
state. At higher energies, decays to $\rho \sigma$ and
$\rho f_2(1270)$ are likely to throw light on the D-wave
content of the initial state.
If the luxury of a linearly polarised photon beam is available,
D-wave amplitudes will appear directly via azimuthal dependence of
the form $\cos \phi$ and $\cos 2\phi$ with repect to the
plane of polarisation. This arises through the
partial wave decomposition
$$ |1,1> = \sqrt {\frac{6}{10}}|2,2>_L|1,-1>_s
                  -\sqrt {\frac {3}{10}}|2,1>_L|1,0>_s
                  +\sqrt {\frac {3}{10}}|2,0>_L|1,1>_s $$
A linearly polarised photon is a superposition of initial states
$|1,1>$ and $|1,-1>$ and $\phi$ dependent terms arise in the
differential cross section from the real part of the interference
between these components.

\section {Glueballs}
The latest Lattice QCD calculations of Morningstar and Peardon [86]
predict a sequence of glueballs beginning with $J^{PC}= 0^{++}$ at
about 1600 MeV, $2^{++}$ around 2100 MeV, then $0^{-+}$ and
a second $0^{++}$ close by in mass.
There are `extra' states which do not fit naturally into $q\bar q$
spectroscopy at 1500, 1980, 2190 and 2100 MeV with just these
quantum numbers. All except the $f_2(1980)$ appear prominently
in $J/\Psi$ radiative decays, the classic place to look for
glueballs; and there is tentative evidence for the $f_2(1980)$ too.

Lattice QCD calculations suffer from a scale error of $\sim (10-15)\%$,
so one should be prepared for some overall shift in mass scale.
Also the Lattice calculations are presently done in the quenched
approximation, where coupling to $q\bar q$ states is omitted.
However, mass {\it ratios} are predicted with smaller errors.
Table 10 compares mass ratios of the four candidate glueballs with
Lattice predictions. There is remarkable agreement, suggesting
this is not an accidental coincidence.
This coincidence was pointed out by Bugg, Peardon and Zou [87].
Here the properties of the candidate glueballs will be reviewed
in detail.
\begin{table}[htb]
\begin {center}
\begin{tabular}{cccc}
\hline
Ratio & Prediction & Experiment\\\hline
$M(2^{++})/M(0^{++})$ & 1.39(4) & 1.32(3) \\
$M(0^{-+})/M(0^{++})$ & 1.50(4) & 1.46(3)\\
$M(0^{*++})/M(0^{++})$ & 1.54(11) & 1.40(2)\\
$M(0^{-+})/M(2^{++})$ & 1.081(12) & 1.043(36)\\\hline
\end{tabular}
\caption{Glueball mass ratios predicted by Morningstar and
Peardon [86]
compared with experimental candidates described in the text;
errors are in parentheses.}
\end {center}
\end{table}

What is clear immediately is that the candidate
states do not decay flavour blind, except perhaps for the $0^{-+}$
state. There is obviously mixing with nearby $q\bar q$ states,
as one would expect from the fact that gluon and quark
jets develop readily at high energies. At present, this mixing
is still being unravelled.

\subsection {$f_0(1500)$ and the $q\bar q$ nonet}
The evidence for the three $0^+$ states $f_0(1370)$, $f_0(1500)$
and $a_0(1450)$ appeared in four papers of Anisovich et al.
early in 1994 [88-91].
In the first of these, it was already realised that $f_0(1370)$
and $f_0(1500)$ lie too close together for them both to be
non-strange $q\bar q$ states.
The $f_0(1370)$ was associated with a nonet containing also
$K_0(1430)$ and $a_0(1450)$; the glueball identification
of $f_0(1500)$ was pointed out.
Ref. [90] gave a detailed exposition of the $N/D$ technique
used in the analysis and Ref. [91] gave extensive details of
results. These results were confirmed 18 months later with
higher statistics and using K-matrix techniques by Amsler et al.
[92].

With the benefit of hindsight, it is clear that $f_0(1370)$
was observed as the $S$ or $\epsilon (1300)$ meson in earlier
experiments on $\pi \pi \to K\bar K$ [93-96], where a small
but definite peak occurs in the S-wave at 1300 MeV. It also
appeared in $4\pi$ in the work of Gaspero [97] on
$\bar pn \to 2\pi ^+ 3\pi ^-$.
Establishing masses and absolute branching ratios of nonet members is
important in isolating the glueball component.
The $f_0(1370)$  decays weakly to $\pi \pi$, $\eta \eta$ and $K\bar K$;
decays to $4\pi$  have a branching ratio at least a factor 5 larger
than to $\pi\pi$, as  demonstrated in the analysis of $5\pi$ data
from Obelix [98] and data  on $\eta 4\pi$ [99] and $5\pi$ [100-102]
from Crystal Barrel.
In view of the dominant $4\pi$ decay, it is unlikely to be an
$s\bar s$ state.

\begin{figure} [b]
\begin{center}
\epsfig{file=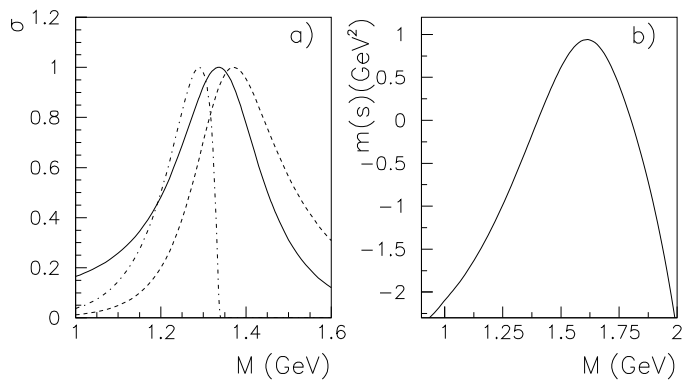,width=10cm}\
\end{center}
\caption{(a) The shape of $f_0(1370)$: in the $2\pi$ channel (chain
curve), in the $4\pi$ channel (dashed); the full curve shows a
Breit-Wigner resonance of constant width. (b) $m(s)$.}
\end{figure}
An important point in analysing data is that the peak observed in
$4\pi$ is shifted up by $\sim 75$ MeV with respect to that in $2\pi$,
because of the rapidly increase of $4\pi$ phase space with mass. This
is illustrated in Fig. 57(a) [99]. The chain curve shows the expected
$2\pi$ cross section and the dashed curve shows that expected for
$4\pi$; all curves are normalised to 1 at their peaks. Also there has
been no full analysis yet including the term $m(s)$ of the full
Breit-Wigner amplitude of eqn. (4) of Section 4.1; it arises through
dispersive effects from the $s$-dependence of $\Gamma (4\pi )$ and is
shown on Fig. 57(b). It has a large effect and consequently all claims
about the mass and width should be  treated with some reserve. The
$f_0(1370)$ does not appear significantly in the Cern-Munich data for
$\pi \pi$ elastic scattering [103,104], where its intensity depends on
$\Gamma ^2_{2\pi }$. It does appears clearly in the new BES II data on
$J/\Psi \to \phi \pi \pi$, Fig. 26, Section 5.5; the mass found there
is $1335 \pm 27$ MeV and the width $265 \pm 60$ MeV [37].

With the benefit of hindsight, it is also clear that the
$f_0(1500)$ was observed by Kalogeropoulos and collaborators [105-7].
It was also observed as $G(1590)$ by GAMS [108-112], but shifted
in mass through interference effects; they appreciated its
significance and claimed it as a glueball because of its $\eta \eta$
decay, regarded then as a touchstone for gluonic content.

Since its discovery, the $f_0(1500)$ has been identified
as a $4\pi$ peak in a re-analysis of Mark III data on
$J/\Psi \to \gamma 4\pi$ [39]. The $f_0(1770)$ and $f_0(2100)$
appear as two further peaks.
The data have been shown earlier in Fig. 25, Section 5.5, together with
E760 data which show a striking resemblance. In the original Mark III
and DM2 analyses [113,114], the peaks were not separated from a large,
slowly varying $\rho \rho$ component with $J^P = 0^-$. The
analysis of $J/\Psi \to \gamma 4\pi$ has been confirmed by BES I [115].
Fig. 58, taken from this analysis, shows the slowly varying $0^-$
signal in (a), and $0^+$ and $2^+$ components in (b) and (c); these are
determined by the spin-parity analysis and are not conspicuous in the
raw data shown as points with errors. The latest BES II data [116] have
7 times higher statistics and confirm the analysis, improving on all
individual components.
\begin{figure}
\centerline{\epsfig{file=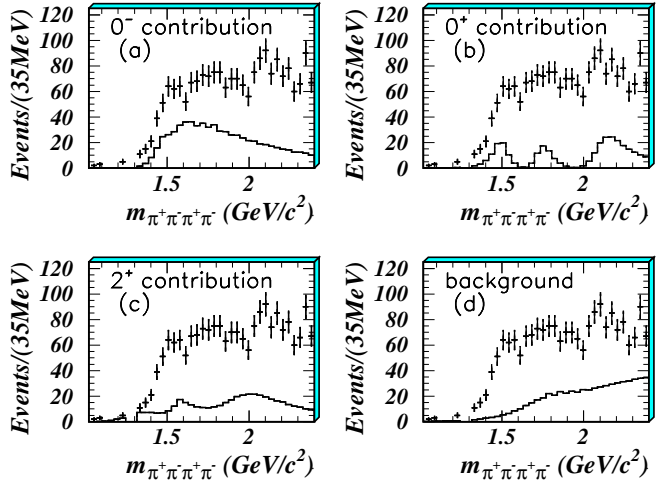,width=9cm}}\
\caption{Components from the BES I analysis of $J/\Psi \to \gamma
 (4\pi )$. }
\end{figure}

From the early analysis of $J/\Psi$ radiative decays, there was
some confusion whether the $f_J$ at 1710 MeV in $K\bar K$ has
$J = 0$ or 2. This has now been established as $J = 0$ by a
re-analysis of those data by Dunwoodie [117].
It has also been confirmed by the Omega group [118] and
from new BES II data on $J/\Psi \to \gamma K\bar K$ [119].

The $a_0(1450)$ has recently been confirmed in decays to
$\omega \rho$ in Crystal Barrel
data on $\bar pp \to \omega \pi ^+ \pi ^- \pi ^0$ at rest [120].
Decays to $\omega \rho$ are a factor $10.7 \pm 2.7$ times stronger
than to $\eta \pi$, where is was originally discovered.
The weak decay to $\eta \pi$ explains why it has not been observed
there in E852 and VES data.
Fig. 59(a) shows the percentage contribution of $a_0(1450)$ as its
mass is scanned; (b) shows the line shape in different channels;
the peak appears quite narrow in the $\omega \rho$ channel because
of the unavoidable effect of the rapidly rising $\Gamma (\rho \omega )$
v. mass.
The percentage contribution fitted to data is shown in
(c) v. width when a Breit-Wigner amplitude of constant width is
used as a test; (d) shows the Argand diagram in the $\eta \pi$ channel.
\begin{figure}
\centerline{\epsfig{file=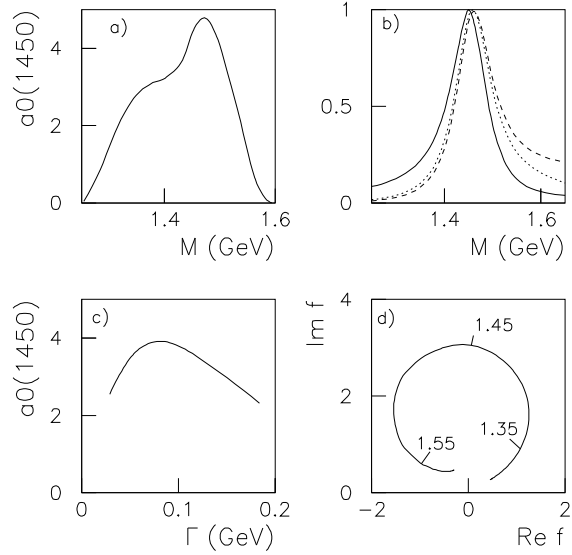,width=9cm}}\
\caption {(a) Percentage contribution of $a_0(1450)$ v. fitted mass
$M$ using a Flatt\' e formula including $\Gamma (\eta \pi)$,
$\Gamma (\rho \omega )$ and $m(s)$.
(b) The line-shapes for the Flatt\' e
amplitude itself (full curve), its contribution to $\omega \rho$
(dashed) and its contribution to $\bar pp \to (\omega \rho )\pi$
(dotted).
(c) Percentage contribution of $a_0(1450)$ v.
fitted width for a Breit-Wigner amplitude of constant width,
(d) the Argand diagram for the $\eta \pi$ channel; numerical values
of mass are shown in GeV.}
\end{figure}

In conclusion, there is available a standard nonet
made up of $f_0(1370)$, $a_0(1450)$, $K_0(1430)$ and $f_0(1710)$,
similar to known $^3P_2$ and $^3P_1$ nonets.
There is some doubt within $\sim 70$ MeV of the precise mass of
$f_0(1370)$: two-body data tend to favour masses of 1300--1335 MeV
and 4-body data favour higher masses close to 1400 MeV.
The mass of $a_0(1450)$ may be affected by its strong coupling
to the $\omega \rho$ threshold. So it would be a mistake to
make too much of mass differences amongst the lowest three members
of this nonet. The $f_0(1710)$ is rather further in mass from
$f_0(1370)$ than $f_2'(1525)$ is from $f_2(1270)$, but the
$f_0(1370)$ and $f_0(1710)$ could be pushed apart by the usual
level repulsion due to an extra state $f_0(1500)$ between them.

\subsection {Mixing of the glueball with $q\bar q$ states}
There have been many studies of the mixing, based on both the
observed masses and on branching ratios of states between
$\pi \pi$, $\eta \eta$, $\eta \eta '$ and $K\bar K$.
The much stronger coupling of $f_0(1500)$ to $\pi \pi$ than
to $K\bar K$ implies
that it is not primarily an $s\bar s$ state.
Amsler and Close [121] consider the mixing of $n\bar n$ and $s\bar s$
components into the glueball.
If the composition of the mixed state $|G>$ is written
\begin {equation}
|G> = \frac {|G_o> + \xi [\sqrt {2}|n\bar n> + \omega |s\bar s>]}
       {1 + \xi ^2(2 + \omega ^2)},
\end {equation}
then lowest order perturbation theory gives
\begin {eqnarray}
\xi    &=& <d\bar d|V|G_o>/(E_{G_o} - E_{n\bar n}) \\
\omega &=& (E_{G_o} - E_{n\bar n}/(E_{G_o} - E_{s\bar s}),
\end {eqnarray}
where E are masses of unmixed states, $G_o$ refers to the
unmixed glueball and $V$ is the interaction responsible for mixing.
If $E_{G_o}$ lies midway between the masses of unmixed $n\bar n$ and
$s\bar s$ states, $|G>$ will acquire extra $d\bar d$ component, losing
$s\bar s$ component. The state $G$ then moves towards a flavour
octet, in agreement with observation for $f_0(1500)$.
If the bare glueball were to lie close to $f_0(1710)$, it would
acquire a large $s\bar s$ component and move towards a flavour
singlet. One does not see two neighbouring $s\bar s$ states.
This suggests that the mass of $G_o>$ lies close to 1500 MeV.

Close and Kirk [122] use observed masses and WA102 branching ratios
and arrive at wave functions as follows:
\begin {eqnarray}
|f_0(1710)> &=& 0.39|G> + 0.91|s\bar s> + 0.14|n\bar n> \\
|f_0(1500)> &=& -0.69|G> + 0.37|s\bar s> - 0.62|n\bar n> \\
|f_0(1370)> &=& 0.60|G> - 0.13|s\bar s> - 0.79|n\bar n>.
\end {eqnarray}
Strohmeier-Presicek et al. [123] arrive at somewhat different
numerical coefficients, but the overall picture is similar.
All authors neglect mixing between $f_0(1710)$ and the nearby
$f_0(1770)$.
Amsler points out that the observed upper limit for $f_0(1500) \to
\gamma \gamma$ is inconsistent with a pure $q\bar q$ state [124].

In a series of papers, Anisovich and Sarantsev examine a somewhat
different scheme [125], using the K-matrix.
They take the view that K-matrix poles represent `bare', unmixed
states. The coupling constants of these poles to $\pi \pi$ etc.
then `dress' these bare states and allow for the effects of
meson decays. The virtue of their approach is that they fit
a very large volume of data on $\pi \pi$ elastic scattering,
$\pi \pi \to K\bar K$ and $\eta \eta$, plus many channels of
$\bar pp$ annihilation at rest. There is also the virtue that
bare $n\bar n$, $s\bar s$ and $|G_o>$ states are fitted directly
to data. They arrive at five K-matrix
poles from 650 to 1830 MeV and identify the lowest nonet
with T-matrix poles as $f_0(980)$, $a_0(980)$,
$K_0(1430)$ and $f_0(1370)$, all strongly mixed with a
very broad glueball with mass 1200--1600 MeV. The $f_0(1500)$,
$a_0(1450)$ and $f_0(1710)$ then fall into the first
radial excitation.

There can be no doubt that the amplitudes fitted to data
reproduce the same narrow T-channel resonances and the same branching
ratios as are found in other analyses.
The difference lies in how to interpret the broad component
which appears so conspicuously in the $\pi \pi$ S-wave
from threshold to 2 GeV and can be built into other channels
also as a broad underlying background [16].
I have had extensive and friendly discussions with Anisovich and
Sarantsev about this scheme.
It has the awkward feature that the lowest
K-matrix pole is largely $s\bar s$;
an obvious question is whether this could be replaced by the
$\sigma$ pole discussed in the next section.
Another point is that the broad structure at high mass may
arise from the opening of the strong $4\pi$ threshold and
associated dispersive effects.
This requires fitting $4\pi$ cross sections to data.
It is rather important to establish whether their scheme is
unique. An alternative is to fit
(i) $\sigma$, $\kappa$, $f_0(980)$ and $a_0(980)$ as the lowest nonet
(not necessarily $q\bar q$), (ii) the conventional
nonet of $f_0(1370)$, $a_0(1450)$, $K_0(1430)$ and $f_0(1770)$ mixed
with $f_0(1500)$, plus
(iii) $f_0(1770)$ as a member of the next nonet.
The question is whether this rearrangement
of the broad background fails to fit the data, and if so how.

In all schemes, there is agreement that the data favour the
presence of a glueball in the mass range close to 1500 MeV.
Given this and the Lattice QCD calculations, the question is:
where are the next three glueballs? This question is addressed
in the following.

\subsection {Features of Central Production Data}
In central production, two protons scatter through a small
amgle with small momentum transfer. The standard hypothesis
for the mechanism of the reaction is shown in Fig. 60.
Two gluons, or a ladder of gluons making the Pomeron, undergo
a collision close to the centre of mass of the system and
produce a meson $R$.
There are extensive data from the Omega and WA102 collaborations
at CERN for many final states.
\begin{figure}
\centerline{\epsfig{file=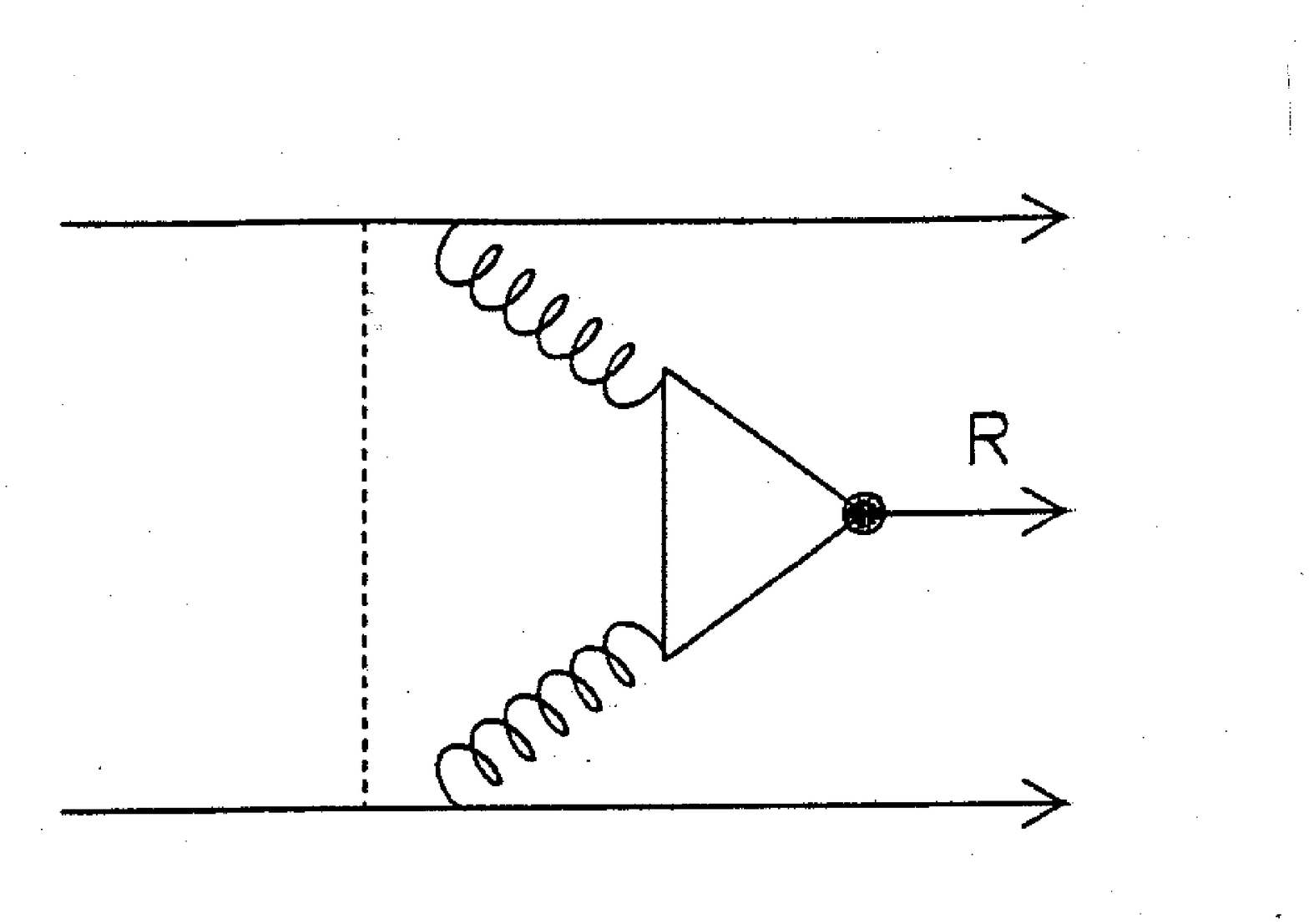,width=4cm}}\
\caption{The mechanism of central production. }
\end{figure}

Close and Kirk found a mechanism for filtering out
well-known $q\bar q$ states by selecting collisions where
the two exchanged gluons have a small difference in
transverse momentum, $\Delta P_T$ [126].
Well known $q\bar q$ states such as $f_1(1285)$, $f_2(1270)$
and $f_0(1370)$ are P-states and have a zero at the origin
of their wave functions. The idea is that they will be
suppressed close to zero transverse momentum in the collision
forming resonance $R$.
It is observed experimentally that this does indeed suppress
well known $q\bar q$ states, leaving exotic states such as
$f_0(1500)$.

\begin{figure}
\centerline{\epsfig{file=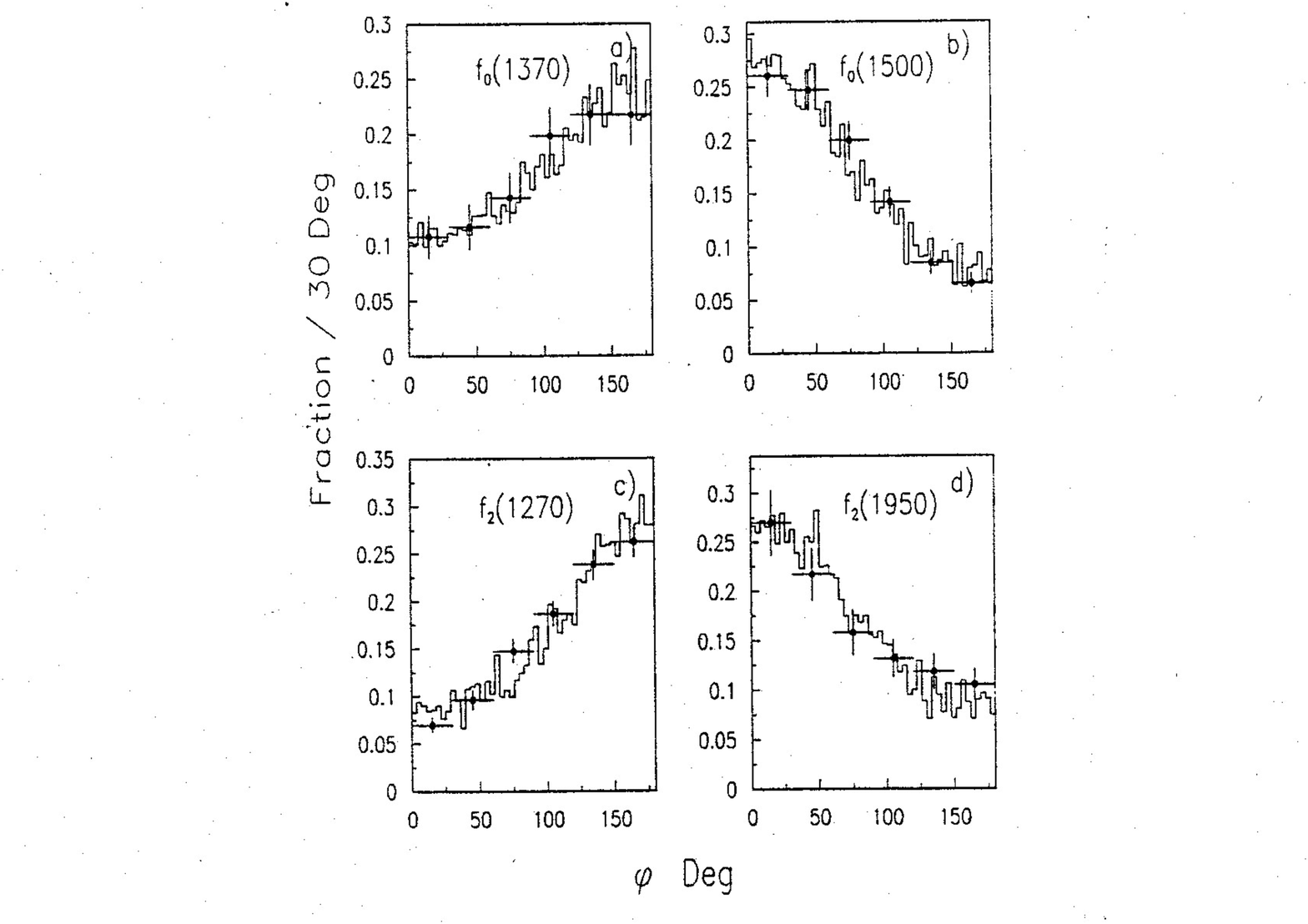,width=7cm}}\
\caption{The $\phi$ dependence for central production
of $f_0(1370)$, $f_0(1500)$, $f_2(1270)$ and $f_2(1950)$. }
\end{figure}
The idea developed further by examining the angular dependence
of the collision.
Treating it in analogy to $e^+e^- \to R$,
Close, Kirk and Schuler studied the azimuthal dependence of
the production process [127--129].
They define $\phi$ as the azimuthal angle between the
transverse momenta of the two exchanged gluons (reconstructed
from the kinematics of the outgoing protons).
From the model, the azimuthal dependence is predicted uniquely
for production of $0^-$, $1^+$ and $2^-$ resonances $R$.
The data agree closely with predictions, vindicating the
analogy between $e^+e^-$ interactions where photons
are exchanged and central production where gluons are
exchanged.

If $R$ has quantum numbers $0^+$ and $2^+$, the azimuthal
dependence is not unique. It depends on two cross-sections which they
denote as TT and LL, standing for transverse or longitudinally
polarised gluons.
For $0^+$ and $2^+$ mesons, results are shown in Fig. 61 and are truly
remarkable.
The well known $q\bar q$ states $f_0(1370)$ and $f_2(1270)$
have one distinctive form of angular dependence;
$f_0(1500)$ and a broad $2^+$ state at $\sim 1950$ MeV
have a quite contrary azimuthal dependence.
Although the origin of this difference is not understood yet,
it suggests something different about $f_0(1500)$
and $f_2(1980)$.
If $f_0(1500)$ has a large glueball content, it suggests
that perhaps $f_2(1950)$ does also.

\begin{figure}
\centerline{\epsfig{file=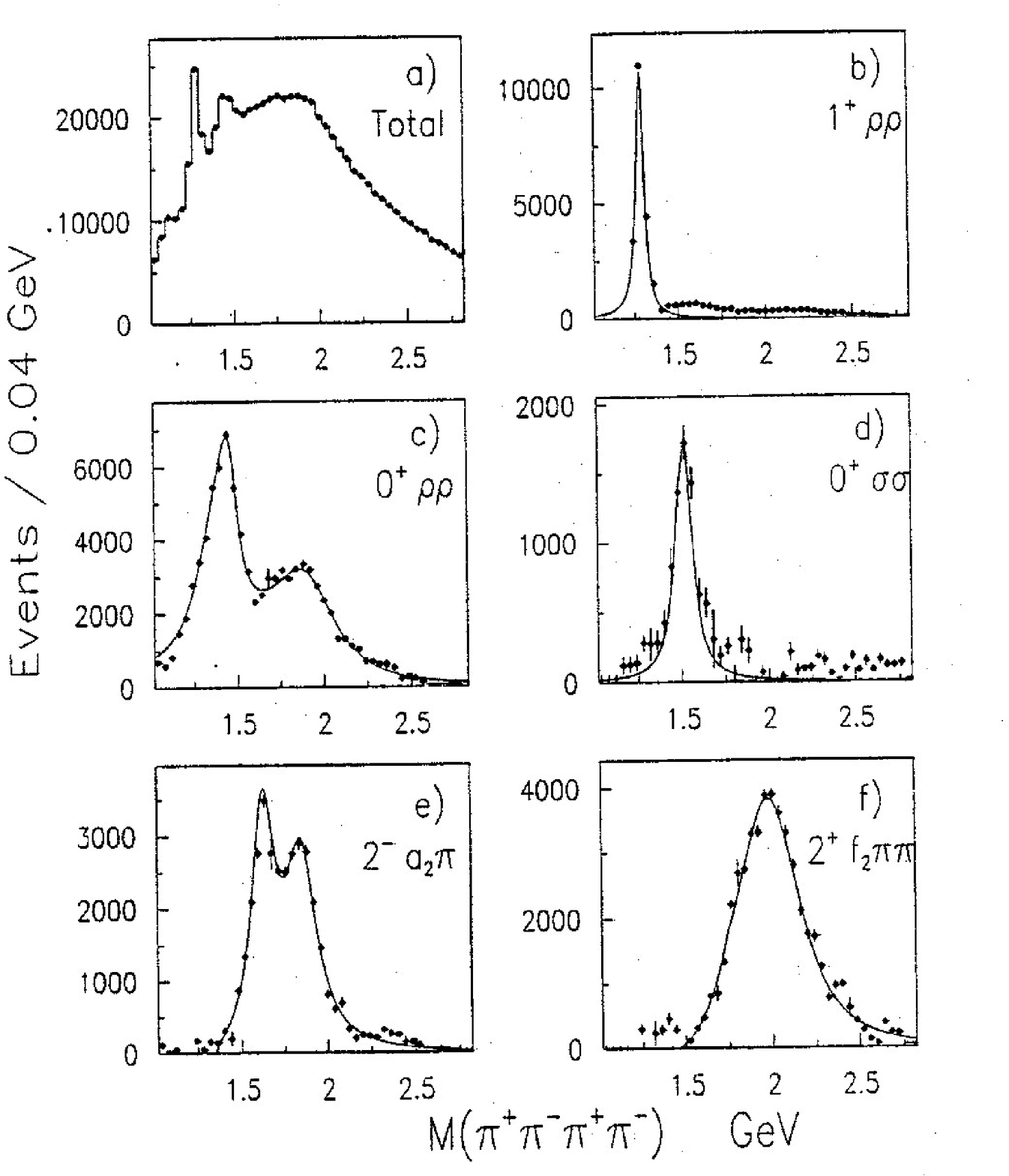,width=9cm}}\
\caption{WA102 data for central production of $\pi ^+\pi ^-\pi ^+\pi
^-$: (a) total mass spectrum, (b) $1^{++}$, (c) $0^{++}$ $\rho
\rho$, $0^+$ $\sigma \sigma$, (e) $2^{-+}$ and $f_2(1270)\pi \pi$
compared with the fits.}
\end{figure}
The broad $f_2(1950)$ in central production data of $4\pi$
was first reported by the Omega collaboration in 1995 [130] with
a width of $390 \pm 60$ MeV.
It may be the same object as reported earlier in $K^*\bar K^*$
by LASS [131] with a width of $250 \pm 50$ MeV.
The WA102 collaboration produced higher statistics in 1997 [132]
and again in 2000 [133] in $4\pi ^0$, $\pi ^+\pi ^-\pi ^0
\pi ^0$ and $\pi ^+\pi ^-\pi ^+\pi ^-$. The latest data are
shown in Fig. 62. Decays are consistent with being purely
to $f_2(1270)\sigma$, where the $\sigma$ has parameters
close to those of the $\sigma$ pole discussed in the next Section.
The $4\pi$ peak has $M = 1995 \pm 22$ MeV, $\Gamma = 436 \pm 42$
MeV for data containing 4 charged pions and $M = 1980 \pm 22$ MeV,
$\Gamma = 520 \pm 50$ MeV for 2 charged pions.
Branching ratios between the three sets of data are accurarely
consistent with $I = 0$.
Note in passing that Fig. 62(e) confirms the $\eta _2(1645)$
and $\eta _2(1860)$ discussed earlier in Section 5.3.
Also note that the $0^{++}$ signal in (d) is at 1500 MeV and
narrow. In (c) there is a broader $0^{++}$ signal peaking
at 1450 MeV and coming from interference between $f_0(1370)$
and $f_0(1500)$; the upper peak in this figure is evidence
for the $0^+$ state called $f_0(2020)$ in Section 5.

\begin{figure} [b]
\centerline{\epsfig{file=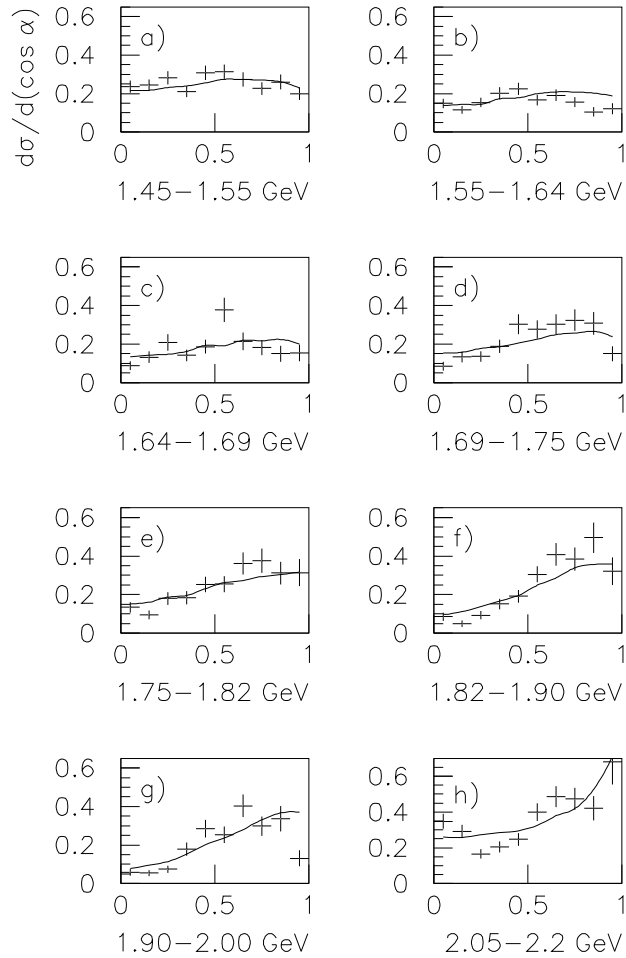,width=10cm}}\
\caption{Decay angular distributions for Crystal Barrel data for
$\bar pp \to \eta \eta \pi ^0$ in slices of $\eta \eta$ mass. }
\end{figure}
Meanwhile Crystal Barrel observed a very similar broad
$2^+$ peak in $\eta \eta$ with $M = 1980 \pm 50$ MeV,
$\Gamma = 500 \pm 100$ MeV, values very close to WA102.
The observation was made in $\bar pp \to \eta \eta \pi ^0$
at momenta 1350 to 1940 MeV/c.
The $\eta \eta$ decay angular distribution is shown
in intervals of $\eta \eta$ mass in Fig. 63.
In the lowest interval, 1.45--1.55 GeV, the process is
dominated by a strong $f_0(1500)$ signal and the
decay angular distribution is almost flat. At higher masses
one sees the gradual onset of a higher partial wave
which is fitted as purely $2^+$; no $4^+$ component
is observed. The gradual onset of the $2+$ amplitude
requires a very broad structure inconsistent with
$f_2(1934)$.

A curious feature of the data was reported in this publication,
but not understood at the time.
The broad $f_2(1950)$ is produced with orbital
angular momentum $L = 1$ in the final state and with component
of spin $\pm 1$ along the beam direction.
Later, when the combined analysis of $I = 1$, $C = +1$ was
carried out, the explanation of this result was uncovered.
The $f_2(1950)$ is produced purely from the decay of the
$a_2(2255)$ in the $\eta \eta \pi ^0$ final state.
The value of the ratio $r_2 = ^3F_2/^3P_2$ for that resonance
fits as $r_2 = -1.9 \pm 0.4$.
Clebsch-Gordan coefficients for coupling of $\bar pp$ to
$^3F_2$ and $^3P_2$ are such that the initial state will have
only helicity 0 if $r_2 = -\sqrt {7/2}$.
This helicity state decays purely to final states with
helicity $\pm 1$, again because of Clebsch-Gordan coefficients.
This explains the curious observation of the helicity of $f_2(1950)$.
However, it also firmly establishes that $a_2(2255)$ is dominantly
$^3F_2$.

The broad $f_2(1950)$ makes a large improvement to $\chi ^2$ in the
combined analysis of $I = 0$, $C = +1$ data [3] in the $\eta \eta$
channel.
Without it, there are awkward interferences between the two
$^3P_2$ and $^3F_2$ interferences, so as to reproduce not
only the narrow states but generate also a broad background.
Adding the broad background explicitly has the effect of
decoupling the narrow states to a large extent.
The broad component fits with $M = 2010 \pm 25$ MeV,
$\Gamma = 495 \pm 35$ MeV.
It appears  also in $\pi \pi$ with $g^2_{\eta \eta }/
g^2_{\pi ^0\pi ^0} = 0.72 \pm 0.06$.
This is to be compared with the prediction 0.41 for $n\bar n$ and
1 for a glueball.
It leads to a flavour mixing angle $\Phi = (23.6 \pm 3.5)^\circ$,
compared with the value 35.6$^\circ$ for a glueball.
Some mixing between a glueball and neighbouring $q\bar q$ states
is plausible.

There does however remain one anomaly which needs further work.
The WA102 collaboration has studied central production of
$\eta \eta$. There they do not find the broad $f_2(1980)$ or
$f_0(2100)$ which are so conspicuous in Crystal Barrel data [134].
Instead they observe an $f_2(2150)$ peak with $M = 2151 \pm 16$ MeV
and width $\Gamma = 280 \pm 70$ MeV; but on close examination
the data actually peak at 2060 MeV because of interference
with the S-wave. The actual moments $t_{LM}$ fitted to the data
are quite noisy, see their Fig. 3. This discrepancy is not
understood at present.

There is one small comment on the mass. The PDF quotes an average
mass of $1934 \pm 12$ MeV.
This average is strongly influenced by the low value reported by
Amsler et al. [49]: $M = 1867 \pm 46$ MeV. The data used in that
publication were at 900 MeV/c where the upper limit available
for $\eta \eta$ mass is 1914 MeV and the broad $2^+$ signal is
much weaker than in the momentum range 1350 to 1940 MeV/c.
If one instead averages the results of Anisovich et al. and the
post-1995 Central Production data, the average mass is $1972 \pm 12$
MeV.

\begin{figure}
\centerline{\epsfig{file=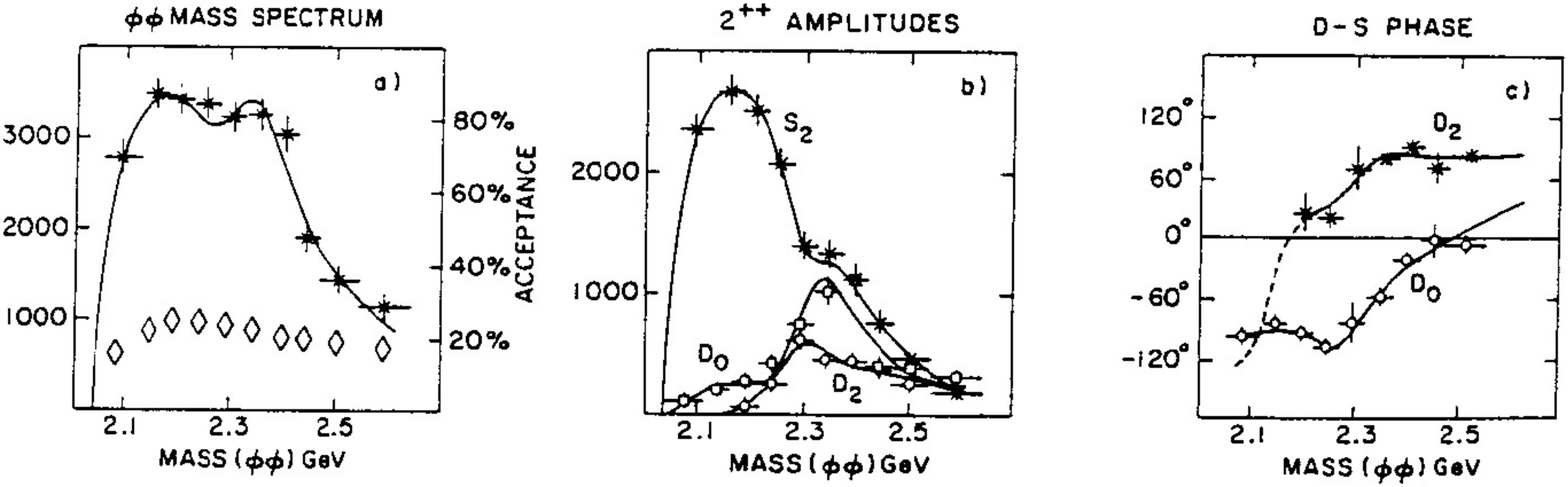,width=10cm}}\
\caption{Etkin et al. data on $\phi \phi$:
(a) acceptance corrected mass spectrum, (b) intensities of partial
waves, (c) relative phases; curves show the fit to three resonances. }
\end{figure}
\subsection {$\pi \pi \to \phi \phi$}
Etkin et al. [135] observe $J^{PC}=2^{++}$ peaks in $\pi \pi \to \phi
\phi$ at 2150 and 2340 MeV, Fig. 64.
This process violates the OZI rule.
The overall observed cross section is a factor 100 stronger than can
be  explained readily by OZI conserving processes.
An obvious suggestion is that the peaks are due to $SU(3)$ singlet
states which mix strongly with a broad $2^+$ glueball.
The narrow states can be produced via an $n\bar n$ component
and decay via the $s\bar s$ component.
An $f_2(2150)$ is listed by the PDG, corresponding to observation
by WA102 in $\eta \eta$ [136] and $K\bar K$ [117] and by GAMS in
$\eta \eta$ [137].
However, remaining entries listed by the PDG were quoted as the
$f_0(2100)$ in the original publications.

An $s\bar s$ $^3P_2$ state is expected $\sim 250$ MeV above
$f_2(1934)$ and could well explain the $f_2(2150)$.
It would fit in with the S-wave decay to $\phi \phi$ of the
2150 MeV peak.
Then a $^3F_2$ $s\bar s$ state is expected $\sim 100$ MeV higher.
Etkin et al. claim two D-wave $\phi \phi$ states at $2297 \pm 28$
MeV (with $\Gamma = 149 \pm 41$ MeV) and at $2339 \pm 55$ MeV
($\Gamma =319^{+81}_{-69})$ MeV.
An obvious question is whether there is really one D-wave state
and an interfering broad $2^+$ background from $f_2(1950)$.
A re-analysis of the data is required to answer this question.

\begin{figure}
\centerline{\epsfig{file=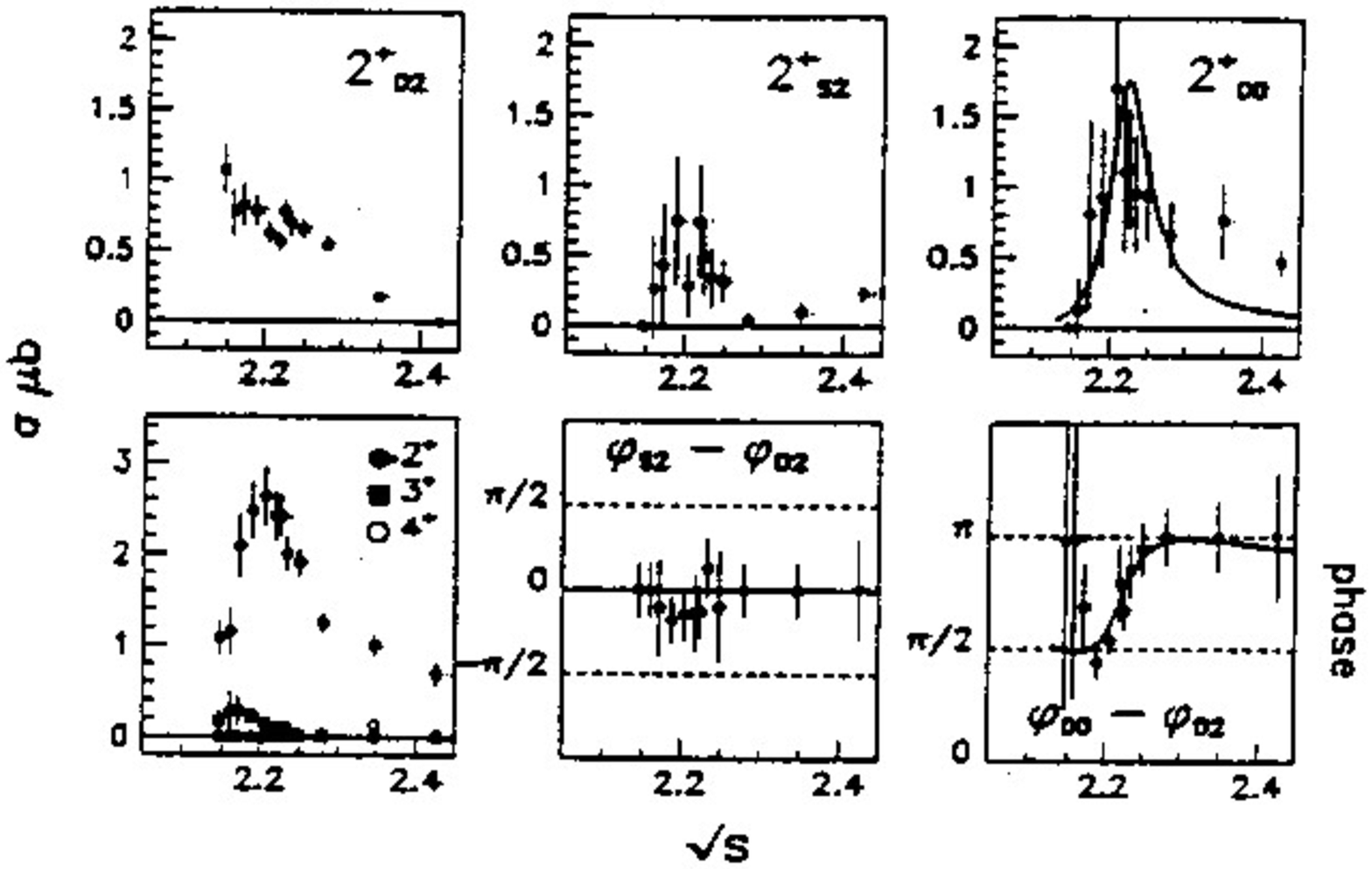,width=8cm}}\
\caption{JETSET data for $\bar pp \to \phi \phi$. }
\end{figure}
Palano [138] has presented JETSET data taken at LEAR on
$\bar pp \to \phi \phi $, Fig. 65.
A peak is observed in the $D0$ wave ($J = 2$, $S = 0$, $L = 2$).
It is fitted with $M = 2231 \pm 2$ MeV, $\Gamma = 70 \pm 10$ MeV;
however there is noise in the signal and Palano remarks that
results could be compatible with the Etkin et al. peak at 2297 MeV.
This is because of interference with a broader stucture in
the $D2$ wave, $J = 2$, $S = 2$, $L=2$, in the first panel.
Phase motion is observed between the $D0$ and $D2$ components.

\subsection {Evidence for the $0^-$ glueball}
This is an entertaining story of unitarity and analyticity at
work.
It has appeared in two publications by Bugg, Dong and Zou [139,140].
However, even in these publications the full ramifications of the
story were not fully appreciated and a further twist
needs to be given here.

Radiative $J/\Psi$ decays are a classic hunting ground for
glueballs.
The $J/\Psi$ has $J^{PC}=1^{--}$.
A process which conserves $C$-parity is the radiation of
one photon and two gluons.
The two gluons will couple preferentially to
glueballs because $\alpha_S(M^2) \sim 0.5$ at these masses, so the
intensity of coupling to $q\bar q$ states is suppressed by a
large factor.

All of the radiative decays to $\eta \pi \pi$, $K\bar K\pi$,
$4\pi$ and $K^*(890)\bar K(890)$ contain a dominant $0^-$ final
state.
This is illustrated in Fig. 66 with data from a variety of
experiments.
In (a), from Mark III [141], attention was focussed on the two
narrow peaks at low mass.
What is not displayed is a
broad $0^-$ signal under the peaks, covering the whole
mass range;
this broad component is shown on Fig.58(a).
Other panels of Fig. 66 show BES I data [115,142,143].
In all three cases there is a broad $0^-$ peak.
Wermes made the first attempt to fit this broad signal [144].
\begin{figure}
\centerline{\epsfig{file=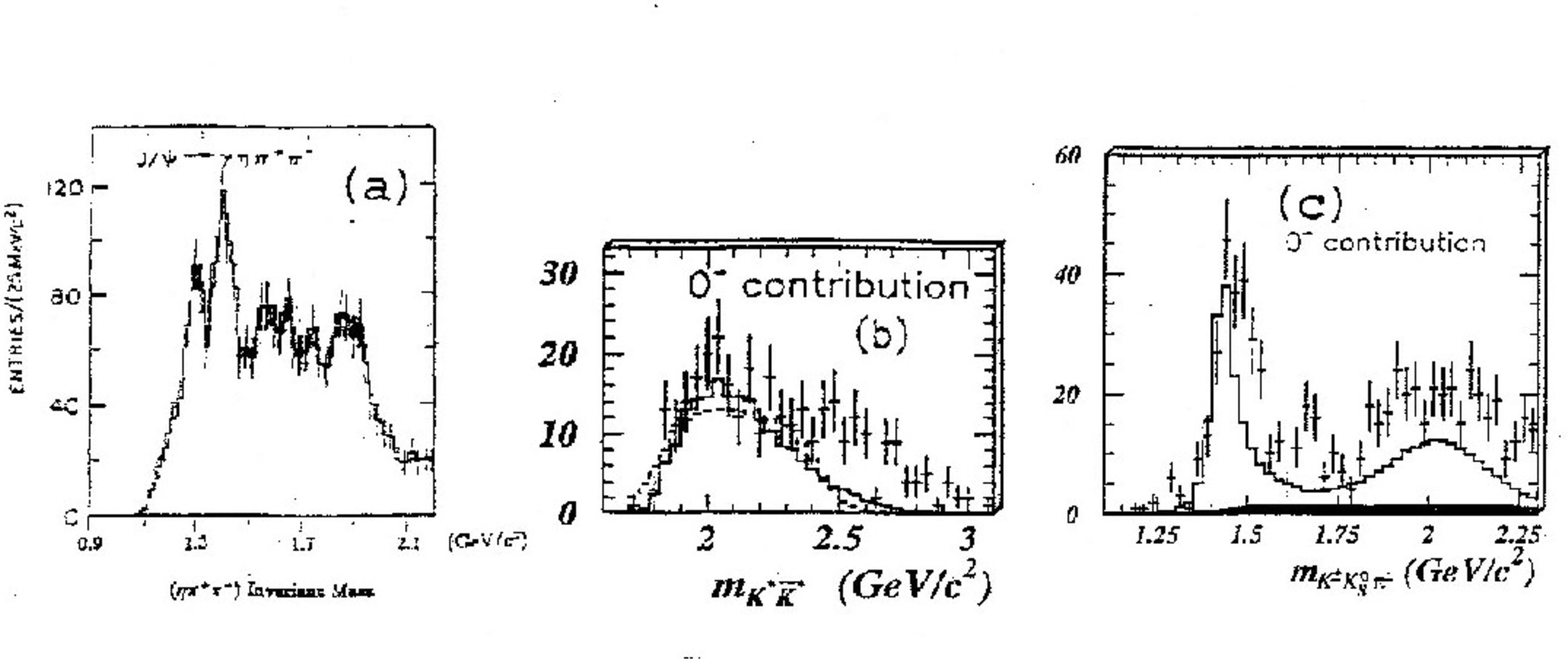,width=15cm}}\
\caption{(a) Mass spectrum for $J/\Psi$ radiative decays to
(a) $(\eta \pi ^+\pi ^-)$ after selecting $a_0(980)$ [141], (b)
$K^*\bar K ^*$ [142] and (c) $K^\pm K^0_S\pi ^\mp$ [143]. }
\end{figure}

Bugg, Dong and Zou [140] showed that it may
be fitted by a single resonance, as will be explained below.
However, their paper failed to point out one important feature.
In the production process $J/\Psi \to \gamma 0^-$, there is one unit of
orbital angular momentum for production.
This leads to an $E^3_\gamma$ factor in the intensity,
where $E_\gamma$ is the photon energy.
This factor inflates the lower side of the resonance.
Fig. 67 shows the line-shape after dividing out the factor $E^3_\gamma$.
It peaks at 2260 MeV with a conventional full width at
half maximum of $\sim 350$ MeV. The shoulder on the lower side
will be explained below.
\begin{figure}
\centerline{\epsfig{file=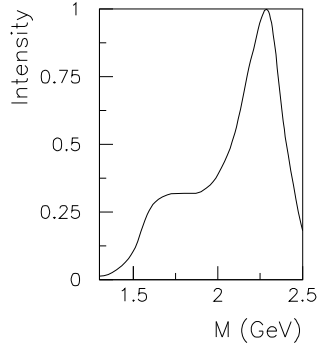,width=6cm}}\
\caption{The line-shape of the broad $0^-$ component
after dividing out the $E^3_\gamma$ factor. }
\end{figure}

A single Breit-Wigner form is used to fit simultaneously
all decay channels $i$:
\begin{equation}
f_{res} = \frac {\Lambda _i}{M^2 - s - m(s) - iM\sum _i \Gamma _i(s)},
\end{equation}
where $\Lambda _i$ are coupling constant for each channel;
$\Gamma _i(s)$ are taken to be proportional
to the phase space available for decay to each channel,
multiplied by a form factor $\exp (-R^2p^2/6)$
with $R = 0.8$ fm.
These widths are shown in Fig. 67 after normalisation to
cross sections in Fig. 66.
The function $m(s)$ is an $s$-dependent dispersive
correction to the mass:
\begin{equation}
m(s) = (s - M^2)
\int ^{\infty}_{4m^2_{\pi}} \frac {ds'}{\pi} \frac {M\Gamma _{tot}(s')}
{(s' - s)(s' - M^2)}.
\end{equation}
The total width $\Gamma _{tot}$ tends to zero at high mass sufficiently
rapidly that the dispersion integral converges well with a subtraction
at $s = M^2$.
\begin{figure}
\begin{center}
\epsfig{file=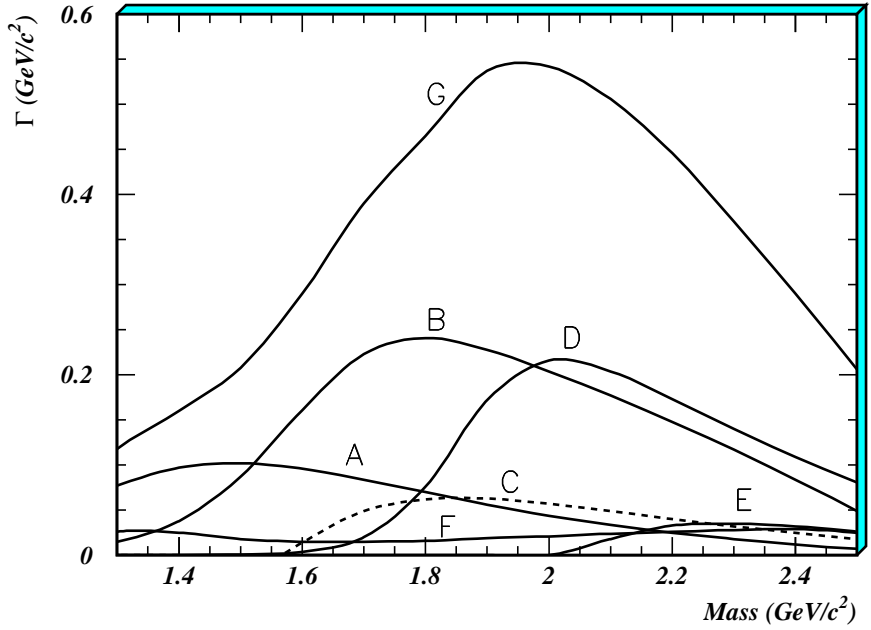,width=12cm}~\
\caption{ Widths $\Gamma _i(s)$ for A) $\eta \pi \pi$,
B) $\rho \rho$, C) $\omega \omega$, D) $K^*\bar K^*$, E) $\phi \phi$,
F) $\kappa K$ and G) total;
C is shown dashed for clarity.}
\end{center}
\end{figure}

On Figs. 69, the magnitude of each $\Gamma _i$ is fitted freely.
The fit gives coupling constants
squared for $\rho \rho$, $\omega \omega$, $K^*\bar K^*$ and
$\phi \phi$ remarkably close to the ratio 3:1:4:1 expected
for flavour-blind decays.
However, the available phase space has quite a significant
impact: obviously, the $\phi \phi$ channel is forbidden
below its threshold, and kaonic channels are significantly suppressed.
The fitted mass is  $2190 \pm 50$ MeV.

Fig. 69 shows the actual fit to data, where the $E^3_\gamma$
factor is included.
This figure shows the essential features of the fit.
At low masses, the only open channel is $\eta \pi \pi$.
Then strong channels $\rho \rho$ and $\omega
\omega$ open at $\sim 1550$ MeV. Their widths in the denominator
of the Breit-Wigner amplitude cut off the $\eta \pi \pi$ channel
rather sharply, as one see in Fig. 69 (a).
This cut-off is responsible for the shoulder in the total
cross section of Fig. 67.
Likewise, branching ratios to $\rho \rho$ and $\omega \omega$ are
cut off at 1750 MeV by the opening of the strong $K^*\bar K^*$
channel.
The resulting line-shape is rather difficult to interpret.
Perhaps one should take a somewhat larger full width than
350 MeV in order to allow for the shoulder at low mass.
The fitted mass $2190 \pm 50$ MeV is somewhat below the
peak value because of the strong effect of the $K^*\bar K^*$
width.
\begin{figure}
\begin {center}
\centerline{\epsfig{file=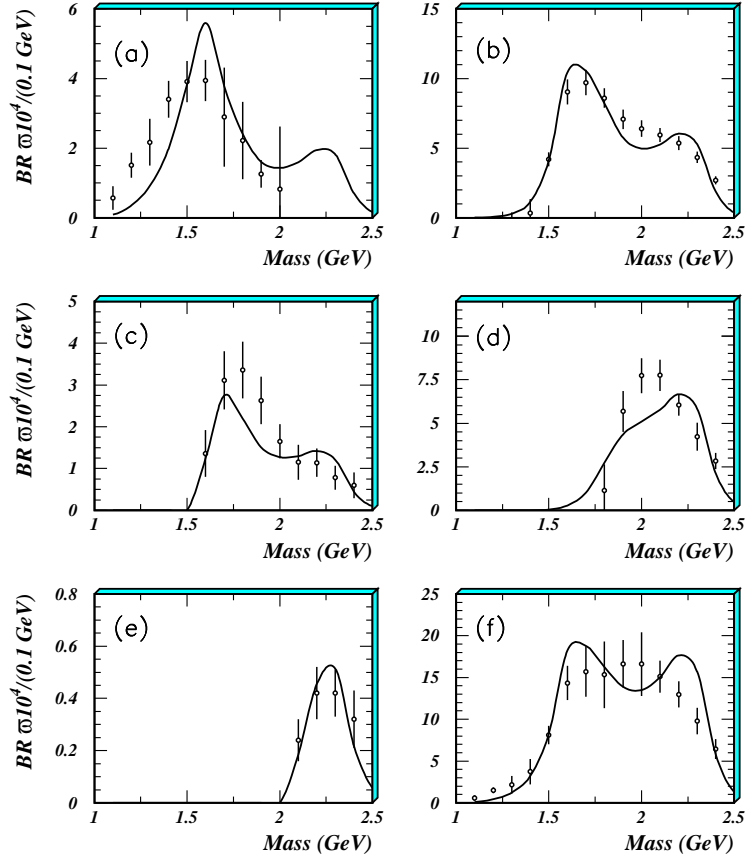,width=10cm}}\
\caption{The fit to the broad component of the $0^-$ signal
in (a) $\eta \pi \pi$, (b) $\rho \rho$, (c) $\omega \omega$,
(d) $K^*\bar K^*$, (f) $\phi \phi$ and (f) the sum of all
channels [140]. }
\end {center}
\end{figure}

\subsection {Sum Rules for $J/\Psi$ radiative decays}
Branching fractions for production of glueballs in $J/\Psi \to \gamma (gg)$
will now be considered.
Close, Farrar and Li [145] make predictions for branching fractions for
individual $J^P$ of $gg$; results depend on masses and
widths of resonances. Using experimental values for these, Table 11
[87] makes a comparison of their predictions with experiment.
\begin{table}[htp]
\begin{center}
\begin{tabular}{ccc}
\hline
State & Prediction & Observation \\\hline
$f_0(1500)$ & $1.2 \times 10^{-3}$ & $(1.03 \pm 0.14)\times 10^{-3}$\\
$\eta (2190)$ & $1.4 \times 10^{-2}$ & $(1.9 \pm 0.3)\times 10^{-2}$\\
$f_0(2105)$ & $3.4 \times 10^{-3}$ & $(6.8 \pm 1.8)\times 10^{-4}$\\
$f_2(1980)$ & $2.3 \times 10^{-2}$ & $(2.6 \pm 0.6)\times 10^{-3}$\\\hline
\end {tabular}
\caption {Branching fractions of glueball candidates in $J/\Psi$ radiative
decays, compared with predictions of Close, Farrar and Li [144].}
\end{center}
\end{table}

The branching ratio for $f_0(1500)$ is close to prediction for a
glueball and far above that predicted for $q\bar q$. The production of
$\eta (2190)$  in $\rho \rho$, $\omega \omega$, $\phi \phi$, $K^*\bar
K^*$, $\eta \pi \pi$ and $K\bar K\pi$ [14,28] is slightly above the
prediction. The $f_0(2105)$ has so far been observed in $J/\Psi$
radiative decays only to $\sigma \sigma$ [14]. For this mode alone, the
observed production is only 20\% of the prediction; decays to $\eta
\eta$ and $\pi \pi$ will also contribute, but their decay branching
ratios relative to $4\pi$ are not presently known.

The prediction for the branching fraction of the $2^+$ glueball is large
if the width is taken to be the 500 MeV fitted to  $f_2(1980)$.
Observed decays to $\sigma \sigma$ and
$f_2(1270)\sigma$ account for $(10 \pm 0.7 \pm 3.6)\times 10^{-4}$ of
$J/\Psi $ radiative decays and $K^*\bar K^*$ decays a further
$(7 \pm 1 \pm 2) \times 10^{-4}$.
If one assumes flavour-blindness for vector-vector final states,
the vector-vector contribution  increases to
$(16 \pm 2 \pm 4.5)\times 10^{-4}$.
The total of $2.6 \times 10^{-3}$
is still a factor 9 less than predicted for a glueball.
This is presently a problem in identifying $f_2(1980)$ with the
$2^+$ glueball.
The total prediction for glueball production in $J/\Psi$ radiative decays
is 4.2\%, compared with $\sim 6\%$ for all radiative decays.
As yet, only half of the products of $J/\Psi$ radiative decays
have been assigned to specific $J^P$.
Data on radiative decays to $\eta \eta$ and $\eta \eta \pi \pi$ would
be particularly valuable.

Longacre [146] has made an interesting suggestion about the
missing $2^+$ component in $J/\Psi$ radiative decays.
He suggests that the glueball may be a ring, like a doughnut,
with a hole in the middle.
If so, it has a D-state wave function.
There would then be a centrifugal barrier depressing its
production rate in $J/\Psi$ radiative decays.
A better place to look would be in $\Psi ''$ decays,
where the initial state has a D-state wave function providing
a better match to the glueball.
The channel to study is $\Psi '' \to \eta (4\pi )$,
but there are no data at present.

However, it is not obvious that this conjecture is consistent with
central production data.
Close and Kirk argue that $q\bar q$ states
are suppressed at small $\Delta P_T$ because they have
P-state wave functions.
A D-state glueball would be suppressed even further.
Possibly there is some other explanation of the
enhancement of $f_0(1500)$ and $f_2(1950)$ at small $\Delta P_T$
in central production.

\section {The light scalars}
There has been much argument whether $\sigma$ and $\kappa$
exist; also  whether $f_0(980)$ and $a_0(980)$ are
part of the ground-state $q\bar q$ family or whether they are
4-quark states or $K\bar K$ `molecules'. Over the last 2 years
valuable new information have appeared.
The experimental situation will be reviewed first. This is
followed by a brief summary of theoretical approaches.
A rough concensus is that the light scalars make a nonet separate
from $q\bar q$ states, but probably with some $q\bar q$ mixed in and
probably with significant breaking of SU(3) due to meson masses.

The E791 collaboration has observed a low mass
$\pi ^+\pi ^-$ peak in $D^+ \to \pi ^+\pi ^-\pi ^+$ [147];
they fit it with a simple Breit-Wigner resonance with $M = 478 ^{+24}_{-25}$ MeV,
$\Gamma = 324 ^{+42}_{-40} \pm 21$ MeV.
A low mass peak was observed earlier in DM2 data for
$J/\Psi \to \omega \pi ^+\pi ^-$ [148].
New BES II data [149] also contain a similar peak shown in Fig. 70(c).
The $\sigma$ appears as a conspicuous band at the upper right-hand edge
of the Dalitz plot in (b).
The fitted $0^+$ and $2^+$ components are shown in (e) and (f);
the cross-hatched area in (c) shows the coherent sum of $\sigma$ and
and a small $f_0(980)$ contribution.
\begin{figure}[htbp]
\begin{center}
\epsfig{file=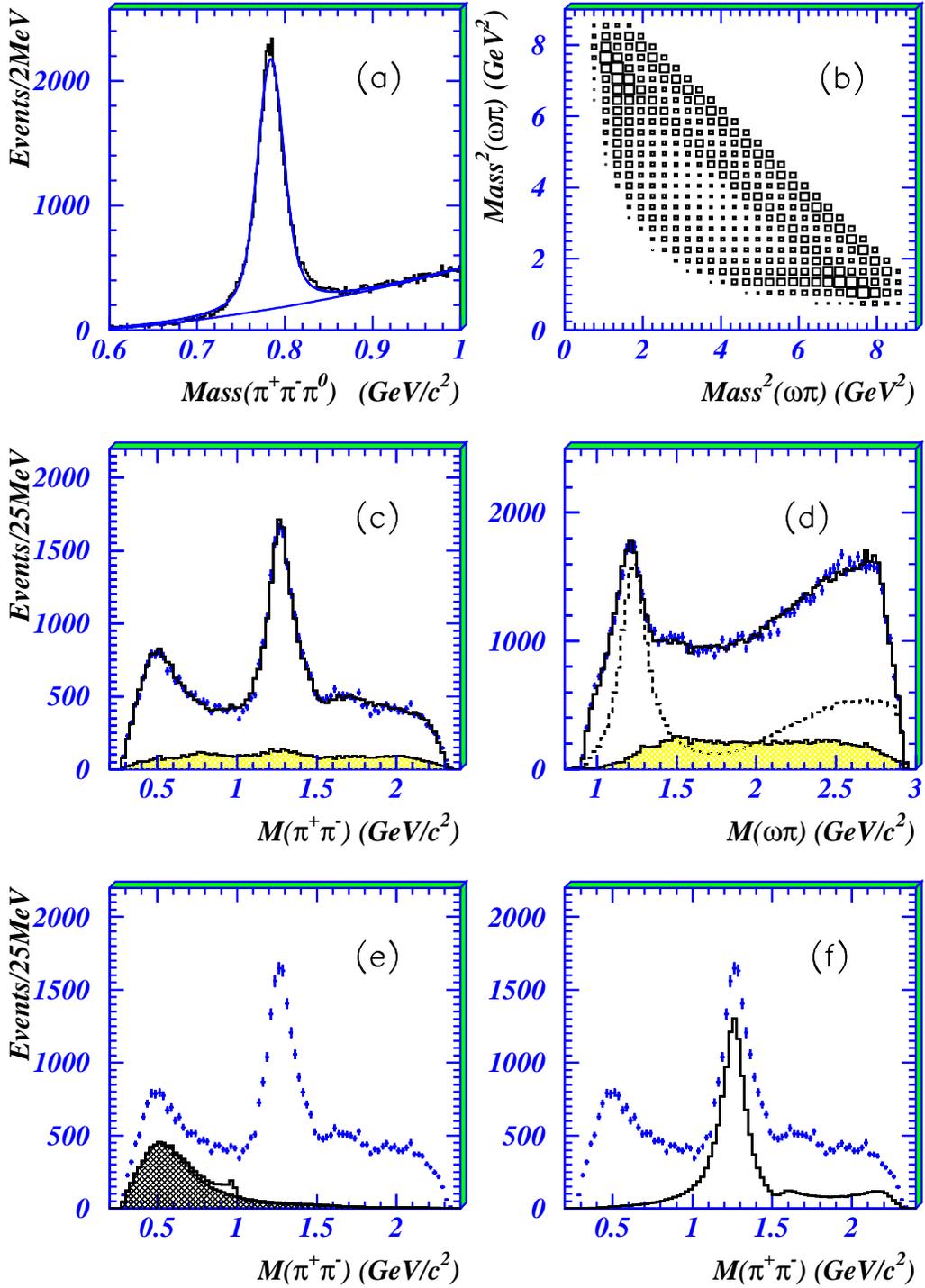,width=14.0cm}\
\caption[]{(a) The $\omega$ peak; the lower curve shows the
quadratic fit to the background. (b) the Dalitz plot for
$\omega \pi ^+\pi ^-$. (c) and (d) projections on to
$\pi \pi $ and $\omega \pi$ mass; histograms show the
fit  and the shaded region the background estimated from
sidebins;  the dashed curve in (d) shows the fitted $b_1(1235)$ signal
(two charge combinations). (e) and (f) show mass projections of $f_0$
and $f_2$ contributions to $\pi ^+ \pi ^-$ from the fit; in (e),
the shaded area shows the $\sigma$ contribution alone, and the full
histogram the coherent sum of $\sigma$ and $f_0(980)$.}
\end{center}
\end{figure}

Why does the $\sigma$ not appear the same way in $\pi \pi$ elastic
scattering? The answer is discussed in a recent paper which refers to
both $\sigma$ and $\kappa$ [150].
In $\pi \pi$ elastic scattering, there is a well known
Adler-Weinberg zero just below the elastic threshold [151]
at $s_A \simeq m^2_\pi /2$.
This zero is today a central feature of Chiral
Perturbation Theory.
To a first approximation, it cancels the pole in the denominator
near threshold, with the result that $\delta (s)$ is rather featureless
at low mass and shows little direct evidence for the pole.

This Adler zero is absent in production processes. The production
amplitude may be taken to be
\begin {equation}
T(J/\Psi \to \omega \sigma) = \frac {\Lambda _\sigma}
{M^2 - s - iM\Gamma (s)},
\end {equation}
where $\Gamma$ is $s$-dependent. For $\pi \pi$ elastic
scattering, the amplitude must lie on the unitarity circle and the
corresponding amplitude is
\begin {equation}
T(\pi \pi \to \pi \pi) = \frac {M\Gamma (s)}{M^2 - s - iM\Gamma (s)}.
\end {equation}
This contains an extra factor $M\Gamma (s)$ in the numerator.
The denominator is common to all $\pi\pi$ reactions.
If a Breit-Wigner amplitude describes the $\pi \pi$ elastic
amplitude, numerator and denominator must be self-consistent.

For elastic scattering, there are other small contributions
from higher resonances $f_0(980)$, $f_0(1370)$,
$f_0(1500)$, etc. They are accomodated by adding their phases to that
of the $\sigma$; this prescription satisfies elastic unitarity. In
principle there could be a further background amplitude. The question
is whether the simple prescription of eqns. (34) and (35) fits the data.
In practice it does.

The rate for a transition is proportional to
$|F|^2$, where $F$ is the matrix element for the transition.
The width of a resonance is proportional to this transition rate.
It is therefore logical to include the Adler zero explicitly
into the width. The idea behind this is that
the matrix element for $\pi \pi$ elastic scattering goes to
zero for  massless pions as $k \to 0$.
In production reactions such as $J/\Psi  \to \omega \sigma$, the pions
are no longer `soft' and no Adler zero is required.

The parametrisation used in Ref. [150] is
\begin {equation}
\Gamma (s)  = \rho (s)\frac { s - s_A }{M^2 - s_A}f(s) \exp
\frac {-(s - M^2)}{A}.
\end {equation}
Here $s_A = m^2_\pi /2$, the position of the Adleer zero.
Empirically, the best form for $f(s)$ is $b_1 + b_2 s$.
A combined fit is made to BES II data, CERN-Munich phase shifts for
$\pi ^-\pi ^+\to \pi ^-\pi ^+$ [104], the most recent $K_{e4}$ data
[152], CP violation in $K^0$ decays [11] and data for $\pi ^- \pi ^+
\to \pi ^0 \pi ^0$ [153,154].
The fit to elastic data is shown by the curve in Fig. 71(a).
There are some systematic discrepancies above 740 MeV
between  phase shifts of Hyams et al. (filled
circles) and $\pi ^0 \pi ^0$ data (open circles)
and the fit runs midway between them.
Fitted parameters are $M = 0.9254$, $A = 1.082$, $b_1 = 0.5843$, $b_2 =
1.6663$, all in units of GeV.
\begin{figure}
\begin{center}
\epsfig{file=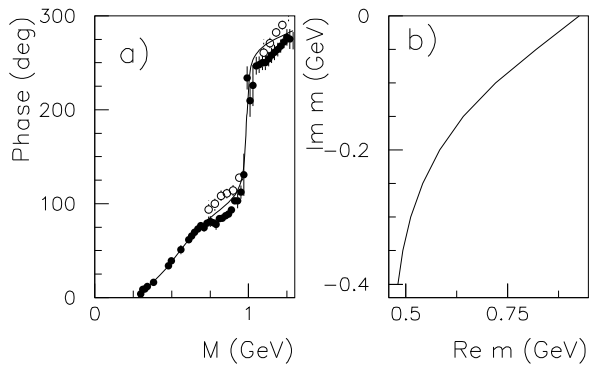,width=10cm}\
\caption{(a) Phase shifts for $\pi\pi$ elastic scattering;
dark points are phase shifts of Hyams et al. [104] above 450 MeV and
from $K_{e4}$ [152] data below 400 MeV; open circles show
$\pi ^0 \pi ^0$ data [153,154]; the curve shows
the fit. (b)The trajectory in the complex $s$-plane where
the real part of the amplitude goes through zero.}
\end{center}
\end{figure}

A consistent fit is found to both elastic and production data.
As a check, the BES data alone have also been fitted with
(i) $\Gamma = \Gamma _0 \rho (s)/\rho (M^2)$, where $\rho (s)$ is
2-body phase space, (ii) a Breit-Wigner amplitude with constant
width. All three pole positions shown in Table 12 are in close
agreement. The first entry is the most
reliable since it fits all data.
The square of the modulus of the
amplitude in eqn. (34) peaks at 460 MeV, slightly below the peak in Fig.
70(c) because of the phase space for $\omega \pi \pi$
at low $\pi\pi$ mass. The pole is in reasonable agreement with the
result of Colangelo et al.: $\rm {M} = (490 \pm 30) - i(290 \pm 20)$
MeV [155]. Their determination has the virtue of fitting the left-hand
cut consistently with the right-hand cut, using the Roy equations.
However, they do not fit data above 800 MeV, so the effect of
$f_0(980)$ is omitted. Markushin and Locher [156] show a table of
earlier determinations which cluster around the values in Table 12,
with quite a large scatter.
\begin {table}[htp]
\begin {center}
\begin {tabular} {|ccc|} \hline Fit & Form & Pole position
(MeV)\\\hline
A &   eqns. (8)-(10) & $533 \pm 25 - i(249 \pm 25 )$ \\
B  &   $\Gamma = \propto \rho (s)$ & $569 \pm 40 - (274 \pm 25 )$ \\
C &   $\Gamma = $ constant & $537 \pm 40 - i(268 \pm 50)$ \\\hline
\end {tabular}
\caption{Pole positions of the $\sigma$ for the three fits.}
\end {center}
\end {table}

A popular misconception is that the pole needs to be close to
the mass $M$ where the phase shift passes through $90^\circ$
on the real axis. That is true only when $\Gamma $ is constant.
Because $\Gamma$ varies with complex $s$ in an explicit way,
the phase of the amplitude varies as one goes off the real $s$ axis to
imaginary values. Fig. 71(b) shows the position on the complex plane
where the phase goes through $90^\circ$; there is a rapid variation
with $Im~m$, related directly to the Adler zero in the amplitude
near $s = 0$.

In BES data, the phase of the $\sigma$ amplitude is determined by
strong interferences with $b_1(1235)$. There are two $b_1(1235)$
bands on the Dalitz plot of Fig. 70(b) and together they account
for 41\% of the intensity. The $\sigma$ accounts for $19\%$
and $f_2(1270)$ for most of the rest.
An independent check is made on the phase of the $\sigma$ amplitude
by dividing up the $\pi \pi$ mass range into bins 100 MeV wide
from 300 to 1000 MeV. In each bin, the magnitude and phase
of the $\sigma$ amplitude are fitted freely, re-optimising
all other components.

Results for the phase are
shown on Fig. 72.
Curves show phases from the three parametrisations described above.
[A detail is that zero phase is taken from fit A at the $\pi \pi$
threshold.
Strong interactions introduce an overall phase for all formulae
and the dashed and dotted curves have been adjusted vertically
to agree with the full curve at 550 MeV; only the mass dependence
of the phase is meaningful.]
There is only small discrimination between the three forms.
The agreement of phases with the curves demonstrates the
correlation of magnitude and phase expected from analyticity.
Magnitudes in individual bins fluctuate from 0.90 to 1.17;
removing those fluctuations reduces errors on phases by
a factor $\sim 1.4$, but does not affect the observed agreement
significantly.
\begin {figure}[htp]
\begin {center}
\epsfig{file=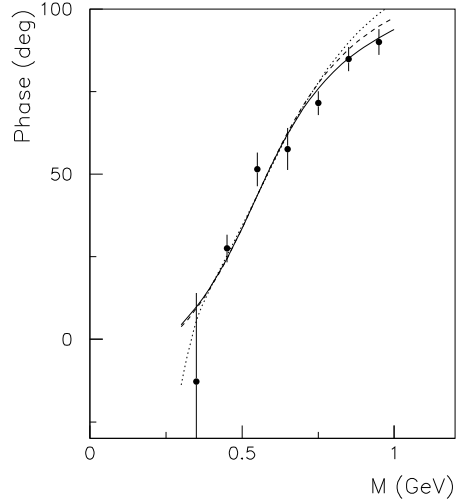,width=8cm}\
\caption{The phase of the $\pi \pi$ S-wave amplitude.
The full curve shows fit A, the dashed curve fit B (Breit-Wigner
amplitude of constant width) and the dotted curve fit C ($\Gamma
\propto \rho (s)$);
points with errors show results fitted to slices of $\pi \pi$ mass
100 MeV wide.}
\end {center}
\end {figure}

There is one technicality concerning the form $\Gamma (s) = \Gamma _0
\rho (s)$. It leads to a false virtual state pole in the mass range
$M = 200-230$ MeV, below the $\pi \pi$ threshold [150, 157].
It is therefore not a good choice of fitting function, though
Table 12 shows it does not have a drastic effect in fitting data.

\subsection {The $\kappa$}
The LASS data for $K\pi$ elastic scattering [158] are fitted with
\begin {equation}
\Gamma (s)  = \rho (s)\frac { s - s_A }{M^2 - s_A}\Gamma _0
\exp {-(\alpha M)};
\end {equation}
$s_A = M^2_K + m^2_\pi /2$.
As for $\pi\pi$, it is assumed that phases of the $\kappa$ and
$K_0(1430)$ add.
For $K_0(1430)$, a Flatt\' e form is
used:
\begin {equation}
\Gamma (s)  = \frac {(s - s_A)}{M^2 - s_A}
[g_1\rho _{K\pi }(s) + g_2\rho _{K\eta '}(s)].
\end {equation}
There is some freedom in fitting the $K_0(1430)$, because
it rides on top of the $\kappa$ `background'.
As other authors have found, there is no improvement if decays to
$K\eta$ are included.

Fitted parameters for the $\kappa$ are
$M = 3.3$ GeV, $b_1 = 24.53$ GeV, $\alpha = 0.4$ GeV$^2$.
These parameters are highly correlated and one can obtain almost
equally good fits over the range $M = 2.1$ to 4 GeV by
re-optimising $b_1$ and $A$.
The $K_0(1430)$ is fitted with $M = 1.535$ GeV, $g_1 =0.361$ GeV,
$g_1/g_1 = 1.0$. Its pole position is $(1433 \pm 30 \pm 10)-
i(181 \pm 10 \pm 12)$ MeV; that is close to the values fitted by Aston
et al: $M = (1412 \pm 6$ MeV, $\Gamma = 294 \pm 23)$ MeV. [The full
width $\Gamma$ is twice the imaginary part of the pole position.]
\vskip 5.1cm
\begin{figure} [h]
\begin{center}
\epsfig{file=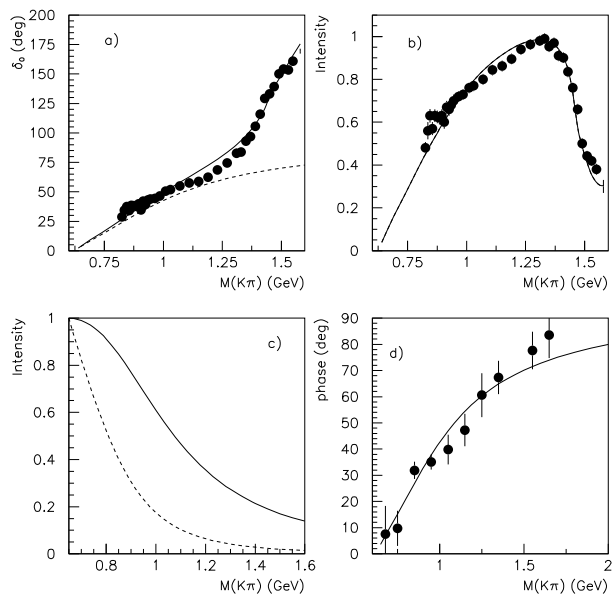,width=8cm}\
\caption{(a) $K\pi$ phase shifts.
Points are from LASS [158]. The full curve shows the fit to LASS data
and the dashed curve the $\kappa$ contribution. (b) The fit to the
intensity of $K\pi$ elastic scattering from LASS data. (c) The
intensity of the $\kappa$ amplitude in production processes, normalised
to 1 at its peak; the full curve is from the fit to LASS data and the
dashed curve from the fit to E791 data. (d) the
$\kappa$ phase fitted to BES II data compared with the curve fitted
to LASS data.}
\end{center}
\end{figure}

The fitted $K\pi$ scattering length is $0.17m_\pi ^{-1}$, very
close to Weinberg's prediction from current algebra of
$0.18m_\pi ^{-1}$.
Fits to $K\pi$ phase shifts are shown in Fig. 73(a) and to the
intensity of the elastic amplitude in
Fig.73(b).
The pole position of the $\kappa$ is at M = $(722 \pm 60) -
i(386 \pm 50)$ MeV.
The fitted value of $M$ for the $\kappa$ is very high;
it is essential that it lies above the $K_0(1430)$ since the
phase shift does not reach $90^\circ$ before that resonance.
As shown on Fig.73(d), the  fitted $\kappa$ phase reaches
$\sim 70 ^\circ$  at 1600 MeV. It is possible that the
phase for real $s$ does not actually reach 90$^\circ$.
A similar case arises for the well-known $NN~^1S_0$ virtual
state, where the phase shift rises to $60^\circ$ and then
falls again.

The full curve in Fig.73(c) shows the intensity of the $\kappa$
amplitude as it will appear in production processes;
the curve is normalised to 1 at the peak at 700 MeV.
In production reactions such as $D^+ \to \pi ^+\kappa$,
the observed intensity will be multiplied by the
available three-body phase space.

If the Adler zero is removed from the parametrisation of the
$K\pi$ S-wave and from the $K_0(1430)$, the $\kappa$ pole
disappears. So the existence of the $\kappa$ is tied closely
to the existence of the Adler zero.
Zheng et al. have reported similar results [159].

The E791 collaboration has also reported an observation of
the $\kappa$ pole in $D^+ \to K^-\pi ^+\pi ^+$ [160].
They parametrise the $\kappa$ width as $\Gamma = \Gamma _0\rho (s)/
\rho (M^2)$,
where $\rho (s)$ is $K\pi$ phase space and find
$M = 797 \pm 19 \pm 43$ MeV, $\Gamma _0 = 410 \pm 43 \pm 87$ MeV.
The corresponding pole position is at $\rm {M} = 721 - i292$ MeV.
The intensity calculated from their Breit-Wigner denominator
is shown by the dashed curve on Fig. 73(c).
Their $\kappa$ is much narrower than from LASS data.

\begin{figure}[t]
\begin {center}
\epsfig{file=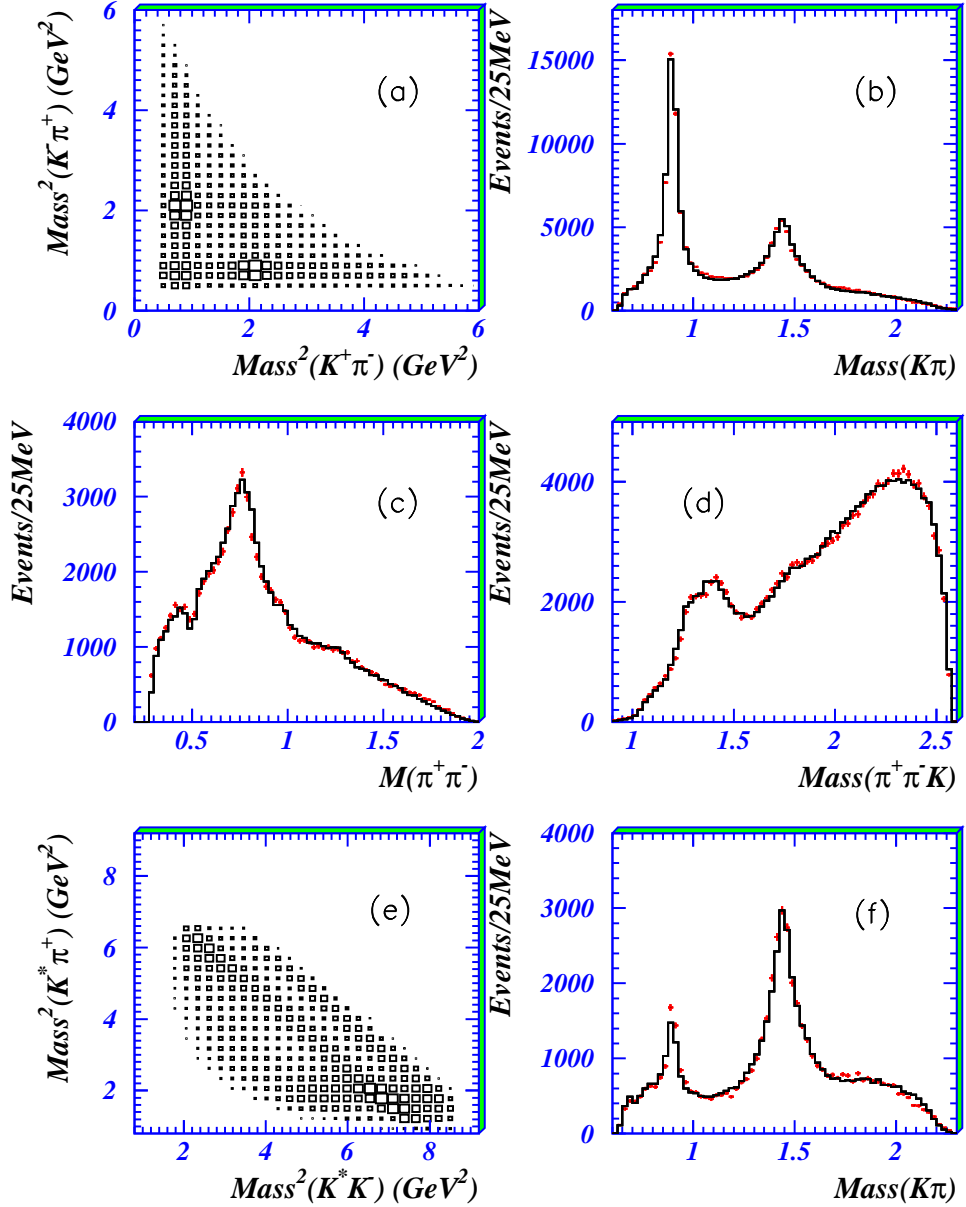,width=13cm}\
\caption[]{(a) The scatter plot of $M(K^+\pi ^-)$ against $M(K^-\pi
^+)$.
(b) The projection on to $K^\pm \pi ^\mp$ mass; the histogram shows
the fit.
(c) The projection on to $\pi \pi $ mass.
(d) The projection on to $K\pi \pi $ mass.
(e) The Dalitz plot for
events where one $K^\pm \pi ^\mp$ combination is in the mass range
$892 \pm 100$ MeV.
(f) The mass projection of the second $K^\mp \pi ^\pm$ pair for
the same selection as (e).}
\end {center}
\end{figure}
BES II data for the $\kappa$ have been presented by Wu
[161] in the channel $J/\Psi \to K^+K^-\pi ^+\pi ^-$. Strong $K^*(890)$
and $K_0(1430)$ bands appear in the scatter plot of Fig. 74(a) and as
peaks in (b). In (c) a peak due to $\rho (770)$ is visible and in (d)
peaks due to $K_1(1270)$ and $K_1(1410)$.
The $\kappa$ appears as a broad low mass peak in the
final state $K^*(890)K\pi$.
It is shown in (f), where one $K^\pm \pi ^\mp$ combination
is chosen and then the mass distribution of the other
$K^\mp \pi ^\pm$ combination is plotted;
(e) shows the Dalitz plot of $K^* K\pi$ for this selection.

In the analysis, it is necessary to show that the $\kappa$ signal
does not arise from decays of $K_1(1270)$ and $K_1(1410)$,
since all their decays give low mass $K\pi$ combinations.
The broad $\kappa$ signal of Fig. 74(e) in fact arises partly
from this source, but mostly from $\kappa$, $K^0(1430)$
and their interferences (and a small amount of $K_2(1430)$).
Two discrete solutions exist, both giving a $\kappa$ pole but
with different widths.
\begin{figure}[htbp]
\begin{center}
\epsfig{file=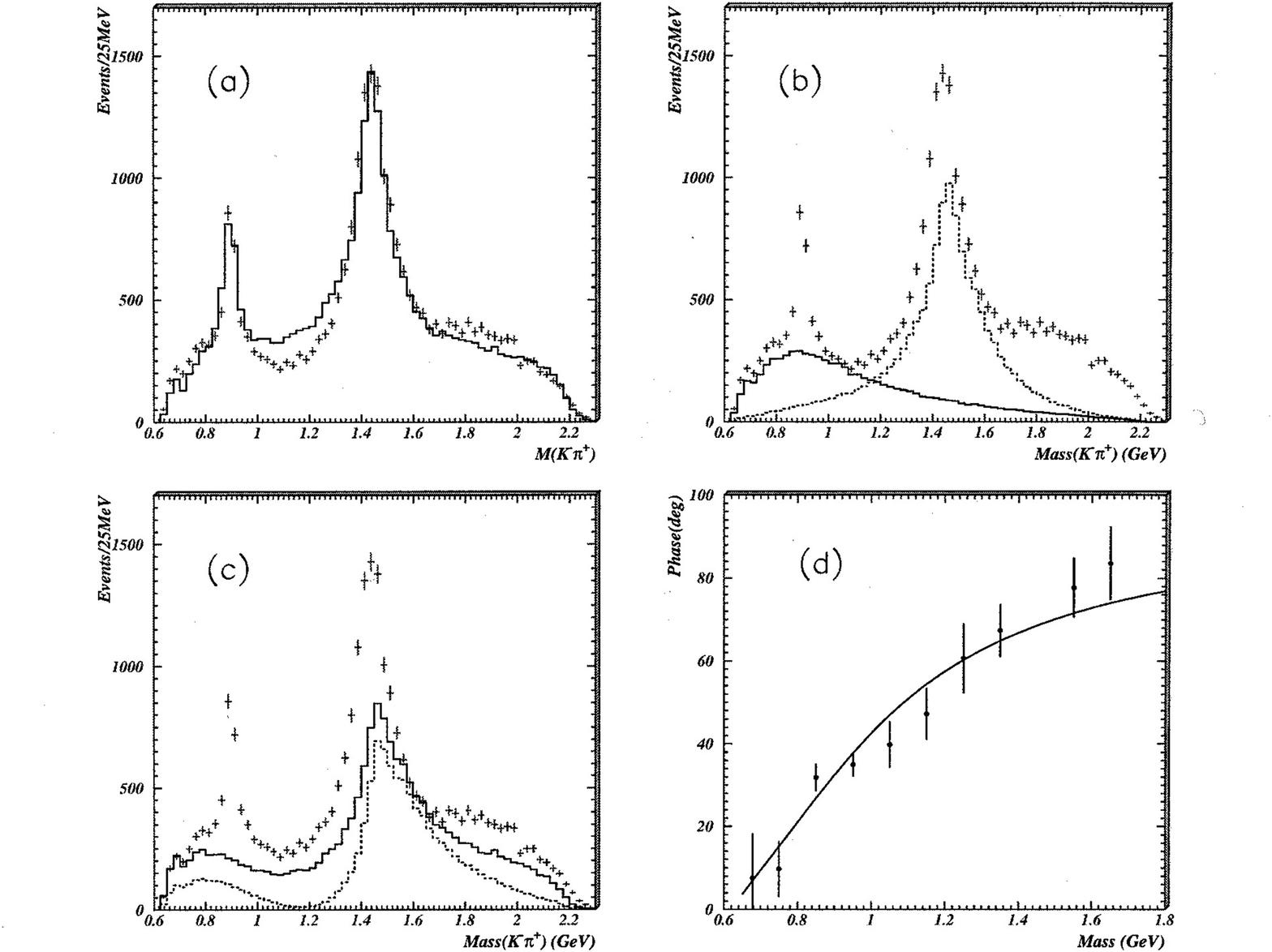,width=13cm}\
  \caption[]{(a) The poor fit when the channel $K^*(890)\kappa$ is
removed; (b) individual contributions from $\kappa$ (full histogram)
and $K_0(1430)$ (dashed); (c) the coherent
sum of $\kappa$ + $K_0(1430)$ (full histogram) and the coherent sum
$\kappa$ + $K_0(1430)+ K_1(1270) + K_1(1410)$ (dashed); (d) the phase
of the $\kappa$ amplitude (full curve) , compared with phases in
individual slices of 100 MeV (points with errors).} \end {center}
\end{figure}

Fig. 75(a) shows that a fit without the $\kappa$ in the $K\pi$
S-wave is bad.
If the Adler zero is included in the formulae, there is a
broad $\kappa$ pole with pole position $M =(760 \pm 20 \pm 40) - i(420
\pm 45 \pm 60)$ MeV;
the $K_0(1430)$ pole position is $(1433 \pm 32) - i(181 \pm 16)$ MeV.
In this solution, there is rather strong
destructive interference between $\kappa$ and $K_0(1430)$.
Contributions from $\kappa$ and $K_0(1430)$ alone are shown in Fig.
75(b) and further combinations in (c).
The phase of the $\kappa$ is determinined accurately from
interference between $\kappa$ and $K_0(1430)$.
Results are shown as points on Fig. 75(d). They agree closely with the
full curve, which shows the fit to LASS data.
Note that in this solution, the coherent sum of $\kappa$ and
$K_0(1430)$ produces a low mass peak with a width of
$\sim 400$ MeV.
This narrower peak is similar to the solution found by E791.
The effect of the destructive interference is also to make
the $K_0(1430)$ peak appear narrower than it really is. The width
fitted to $K_0(1430)$ by E791 is low: $175 \pm 12 \pm 12$ MeV.

In the second type of solution, the Adler zero is omitted
and a simple Breit-Wigner amplitude is fitted with
$\Gamma \propto \rho (s)$.
In this case, there is weak constructive interference between
$\kappa$ and $K_0(1430)$. The width fitted to the $\kappa$
is smaller, 450--600 MeV,
in agreement with the rather narrow $\kappa$ peak in Fig. 75(b).
In order to fit this width of $\kappa$ to LASS data it is
necessary to add a strong background amplitude producing a negative
phase; the sum of this repulsive contribution and the $\kappa$
generates the Adler zero [162].

Both approaches require a $\kappa$ pole.
The difference between the two approaches reduces
to one  question.
Should the Adler zero be included directly into the width of
the $\kappa$ or should it appear as a destructive
interference between a conventional Breit-Wigner resonance
and a background amplitude arising from the left-hand cut?
Fitting both left and right-hand cuts together
might resolve this question; but it is also
possible that uncertainties about distant
parts of the left-hand cut (how to include Regge poles)
will obstruct this way of resolving the issue.
Crystal Barrel data for $\bar pp \to K\bar K\pi$ at rest find
$K_0(1430)$ parameters $M = 1415 \pm 25 - i(165 \pm 25)$ MeV.

\subsection {Theoretical approaches}
Several theory groups have tackled the question of fitting
Chiral Perturbation Theory (ChPT) to data.
Ollet, Oset, Pel\' aez and collaborators
have published a series of papers where they use ChPT for the second
order driving force and then unitarise the theory by including fourth
order terms [163--168]. Some parameters of ChPT are not known
accurately. They vary these and fit to data. A virtue of these
calculations is that they also fit repulsive phase shifts in the $I =
2$ $\pi \pi$ channel and the $I = 3/2$ $K\pi$ system. Poles are
required for $\sigma$, $\kappa$, $f_0(980)$ and $a_0(980)$. A
representative set of pole positions is quoted and fits to data are
shown in a recent review by Pel\' aez [169]. Their $\sigma$ pole tends
to come out around $440 - i215$ MeV, slightly  narrower and lower than
values given above; their $\kappa$ is typically $750-i226$ MeV, again
fairly narrow.

Schechter and collaborators adopt an effective Lagrangian
based on ChPT [170--173]. Their conclusion is that there is mixing
between $q\bar q$ and a 4-quark system, similar to that proposed by
Jaffe [174]. Beisert and Borasoy also adopt an effective
Lagrangian approach [175] fitting experimental phase shifts but
including $\eta '$ as an explicit degree of freedom.

An approach based on meson exchanges was pioneered by the
J\" ulich group of Speth et al. [176--8].
Zou and collaborators adopt this approach fitting data with K-matrix
unitarisation [179,180].

Morgan, Pennington and collaborators emphasise unitarity
and analyticity as a general approach to fitting data [54,181,182].
Kaminski et al. likewise fit data imposing constraints of
analyticity and crossing symmetry [183].
The approach of T\" ornqvist is based on the linear $\sigma$
model, broken SU(3) and analyticity [184,185].
Fits by the Ishida group are reviewed in
Ref. [186].

Important insight about the connection between mesons
and their short-range content appears in the approach of
Van Beveren and Rupp [187].
They write down two relativistic coupled Sch\" odinger
equations. One describes the $q\bar q$ system at short
range.
The second describes mesons at long range.
There is a transition potential between them, based on
the $^3P_0$ model for transitions between $q\bar q$ and
4 quarks.
This models deconfinement of $q\bar q$ into mesons.
Model parameters are fitted to observed spectra using a
simple harmonic oscillator potential for confinement.
Although the model is not precise, it gives considerable
insight into the general picture.
The Adler zero is built into the equations to a good
approximation.
The transition potential has the shape of the $q\bar q$
P-state and peaks at $\sim 0.7 $ fm.

The transition potential acts as a partially reflecting
barrier, like the surface of a sheet of glass.
The $q\bar q$ states sit in a confining potential,
but can leak through the barrier.
The interesting point is that $\sigma$, $\kappa$, $f_0(980)$
and $a_0(980)$ appear as weakly bound states of the
transition potential itself.
They are mixed states of mesons and $q\bar q$.
Only S-wave mesons are bound by the transition potential.
In this model, SU(3)-breaking appears directly through
the masses of the mesons.

I will risk offending everyone by summarising present problems.
It is a marriage between theory and experiment:
Lattice QCD calculations cannot yet contribute with any
accuracy.
Experimental approaches often fit only the physical
region and ignore constraints from
the left-hand cut; the work of Leutwyler et al. and Kaminski
et al. are exceptions. All experimental fits have trouble
with the opening of inelastic thresholds above 1 GeV.
More work on understanding the $4\pi$ threshold could be
profitable.

Chiral Perturbation Theory is rigorously connected to QCD
in principle, but in practice has a large number of
parameters.
The Adler zero clearly plays an essential role, but it is
uncertain whether this should be built directly into
resonance amplitudes or whether it appears as a cancellation
between a conventional resonance and background from the left-hand cut.
ChPT requires some unitarisation procedure and fitting parameters to
data; its success verifies fundamental theoretical ideas.
Meson exchange models suffer from uncertainties about heavy
exchanges and Reggeisation.
Analyticity and unitarity are sancrosanct, but in reality
one is always battling with the effects of distant
singularities, not knowing quite what effects they have;
Occam's razor is needed to get results.
Theory does not know how to build in $q\bar q$ states and
decays, except in the empirical approach of van Beveren
and Rupp; that is where their scheme has much to offer,
even if it is empirical.

Despite rough edges between different approaches, the
general consensus is in favour of a nonet of light scalar
mesons, with SU(3) breaking due to explicit mesons masses.
The light mesons form a family of abnormal states resembling
the deuteron and the virtual $pp$ $^1S_0$ state.

\subsection {$f_0(980)$}
Recent BES II data on $J/\Psi \to \phi \pi ^+\pi ^-$ and $\phi K^+K^-$
provide a major improvement in the parameters of $f_0(980)$.
It appears as a conspicous peak in Fig. 26, Section 5.5, interfering on
its lower side with the $\sigma$. The important point is that it is
also observed clearly in decays to $K\bar K$, Fig. 27. It is fitted
with the Flatt\' e formula:
\begin  {equation} f = 1/[M^2 - s - i(g^2_{\pi \pi}\rho (s)_{\pi \pi}
+ g^2_{K\bar K}\rho (s)_{K\bar K})].
\end {equation}
The branching ratio between $K\bar K$ and $\pi \pi$
determines $g_2^2/g_1^2 = 4.40 \pm 0.25$.
In much earlier work, this ratio was unclear because of
lack of information on $K\bar K$.
The value of $g_2$ correlates rather strongly with $M$ and $g_1$
because the last term in eqn. (39) needs to be continued analytically
below threshold so as to satisfy analyticity explicitly.
From BES data, $M = 974 \pm 4$ MeV and $g^2_1 = 142 \pm 10$ MeV.
The paper gives the correlations.

\subsection {$a_0(980)$}
Crystal Barrel data at rest provide accurate parameters for
$a_0(980)$.
Data on $\bar pp \to \omega \eta \pi ^0$ [188] determine the
full-width at half-maximum very precisely:
$\Gamma = 54.12 \pm 0.34 \pm 0.12$ MeV.
This value is used as a constraint in fitting data on
$\bar pp \to \eta \pi ^0 \pi ^0$, where the $a_0(980)$
is conspicuous [90].
On the Dalitz plot, the $K\bar K$ threshold appears as a very clear
`edge'.
The intensity of the signal changes abruptly there.
From the change in slope at this edge, the relative coupling
to $\eta \pi$ and $K\bar K$ is determined rather accurately
compared with most other experiments.
The $a_0(980)$ in fitted  with a Flatt\' e
formula like eqn. (39) with
$M = 999 \pm 5$ MeV, $g^2_{\eta \pi } = 0.221 \pm 0.020$ GeV,
$g^2_{K\bar K}/g^2_{\eta \pi } = 1.16 \pm 0.08$.

\subsection {$\phi \to \gamma \pi \pi$ and $\gamma \eta \pi$ }
Data for $\phi (1020) \to \gamma \eta \pi ^0$ and
$\gamma \pi ^0 \pi ^0$ from Novosibirsk [189-191] and the Kloe group
at Daphne [192,193] are a topic of great interest.
Achasov and Ivanchenko pointed out in 1989 that these data
may throw light on $a_0(980)$ and $f_0(980)$ [194].
A vigorous debate has followed the publication of data;
Boglione and Pennington [195] give
an extensive list of references.

I have myself made unpublished fits to these data and wish
to draw upon one important conclusion, which supports
work of Achasov and collaborators.
The fits are made using parameters of $f_0(980)$ and $a_0(980)$
given in the previous subsections; formulae are
taken from the $K\bar K$-loop model of Achasov and equations
of Achasov and Gubin [196].
Apart from the absolute normalisation, there are no free
parameters.
The results of the fit to Kloe data for the $\eta \pi \gamma$
final states are shown in Fig. 76(a) and (b).
The good fit immediately verifies the $K\bar K$ loop as the
correct model for the process.
This loop gives rise to a violent energy dependent factor
$|I(a,b)|^2$ shown in (d).
This multiplies the factor $E^3_\gamma$ for the M1 transition;
the product of all energy dependent factors in shown in (c).
This leaves little doubt that the basic mechanism is
understood.
\begin {figure}
\begin {center}
\epsfig{file=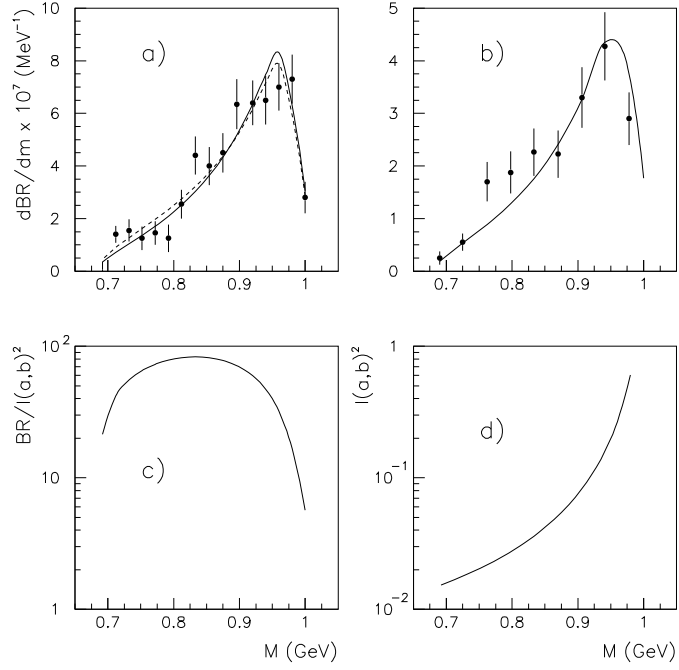,width=10cm}\
\caption{Kloe data for $\phi \to \gamma \eta \pi ^0$ for
(a) $\eta \to \gamma \gamma$, (b) $\eta \to \pi ^+\pi ^- \pi ^0$;
full curves show the fit. The dashed curve in (a) illustrates
uncertainties from a possible form factor and the Adler zero;
(c) the energy dependent factor appearing in $d\Gamma /dm$, but
omitting $|I(a,b)|^2$, (d) $|I(a,b)|^2$ itself.}
\end {center}
\end {figure}

The situation is not quite so good for $\phi \to \gamma \pi ^0 \pi ^0$.
Fig. 77 shows the fit.
However, there are two complications which affect the mass range
below 0.6 GeV (not important for present conclusion).
The first is that the final state $\rho \pi ^0$, $\rho \to \pi ^0
\gamma$ contributes there and is poorly known in normalisation.
The second is that there may be a small $\sigma$ contribution there.
The fit to the $f_0(980)$ peak using BES parameters looks slightly
narrow, but can be optimised by increasing $g^2_{\pi \pi}$ slightly to
160 MeV. With those mild
reservations, the fit is again good to the $f_0(980)$ peak and verifies
the $K\bar K$-loop model.
\begin {figure} \begin {center}
\epsfig{file=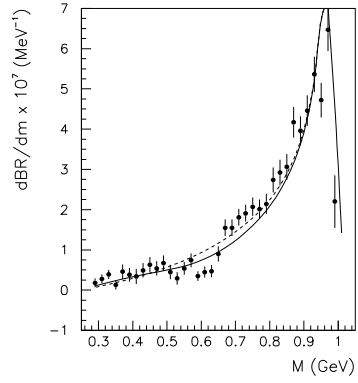,width=6cm}\
\caption{Kloe data for $\phi \to
\gamma \pi ^0 \pi ^0$; the full curve shows the fit and the
dashed curve the $f_0(980)$ intensity alone. }
\end {center}
\end {figure}

Achasov argues that these data
reveal the origin of $a_0(980)$ and $f_0(980)$ [197].
The transition $\phi \to a_0(980)$ is OZI-violating,
since $s\bar s$ must change into $n\bar n$.
Physically, the argument is that $s\bar s$ states
like $\phi$ and $f_2(1525)$ do not decay readily to
non-strange mesons.
His conclusion is that $a_0(980)$ must be a state such as Jaffe [174]
has proposed, made up of $(u\bar s s \bar u - d\bar s s \bar d)/\sqrt
{2}$.
The $u$ and $d$ quarks in the $K\bar K$-loop appear in the
$f_0(980)$ and $a_0(980)$ final states.

Jaffe's state involves 4 quarks with a small spatial extent;
the binding energy arises from pairing of quarks.
Pairs of $q$ and $\bar q$ form an SU(3) flavour 3 and $\bar 3$
configurations; these combine to a flavour singlet.
Note that all these configurations are confined to rather short
ranges.

A $K\bar K$ molecule contains the same quarks, but configured into
$K\bar K$. It is not obvious how to separate these
two possibilities.
The $f_0(980)$ and $a_0(980)$ have small binding energies.
They must therefore be surrounded by
long-range clouds of kaons.
For a given binding energy $B$, the wave function of the kaon
pair is just $\propto (1/r)\exp (-\mu r)$, where $\mu \propto
\sqrt {MB}$; $M$ is the reduced mass.
Figs. 78 and 79 shows results obtained by weighting $|\psi |^2$
with the line-shape of $f_0(980)$ or $a_0(980)$.
\begin {figure}
\begin {center}
\epsfig{file=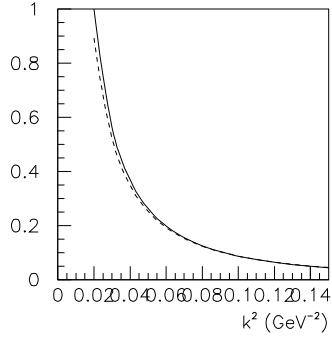,width=6cm}\
\caption{Momentum-space wave functions of $f_0(980)$, full curve
and $a_0(980)$, dashed.}
\end {center}
\end {figure}
Wave functions of $a_0(980)$ and $f_0(980)$ are remarkably similar.
This is illustrated in momentum space $k$  in Fig. 78. It need not be
so. The Flatt\' e formulae describing them have very different
parameters at face value. Nonetheless, the wave functions of $f_0(980)$
and $a_0(980)$ are within 10 percent of one another for all $k$.
That suggests they have a common character.

\begin {figure} [b]
\begin {center}
\epsfig{file=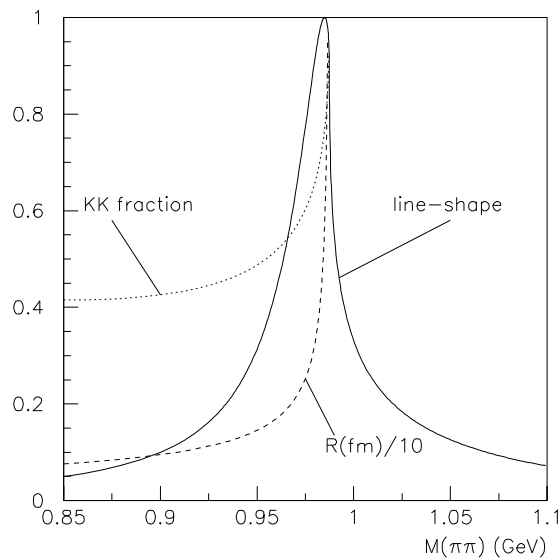,width=9cm}\
\caption{The line-shape of $f_0(980)$, full curve;
the $K\bar K$ fraction in the wave function, dotted curve;
and 0.1*RMS radius, dashed curve.}
\end {center}
\end {figure}
T\" ornqvist [18] points out that both $a_0(980)$ and $f_0(980)$
{\it must} contain a $K\bar K$ component.
His eqn. (15) includes mesonic channels $A_iB_i$:
\begin {equation}
\psi = \frac {|q\bar q> + \sum _i [-(d/ds)Re~\Pi _i(s)]^{1/2}|A_iB_i>}
         {1 - \sum _i (d/ds) Re ~ \Pi _i(s)}.
\end {equation}
Here $Re~\Pi (s) = g^2_{K\bar K}\sqrt {4m^2_K/s - 1}$ for $s < 4m^2_K$.
This formula is easily evaluated to find the $K\bar K$
components in $a_0(980)$ and $f_0(980)$ as functions of $s$. At the
$K\bar K$ threshold, these components are of course 100\%. But they are
$>40\%$ for all masses above 900 MeV.
This is illustrated in Fig. 79 by the dotted curve.
This figure also shows as the full curve the line-shape of
$f_0(980)$ v. mass.
The RMS radius of the kaonic wave function is shown by the dashed
curve in Fig. 79, scaled down by a factor 10 for display purposes.
When the intensity of $f_0(980)$ is above half-height, the
RMS radius is $>1.7$ fm.

The photon in $\phi$ radiative decays couples to
kaons dominantly at large $r$, where the E1 matrix element
has a large dipole moment $e|r|$.
Note too that the kaons couple with charge 1,
unlike the quarks which have charges 1/3 or $-2/3$.
Achasov argues that $K\bar K$ `molecules' cannot explain
the observed large rates of $\phi$ radiative dedcays.
However, a quantitative assessment of the rates is needed,
taking into account the large radii of the kaonic clouds.
It is hard to believe that the matrix element
for $\phi$ radiative decays can be dominated
by small radii.

\begin{figure}
\begin{center}
\epsfig{file=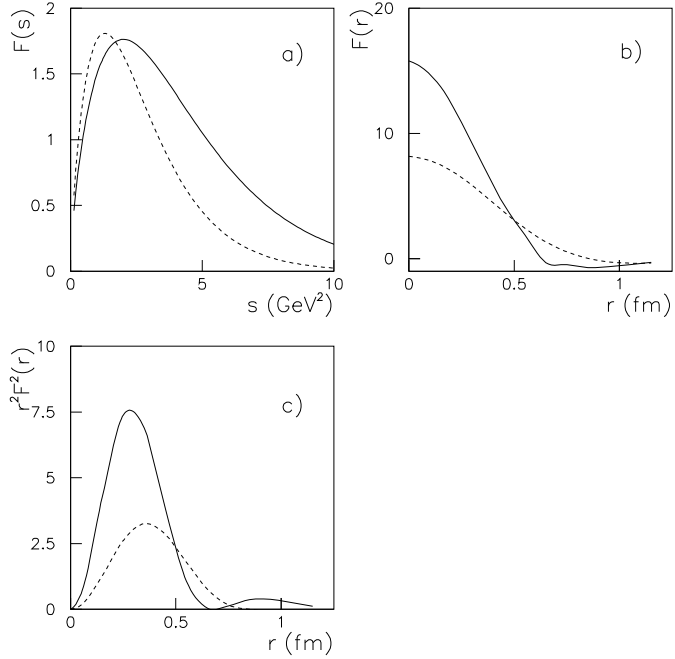,width=10cm}\
\caption{Matrix
elements $F$ for $\pi \pi \to \pi \pi$ (a) in momentum space, $F(s)$,
(b) $F(r)$ and (c) $r^2F^2(r)$. The full curves  show the optimum fit
to elastic data and dashed curves show results with a stronger
exponential cut-off in $s$ above 1 GeV$^2$.}
\end{center}
\end{figure}
However, the matrix element for $K\bar K \to f_0(980)$ or
$a_0(980)$ will be concentrated at small $r$. This matrix element
may go via a short-range interaction at the quark level.
Or it may go through $t$-channel mesons exchange;
these will be dominated by exchange of $K^*(890)$.
This exchange will be responsible for $\pi \pi \to K\bar K$.
It is then `natural' that $a_0(980)$ and $f_0(980)$
lie close to the $K\bar K$ threshold.
In Ref. [150], the matrix element for $\pi \pi \to \pi \pi$
was found from the Fourier transform of the matrix element
in momentum space.
Although this matrix element is uncertain above 1 GeV, as
shown by the dashed curve in Fig. 80(a), $F(r)$ is definitely
concentrated at a radius $ < 0.4$ fm. It appears that binding of
$f_0(980)$ and $a_0(980)$ could come either from interactions between
quarks or from meson exchanges. These states must have large inert
kaonic clouds, as T\" ornqvist emphasises.

Kalashnikova and collaborators [198] re-examine an old idea
of Weinberg to decide whether the $a_0(980)$ and
$f_0(980)$ are composites of elementary particles
or are `bare' states.
The choice between these possibilities rests upon whether
the resonances have a single pole close to the $K\bar K$
threshold or a pair of poles on different sheets.
A single pole favours a composite, like the deuteron.
Morgan made this point in a different way in 1992 [199].
The BES data give a clear answer for $f_0(980)$. There is
a single nearby second-sheet pole at $M = 0.999-i0.014$ GeV and
a distant third-sheet pole at $M = 0.828-i0.345$ GeV.
For $a_0(980)$, the
situation is not so clear-cut. There is a second-sheet pole at $1.032
-i0.085$ GeV and a third-sheet pole at $M = 0.982 - i0.204$ GeV.

\begin{figure}  [b]
\begin{center}
\epsfig{file=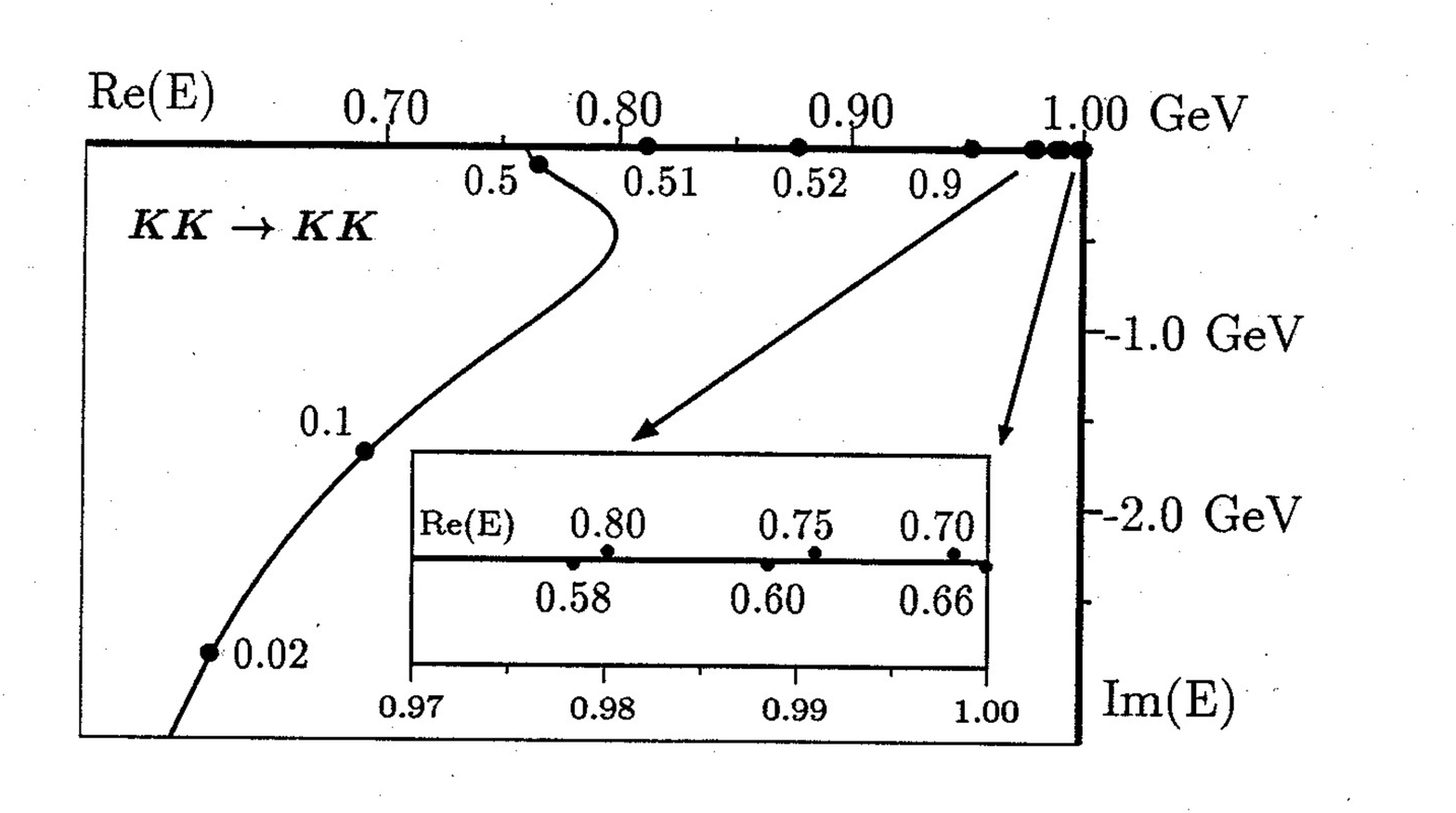,width=10cm}\
\caption{From Ref. [203]. The movement of the $a_0(980)$ pole is
shown as a
function of the coupling constant for the $I = 1$ $K\bar K$  S-wave.
Values of coupling constants are indicated.
The inset shows a magnified view near the $K\bar K$ threshold.}
\end{center}
\end{figure}
\subsection {$D_s(2317)$}
The discovery of the narrow $D_s(2317)$ by Babar [200] and
Cleo [201] has generated a flurry of interest.
The predicted mass for a $0^+$ $c\bar s$ state in typical
quark potential models is 2.48--2.49 GeV, above the
$DK$ threshold at 2360 MeV.
Van Beveren and Rupp [202] argue that the $D_s(2317)$ is
a cousin of the light scalar nonet associated with the $DK$
threshold.
They present a calculation following an earlier one of
$a_0(980)$ [203].

There, they model the attraction in the $I=1$ $K\bar K$ S-wave,
ignoring coupling to $\eta \pi$.
Fig. 81 shows the movement of the pole as the coupling
constant to $K\bar K$ is varied.
A virtual state appears first at 0.76 GeV.
As the coupling to $K\bar K$ increases, the pole migrates along the
lower side of the real axis to the $K\bar K$ threshold.
For still larger coupling, it becomes a bound state and slowly
moves away from the $K\bar K$ threshold on the upper side of the cut.
The pole remains in the vicinity of the $K\bar K$ threshold for quite
a wide  range of coupling constants.
This leaves little doubt that the $K\bar K$  threshold plays a strong
role in the dynamics.
If the mass of the kaon  were to change, the pole would move with it.
This does not necessarily  mean that $a_0(980)$ is a bound-state
of $K\bar K$.
The attraction in this  channel may arise from a variety of
processes generating attraction,  either at the meson level or
at the quark level.

Likewise, the $D_s(2317)$ appears as either a virtual or bound
state of $DK$ with some $c\bar s$ component mixed in.
The regular $c\bar s$ state is predicted to move up in mass above
the mass of simple quark models through level repulsion.
A similar phenomena near the $BK$ threshold is of course predicted.
However, there are many other competing ideas to explain
$D_s(2317)$ and other narrow states.

The $\sigma$ and $\kappa$ appears close to $\pi \pi$
and $K\pi$ thresholds, but are forced away from those
thresholds by the Adler zeros.
It is good for us that this happens. If they were
bound states below threshold, the Universe as we know
it would be totally different because of the long-range
$\sigma$ exchange binding nuclei.
Van Beveren and Rupp predict [202] that there will be a $D\pi$ $0^+$
state not far above the $D\pi$ threshold.
This could be the $D^*_0(2290)$  reported by Belle with
a width of 305 MeV [204].

\section {Future Prospects}

It is possible to check experimentally whether $f_0(1710)$ is
a ground-state $q\bar q$ state or a radial excitation.
This may be done by checking its production rate in
$D_s^+ \to f_0(1710)\pi$.
Van Beveren, Rupp and Scadron argue [205] that $f_0(980)$ is dominantly
$s\bar s$ because of its absolute production rate.
This is calculated assuming that the $D_s^+$ decays by
$c \to sW^+$, followed by $W^+$ turning into $\pi ^+$.
Their argument could be perturbed by other diagrams, but these
will only change the rate by small factors.
Implicit in their calculation is that there is good overlap
between the radial wave functions of $D_s$ and the radial
wave function of the $f_0$.
This is true for the mesonic ground-state.
But if the $f_0(1710)$ is a radially excited $s\bar s$ state,
as Anisovich and Sarantsev would have us believe,
the decay rate will be suppressed by at least a factor 100.

E852 and VES have data which could locate the missing $q\bar q$
states in the mass range 1600--1800 MeV. The $I=0$ hybrid
partner of $\pi _1(1600)$ is also missing.

CLEO C is already taking data on $\Psi ''$; data on
$J/\Psi$ and $\Psi '$ will follow during the next 5 years
with better angular acceptance than BES II and excellent
$\gamma$ detection.
A high priority should be looking for the $2^+$ glueball
in $\psi '' \to \eta (4\pi )$
as suggested above. Decays $J/\Psi \to \gamma \eta \eta $
and $\gamma \phi \phi$ are important for full information
on $0^+$, $2^+$ and $0^-$ glueballs and sorting out mixing
schemces with $q\bar q$.
A search for $f_0(2100) \to K\bar K$ and $\eta \eta$ would be simple
and valuable. Radiative decays of $\Psi '$ are almost unknown
at present; the node in its radial wave function will throw
light on wave functions of resonances produced in $\Psi '$
decays. High quality data on $\chi$ states should also
appear. This looks an exciting program of physics over
the next few years.

The way forward with $1^{--}$ $q\bar q$ states is to study
diffractive dissociation of the photon.

What would complete the picture of $q\bar q$ states and make it more
secure is to rerun the Crystal Barrel experiment
using a frozen spin polarised target.
Study of $\pi ^0 \eta$ and $3\pi ^0$ would
remove ambiguities and tighten the analysis for $I = 1$, $C = +1$.
Study of $\omega \pi ^0$ and $\omega \eta$ channels would
likewise improve and hopefully complete $I = 1$ and $0$, $C = -1$
states. Such an experiment is technically straightforward. Data are
needed right down to 360 MeV/c to cover the tower of resonances around
2 GeV. This is a possible program using the new $\bar p$ ring proposed
for GSI. It could bring the long program of work on light non-strange
mesons to a definitive conclusion.

\section {Acknowledgments}
It is a pleasure to thank the St. Petersburg group of Volodya
Anisovich, Andrei Sarantsev, Viktor Nikonov, Viktor Sarantsev
and Alexei Anisovich for their keen cooperation in analysing the
Crystal Barrel data in flight, and for extensive discussions.
Thanks go also to the Crystal Barrel collaboration for efforts
over many years in producing the data; Martin Suffert and Dieter
Walter deserve special praise for their patient work perecting
the performance of the CsI crystals.
The work has been funded by the British Particle Physics and
Astronomy Committee. Special thanks go to the Royal Society
for supporting collaboration between Queen Mary and the BES
collaboration over a six year period.


\begin{thebibliography}{999}
\bibitem{1} A. Hasan et al., Nucl. Phys. B378 (1992) 3
\bibitem{2} E. Eisenhandler et al., Nucl. Phys. B98 (1975) 109.
\bibitem{3} A.V. Anisovich et al., Phys. Lett. B491 (2000) 47
($I = 0$, $C = +1$).
\bibitem{4} A.V. Anisovich et al., Phys. Lett. B517 (2001) 261
($I = 1$,$C = +1$).
\bibitem{5} A.V. Anisovich et al., Phys. Lett. B517  (2001) 273
($\eta \eta \pi ^0$).
\bibitem{6} A.V. Anisovich et al., Phys. Lett. B542 (2002) 8
($I = 1$, $C = -1$).
\bibitem{7} A.V. Anisovich et al., Phys. Lett. B542 (2002) 19
($I = 0$, $C = -1$).
\bibitem{8} E. Aker et al, Nucl. Instr. A321 (1992) 69.
\bibitem{9} C.A. Baker, N.P. Hessey, C.N. Pinder and C.J. Batty,
Nucl. Instr. A394 (1997) 180.
\bibitem{10} A.V. Anisovich et al., Nucl. Phys. A 662 (2000) 344.
 ($\pi ^0 \pi ^0$, $\eta \eta$ and $\eta \eta '$.)
\bibitem{11} Particle Data Group, Phys. Rev. D66 (2002) 1.
\bibitem{12} A.V. Anisovich et al., Phys. Lett. B452 (1999) 173.
($\pi ^0 \eta$ and $\pi ^0 \eta ')$.
\bibitem{13} A.V. Anisovich et al., Phys. Lett. B471 (1999) 271
and Nucl. Phys. A662 (2000) 319.
($\pi ^+ \pi ^-$, $\pi ^0 \pi ^0$, $\eta \eta$ and $\eta \eta '$).
\bibitem{14} A.V. Anisovich et al., Phys. Lett. B452 (1999) 173.
($\eta \pi ^0 \pi ^0$).
\bibitem{15} A.V. Anisovich et al., Nucl. Phys. A651 (1999) 253.
($\eta \pi ^0 \pi ^0)$.
\bibitem{16} D.V. Bugg, A.V. Sarantsev and B.S. Zou, Nucl. Phys. B471
(1996) 59.
\bibitem{17} S.M. Flatt\' e, Phys. Lett. 63B (1976) 224.
\bibitem{18} N.A. T\" ornqvist, Z. Phys. C. 68 (1995) 647.
\bibitem{19} S.U. Chung,
\bibitem{20} G.F. Chew and F.E. Low, Phys. Rev. 101 (1956) 1570.
\bibitem{21} D.M. Alde et al., Phys. Lett. B216 (1989) 447 and
B241 (1990) 600.
\bibitem{22} G.M. Beladidze et al., Z. Phys. C54 (1992) 367 and
C57 (1992) 13.
\bibitem{23} A.V. Anisovich et al., Phys. Lett. B491 (2000) 40
($\eta '\pi ^0 \pi ^0$).
\bibitem{24} A.V. Anisovich et al., Phys. Lett. B496 (2000) 145.
($\eta \eta \eta $)
\bibitem{25} J. Adomeit et al., Zeit. Phys. C71 (1996) 227.
\bibitem{26} D. Barberis et al., Phys. Lett. B413 (1997) 217 and 225.
\bibitem{27} A.V. Anisovich et al., Phys. Lett. B477 (2000) 19
($\eta \pi ^0 \pi ^0 \pi ^0$).
\bibitem{28} N. Isgur, R. Kokoski and J. Paton, Phys. Rev. Lett. 54
   (1985) 869.
\bibitem{29} T. Barnes, F.E. Close, P.R. Page and E.S. Swanson,
   Phys. Rev. D55 (1997) 2147.
\bibitem{30} P.R. Page, E.S. Swanson and A.P. Szczepaniak, Phys.
Rev. D50 (1999) 034016.
\bibitem{31} H. Baher, {\it Analog and Digital Signal Processing},
Wiley, Chichester, 1990, p243.
\bibitem{32} A.D. Martin and M.R. Pennington, Nucl. Phys. B169 (1980)
216.
\bibitem{33} B.R. Martin and D. Morgan, Nucl. Phys. B176 (1980) 355.
\bibitem{34} A.A. Carter et al., Phys. Lett. 67B (1977) 117.
\bibitem{35} D. Ryabchikov, private communication.
\bibitem{36} D. Ryabchikov, AIP Conf. Proc. 432 (1998) 603.
\bibitem{37} L.Y. Dong, {\it Study of $J/\Psi \to \gamma 4\pi$},
private communication).
\bibitem{38} A. Abele et al., Euro. Phys. J C8 (1999) 67.
\bibitem{39} D.V. Bugg et al., Phys. Lett. B353 (1995) 378.
\bibitem{40} T.A. Amstrong et al., Phys. Lett. B307 (1993) 394.
\bibitem{41} D. Barberis et al., Phys. Lett. B413 (1997) 217,
B471 (2000) 440 and B484 (2000) 198.
\bibitem{42} R.M. Baltrusaitis et al., Phys. Rev. D35 (1987) 2077.
\bibitem{43} J. E. Augustin et al., Phys. Rev. Lett. 60 (1988) 2238.
\bibitem{44} G.J. Feldman et al., Phys. Rev. Lett. 33C (1977) 285.
\bibitem{45} A. Falvard et al., Phys. Rec. D38 (1988) 2706.
\bibitem{46} L. K\" opke and N. Wermes, Phys. Rep., 174 (1989) 67.
\bibitem{47} L.Y. Dong (private communication).
\bibitem{48} A.V. Anisovich et al., Phys. Lett. B449 (1999) 154
($f_0(1770)$).
\bibitem{49} C. Amler et al., Euro. Phys. J C23 (2002) 29.
\bibitem{50} D. Alde et al., Phys. Lett. B182 (1986) 105.
\bibitem{51} D. Alde et al., Phys. Lett. B284 (1992) 457.
\bibitem{52} V.V. Anisovich, Phys. Lett. B364 (1995) 195.
\bibitem{53} V.V. Anisovich, Phys. Lett. B452 (1999) 187.
($f_2(1270)\pi )$
\bibitem{54} K.L. Au, D. Morgan and M.R.Pennington, Phys. Rev. D35
(1987) 1633.
\bibitem{55} R.S. Dulude et al., Phys. Lett. B79 (1978) 329.
\bibitem{56} A.V. Anisovich et al., Phys. Lett. B472 (2000) 168.
\bibitem{57} A.V. Popov, AIP Conf. Proc 619 (2002) 565.
\bibitem{58} J. Kuhn et al., hep-ex/0401004.
\bibitem{59} A.V. Anisovich et al., Phys. Lett. B500 (2001) 222.
($\eta \eta \pi ^0 \pi ^0)$.
\bibitem{60} D. Amelin et al., Phys. At. Nucl. 59 (1996) 976.
\bibitem{61} D. Ryabchikov, Ph. D. thesis, IHEP, Protvino, 1999.
\bibitem{62} C. Daum et al., Nucl. Phys. B182 (1981) 269.
\bibitem{63} P. Eugenio, AIP Conf. Proc 619 (2002) 573.
\bibitem{64} A.V. Anisovich et al., Phys. Lett. B508 (2001) 6.
($\omega \pi ^0,~\omega \to \pi ^0 \gamma$).
\bibitem{65} A.V. Anisovich et al., Phys. Lett. B513 (2001) 281.
($\omega \eta \pi ^0$).
\bibitem{66} A.V. Anisovich et al., Phys. Lett. B542 (2002) 8.
($I = 1$, $C = -1$).
\bibitem{67} F. Foroughi, J. Phys. G: Nucl. Phys., 8 (1982) 345.
\bibitem{68} D. Alde et al., Z. Phys. C 66 (1995) 379.
\bibitem{69} M.E. Biagini et al., Nuovo Cim. 104A (1991) 363.
\bibitem{70} A.B. Clegg and A. Donnachie, Z. Phys. C45 (1990) 677.
\bibitem{71} M. Matveev, Proc. Hadron03 (to be published).
\bibitem{72} A.V. Anisovich et al., Phys. Lett. B476 (2000) 15.
($\omega \pi ^0 \pi ^0$).
\bibitem{73} A.V. Anisovich et al., Phys. Lett. B507 (2000) 23.
($\omega \eta$).
\bibitem{74} A.V. Anisovich et al., Phys. Lett. B542 (2001) 19.
(Combined $I = 0$, $C = -1$).
\bibitem{75} G.S. Adams et al., Phys. Rev. Lett. 81 (1998) 5760.
\bibitem{76} D.R. Thompson et al., Phys. Rev. Lett. 79 (1997) 1630.
\bibitem{77} A. Abele et al., Phys. Lett. B423 (1998) 175.
\bibitem{78} C.A. Baker et al., Phys. Lett. B467 (1999) 147.
\bibitem{79} D.V. Bugg et al., J. Phys. G: Nucl. Phys. 4 (1978) 1025.
\bibitem{80} S. Godfrey and N. Isgur, Phys. Rev. D32 (1985) 189.
\bibitem{81} L. Ya. Glozman, Phys. Lett B541 (2002) 115.
\bibitem{82} L. Ya. Glozman, hep-ph/0205072 and 0301012.
\bibitem{83} Y. Nambu, Phys. Rev. D210 (1974) 4262.
\bibitem{84} L.M. Kuradadze et al., JETP Lett. 37 (1983) 733
\bibitem{85} A. Bertin et al., Phys. Lett. B414 (1997) 220.
\bibitem{86} C.J. Morningstar and M. Peardon, Phys. Rev. D60 (1999)
034509.
\bibitem{87} D.V. Bugg, M. Peardon and B.S. Zou, Phys. Lett. B486
(2000) 49.
\bibitem{88} V.V. Anisovich et al., Phys. Lett. B323 (1994) 233.
\bibitem{89} C. Amsler et al., Phys. Lett. B333 (1994) 277.
\bibitem{90} D.V. Bugg, V.V. Anisovich, A. Sarantsev and B.S. Zou,
Phys. Rev. D50 (1994) 4412.
\bibitem{91} V.V. Anisovich, D.V. Bugg, A.V. Sarantsev and B.S. Zou,
Phys. Rev. D50 (1994) 1972.
\bibitem{92} C. Amsler et al., Phys. Lett. B355 (1995) 425.
\bibitem{93} V.A. Polychronakos et al., Phys. Rev. D19 (1979) 1317.
\bibitem{94} A.B. Wicklund et al., Phys. Rev. Lett. 45 (1980) 1469.
\bibitem{95} A. Etkin et al., Phys. Rev. D25 (1982) 1786.
\bibitem{96} N.M. Cason et al., Phys. Rev. D28 (1983) 1586.
\bibitem{97} M. Gaspero, Nucl. Phys.  A562 (1993) 407.
\bibitem{98} A. Adamo et al., Nucl. Phys. A558 (1994) 130.
\bibitem{99} A.V. Anisovich et al., Nucl. Phys. A690 (2001) 567.
\bibitem{100} C. Amsler et al., Phys. Lett. B322 (1994) 431.
\bibitem{101} A. Abele et al., Euro. Phys. J C19 (2001) 667.
\bibitem{102} A. Abele et al., Euro. Phys. J C31 (2001) 261.
\bibitem{103} B.D. Hyams et al., Nucl. Phys. B64 (1973) 134.
\bibitem{104} W. Ochs, Ph.D. Thesis, Univeristy of Munich (1973).
\bibitem{105} L. Gray et al., Phys. Rev. D27 (1983) 307.
\bibitem{107} D. Bridges, I. Daftari and T.E. Kalogeropoulos,
      Phys. Rev. Lett. 57 (1986) 1534.
\bibitem{107} I. Daftari et al., Phys. Rev. Lett. 58 (1987) 859.
\bibitem{108} F.G. Binon et al., Nu. Cim. 78A (1983) 313.
\bibitem{109} F.G. Binon et al., Nu. Cim. 80A (1984) 363.
\bibitem{110} D.M. Alde et al., Nucl. Phys. B269 (1986) 485.
\bibitem{111} D.M. Alde et al., Phys. Lett. B198 (1987) 286.
\bibitem{112} D.M. Alde et al., Phys. Lett. B201 (1988) 160.
\bibitem{113} R.M. Baltrusaitis et al., Phys. Rev. Lett. 55 (1985) 1723
   and Phys. Rev. D33 (1986) 1222.
\bibitem{114} D. Bisello  et al., Phys. Lett. B192 (1987) 239
   and Phys. Rev. D39 (1989) 707.
\bibitem{115} J.Z. Bai et al., Phys. Lett. B472 (2000) 207.
\bibitem{116} L.Y. Dong (private communication)
\bibitem{117} W. Dunwoodie, AIP Conf. Proc. 432 (1998) 753.
\bibitem{118} D. Barberis et al., Phys. Lett. B453 (1999) 305 and 316
   and B462 (1999) 462.
\bibitem{119} J.Z. Bai et al., Phys. Rev. D68 (2003) 052003.
\bibitem{120} C.A. Baker et al., Phys. Lett. B563 (2003) 140.
\bibitem{121} C. Amsler and F.E. Close, Phys. Lett. B353 (1995) 385.
\bibitem{122} F.E. Close and A. Kirk, Phys. Lett. B483 (2000) 345
and Euro. Phys. J C21 (2001) 531.
\bibitem{123} M. Strohmeier-Presicek et al., Phys. Rev. D60 (1999)
54010.
\bibitem{124} C. Amsler, Phys. Lett. B541 (2002) 22.
\bibitem{125} V.V. Anisovich and A.V. Sarantsev, Euro. Phys. J
A16 (2003) 229 and references cited there.
\bibitem{126} F.E. Close and A. Kirk, Phys. Lett. B397 (1997) 333.
\bibitem{127} F.E. Close and G.A. Schuler, Phys. Lett. B458 (1999) 127.
\bibitem{128} F.E. Close,A. Kirk and G.A. Schuler, Phys. Lett. B477
(1999) 13.
\bibitem{129} A. Kirk, Phys. Lett. B489 (2000) 29.
\bibitem{130} F. Antinori et al., Phys. Lett. B353 (1995) 589.
\bibitem{131} D. Aston et al., Nucl. Phys. B21 (1991) 5???
\bibitem{132} D. Barberis et al., Phys. Lett. B413 (1997) 217.
\bibitem{133} D. Barberis et al., Phys. Lett. B471 (2000) 440.
\bibitem{134} A.V. Anisovich et al., Phys. Lett. B449 (1999) 154.
($\eta \eta \pi ^0$).
\bibitem{135} A, Etkin et al., Phys. Lett. B201 (1988) 568.
\bibitem{136} D. Barberis et al., Phys. Lett. B479 (2000) 59.
\bibitem{137} A.V. Singovsky et al., Nuc. Cim. 107A (1994) 1911.
\bibitem{138} A. Palano, Frascati Physics Series 15, {\it Workshop on
Hadron Spectroscopy} (1999) 363.
\bibitem{139} D.V. Bugg and B.S. Zou, Phys. Lett. B396 (1997) 295.
\bibitem{140} D.V. Bugg, L.Y. Dong and B.S. Zou, Phys. Lett. B459
   (1999) 511.
\bibitem{141} T. Bolton et al., Phys. Rev. Lett. 69 (1992) 1328.
\bibitem{142} J.Z. Bai et al., Phys. Lett. B472 (2000) 200.
\bibitem{143} J.Z. Bai et al., Phys. Lett. B476 (2000) 25.
\bibitem{144} N. Wermes, SLAC-PUB-3730 (1095) and
L. K\" opke and N. Wermes, Phys.Rep. 174 (1989) 67, p 177.
\bibitem{145} F.E. Close, G.R. Farrar and Z. Li, Phys. Rev. D55 (1997)
5749.
\bibitem{146} R. Longacre (private communication).
\bibitem{147} E.M. Aitala et al., Phys. Rev. Lett. 86 (2001) 770.
\bibitem{148} J.E. Augustin et al., Nucl. Phys. B320 (1989) 1.
\bibitem{149} J.Z. Bai et al., {\it The $\sigma$ pole in $J/\Psi \to
\omega \pi ^+\pi ^-$}, submitted to Phys. Lett. B.
\bibitem{150} D.V. Bugg, Phys. Lett. B572 (2003) 1.
\bibitem{151} S. Weinberg, Phys. Rev. Lett. 17 (1966) 616.
\bibitem{152} S. Pislak et al., Phys. Rev. Lett. 87 (2001) 221801.
\bibitem{153} K. Takamatsu et al., Nucl. Phys. A675 (2000) 312c and
Progr. Theor. Phys. 102 (2001) E52.
\bibitem{154} J. Gunter et al., Phys. rev. D64 (2001) 072003.
\bibitem{155} G. Colangelo, J. Gasser and H. Leutwyler,
Nucl. Phys. B603 (2001) 125.
\bibitem{156} V.E. Markushin and M.P. Locher,
Frascati Phys. Ser. 15 (1999) 229.
\bibitem{157} H.Q. Zheng, hep-ph/0304173.
\bibitem{158} D. Aston et al., Nucl. Phys. B296 (1988) 253.
\bibitem{159} H.Q. Zheng et al., hep-ph/0309242 and 0310293.
\bibitem{160} E.M. Aitala et al., Phys. Rev. Lett. 89 (2002)
121801.
\bibitem{161} N. Wu, {\it Int. Symp. on Hadron Spectroscopy, Chiral
Symmetry and Relativistic Description of Bound Systems}, Tokyo, Feb
24-26, 2003.
\bibitem{162} S. Ishida et al., Progr. Theor. Phys. 98 (1997) 621.
\bibitem{163} J.A. Oller and E. Oset, Nucl. Phys. A620 (1997) 438.
\bibitem{164} J.A. Oller, E. Oset and J.R. Pel\' aez, Phys. Rev.
D59 (1999) 074001.
\bibitem{165} J.A. Oller and E. Oset, Phys. Rev D60 (1999) 074023.
\bibitem{166} M. Jamin, J.A. Oller and A. Pich, Nucl. Phys. B587 (2000)
331.
\bibitem{167} A. G\' omez Nicola and J.R. Pel\' ez, Phys. Rev. D65
(2002) 054009.
\bibitem{168} J.A. Oller, Nucl. Phys. A727 (2003) 353.
\bibitem{169} J.R. Pel\' ez, hep-ph/0307018.
\bibitem{170} D. Black et al., Phys. Rev. D58 (1998) 054012.
\bibitem{171} D. Black et al., Phys. Rev. D59 (1999) 074026.
\bibitem{172} D. Black, A.H. Fariborz and J. Schechter (2000) 074001.
\bibitem{173} D. Black et al., Phys. Rev. D64 (2001) 014031.
\bibitem{174} R.L. Jaffe, Phys. Rev. D15 (1977) 281.
\bibitem{175} N. Beiset and B. Borasoy, Phys. Rev. D67 (2003) 074007.
\bibitem{176} D. Lohse, J.W. Durso, K. Holinde and J. Speth,
Phys. Lett. B234 (1990) 235.
\bibitem{177} G. Janssen, B.C. Pearce, K. Holinde and J. Speth,
Phys. Rev. D52 (1995) 2690.
\bibitem{178} F.P.Sassen, S. Krewald, J. Speth and A.W. Thomas,
Phys.  Rev. D68 (2003) 036003.
\bibitem{179} B.S. Zou and D.V. Bugg, Phys. Rev. D48 (1994) R3948 and
D50 (1994)  591.
\bibitem{180} L. Li, B.S. Zou and G.L. Li, Phys. Rev. D67 (2003)
034025.
\bibitem{181} D. Morgan and M.R. Pennington, Phys. Rev. D48 (1993)
1185.
\bibitem{182} S.N. Cherry and M.R. Pennington, Nucl. Phys. A688
(2001) 823.
\bibitem{183} R. Kaminski, L. Lesniak and B. Loiseau, Phys. Lett. B551
(2003) 241.
\bibitem{184} N.A. T\" ornqvist and M. Roos, Phys. Rev. Lett. 76
(1996) 1575.
\bibitem{185} N.A. T\" ornqvist, Euro. Phys. J. C11 (1999) 359.
\bibitem{186} M. Ishida, hep-ph/0212383.
\bibitem{187} E.van Beveren and G. Rupp, Euro. Phys. J C22 (2001) 493
and hep-ph/0201006.
\bibitem{188} C, Amsler et al., Phys. Lett. B327 (1994) 425.
\bibitem{189} M.N. Achasov et al., Phys. Lett. B438 (1998) 441 and
B479 (2003) 53.
\bibitem{190} R.R. Akhmetshin et al., Phys. Lett. B462 (1998) 380.
\bibitem{191} M.N. Achasov et al., Phys. Lett. B440 (1998) 442 and
B485 (2000) 349.
\bibitem{192} A. Aloisio et al., Phys. Lett. B536 (2002) 209.
\bibitem{193} A. Aloisio et al., Phys. Lett. B537 (2002) 21.
\bibitem{194} N.N. Achasov and V.N. Ivanchenko, Nucl. Phys. B315 (1989)
  465.
\bibitem{195} M. Boglione and M.R. Pennington, Euro. Phys. J C30 (2003)
503.
\bibitem{196} N.N. Achasov and V.V. Gubin, Phys. Rev. D63 (2001) 094007.
\bibitem{197} N.N. Achasov, Nucl. Phys. A728 (2003) 425.
\bibitem{198} Yu.S. Kalashnikova et al., hep-ph/0311302.
\bibitem{199} D. Morgan, Nucl. Phys. A543 (1992) 632.
\bibitem{200} B. Aubert et al., Phys. Rev. Lett. 90 (2003) 242001.
\bibitem{201} D. Besson et al., hep-ex/0305100 and 0305017,
\bibitem{202} E. Van Beveren and G. Rupp, Phys. Rev. Lett. 91 (2003)
012003.
\bibitem{203} E. Van Beveren and G. Rupp, hep-ph/0211411.
\bibitem{204} P.Krokokovny, for the Belle Collaration, hep-ex/0210037.
\bibitem{205} E. van Beveren, G. Rupp and M.D. Scadron, Phys.
Lett. B495 (2000) 300 and B509 (2001) 365.






\end{thebibliography}
\end {document}